\documentclass[11pt]{article}
\usepackage[utf8]{inputenc}
\usepackage{textcomp}
\usepackage{amsfonts}
\usepackage{amsmath}
\usepackage{amssymb}
\usepackage{mathrsfs}
\usepackage{bbold}
\usepackage[margin=2.2cm]{geometry}
\usepackage{longtable}
\usepackage{lscape,graphicx}
\usepackage{graphics}
\usepackage{color}
\usepackage{array}
\usepackage{makecell}
\usepackage{upgreek}
\usepackage{physics}
\usepackage{bm}
\usepackage{comment}
\usepackage{multicol}
\usepackage{blindtext}
\usepackage{float}

\usepackage{hyperref}
\usepackage{cleveref}

\usepackage[numbers, sort&compress]{natbib}

\usepackage{slashed}

\usepackage{sidecap}
\usepackage{caption}
\usepackage{subcaption}

\newcommand{\ba}{\begin{array}{c}}
\newcommand{\ea}{\end{array}}

\usepackage[dvipsnames]{xcolor}

\begin{document}
\begin{titlepage}
\vspace*{-1cm}
\phantom{hep-ph/***}
\flushright
\hfil{CP3-22-23}
\hfil{IFIC/21-49}
\hfil{FTUV-21-1125.6337}

\vskip 1.5cm
\begin{center}
\mathversion{bold}
{\LARGE\bf
Low-scale leptogenesis with flavour and CP symmetries
\mathversion{normal}
}
\vskip .3cm
\end{center}
\vskip 0.5  cm
\begin{center}
{\large M.~Drewes}$^{1}$,
{\large Y.~Georis}$^{1}$,
{\large C.~Hagedorn}$^{2,3}$,
{\large J.~Klarić}$^{1,4}$
\\
\vskip .7cm
{\footnotesize
$^{1}$ Centre for Cosmology, Particle Physics and Phenomenology, Université catholique de Louvain, Louvain-la-Neuve B-1348, Belgium\\[0.3cm]
$^{2}$ Instituto de F\'isica Corpuscular, Universidad de Valencia and CSIC,
Edificio Institutos Investigaci\'on, Catedr\'atico Jos\'e Beltr\'an 2, 46980 Paterna, Spain\\[0.3cm]
$^{3}$ Istituto Nazionale di Fisica Nucleare, Sezione di Padova, Via F. Marzolo 8, 35131 Padua, Italy\\[0.3cm]
$^{4}$ Institute of Physics, École Polytechnique Fédérale de Lausanne, CH-1015 Lausanne, Switzerland
\vskip .5cm
\begin{minipage}[l]{.9\textwidth}
\begin{center}
\textit{E-mail:}
\tt{marco.drewes@uclouvain.be}, \tt{yannis.georis@uclouvain.be}, \tt{hagedorn@ific.uv.es}, \tt{juraj.klaric@uclouvain.be}
\end{center}
\end{minipage}
}
\end{center}
\vskip 1cm
\begin{abstract}
We consider a type-I seesaw framework endowed with a flavour symmetry, belonging to the series of non-abelian groups $\Delta (3 \, n^2)$ and $\Delta (6 \, n^2)$, 
and a CP symmetry. Breaking these symmetries in a non-trivial way results in the right-handed neutrinos being degenerate in mass up to possible (further symmetry-breaking) splittings $\kappa$ and $\lambda$, 
while the neutrino Yukawa coupling matrix encodes the entire flavour structure in the neutrino sector. For a fixed combination of flavour and CP symmetry and residual groups, 
 this matrix contains five real free parameters. Four of them are determined by the light neutrino mass spectrum and by
 accommodating experimental data on lepton mixing well, while the angle $\theta_R$ is related to right-handed neutrinos. We scrutinise for all four lepton mixing patterns, grouped into 
 Case 1) through Case 3 b.1), the potential to generate the baryon asymmetry of the Universe through low-scale leptogenesis numerically and analytically. The main results are: $a)$ 
 the possible correlation of the baryon asymmetry and the Majorana phases, encoded in the Pontecorvo-Maki-Nakagawa-Sakata mixing matrix, in certain instances; $b)$  
 the possibility to generate the correct amount of baryon asymmetry for vanishing splittings $\kappa$ and $\lambda$ among the right-handed neutrinos as well as for 
 large $\kappa$, depending on the case and the specific choice of group theory parameters; $c)$ the chance to produce sufficient baryon asymmetry 
 for large active-sterile mixing angles, enabling direct experimental tests at current and future facilities, if $\theta_R$ is close to a special value, potentially protected by an enhanced residual symmetry. 
 We elucidate these results with representative examples of flavour and CP symmetries, which all lead to a good agreement with the measured values of the lepton mixing angles and, possibly, the current indication of the CP phase $\delta$.
 We identify the CP-violating combinations relevant for low-scale leptogenesis, and show that
 the parametric dependence of the baryon asymmetry found in the numerical study
 can be understood well with their help.
\end{abstract}
\end{titlepage}
\setcounter{footnote}{0}
\tableofcontents

\section{Introduction}
\label{intro}

The Standard Model (SM) of particle physics has proven to be of unprecedented success. However, it fails to explain important
observations such as non-vanishing neutrino masses, the peculiar flavour structure in the quark and lepton sector as well as the baryon asymmetry 
of the Universe (BAU), $Y_B = (8.61 \pm 0.05) \cdot 10^{-11}$~\cite{ParticleDataGroup:2020ssz}.  

Many mechanisms have been proposed in order to generate neutrino masses, see e.g.~\cite{Cai:2017jrq}. Most of them predict neutrinos to be Majorana particles. 
In the case of the three different types of seesaw mechanisms, type-I seesaw, type-II seesaw and type-III seesaw, only one new species of particles is
added to the SM, right-handed (RH) neutrinos, Higgs triplets and fermion triplets, respectively. The smallness of neutrino masses is usually related to the
heaviness of the new particle species for couplings of order one. It is, however, also plausible that these particles have masses below the TeV scale, if
some of the couplings are suitably suppressed. An elegant and minimalistic example is the \emph{Neutrino Minimal Standard Model} ($\nu$MSM)~\cite{Asaka:2005pn,Asaka:2005an} 
with masses of the heavy neutrinos being (much) smaller than the electroweak scale. The important advantage of frameworks such as the $\nu$MSM is the fact that these can be comprehensively tested in different types of 
terrestrial experiments, e.g.~colliders, precision tests of flavour physics observables and fixed-target experiments \cite{Chun:2017spz,Abdullahi:2022jlv}.

Likewise several approaches have been studied to predict the number of fermion generations and the peculiar flavour structure in the quark and lepton sector, see 
e.g.~\cite{Ishimori:2010au,King:2013eh,Feruglio:2019ybq,Grimus:2011fk}.  
A symmetry acting on flavour space, a so-called flavour symmetry, which possesses at least one irreducible three-dimensional representation can explain (some of) these features. 
Several different types of symmetries have been investigated and discrete non-abelian groups have shown
to be most useful in the description of the features of the lepton sector. If such a symmetry is broken in a specific way, i.e.~to residual symmetries in different parts of 
the theory (e.g.~the charged lepton mass matrix and the matrices responsible for neutrino masses are governed by different residual groups), this approach turns out to be very constraining regarding the lepton 
mixing parameters accessible in neutrino oscillation experiments. The predictive power can be further enhanced, if a CP symmetry, also acting non-trivially on flavour
space in general, is involved~\cite{Feruglio:2012cw,Holthausen:2012dk,Chen:2014tpa,Grimus:1995zi,Ecker:1983hz,Ecker:1987qp,Neufeld:1987wa,Harrison:2002kp,Grimus:2003yn}. 
Most successful is the combination of a non-abelian flavour group $G_f$ and a CP symmetry
 that are broken to the residual symmetry $G_l$ among charged leptons,
contained in $G_f$, and to $G_\nu=Z_2 \times CP$, with $Z_2$ in $G_f$, among neutrinos. 
In this approach, the three lepton mixing angles $\theta_{ij}$ and the three CP phases, the Dirac phase $\delta$ and the two Majorana phases $\alpha$ and $\beta$, only depend
on one real parameter, up to permutations of the rows and columns of the Pontecorvo-Maki-Nakagawa-Sakata (PMNS) mixing matrix~\cite{Feruglio:2012cw}. Suitable candidates of a flavour group are found 
among the series of groups $\Delta (3 \, n^2)$~\cite{Luhn:2007uq} and $\Delta (6 \, n^2)$~\cite{Escobar:2008vc}, $n\geq 2$. 
The automorphisms of the flavour group then determine the possible CP symmetries. See~\cite{Hagedorn:2014wha} for resulting lepton mixing patterns.

The generation of the BAU in the SM could in principle be successful, if the electroweak phase transition were first order and a sufficient amount of CP violation could be achieved.
Given that this is impossible without an appropriate extension of the SM, e.g.~adding further scalars to the SM, leptogenesis~\cite{Fukugita:1986hr} is an interesting alternative. In leptogenesis the  BAU is produced from a generated lepton asymmetry which is converted into a baryon asymmetry through sphaleron processes~\cite{Kuzmin:1985mm}. Indeed, many mechanisms leading to neutrino masses
also offer the possibility to have viable leptogenesis, for a review see e.g.~\cite{Davidson:2008bu}.
We focus in the following on low-scale leptogenesis~\cite{Akhmedov:1998qx,Pilaftsis:2003gt,Pilaftsis:2005rv,Asaka:2005pn} in the context of the type-I seesaw framework with three RH neutrinos.

In this work, we consider the aforementioned framework, endowed with flavour and CP symmetries.\footnote{Leptogenesis has been discussed in several scenarios with flavour (and CP) symmetries, see e.g.~\cite{Hagedorn:2017wjy} for a concise review as well as the following publications for explicit examples~\cite{Jenkins:2008rb,Bertuzzo:2009im,Hagedorn:2009jy,AristizabalSierra:2009ex,Mohapatra:2015gwa,Hagedorn:2016lva,Chen:2016ptr,Fong:2021tqj,Chauhan:2021xus}.} The three generations of left-handed (LH) lepton doublets and of the RH neutrinos are
assigned to irreducible three-dimensional representations of the flavour group, whereas the RH charged leptons transform as singlets in order to ensure the possibility to achieve the hierarchy
among charged lepton masses without fine-tuning. The RH neutrinos are degenerate in mass, since their mass matrix does not break the flavour and CP
symmetry, while the neutrino Yukawa coupling matrix $Y_D$ relating LH and RH neutrinos instead preserves the residual symmetry $G_\nu=Z_2 \times CP$. For fixed $G_f$ and CP as well as residual groups $G_l$ and $G_\nu$, the matrix $Y_D$ contains five real free parameters, $y_f$, $f=1,2,3$, $\theta_L$ and $\theta_R$, which correspond to the three light neutrino masses, the free parameter
adjusted to accommodate well the experimental data on lepton mixing and the free parameter $\theta_R$ related to the RH neutrinos.
The residual symmetry $G_l$ among charged leptons is taken to be a $Z_3$ group, the minimal symmetry which allows to distinguish between the three generations. 
Indeed, we can without loss of generality assume that the $Z_3$ group is always generated by the same element of the flavour group.
Depending on the choice of $G_\nu$, we obtain different lepton mixing patterns. These can be grouped into four different cases, Case 1) through Case 3 b.1).
As is well-known, successful leptogenesis requires a non-vanishing mass splitting among the heavy neutrinos. One possible source is the (small) breaking of the flavour
and CP symmetry in the RH neutrino mass matrix, either arising from the symmetry breaking present in the charged lepton sector, encoded in the splitting $\kappa$,
or from other sources, as exemplified with the splitting $\lambda$.
To calculate the BAU we numerically solve the quantum kinetic equations from~\cite{Klaric:2020phc},
supplemented with the interaction rates from~\cite{Ghiglieri:2017csp}. Although these equations can in general not be solved analytically, we can gain significant insight into the parametric dependence of the BAU by identifying and evaluating the relevant CP-violating combinations that arise when perturbatively solving the equations.

The paper is organised as follows: in section~\ref{sec2} we first discuss the low-scale type-I seesaw framework and then its extension with flavour and CP symmetries. In section~\ref{sec3}
we elaborate on the results for light neutrino masses and lepton mixing, depending on the residual symmetry in the neutrino sector. Section~\ref{sec50} is dedicated to the numerical analysis of 
the lepton mixing parameters for the different cases, Case 1) through Case 3 b.1), characterised by different $G_\nu$. We give representative examples of the index $n$ of the flavour group and
the parameters describing the CP symmetry which give rise to an acceptable agreement with experimental data~\cite{Esteban:2020cvm} for at least one value of the free parameter $\theta_L$. We scrutinise the possibility to generate a sufficient amount of BAU for a subset of these representative examples in section~\ref{sec:LowScalelepto}. In section~\ref{sec6} we discuss analytic expressions for source
and washout terms which are very useful in order to understand the dependence of the generated BAU on the different parameters of the scenario at hand, e.g.~the choice of the 
CP symmetry and the size of the splitting $\kappa$ of the heavy neutrino masses.
We show in section~\ref{sec4} that special values of the free parameters $\theta_L$ and $\theta_R$
can be related to an enhancement of the residual symmetry in the neutrino Yukawa coupling matrix $Y_D$. Section~\ref{summ}
contains the summary of the main results and an outlook. The appendices, appendix~\ref{appA}-\ref{appG}, contain details on the group theory of the flavour symmetries, the form of the representation matrices of the residual
symmetries and of the CP transformations, the conventions of lepton mixing, neutrino masses and the corresponding experimental data, as well as supplementary information such as
further tables belonging to section~\ref{sec50}, additional plots for section~\ref{sec:LowScalelepto} and formulae for the CP-violating combinations found in section~\ref{sec6}.

\section{Scenario}
\label{sec2}

In this section we first summarise the main features of the low-scale  type-I seesaw which we use as framework. We then endow this framework
with flavour and CP symmetries and define the scenario we investigate in detail. In doing so, we fix our notation and specify the used symmetries and their residual groups.

\subsection{Low-scale type-I seesaw framework}
\label{sec:LowScaleSeesaw}

One of the most straightforward and commonly employed explanations for neutrino masses is the type-I seesaw mechanism~\cite{Minkowski:1977sc,Glashow:1979nm,Gell-Mann:1979vob,Mohapatra:1979ia,Yanagida:1980xy,Schechter:1980gr}, which requires complementing the SM field content with RH neutrinos $\nu_R$. This mechanism is not only appealing because all other fermions in the SM exist with both chiralities, but also because the CP-violating interactions of the RH neutrinos in the early Universe can generate a matter-antimatter asymmetry in the primordial plasma through leptogenesis~\cite{Fukugita:1986hr}. In this way, another important question in cosmology is simultaneously addressed, namely the origin of ordinary matter in the observable Universe.\footnote{See e.g.~\cite{Canetti:2012zc} for a detailed discussion of the empirical evidence for a matter-antimatter asymmetry.}

Being SM gauge singlets RH neutrinos can have a Majorana mass term $\overline{\nu^c_R} \, M_R \, \nu_R$. The implications of the existence of RH neutrinos for particle physics and cosmology strongly depend on the magnitude of the eigenvalues $M_i$ of the Majorana mass matrix $M_R$~\cite{Drewes:2013gca}. While the original proposals of both the type-I seesaw mechanism and leptogenesis assume that the \emph{seesaw scale} is several orders of magnitude larger than the electroweak scale, it has meanwhile become clear that low-scale seesaw models can successfully explain current neutrino oscillation data as well as the BAU. This does not only lead to a variety of leptogenesis mechanisms even in minimal models~\cite{Garbrecht:2018mrp,Bodeker:2020ghk}, but also opens up the possibility to discover the heavy neutrinos -- which are often referred to as \emph{heavy neutral leptons}  -- in direct searches
at the LHC~\cite{Atre:2009rg,Deppisch:2015qwa,Cai:2017mow,Agrawal:2021dbo,Abdullahi:2022jlv}, future colliders~\cite{Antusch:2016ejd,FCC:2018evy,CEPCStudyGroup:2018ghi,Abdullahi:2022jlv} or fixed target experiments~\cite{Agrawal:2021dbo,Abdullahi:2022jlv},
and thus to potentially test the origin of the BAU~\cite{Chun:2017spz}. For
theoretical motivations of a low seesaw scale see e.g.~section 5 of~\cite{Agrawal:2021dbo}.

In the following, we consider the renormalisable Lagrangian
\begin{eqnarray}
	\label{L}
	\mathcal L \supset
	\mathrm{i} \, \overline{\nu_R} \, \slashed\partial \, \nu_{R}
	- \frac{1}{2}
	\overline{\nu^c_R}\, M_R\, \nu_{R}
	- \overline{l_{L}} \, Y_D \, \varepsilon H^* \, \nu_{R}
	+ {\rm h. c.} \, ,
\end{eqnarray}
where $M_R$ is the Majorana mass matrix of the RH neutrinos $\nu_R$ (with the index $i$), $Y_D$ the neutrino Yukawa coupling matrix, $H$ the SM Higgs doublet and $l_L$ are the three LH lepton doublets (with the lepton flavour index $\alpha=e,\, \mu,\, \tau$).

We focus on the case of three generations of RH neutrinos $\nu_{R \, i}$, $i=1,2,3$.
Studies of low-scale leptogenesis in this case~\cite{Akhmedov:1998qx,Drewes:2012ma,Canetti:2014dka,Garbrecht:2014bfa,Shuve:2014zua,Hernandez:2015wna,Abada:2018oly,Drewes:2021nqr} have been either limited to specific parameter choices or performed by scanning over all possible values of the elements of the matrices $M_R$ and $Y_D$ that are consistent with current neutrino oscillation data (and constraints on light neutrino masses). Taking the latter approach leads to a large available parameter space~\cite{Drewes:2021nqr} which
does not only pose a practical computational challenge for a complete exploration of the parameter space due to its high dimensionality (without additional assumptions on their form $M_R$ and $Y_D$ encode 18 new parameters in the case of three RH neutrinos), it also severely limits the predictive power of the framework.\footnote{Note that one can strongly constrain the parameter space, if the framework with three RH neutrinos can not only explain current neutrino oscillation data and leptogenesis, but also provide a Dark Matter candidate~\cite{Dodelson:1993je,Shi:1998km}. In this minimal scenario, known as $\nu$MSM~\cite{Asaka:2005pn,Asaka:2005an}, the constraints on the Dark Matter candidate are so strong~\cite{Drewes:2016upu,Boyarsky:2018tvu} that this particle plays practically no role for leptogenesis and the generation of light neutrino masses. The reduced parameter space of the remaining two heavy neutrinos is then sufficiently small to be predictive and testable, cf.~e.g.~\cite{Hernandez:2016kel,Drewes:2016jae}.} In the present work, the (main) structure of the matrices $M_R$ and $Y_D$ (as well as of the charged lepton mass matrix $m_l$) is  determined by the flavour and CP symmetries and their residual groups. This drastically reduces the number of free parameters, see section~\ref{sec:symmetries}.

Before closing this section we briefly define the quantities necessary to characterise the mass eigenstates of the neutral fermions and their mixing.
After electroweak symmetry breaking there are two sets of neutrino mass eigenstates, the light neutrinos $\upnu_i$ and the heavy neutrinos $N_i$, which can be described by the Majorana spinors
\begin{align}
	\upnu_i &
	= \left[
		V_{\nu}^{\dagger}\nu_L-U_{\nu}^{\dagger}\uptheta \nu_R^c+V_{\nu}^T\nu_L^c-U_{\nu}^T\uptheta^{\ast} \nu_R
	\right]_i
	\ ,
	& N_i &
	= \left[
		V_N^\dagger\nu_R+\Uptheta^T \nu_L^c + V_N^T\nu_R^c+\Uptheta^{\dagger}\nu_L
	\right]_i
	\ . \label{HeavyMassEigenstates}
\end{align}
Here the matrix $\uptheta$, $\uptheta=\langle H \rangle \, Y_D \, M_R^{-1}$, encodes the mixing between LH and RH neutrinos.
$\langle H \rangle$ is the vacuum expectation value (VEV) of the SM Higgs doublet $H$, $\langle H \rangle \approx 174 \, \mathrm{GeV}$. The matrix $V_\nu$,
$V_\nu = (1 - \frac{1}{2}\uptheta\uptheta^\dagger ) U_\nu$, is the light neutrino mixing matrix, while the matrix $V_N$, $V_N = (1 - \frac{1}{2} \uptheta^T \uptheta^*) U_N$, is its equivalent in the heavy neutrino sector.
The unitary matrices $U_\nu$ and $U_N$ diagonalise the matrices
\begin{eqnarray}
\label{eq:blocks_mass_matrix}
m_\nu
= - \uptheta M_R \uptheta^T
\;\; \mbox{and} \;\;
M_N
= M_R + \frac{1}{2} (\uptheta^\dagger \uptheta M_R + M_R^T \uptheta^T \uptheta^{*})
\end{eqnarray}
as $ U_\nu^\dagger m_\nu U_\nu^*$ and $ U_N^T M_N U_N$, respectively.\footnote{The matrix $U_\nu$ coincides with the PMNS mixing matrix in Eq.~\eqref{eq:UPMNSdef} in appendix~\ref{appD1} in the charged lepton mass basis.
In this work, we ignore the difference between $V_\nu$ and $U_\nu$.
}
The squared masses of the light neutrinos $\upnu_i$ and of the heavy neutrinos $N_i$  are given at tree level by the eigenvalues of the matrices $m_\nu m_\nu^\dagger$ and $M_N M_N^\dagger$, respectively.
The eigenvalues of the matrix $M_N$ are close to those of $M_R$,
but in the regime of quasi-degenerate masses discussed in the following,
corrections of the order $\mathcal{O}(\uptheta^2)$ to the splitting between them can impact both, leptogenesis~\cite{Shaposhnikov:2008pf} and lepton number violating (LNV) event rates at colliders~\cite{Drewes:2019byd}.
The suppression of the  weak interactions of the heavy neutrinos, relative to those of the light neutrinos, is given by the elements of the so-called active-sterile neutrino mixing matrix
\begin{equation}
\Uptheta = \uptheta U_N^*.
\end{equation}
For the discussion of event rates at accelerator based experiments it is convenient to introduce the quantities
\begin{eqnarray}
\label{U2defs}
U_{\alpha i}^2 = |\Uptheta_{\alpha i}|^2 \ , \ U_i^2 = \sum_\alpha U_{\alpha i}^2 \ , \ U_\alpha^2 = \sum_i U_{\alpha i}^2 \;\; \mbox{and} \;\; U^2 = \sum_i U_i^2.
\end{eqnarray}
If one were to estimate the expected magnitude of the active-sterile mixing angles from Eq.~\eqref{eq:blocks_mass_matrix} without taking into account the matrix structure in flavour space, the smallness of the light neutrino masses would constrain these to be extremely small
\begin{equation}
	\label{eq:naiveseesawformula}
	U_i^2 \sim \frac{\sqrt{\sum_j m_{j}^2}}{M_i} \lesssim 10^{-10} \frac{\mbox{GeV}}{M_i},
\end{equation}
where $m_j$ denote the light neutrino masses and we have ignored the difference between the eigenvalues of the matrices $M_R$ and $M_N$.
The estimate in Eq.~(\ref{eq:naiveseesawformula}) is called the naive seesaw limit and it would strongly constrain
the possibility of any direct detection of the heavy neutrinos in the near future. However, in low-scale seesaw models the smallness of the light neutrino masses is usually not the result of a suppression by the seesaw scale, but is explained by an approximate $B-\bar{L}$ symmetry~\cite{Shaposhnikov:2006nn,Kersten:2007vk,Moffat:2017feq}, where $\bar{L}$ is a generalised lepton number. This, in principle, allows for large active-sterile mixing angles  (meaning order-one Yukawa couplings) and heavy neutrinos with sub-TeV masses without fine-tuning,
as the $B-\bar{L}$ symmetry dictates cancellations in the product of the matrices in Eq.~\eqref{eq:blocks_mass_matrix}.

\subsection{Type-I seesaw framework with flavour and CP symmetries}
\label{sec:symmetries}

In the following, we endow the low-scale type-I seesaw framework with a flavour and a  CP symmetry. These determine together with their residual groups $G_\nu$ and $G_l$
the form of the matrices $M_R$ and $Y_D$ as well as the charged lepton mass matrix $m_l$. In particular, they lead to the heavy neutrinos being (almost) degenerate in mass.

In the present study, we impose the flavour and CP symmetries as well as $G_\nu$ and $G_l$ directly on the matrices $M_R$, $Y_D$ and $m_l$. In concrete models, additional motivation for the choice of a particular flavour and CP symmetry can exist, for example in certain string theory inspired models, see e.g.~\cite{Baur:2019iai}. Furthermore, the breaking of these symmetries to the residual groups $G_\nu$ and $G_l$ often occurs spontaneously. For this to work, certain flavour (and CP) symmetry breaking fields are introduced that acquire peculiar VEVs such that $G_l$ and $G_\nu$ remain as residual symmetries~\cite{Ishimori:2010au,King:2013eh,Feruglio:2019ybq}.
Numerous realisations of such a spontaneous symmetry breaking of a flavour symmetry, belonging to the series of groups $\Delta (3 \, n^2)$ and $\Delta (6 \, n^2)$ with $n$ integer, (and CP) to residual groups can be found in the literature, for examples see~\cite{Altarelli:2005yx,deMedeirosVarzielas:2005qg,Lin:2009bw,Ding:2012xx,Feruglio:2013hia,Luhn:2013lkn,Hagedorn:2016lva,Hagedorn:2018gpw,Hagedorn:2018bzo}.\footnote{Another option is to consider the (explicit) breaking of the flavour symmetry at the boundaries  of an extra dimension very much like the breaking of the gauge symmetry, see e.g.~\cite{Hagedorn:2011un,Hagedorn:2011pw}.} 

We leave aside these specifications and assume in the following as relevant degrees of freedom only those mentioned in section~\ref{sec:LowScaleSeesaw}, meaning that additional degrees of freedom are either sufficiently heavier than the involved scales (electroweak scale, seesaw scale, temperature during leptogenesis) such that they can be ignored, or their abundances and strength of their interactions with the SM particles and RH neutrinos are sufficiently small so that they do not have to be treated dynamically during leptogenesis.\footnote{See e.g.~\cite{Fischer:2021nha,Flood:2021qhq} for some recent studies that investigate the impact of additional degrees of freedom on leptogenesis.}

As flavour symmetry $G_f$ we use a group which belongs to the series $\Delta (6 \, n^2)$ with $3 \nmid n$ ($n$ is not divisible by three)~\cite{Escobar:2008vc}. This group can be generated
by four generators, called $a$, $b$, $c$ and $d$.\footnote{We could also consider a group of the form $\Delta (3 \, n^2)$~\cite{Luhn:2007uq} in some of the cases. It is, however, contained in the corresponding group $\Delta (6 \, n^2)$ so that we can stick, without loss of generality, to the latter only.} They are given in the relevant representations
in appendix~\ref{appA}. The groups $\Delta (6 \, n^2)$ for $n > 2$ are interesting, as they possess at least one irreducible, faithful, complex three-dimensional
representation ${\bf 3}$.\footnote{For $n=2$ the irreducible three-dimensional representations are real.} In the following, we assign the three generations of LH lepton doublets $l_{L \, \alpha}$, $\alpha=e , \, \mu, \, \tau$, to ${\bf 3}$. For concreteness, we choose the representation
as ${\bf 3}_{\bf{1} \, (1)}$ according to the nomenclature used in~\cite{Escobar:2008vc}. The RH charged leptons are assigned to the representation
${\bf 1}$, the trivial singlet, of $G_f$, while the three generations of RH neutrinos $\nu_{R \, i}$, $i=1, 2, 3$, are unified into an irreducible, in general unfaithful, real representation ${\bf 3^\prime}$
of $G_f$.\footnote{Only for $n=2$ this representation is faithful.} The latter requires the index $n$ of the group $\Delta (6 \, n^2)$ to be even, see appendix~\ref{appA} for details.
In the nomenclature of~\cite{Escobar:2008vc} the representation ${\bf 3^\prime}$ is denoted as ${\bf 3}_{\bf{1} \, (n/2)}$.
Assigning LH lepton doublets and RH neutrinos to these different
three-dimensional representations of $G_f$ is crucial, as the assignment $l_{L \, \alpha} \sim {\bf 3}$ allows to fully explore the predictive power of $G_f$ (and not only of one of its subgroups),
while $\nu_{R \, i} \sim {\bf 3^\prime}$ permits the RH neutrinos to have a (flavour-universal) mass term without breaking $G_f$ and the CP symmetry. In addition, we assume the existence of a $Z_3$
symmetry, called $Z_3^{(\mathrm{aux})}$, which is employed in order to distinguish the three RH charged leptons $e_R$, $\mu_R$ and $\tau_R$. These are assigned
to $1$, $\omega$ and $\omega^2$ with $\omega=e^{2 \pi i/3}$, respectively, whereas LH lepton doublets and RH neutrinos are invariant under $Z_3^{(\mathrm{aux})}$.

The CP symmetry
imposed on the theory corresponds to an automorphism of $G_f$~\cite{Feruglio:2012cw,Holthausen:2012dk,Chen:2014tpa,Grimus:1995zi,Ecker:1983hz,Ecker:1987qp,Neufeld:1987wa,Harrison:2002kp,Grimus:2003yn}, belonging to those studied in~\cite{Hagedorn:2014wha}. They are represented by the CP transformation $X (\mathrm{\mathbf{r}})$
in the different (irreducible) representations ${\mathrm{\mathbf{r}}}$ of $G_f$ and depend on the parameter(s)  specifying the automorphism. Note that we
always choose $X (\mathrm{\mathbf{r}})$ symmetric and that it fulfils $X (\mathrm{\mathbf{r}}) \, X (\mathrm{\mathbf{r}})^*= X (\mathrm{\mathbf{r}})^* \, X (\mathrm{\mathbf{r}})= 1$.
For completeness, we show the form of the automorphisms and of $X (\mathrm{\mathbf{r}})$
for the relevant representations in appendix~\ref{appC}.

The residual symmetry $G_l$ is chosen as $Z_3^{(\mathrm{D})}$, which is the diagonal subgroup of the group, generated by $a$ of $G_f$, see appendix~\ref{appA}, and the auxiliary symmetry $Z_3^{(\mathrm{aux})}$, while $G_\nu$ is $Z_2 \times CP$, where the $Z_2$ symmetry is a subgroup of $G_f$.
Its generator $Z$ is denoted as $Z (\mathrm{\mathbf{r}})$ in the different representations $\mathrm{\mathbf{r}}$.
The $Z_2$ symmetry and CP commute, i.e. they satisfy
\begin{equation}
X (\mathrm{\mathbf{r}}) \, Z (\mathrm{\mathbf{r}})^* - Z (\mathrm{\mathbf{r}}) \, X (\mathrm{\mathbf{r}}) = 0
\end{equation}
for all representations $\mathrm{\mathbf{r}}$ of $G_f$.
The mismatch of
the residual symmetries $G_l$ and $G_\nu$ determines the form of the PMNS mixing matrix, as has been discussed in general in~\cite{Feruglio:2012cw} and, in particular for the groups $\Delta (3 \, n^2)$ and $\Delta (6 \, n^2)$, in~\cite{Hagedorn:2014wha}
as well as in~\cite{Ding:2014ora}. It has been found in~\cite{Hagedorn:2014wha} that the patterns, potentially compatible with experimental data on lepton mixing parameters, can be classified according to four types, called Case 1), Case 2), Case 3 a) and Case 3 b.1), that  have different
features. The form of the PMNS mixing matrix for the four different types is shown in section~\ref{sec3}.

The form of the charged lepton mass matrix $m_l$, the neutrino Yukawa coupling matrix $Y_D$ and the Majorana mass matrix $M_R$ of the RH neutrinos is determined by $G_l$ and $G_\nu$, at least at leading order.
In the chosen basis, see appendix~\ref{appA}, the mass matrix $m_l$ is diagonal and contains three independent parameters that correspond to the three different charged lepton
masses. As $m_l$ is diagonal, there is no contribution to the PMNS mixing matrix from the charged lepton sector.\footnote{In principle, the masses of charged leptons could be ordered non-canonically, leading to a permutation matrix
entering the PMNS mixing matrix. However, we assume them to be ordered correctly in the following. For a complete discussion of  mixing patterns, taking also into account such permutations, see~\cite{Hagedorn:2014wha}.} As regards the
neutrino sector, we take the neutrino Yukawa coupling matrix $Y_D$
to be invariant under $G_\nu$, whereas the matrix $M_R=M_R^0$ does neither break $G_f$ nor CP.  Being invariant under $Z_2 \times CP$, the matrix $Y_D$, in the basis in which LH fields are on the left and RH
ones on the right, fulfils the following relations
\begin{equation}
\label{eq:YDZX}
Z ({\bf 3})^\dagger \, Y_D \, Z ({\bf 3^\prime}) = Y_D \;\;\; \mbox{and} \;\;\; X ({\bf 3})^* \, Y_D \, X ({\bf 3^\prime}) = Y_D^* \; .
\end{equation}
The form of $Y_D$ is thus\footnote{We can re-write the conditions in Eq.~(\ref{eq:YDZX}) using the matrices
$\Omega({\bf 3})$ and $ \Omega({\bf 3^\prime})$, see Eq.~(\ref{eq:XOmega}), and find
\begin{equation}
\nonumber
\Omega({\bf 3})^\dagger \, Y_D \,  \Omega({\bf 3^\prime})
\end{equation}
is real and can be diagonalised by two rotation matrices from the left and right, respectively,
\begin{equation}
\nonumber
\Omega({\bf 3})^\dagger \, Y_D \,  \Omega({\bf 3^\prime}) = R_{ij} (\theta_L) \,  \mbox{diag} (y_1, y_2, y_3) \, P^{ij}_{kl} \, R_{kl} (-\theta_R) \, .
\end{equation}}
\begin{equation}
\label{eq:YDgen}
Y_D =  \Omega ({\bf 3}) \, R_{ij} (\theta_L) \, \mbox{diag} (y_1, y_2, y_3) \, P^{ij}_{kl} \, R_{kl} (-\theta_R) \, \Omega ({\bf 3^\prime})^\dagger \; .
\end{equation}
The matrices $\Omega ({\bf 3})$ and $\Omega ({\bf 3^\prime})$ are unitary and determined by the form of the CP transformations $X ({\bf 3})$ and $X ({\bf 3^\prime})$ in the
representations of LH lepton doublets and RH neutrinos, i.e. they satisfy
\begin{equation}
\label{eq:XOmega}
X ({\bf 3})= \Omega ({\bf 3}) \, \Omega ({\bf 3})^T \;\; \mbox{and} \;\; X ({\bf 3^\prime})= \Omega ({\bf 3^\prime}) \, \Omega ({\bf 3^\prime})^T \, .
\end{equation}
As the choice of the CP symmetry and thus the corresponding CP transformation $X ({\bf 3})$ and $X ({\bf 3^\prime})$ is in general indicated by natural numbers, see e.g. the parameter $s$ in Eqs.~(\ref{eq:Xs3Case1}) and (\ref{eq:Xs3primeCase1}), also the matrices  $\Omega ({\bf 3})$ and $\Omega ({\bf 3^\prime})$ (can) depend on these parameters. The matrices $R_{ij} (\theta_L)$ and $R_{kl} (\theta_R)$ denote
rotations in the $(ij)$- and $(kl)$-plane, $i,j,k,l=1,2,3$ with $i< j$ and $k< l$, through the angles $\theta_L$ and $\theta_R$, respectively.\footnote{We define
the rotations $R_{ij}$, $i< j$, through the angle $\theta$ in the $(ij)$-plane as follows
\begin{equation}
\nonumber
R_{12} (\theta) = \left(
\begin{array}{ccc}
\cos\theta & \sin\theta & 0\\
-\sin\theta & \cos\theta & 0\\
0 & 0 & 1
\end{array}
\right) \, , \;
R_{13} (\theta) = \left(
\begin{array}{ccc}
\cos\theta & 0 & \sin\theta\\
0 & 1 & 0\\
-\sin\theta & 0 & \cos\theta
\end{array}
\right) \, , \;
R_{23} (\theta) = \left(
\begin{array}{ccc}
1 & 0 & 0\\
0 &  \cos\theta & \sin\theta\\
0 & -\sin\theta & \cos\theta
\end{array}
\right) \, .
\end{equation}}
These angles are free parameters,
i.e. not fixed by the residual symmetry $G_\nu$, and can take values in the range $0 \leq \theta_L \leq \pi$ and $0 \leq \theta_R \leq 2 \, \pi$.
The planes, in which the rotations $R_{ij} (\theta_L)$ and $R_{kl} (\theta_R)$ act, are determined
by the $(ij)$- and $(kl)$-subspaces of degenerate eigenvalues of the generator $Z$ in the representation ${\bf 3}$ and ${\bf 3^\prime}$, when transformed with the matrices $\Omega ({\bf 3})$
and $\Omega ({\bf 3^\prime})$, respectively; for examples see section~\ref{sec3}. If the $(ij)$- and $(kl)$-planes do not coincide, the permutation matrix $P^{ij}_{kl}$ is needed. It  acts in the plane which is determined
by the two indices of $(ij)$ and $(kl)$ that are different, e.g.~if $(ij)=(13)$ and $(kl)=(12)$ the permutation matrix $P^{ij}_{kl}$ must act in the $(23)$-plane.\footnote{The permutation matrix $P^{ij}_{kl}$ is only relevant
in Case 3 a) and Case 3 b.1).}
In addition to the two angles $\theta_L$ and $\theta_R$, the neutrino Yukawa coupling matrix $Y_D$ contains further three real parameters, namely the
couplings $y_f$, $f=1,2,3$. This has also been pointed out in~\cite{Hagedorn:2016lva}. The Dirac neutrino mass matrix $m_D$ is in turn given by
\begin{equation}
\label{eq:mD}
m_D = Y_D \, \langle H \rangle \; .
\end{equation}
As $M_R=M_R^0$ leaves $G_f$ and CP invariant, its form is simply
\begin{equation}
\label{eq:MR}
M_R = M_R^0 = M \, \left(
\begin{array}{ccc}
1 & 0 & 0\\
0 & 0 & 1\\
0 & 1 & 0
\end{array}
\right)
\end{equation}
with $M > 0$ which sets the mass scale of the RH neutrinos
\begin{equation}
\label{eq:defMi}
M_i=M \, .
\end{equation}
The matrix $M_R$ can be made diagonal with\footnote{If the difference between the matrices $M_R$ and $M_N$, see Eq.~(\ref{eq:blocks_mass_matrix}), is negligible, the matrix $U_N$ can be identified with $U_R$.}
\begin{equation}
\label{eq:MRUR}
U_R = \frac{1}{\sqrt{2}} \,
\left( \begin{array}{ccc}
\sqrt{2} & 0 & 0\\
0 & 1 & i\\
0 & 1 & -i
\end{array}
\right) \; , \;\; \mbox{i.e. } \, \hat{M}_R = U_R^T \, M_R \, U_R \, .
\end{equation}
The light neutrino mass matrix $m_\nu$ follows from the type-I seesaw mechanism~\cite{Minkowski:1977sc,Glashow:1979nm,Gell-Mann:1979vob,Mohapatra:1979ia,Yanagida:1980xy,Schechter:1980gr}, compare Eq.~(\ref{eq:blocks_mass_matrix}),
\begin{equation}
\label{eq:mnu}
m_\nu = - \, m_D \, M_R^{-1} \, m_D^T \, .
\end{equation}
As the charged lepton mass matrix $m_l$ is diagonal, the PMNS mixing matrix arises from the diagonalisation of $m_\nu$ only. In general, the resulting lepton mixing parameters involve
a combination of all quantities, appearing in $Y_D$. However, if
\begin{equation}
\label{eq:mnumiddle}
\mbox{diag} \, (y_1, y_2, y_3) \, P^{ij}_{kl} \, R_{kl} (-\theta_R)  \, \Omega ({\bf 3^\prime})^\dagger \, M_R^{-1} \, \Omega ({\bf 3^\prime})^* \, R_{kl} (\theta_R) \, \Big(P^{ij}_{kl}\Big)^T \, \mbox{diag} \, (y_1, y_2, y_3) \;\; \mbox{is diagonal,}
\end{equation}
see section~\ref{sec3},\footnote{Whether the combination in Eq.~(\ref{eq:mnumiddle}) is diagonal or not, does not depend on the presence of the permutation matrix $P^{ij}_{kl}$.}
the lepton mixing parameters only depend on $\theta_L$ and the quantities, describing the flavour and CP symmetries as well as the residual group $G_\nu$, i.e.~we find
\begin{equation}
\label{eq:PMNSspecial}
U_{\mbox{\scriptsize{PMNS}}} = \Omega ({\bf 3}) \, R_{ij} (\theta_L) \, K_\nu
\end{equation}
which fulfils
\begin{equation}
U_{\mbox{\scriptsize{PMNS}}}^\dagger \, m_\nu \, U_{\mbox{\scriptsize{PMNS}}}^* = \mathrm{diag} \, (m_1, m_2, m_3) \; .
\end{equation}
The mass spectrum of the light neutrinos is determined by the couplings $y_f$, $f=1,2,3$ (and
the angle $\theta_R$).
The matrix $K_\nu$ accounts for the CP parities of the light neutrino mass eigenstates and is a diagonal matrix whose (non-vanishing) elements can take the values $\pm 1$ and $\pm i$. Its
explicit form is given for each of the different cases in section~\ref{sec3}. As the couplings $y_f$, $f=1,2,3$, are not constrained other than being real, the scenario can accommodate both neutrino mass orderings, normal ordering (NO) and inverted ordering (IO), as well as a quasi-degenerate light neutrino mass spectrum.
The resulting PMNS mixing matrix in Eq.~(\ref{eq:PMNSspecial}) coincides with the PMNS mixing matrix, obtained in a scenario with three RH neutrinos~\cite{Hagedorn:2016lva}, in which the  Dirac neutrino mass matrix $m_D$
is invariant under the entire flavour and CP symmetry, while the  Majorana mass matrix $M_R$ of the RH neutrinos possesses the residual symmetry $G_\nu$.
The requirement to accommodate the measured lepton mixing angles well further constrains the index $n$ of $G_f$ and the CP symmetry as well as the combination of the residual groups $G_l$ and $G_\nu$, as discussed in detail in~\cite{Hagedorn:2014wha} and in section~\ref{sec50}. In section~\ref{sec3} we also comment on situations in which the condition in Eq.~(\ref{eq:mnumiddle}) is not fulfilled and the resulting PMNS mixing
matrix depends on the couplings $y_f$, $f=1,2,3$, and the angle $\theta_R$ as well.

The form of the neutrino Yukawa coupling matrix $Y_D$ in the basis, in which the Majorana mass matrix $M_R$ of the RH neutrinos is diagonal, is given by
\begin{equation}
\label{eq:YDhat}
\hat{Y}_D = Y_D \, U_R \, .
\end{equation}
Corrections to the displayed form of the matrices $M_R$, $Y_D$ and $m_l$ are expected to arise, at higher order, in concrete models~\cite{Ishimori:2010au,King:2013eh,Feruglio:2019ybq}. These can, indeed, be welcome, since the successful generation of the BAU requires the masses of the heavy neutrinos to be, at least partly, (slightly) non-degenerate~\cite{Dev:2017wwc}.
One way to achieve this is to consider a correction to the Majorana mass matrix $M_R$ of the RH neutrinos.
While the form of such a correction in general strongly depends on the concrete model, we advocate in the following an instance in which this correction is
invariant under the residual symmetry $G_l$. The generator of $G_l$ is given as
\begin{equation}
a ({\bf 3^\prime}) =  \left( \begin{array}{ccc}
1 & 0 & 0\\
0 & \omega & 0\\
0 & 0 & \omega^2
\end{array}
\right)
\end{equation}
in the representation ${\bf 3^\prime}$ of the RH neutrinos $\nu_{R \, i}$,
since these are not charged under the auxiliary symmetry $Z_3^{(\mathrm{aux})}$. The correction $\delta M_R$ must thus satisfy
\begin{equation}
a({\bf 3^\prime})^T \, \delta M_R \, a ({\bf 3^\prime}) = \delta M_R \; ,
\end{equation}
meaning that it is of the form
\begin{equation}
\label{dMRtilde}
\delta M_R = \kappa \, M \, \left( \begin{array}{ccc}
2 & 0 & 0\\
0 & 0 & -1\\
0 & -1 & 0
\end{array}
\right)
\end{equation}
with the splitting $\kappa$ being a small symmetry-breaking parameter. The RH neutrino masses $M_i$, $i=1,2,3$, acquire a (small) correction
\begin{equation}
\label{eq:RHmassspectrum}
M_1 = M \, (1+ 2 \, \kappa) \;\; \mbox{and} \;\; M_2=M_3= M \, (1-\kappa) \, .
\end{equation}
As one can see from Eq.~(\ref{eq:RHmassspectrum}), the masses of the second and third RH neutrino are still degenerate. A way to split these as well is to consider a further correction $\Delta M_R$ that is not (necessarily) invariant under any residual symmetry. A possible (rather minimal) choice is
\begin{equation}
\Delta M_R = \lambda \, M \, \left( \begin{array}{ccc}
0 & 0 & 0 \\
0 & 1 & 0\\
0 & 0 & 1
\end{array}
\right) \, ,
\label{lambda definition}
\end{equation}
where the splitting $\lambda$ is a small parameter, expected to be of the size of $\kappa$ (or smaller).
The Majorana mass matrix $M_R$ of the RH neutrinos then reads
\begin{equation}
M_R = M_R^0 + \delta M_R + \Delta M_R
\end{equation}
and we find for the RH neutrino masses 
\begin{equation}
\label{lambda mass spectrum}
M_1 = M \, (1+ 2 \, \kappa) \; , \;\; M_2=M \, (1-\kappa+ \lambda) \;\; \mbox{and} \;\; M_3= M \, (1-\kappa- \lambda) \, .
\end{equation}
In general, we expect that such (residual) symmetry-breaking effects also correct the form of the
neutrino Yukawa coupling matrix $Y_D$ and of the charged lepton mass matrix $m_l$. However, we assume in the following that their size is small enough so that they have no relevant impact on the shown results.

A concrete model incorporating the presented type-I seesaw framework with a flavour and CP symmetry should, thus, contain means to efficiently control the size and form of such symmetry-breaking effects, e.g.~additional symmetries to suppress these, a judicious choice of flavour symmetry breaking fields, sufficiently small symmetry-breaking parameters.

\section{Residual symmetries, neutrino masses and PMNS mixing matrix}
\label{sec3}

We first present the residual symmetry $G_\nu$ and the form of the corresponding representation matrices for the different cases, called Case 1), Case 2), Case 3 a) and Case 3 b.1) in~\cite{Hagedorn:2014wha}.
We then discuss additional constraints imposed from light neutrino masses on the choice of $G_\nu$ as well as the constraints on the light neutrino mass spectrum arising from imposing the condition in Eq.~(\ref{eq:mnumiddle}).
Furthermore, we mention for each case the form of the PMNS mixing matrix.
In the end, we briefly comment on the possible impact of the splitting $\kappa$ (and $\lambda$) among the RH neutrino masses on the light neutrino mass spectrum and the PMNS mixing matrix.

\subsection{Case 1)}
\label{sec31}

\subsubsection{Residual symmetries}
\label{subsec:case1ressymm}

The residual $Z_2$ symmetry in the neutrino sector is generated by
\begin{equation}
\label{Z2Case1}
Z= c^{n/2}
\end{equation}
which requires the index $n$ of the flavour symmetry to be even. The explicit form of $Z$ in the irreducible, faithful, complex three-dimensional representation ${\bf 3}$
and in the irreducible, unfaithful, real three-dimensional representation ${\bf 3^\prime}$ can be found in appendix~\ref{appB}.
As we see in section~\ref{subsec:case1numass}, due to the form of the generator $Z$ in ${\bf 3^\prime}$ for $n$ divisible by four the neutrino Yukawa coupling matrix $Y_D$ becomes singular
so that the resulting light neutrino mass spectrum is not viable. For this reason, we focus in the following on $n$ not divisible by four.

The CP symmetry corresponds to the automorphism, given in Eq.~(\ref{eq:XP23auto}) in appendix~\ref{appC},
conjugated with the inner automorphism induced by the group transformation $a \, b \, c^s \, d^{2 s}$ with $s=0,1,...,n-1$. The CP transformation $X (s)$ reads
in ${\bf 3}$
\begin{equation}
\label{eq:Xs3Case1}
X(s) ({\bf 3}) = a ({\bf 3}) \, b({\bf 3}) \, c ({\bf 3})^s \, d({\bf 3})^{2 s} \, X_0 ({\bf 3})
\end{equation}
and in ${\bf 3^\prime}$
\begin{equation}
\label{eq:Xs3primeCase1}
X(s) ({\bf 3^\prime}) = a ({\bf 3^\prime}) \, b({\bf 3^\prime}) \, c ({\bf 3^\prime})^s \, d({\bf 3^\prime})^{2 s} \, X_0 ({\bf 3^\prime})
\end{equation}
with $X_0 ({\bf 3})$ and $X_0 ({\bf 3^\prime})$ representing the CP transformation corresponding to the automorphism in Eq.~(\ref{eq:XP23auto}).
Their explicit form can be found in appendix~\ref{appC}.

The matrix $\Omega(s) ({\bf 3})$, derived from $X (s) ({\bf 3})$, given in Eq.~(\ref{eq:case1Xin3}) in appendix~\ref{appC}, can be chosen as
\begin{equation}
\label{case1Omegain3}
\Omega(s) ({\bf 3}) = e^{i \, \phi_s} \, U_{\mbox{\scriptsize{TB}}} \,
\left( \begin{array}{ccc}
1 & 0 & 0 \\
0 & e^{-3 \, i \, \phi_s} & 0\\
0 & 0 & -1
\end{array}
\right)
\end{equation}
with $U_{\mbox{\scriptsize{TB}}}$ describing tri-bimaximal (TB) mixing
\begin{equation}
U_{\mbox{\scriptsize{TB}}} =
\left( \begin{array}{ccc}
\sqrt{2/3} & \sqrt{1/3} & 0\\
-\sqrt{1/6} & \sqrt{1/3} & \sqrt{1/2} \\
-\sqrt{1/6} & \sqrt{1/3} & -\sqrt{1/2}
\end{array}
\right)
\end{equation}
and
\begin{equation}
\label{phi definition case I}
\phi_s=\frac{\pi \, s}{n} \, .
\end{equation}
The form of $\Omega(s) ({\bf 3^\prime})$ only depends on whether $s$ is even or odd
\begin{equation}
\label{Omega3p_seven}
\Omega(s \, \mbox{even}) ({\bf 3^\prime}) =  U_{\mbox{\scriptsize{TB}}}
\end{equation}
and
\begin{equation}
\label{Omega3p_sodd}
\Omega(s \, \mbox{odd}) ({\bf 3^\prime}) =  U_{\mbox{\scriptsize{TB}}} \, \left(
\begin{array}{ccc}
i & 0 & 0\\
0 & 1 & 0\\
0 & 0 & i
\end{array}
\right)
\; .
\end{equation}

Comparing these forms to the form of $\Omega(s) ({\bf 3})$ we see that they have the same
structure and the crucial difference lies in the phase matrix multiplied from the right.

In order to determine the plane in which the rotation $R_{ij} (\theta_L)$ acts, we look at
\begin{equation}
\label{eq:ZOmega3Case1}
\Omega(s) ({\bf 3})^\dagger \, Z ({\bf 3}) \,  \Omega(s) ({\bf 3}) = \left( \begin{array}{ccc}
-1 & 0 & 0\\
0 & 1 & 0\\
0 & 0 & -1
\end{array}
\right) \, ,
\end{equation}
meaning that the rotation through $\theta_L$ is in the $(13)$-plane~\cite{Feruglio:2012cw}.
Similarly, we can find the plane in which the rotation $R_{kl} (\theta_R)$ acts.
The representation matrix $Z ({\bf 3^\prime})$ for $n$ not divisible by four reads after the transformation with $\Omega (s) ({\bf 3^\prime})$
for both, $s$ even as well as $s$ odd,
\begin{equation}
\label{eq:ZOmega3prCase1}
\Omega(s) ({\bf 3^\prime})^\dagger \, Z ({\bf 3^\prime})  \,  \Omega(s) ({\bf 3^\prime}) = \left( \begin{array}{ccc}
-1 & 0 & 0\\
0 & 1 & 0\\
0 & 0 & -1
\end{array}
\right)
\; ,
\end{equation}
meaning that also $R_{kl} (\theta_R)$ acts in the $(13)$-plane.

\subsubsection{Constraints from and on light neutrino mass spectrum}
\label{subsec:case1numass}

First, we discuss constraints on the possible choices of the residual symmetry $G_\nu$ arising from the light neutrino mass spectrum.
In order to find these we consider the form of the neutrino Yukawa coupling matrix $Y_D$ fulfilling the conditions in Eq.~(\ref{eq:YDZX}).
For $n$ divisible by four $Z ({\bf 3^\prime})$ is given by Eq.~(\ref{eq:Zcn2_n2even}) in appendix~\ref{appB} and we find that the form of $Y_D$ needs to be
\begin{equation}
\label{YDnotworking}
Y_D = \left(
\begin{array}{ccc}
y_{11} & y_{12} & y_{13}\\
y_{11} & y_{12} & y_{13}\\
y_{11} & y_{12} & y_{13}
\end{array}
\right)
\end{equation}
with $y_{1i}$ complex, $i=1,2,3$. The determinant of $Y_D$ vanishes and this matrix
has two zero eigenvalues. As a consequence, the light neutrino mass matrix arising from the type-I seesaw mechanism, see Eq.~(\ref{eq:mnu}),
has two zero eigenvalues.\footnote{Indeed, if $Z({\bf 3^\prime})$ is the identity matrix and $Z ({\bf 3})$ is any generator of a $Z_2$
symmetry contained in $SU(3)$, i.e. it can be represented by a matrix $Z ({\bf 3})$ that fulfils $V^\dagger \, Z({\bf 3}) \, V = \mbox{diag} \, (1,-1,-1)$
with $V$ being a unitary matrix, we can show that
\begin{equation}
Z ({\bf 3})^\dagger \, Y_D = V \, \mbox{diag} \, (1,-1,-1) \, V^\dagger \, Y_D = Y_D \, ,
\end{equation}
meaning we can re-write this condition as
\begin{equation}
\mbox{diag} \, (1,-1,-1) \, \left[ V^\dagger \, Y_D \right] = \left[ V^\dagger \, Y_D \right] \; .
\end{equation}
Consequently, the combination $V^\dagger \, Y_D$ must have two vanishing rows, namely the second and the third one.
The determinant of $V^\dagger \, Y_D$ vanishes and the matrix has two zero eigenvalues. Thus, $Y_D$ itself must have these  properties. So, in general knowing
that $Z ({\bf 3^\prime})$ is given by the identity matrix is sufficient in order to discard a case as realistic without corrections which can induce, at least,
one further non-vanishing neutrino mass.
}
Furthermore, we can check that the non-zero eigenvalue has to correspond to the
light neutrino mass $m_2$, since it is always associated with the eigenvector proportional to $\left( 1,1,1 \right)^T$ which can only be identified
with the second column of the PMNS mixing matrix. It is, however, experimentally highly disfavoured that such a form can be the dominant contribution to
light neutrino masses. We thus do not discuss this case further.

For $n$ not divisible by four the form of the matrix $Z ({\bf 3^\prime})$ is shown in Eq.~(\ref{eq:Zcn2_n2odd}) in appendix~\ref{appB}. Again, we can compute the constraints on $Y_D$,
arising from imposing the conditions in Eq.~(\ref{eq:YDZX}). In particular, we see that the first condition in the latter equation reduces the number of free
(complex) parameters in $Y_D$ to five, meaning the other four ones can be expressed in these, e.g.~
\begin{eqnarray}
&& y_{23} = y_{11} + y_{12} + y_{13} - y_{21} - y_{22} \;\; , \;\; y_{31} = y_{12} + y_{13} - y_{21} \; ,\\
&& y_{32} = y_{11} + y_{13} - y_{22} \;\; \mbox{and} \;\; y_{33} = -y_{13} + y_{21} + y_{22} \; .
\end{eqnarray}
The five free complex parameters in $Y_D$ are further constrained by requiring that also the second condition in Eq.~(\ref{eq:YDZX}) is fulfilled.
As a consequence, these parameters are reduced to five real parameters. This is consistent with the findings in the general case where $Y_D$ contains three real couplings $y_f$, $f=1,2,3$,
and two angles $\theta_L$ and $\theta_R$. In general, such a matrix $Y_D$ has a non-vanishing determinant and three different eigenvalues. 

Focusing on $n$ not divisible by four we continue with considering the form of the light neutrino mass matrix and the corresponding mass spectrum.
For all choices of $s$ the combination in Eq.~(\ref{eq:mnumiddle}) is non-diagonal and the $(13)$- and $(31)$-elements are proportional to $y_1$, $y_3$ and
$\sin 2 \, \theta_R$. Thus, in general only the light neutrino mass $m_2$ is given as
\begin{equation}
\label{eq:m2Case1}
m_2 = \frac{y_2^2 \, \langle H \rangle^2}{M} \; ,
\end{equation}
while the masses $m_1$ and $m_3$ are determined by both the couplings $y_1$ and $y_3$ as well as $\theta_R$. One can achieve a diagonal form of the combination in Eq.~(\ref{eq:mnumiddle}) for
\begin{itemize}
\item[$(i)$] $y_1=0$ which corresponds to strong NO, i.e.~light neutrino masses with NO and the lightest neutrino mass $m_0=0$, with
\begin{equation}
\label{eq:strongNOCase1}
m_1=0 \;\; \mbox{and} \;\; m_3=\frac{y_3^2 \, \langle H \rangle^2}{M}  \, |\cos 2 \, \theta_R|
\end{equation}
in addition to $m_2$ in Eq.~(\ref{eq:m2Case1}). For a realistic light neutrino mass
spectrum $m_3$ has to be non-vanishing and thus $\cos 2 \, \theta_R \neq 0$.
The matrix $K_\nu$ is of the form
\begin{equation}
\label{eq:strongNOCase1Knu}
K_\nu = \left( \begin{array}{ccc}
		1 & 0 & 0\\
		0 & \pm 1 & 0\\
		0 & 0 & \pm i^{(s \, \scriptsize \mbox{mod} \normalsize \, 2) + c_R + 1}
	\end{array}
\right)
\;\; \mbox{(strong NO)}
\end{equation}
with $c_R=0$ for $\cos 2 \, \theta_R$ positive and $c_R=1$ for negative $\cos 2 \, \theta_R$.
Since $m_1$ vanishes, the $(11)$-element of $K_\nu$ is not constrained and is set for concreteness
to $+1$.
\item[$(ii)$] $y_3=0$ corresponding to strong IO, i.e.~light neutrino masses with IO and the lightest neutrino mass $m_0=0$, with
\begin{equation}
\label{eq:strongIOCase1}
m_1=\frac{y_1^2 \, \langle H \rangle^2}{M}  \, |\cos 2 \, \theta_R| \;\; \mbox{and} \;\; m_3=0
\end{equation}
with $m_2$ given in Eq.~(\ref{eq:m2Case1}). For a realistic light neutrino mass
spectrum $m_1$ has to be non-vanishing and thus $\cos 2 \, \theta_R \neq 0$.
The matrix $K_\nu$ is of similar form as in Eq.~(\ref{eq:strongNOCase1Knu}) with the roles of the $(11)$- and $(33)$-elements
exchanged,
\begin{equation}
\label{eq:strongIOCase1Knu}
K_\nu = \left( \begin{array}{ccc}
		\pm i^{(s \, \scriptsize \mbox{mod} \normalsize \, 2) + c_R} & 0 & 0\\
		0 & \pm 1 & 0\\
		0 & 0 & 1
	\end{array}
\right)
\;\; \mbox{(strong IO)} \, .
\end{equation}
\item[$(iii)$] $\sin 2 \, \theta_R=0$ where no constraints on the light neutrino masses arise. We do not consider this possibility, as we would like to explore
the effect of the parameter $\theta_R$ on the various phenomenological aspects of the scenario.
\end{itemize}

\subsubsection{PMNS mixing matrix}
\label{subsec:case1mixing}

For light neutrino masses with strong NO or strong IO as well as for $\sin 2 \, \theta_R=0$, the form of the PMNS mixing matrix is given by
\begin{equation}
\label{eq:PMNSspecialcase1}
U_{\mbox{\scriptsize{PMNS}}} = \Omega (s) ({\bf 3}) \, R_{13} (\theta_L) \, K_\nu \, ,
\end{equation}
as deduced from Eq.~(\ref{eq:PMNSspecial}).

In the case in which none of the light neutrinos is massless and the condition in Eq.~(\ref{eq:mnumiddle}) is not fulfilled, we have to take into account an additional rotation in the $(13)$-plane which depends on the couplings $y_1$ and $y_3$ and the angle
$\theta_R$. Since this rotation is in the same plane as the rotation induced by the angle $\theta_L$, the structure
of the PMNS mixing matrix in Eq.~(\ref{eq:PMNSspecialcase1}) does not change, but the angle $\theta_L$ becomes replaced by an effective angle  $\widetilde{\theta}_L$ that is the sum 
of $\theta_L$ and the additional angle.

In the numerical analysis of lepton mixing data, presented in section~\ref{sec50}, we always refer to the free angle, contained in the PMNS mixing matrix, as $\theta_L$.
Nevertheless, all results are also applicable in the case of three massive light neutrinos, where the angle $\theta_L$ has to be read as $\widetilde{\theta}_L$.

\subsection{Case 2)}
\label{sec32}

\subsubsection{Residual symmetries}
\label{subsec:case2ressymm}

The residual $Z_2$ symmetry in the neutrino sector is generated by the same element,
\begin{equation}
\label{Z2Case2}
Z=c^{n/2} \, ,
\end{equation}
as in Case 1). Thus, all comments made, in particular the forms of $Z ({\bf 3})$ and $Z ({\bf 3^\prime})$ in Eqs.~(\ref{eq:Zcn2in3}-\ref{eq:Zcn2_n2odd}) in appendix~\ref{appB}, 
apply.

The CP symmetry is given by the automorphism in Eq.~(\ref{eq:XP23auto}) and the inner
automorphism $h= c^s d^t$ with $0 \leq s, t \leq n-1$. In the three-dimensional representations ${\bf 3}$ and ${\bf 3^\prime}$ the CP transformation $X (s,t)$ is given by
\begin{equation}
X (s,t) ({\bf 3}) = c ({\bf 3})^s \, d ({\bf 3})^t \, X_0 ({\bf 3}) \;\; \mbox{and} \;\; X (s,t) ({\bf 3^\prime}) = c ({\bf 3^\prime})^s \, d ({\bf 3^\prime})^t \, X_0 ({\bf 3^\prime})
\end{equation}
and the explicit forms can be found in appendix~\ref{appC}, see Eqs.~(\ref{eq:case2Xin3}-\ref{eq:case2Xin3prime_stodd}).

In the discussion of the residual symmetries and the corresponding representation matrices we use $s$ and $t$ as parameters unlike in the phenomenological analysis
(e.g.~of lepton mixing in section~\ref{sec50}), where it turns out to be more convenient to use the parameters $u$ and $v$ that are linearly
related to $s$ and $t$
\begin{equation}
\label{defuv}
u= 2 \, s -t \;\; \mbox{and} \;\; v= 3 \, t \; .
\end{equation}
According to the findings in \cite{Hagedorn:2014wha}, a suitable choice of the matrix $\Omega (s,t) ({\bf 3})$ is given by
\begin{equation}
\Omega (s,t) ({\bf 3}) = \Omega (u,v) ({\bf 3}) =  e^{i \phi_v/6} \, U_{\mbox{\scriptsize{TB}}} \, R_{13} \left( -\frac{\phi_u}{2} \right) \, \left( \begin{array}{ccc}
1 & 0 & 0\\
0 & e^{- i \phi_v/2} & 0\\
0 & 0 & -i
\end{array}
\right)
\end{equation}
with
\begin{equation}
\label{defphiuphiv}
\phi_u=\frac{\pi \, u}{n} \;\; \mbox{and} \;\; \phi_v=\frac{\pi \, v}{n} \, .
\end{equation}
The form of the matrix  $\Omega (s,t) ({\bf 3^\prime})$, derived from $X (s,t) ({\bf 3^\prime})$, depends like the latter on whether $s$ and $t$ are even or odd.
The explicit form of $\Omega (s,t) ({\bf 3^\prime})$, however, does neither contain $s$ nor $t$ as parameter. For $s$ and $t$ even we can use
\begin{equation}
\Omega (s \, \mbox{even},t \, \mbox{even}) ({\bf 3^\prime}) = U_{\mbox{\scriptsize{TB}}} \, \left( \begin{array}{ccc}
1 & 0 & 0\\
0 & 1 & 0\\
0 & 0 & i
\end{array}
\right) \; ,
\end{equation}
for $s$ even and $t$ odd a possible choice is
\begin{equation}
\Omega (s \, \mbox{even},t \, \mbox{odd}) ({\bf 3^\prime}) = e^{-i \pi/4} \, U_{\mbox{\scriptsize{TB}}} \, R_{13} \left( \frac{\pi}{4} \right) \, \left(
\begin{array}{ccc}
-i & 0 & 0\\
0 & e^{- i \pi/4} & 0\\
0 & 0 & 1
\end{array}
\right) \; ,
\end{equation}
for $s$ odd and $t$ even we can choose
\begin{equation}
\Omega (s \, \mbox{odd},t \, \mbox{even}) ({\bf 3^\prime}) = U_{\mbox{\scriptsize{TB}}} \, \left( \begin{array}{ccc}
i & 0 & 0\\
0 & 1 & 0\\
0 & 0 & 1
\end{array}
\right)
\end{equation}
and for $s$ and $t$ odd we use
\begin{equation}
\Omega (s \, \mbox{odd},t \, \mbox{odd}) ({\bf 3^\prime}) = e^{- 3 \, i \, \pi/4} \, U_{\mbox{\scriptsize{TB}}} \, R_{13} \left( \frac{\pi}{4} \right) \, \left(
\begin{array}{ccc}
-i & 0 & 0\\
0 & e^{i \, \pi/4} & 0\\
0 & 0 & 1
\end{array}
\right) \; .
\end{equation}
In order to determine the planes in which the rotations $R_{ij} (\theta_L)$ and $R_{kl} (\theta_R)$, see Eq.~(\ref{eq:YDgen}), act, we consider the
combination
\begin{equation}
\label{eq:ZOmega3Case2}
\Omega(s, t) ({\bf 3})^\dagger \, Z ({\bf 3}) \,  \Omega(s, t) ({\bf 3}) = \left( \begin{array}{ccc}
-1 & 0 & 0\\
0 & 1 & 0\\
0 & 0 & -1
\end{array}
\right) \, ,
\end{equation}
meaning that, like in Case 1), the rotation associated with LH leptons in the representation ${\bf 3}$
is always in the $(13)$-plane, i.e.~$R_{13} (\theta_L)$. Similarly, we find for all possible choices of $s$ and $t$ that $\Omega (s,t)({\bf 3^\prime})$ and $Z ({\bf 3^\prime})$ for $n/2$ odd, see Eq.~(\ref{eq:Zcn2_n2odd}) in appendix~\ref{appB}, fulfil
\begin{equation}
\Omega (s,t) ({\bf 3^\prime})^\dagger \, Z ({\bf 3^\prime}) \, \Omega (s,t) ({\bf 3^\prime}) = \left( \begin{array}{ccc}
-1 & 0 & 0\\
0 & 1 & 0\\
0 & 0 & -1
\end{array}
\right) \;\; \mbox{for} \;\; n/2 \;\; \mbox{odd}.
\end{equation}
Hence, also for RH neutrinos in ${\bf 3^\prime}$ the relevant rotation is in the $(13)$-plane, namely $R_{13} (\theta_R)$.
We remind that we only consider $n/2$ odd, since it is shown in section~\ref{subsec:case1numass} that for $n$ divisible by four the form of $Z ({\bf 3^\prime})$ enforces $Y_D$ to have two vanishing eigenvalues,
not allowing for a realistic light neutrino mass spectrum without corrections.

\subsubsection{Constraints from and on light neutrino mass spectrum}
\label{subsec:case2numass}

Leaving aside the choice $n$ divisible by four, the other admitted choices of $n$ and all possible choices of the parameters $s$ and $t$, describing the CP transformation $X (s,t)$,
lead to $Y_D$ having five real parameters which can be identified with $y_f$, $f=1,2,3$, $\theta_L$ and $\theta_R$.

Next, we analyse the form of the combination in Eq.~(\ref{eq:mnumiddle}), appearing in the type-I seesaw formula, for the different choices of $s$ and $t$. For $t$ even and all possible values
of $s$ this combination turns out to be diagonal. In particular, for $s$ even it reads
\begin{equation}
\label{eq:combmnuCase2steven}
\frac{1}{M} \, \left(
\begin{array}{ccc}
y_1^2 & 0 & 0\\
0 & y_2^2 & 0\\
0 & 0 & y_3^2
\end{array}
\right) \; ,
\end{equation}
while for $s$ odd we find
\begin{equation}
\label{eq:combmnuCase2tevensodd}
\frac{1}{M} \, \left(
\begin{array}{ccc}
- y_1^2 & 0 & 0\\
0 & y_2^2 & 0\\
0 & 0 &- y_3^2
\end{array}
\right) \; .
\end{equation}
Thus, for $t$ and $s$ both even the matrix $K_\nu$ contains only $\pm 1$, whereas we have to consider
\begin{equation}
\label{eq:tevensoddCase2Knu}
K_\nu = \left( \begin{array}{ccc}
\pm i & 0 & 0\\
0 & \pm 1 & 0\\
0 & 0 & \pm i
\end{array}
\right) \;\; \mbox{for $t$ even and $s$ odd} \, .
\end{equation}
The three light neutrino masses $m_f$ are given by
\begin{equation}
\label{eq:tevenCase2numasses}
m_f = \frac{y_f^2 \, \langle H \rangle^2}{M} \;\; \mbox{for} \;\; f=1,2,3 \; .
\end{equation}
As one can see, for these choices of $s$ and $t$ no assumptions on the light neutrino mass spectrum have to be made in order to achieve a diagonal form of the combination
in Eq.~(\ref{eq:mnumiddle}).
Instead for $t$ odd the combination contains off-diagonal elements, since the $(13)$- and $(31)$-elements are proportional to $y_1$, $y_3$ and $\cos 2 \, \theta_R$.
Like for Case 1), the light neutrino mass $m_2$ is always directly related to $y_2$ via the relation in Eq.~(\ref{eq:m2Case1}), while the masses $m_1$ and $m_3$ are determined by the couplings $y_1$ and $y_3$ and the angle $\theta_R$. The combination in Eq.~(\ref{eq:mnumiddle}) becomes diagonal in the following three occasions
\begin{itemize}
\item[$(i)$] $y_1=0$ which corresponds to strong NO with
\begin{equation}
\label{eq:strongNOCase2}
m_1=0 \;\; \mbox{and} \;\; m_3=\frac{y_3^2 \, \langle H \rangle^2}{M}  \, |\sin 2 \, \theta_R|
\end{equation}
in addition to $m_2$ in Eq.~(\ref{eq:m2Case1}). For a realistic light neutrino mass
spectrum $m_3$ has to be non-vanishing and thus $\sin 2 \, \theta_R \neq 0$.
The matrix $K_\nu$ is of the form
\begin{equation}
\label{eq:strongNOCase2Knu}
K_\nu = \left( \begin{array}{ccc}
1 & 0 & 0\\
0 & \pm i & 0\\
0 & 0 & \pm i^{(s \, \scriptsize \mbox{mod} \normalsize \, 2) + s_R + 1}
\end{array}
\right)
\;\; \mbox{(strong NO)}
\end{equation}
with $s_R=0$ for $\sin 2 \, \theta_R$ positive and $s_R=1$ for negative $\sin 2 \, \theta_R$.
Since $m_1$ vanishes, the $(11)$-element of $K_\nu$ is not constrained and set for concreteness
to $+1$.
\item[$(ii)$] $y_3=0$ corresponding to strong IO with
\begin{equation}
\label{eq:strongIOCase2}
m_1=\frac{y_1^2 \, \langle H \rangle^2}{M}  \, |\sin 2 \, \theta_R| \;\; \mbox{and} \;\; m_3=0
\end{equation}
and $m_2$ given in Eq.~(\ref{eq:m2Case1}). For a realistic light neutrino mass
spectrum $m_1$ has to be non-vanishing and thus $\sin 2\, \theta_R \neq 0$.
The matrix $K_\nu$ is of similar form as in Eq.~(\ref{eq:strongNOCase2Knu}) with the roles of the $(11)$- and $(33)$-elements
exchanged,
\begin{equation}
\label{eq:strongIOCase2Knu}
K_\nu = \left( \begin{array}{ccc}
\pm i^{(s \, \scriptsize \mbox{mod} \normalsize \, 2) + s_R} & 0 & 0\\
0 & \pm i & 0\\
0 & 0 & 1
\end{array}
\right)
\;\; \mbox{(strong IO)} \, .
\end{equation}
\item[$(iii)$] $\cos 2 \, \theta_R=0$ where no constraints on the light neutrino masses arise. We do not consider this possibility, as we would like to explore
the effect of the parameter $\theta_R$ on the various phenomenological aspects of the scenario.
\end{itemize}

\subsubsection{PMNS mixing matrix}
\label{subsec:case2mixing}

If $t$ is even or if $t$ is odd and light neutrino masses follow strong NO or strong IO or $\cos 2 \, \theta_R=0$,
the resulting PMNS mixing matrix reads
\begin{equation}
\label{eq:PMNSmixingCase2}
U_{\mbox{\scriptsize{PMNS}}} = \Omega (u,v) ({\bf 3}) \, R_{13} (\theta_L) \, K_\nu \, .
\end{equation}
For $t$ odd and all three light neutrinos being massive, the form of the PMNS mixing matrix is still the same. However, the angle $\theta_L$ has to be replaced by the effective angle $\widetilde{\theta}_L$ that depends on $\theta_L$, the couplings $y_1$ and $y_3$ and $\theta_R$, as described for Case 1) in section~\ref{subsec:case1mixing}.

\subsection{Case 3 a) and Case 3 b.1)}
\label{sec33}

\subsubsection{Residual symmetries}
\label{subsec:case3ressymm}

In Case 3 a) and Case 3 b.1) the residual $Z_2$ symmetry in
the neutrino sector is generated by
\begin{equation}
Z= b \, c^m d^m \;\; \mbox{with} \;\; m=0, 1, ..., n-1 \, .
\end{equation}
Since $Z$ involves the generator $b$, Case 3 a) and Case 3 b.1)  can only be achieved with the help of the flavour symmetries $\Delta (6 \, n^2)$.
We have in general $n$ different choices for the generator $Z$. However, as  shown in~\cite{Hagedorn:2014wha}, preferred values of $m$ are either around
$m\approx 0$ and $m\approx n$ for Case 3 a) or $m\approx n/2$ for Case 3 b.1), as long as the charged lepton masses are not permuted.
The form of $Z$ in the representations ${\bf 3}$ and ${\bf 3^\prime}$, $Z ({\bf 3})$ and $Z ({\bf 3^\prime})$, can be found in appendix~\ref{appB}, see Eqs.~(\ref{eq:Case3Zm3}-\ref{Z3pmodd}).

The CP symmetry, used in Case 3 a) and Case 3 b.1), is induced by the automorphism, shown in Eq.~(\ref{eq:XP23auto}) in appendix~\ref{appC}, conjugated with the
inner one, represented by the group transformation $h= b \, c^s \, d^{n-s}$, $s=0, 1, ...,n-1$. The corresponding CP transformation $X (s)$ in ${\bf 3}$ and ${\bf 3^\prime}$
is given by
\begin{equation}
X (s) ({\bf 3}) = b ({\bf 3}) \, c ({\bf 3})^s \, d ({\bf 3})^{n-s} \, X_0 ({\bf 3})
\end{equation}
and
\begin{equation}
X (s) ({\bf 3^\prime}) =b ({\bf 3^\prime}) \, c ({\bf 3^\prime})^s \, d ({\bf 3^\prime})^{n-s} \, X_0 ({\bf 3^\prime}) \, ,
\end{equation}
respectively.
The explicit forms of $X (s) ({\bf 3})$ and $X (s) ({\bf 3^\prime})$ can be found in appendix~\ref{appC}, see Eqs.~(\ref{eq:Case3Xs3}-\ref{eq:Case3Xsodd3prime}).

The form of the matrix $\Omega (s, m) ({\bf 3})$, derived from $X  (s) ({\bf 3})$ in Eq.~(\ref{eq:Case3Xs3}) in appendix~\ref{appC}, can be chosen as~\cite{Hagedorn:2014wha}
\begin{equation}
\Omega (s, m) ({\bf 3}) =e^{i \, \phi_s} \,  \left( \begin{array}{ccc}
1 & 0 & 0\\
0 & \omega & 0 \\
0 & 0 & \omega^2
\end{array}
\right) \, U_{\mbox{\scriptsize{TB}}} \,
\left( \begin{array}{ccc}
1 & 0 & 0\\
0 & e^{-3 \, i \, \phi_s} & 0 \\
0 & 0 & -1
\end{array}
\right) \, R_{13} \left( \phi_m \right)
\end{equation}
with
\begin{equation}\label{phi s and phi m def}
\phi_s=\frac{\pi s}{n} \;\;\; \mbox{and} \;\;\; \phi_m=\frac{\pi \, m}{n} \, .
\end{equation}
The form of the matrix $\Omega (s) ({\bf 3^\prime})$ only depends on whether $s$ is even or odd and is independent of the choice of the parameter $m$.
In particular, we can use for $s$ even
\begin{equation}
\Omega (s \, \mbox{even}) ({\bf 3^\prime}) = \left(
\begin{array}{ccc}
1 & 0 & 0\\
0 & \omega & 0\\
0 & 0 & \omega^2
\end{array}
\right) \, U_{\mbox{\scriptsize{TB}}} \, \left(
\begin{array}{ccc}
1 & 0 & 0\\
0 & 1 & 0\\
0 & 0 & -1
\end{array}
\right)
\end{equation}
and for $s$ odd
\begin{equation}
\Omega (s \, \mbox{odd}) ({\bf 3^\prime}) = \left(
\begin{array}{ccc}
1 & 0 & 0\\
0 & \omega & 0\\
0 & 0 & \omega^2
\end{array}
\right) \, U_{\mbox{\scriptsize{TB}}} \, \left(
\begin{array}{ccc}
i & 0 & 0\\
0  & -1 & 0\\
0 & 0 & -i
\end{array}
\right) \; .
\end{equation}
We note that the form of $\Omega (s \, \mbox{even}) ({\bf 3^\prime})$ coincides with
$\Omega (s, m) ({\bf 3})$ for the special choices $s=0$ and $m=0$ as well as that
$\Omega (s \, \mbox{odd}) ({\bf 3^\prime})$ coincides with
a special form of $\Omega (s, m) ({\bf 3})$, namely for $s=n/2$ and $m=0$. Since $n$ is always even, the choice $s=n/2$
necessarily corresponds to an integer as well.

For all choices of $n$, $m$ and $s$ the form of the matrix $Z (m) ({\bf 3})$, see Eq.~(\ref{eq:Case3Zm3}) in appendix~\ref{appB}, in the basis transformed with $\Omega (s, m) ({\bf 3})$ reads
\begin{equation}
\Omega (s, m) ({\bf 3})^\dagger \, Z (m) ({\bf 3}) \, \Omega (s, m) ({\bf 3}) =
\left( \begin{array}{ccc}
1 & 0 & 0\\
0 & 1 & 0\\
0 & 0 & -1
\end{array}
\right) \; ,
\end{equation}
meaning that the rotation associated with LH leptons in the representation ${\bf 3}$ is always $R_{12} (\theta_L)$
in the $(12)$-plane. In order to find the exact form of $R_{kl} (\theta_R)$, we have to separately consider
the choices $m$ even and $m$ odd. We find for $m$ even
\begin{equation}
\label{eq:Z3pmevenrotOmegaCase3}
\Omega  (s) ({\bf 3^\prime})^\dagger \, Z (m \, \mbox{even}) ({\bf 3^\prime}) \, \Omega  (s) ({\bf 3^\prime}) =
\left(
\begin{array}{ccc}
1 & 0 & 0\\
0 & 1 & 0\\
0 & 0 & -1
\end{array}
\right) \, ,
\end{equation}
i.e.~the rotation associated with the RH neutrinos
in ${\bf 3^\prime}$ is $R_{12} (\theta_R)$. For $m$ odd we have instead
\begin{equation}
\label{eq:Z3pmoddrotOmegaCase3}
\Omega  (s) ({\bf 3^\prime})^\dagger \, Z (m \, \mbox{odd}) ({\bf 3^\prime}) \, \Omega  (s) ({\bf 3^\prime}) =
\left(
\begin{array}{ccc}
-1 & 0 & 0\\
0 & 1 & 0\\
0 & 0 & 1
\end{array}
\right)
\end{equation}
corresponding to the rotation $R_{23} (\theta_R)$ associated with the RH neutrinos. This rotation acts thus in a different plane than the one of LH leptons and hence makes the presence of the
permutation matrix $P^{ij}_{kl}$, see Eq.~(\ref{eq:YDgen}), necessary, that has to act in the $(13)$-plane, i.e.
\begin{equation}
\label{eq:permutation}
P^{12}_{23}=P_{13}= \left(
\begin{array}{ccc}
0 & 0 & 1\\
0 & 1 & 0\\
1 & 0 & 0
\end{array}
\right) \, .
\end{equation}
Note that Eqs.~(\ref{eq:Z3pmevenrotOmegaCase3}) and (\ref{eq:Z3pmoddrotOmegaCase3}) hold independently of the choice of $s$.

\subsubsection{Light neutrino mass spectrum and PMNS mixing matrix for Case 3 a)}
\label{subsec:case3numass}

In contrast to Case 1) and Case 2) there are no constraints from the light neutrino mass spectrum on the choice of the residual symmetries and thus the
parameters $n$, $m$ and $s$.

In the following, we first focus on Case 3 a) and outline the changes to be made for Case 3 b.1) afterwards.
Like for the other cases, we investigate the combination in Eq.~(\ref{eq:mnumiddle}) for the different choices of $m$ and $s$.
For $m$ and $s$ either both even or both odd we see that this combination is diagonal and, in particular, reads for both combinations
\begin{equation}
\label{eq:combmnuCase3amsevenodd}
\frac{1}{M} \, \left(
\begin{array}{ccc}
y_1^2 & 0 & 0\\
0 & y_2^2 & 0\\
0 & 0 & -y_3^2
\end{array}
\right) \, .
\end{equation}
Thus, in both these cases we have
\begin{equation}
m_f = \frac{y_f^2 \, \langle H \rangle^2}{M} \;\; \mbox{for} \;\;  f=1,2,3
\end{equation}
for Case 3 a) and the matrix $K_\nu$ is
\begin{equation}
\label{eq:msevenoddCase3aKnu}
K_\nu = \left( \begin{array}{ccc}
\pm 1 & 0 & 0\\
0 & \pm 1 & 0\\
0 & 0 & \pm i
\end{array}
\right) \, .
\end{equation}
For the choices $m$ even and $s$ odd as well as $m$ odd and $s$ even the combination in Eq.~(\ref{eq:mnumiddle}) is not diagonal
in general. For $m$ even and $s$ odd we have
\begin{equation}
\label{eq:Case3matmevensodd}
\frac{1}{M} \, \left(
\begin{array}{ccc}
-y_1^2 \, \cos 2\,\theta_R & - y_1 \, y_2 \, \sin 2 \, \theta_R & 0\\
- y_1 \, y_2 \, \sin 2 \, \theta_R & y_2^2 \, \cos 2 \, \theta_R & 0\\
0 & 0 & y_3^2
\end{array}
\right) \, ,
\end{equation}
meaning that only the light neutrino mass $m_3$ is directly related to a coupling
\begin{equation}
\label{eq:m3Case3a}
m_3 =\frac{y_3^2 \, \langle H \rangle^2}{M} \, .
\end{equation}
Like for Case 1) and Case 2),
we consider situations in which the off-diagonal elements in the matrix in Eq.~(\ref{eq:Case3matmevensodd}) are zero, i.e.~
\begin{itemize}
\item[$(i)$] $y_1=0$ which corresponds to strong NO with
\begin{equation}
\label{eq:strongNOmevensoddCase3a}
m_1=0 \;\; \mbox{and} \;\; m_2=\frac{y_2^2 \, \langle H \rangle^2}{M}  \, |\cos 2 \, \theta_R|
\end{equation}
in addition to $m_3$ in Eq.~(\ref{eq:m3Case3a}). For a realistic light neutrino mass
spectrum $m_2$ has to be non-vanishing and thus $\cos 2 \theta_R \neq 0$.
The matrix $K_\nu$ is of the form
\begin{equation}
	\label{eq:strongNOmevensoddCase3aKnu}
	K_\nu = \left( \begin{array}{ccc}
			1 & 0 & 0\\
			0 & \pm i^{c_R} & 0\\
			0 & 0 & \pm 1
		\end{array}
	\right)
	\;\; \mbox{(strong NO)}
\end{equation}
with $c_R=0$ for $\cos 2 \, \theta_R$ positive and $c_R=1$ for negative $\cos 2\, \theta_R$.
Since $m_1$ vanishes, the $(11)$-element of $K_\nu$ is not constrained and set to $+1$ for concreteness.
\item[$(ii)$] $\sin 2 \, \theta_R=0$ where no constraints on the light neutrino masses arise. This possibility, however,
we do not consider, since we wish not to constrain the value of $\theta_R$.
\end{itemize}
Note that in contrast to Case 1) and Case 2) for the choice $m$ even and $s$ odd for Case 3 a) strong IO would not
lead to a diagonal form for the combination in Eq.~(\ref{eq:mnumiddle}) and that setting $y_2=0$ is not admitted, since it is experimentally
known that the light neutrino mass $m_2$ cannot vanish.
For $m$ odd and $s$ even the combination in Eq.~(\ref{eq:mnumiddle}) looks very similar to the one in Eq.~(\ref{eq:Case3matmevensodd}), i.e.
\begin{equation}
\label{eq:Case3matmoddseven}
\frac{1}{M} \, \left(
\begin{array}{ccc}
-y_1^2 \, \cos 2\,\theta_R & y_1 \, y_2 \, \sin 2 \, \theta_R & 0\\
y_1 \, y_2 \, \sin 2 \, \theta_R & y_2^2 \, \cos 2 \, \theta_R & 0\\
0 & 0 & y_3^2
\end{array}
\right) \, .
\end{equation}
The discussion of light neutrino masses and the mass ordering, leading to a diagonal form of the matrix in Eq.~(\ref{eq:Case3matmoddseven}), is thus the
same as for the combination $m$ even and $s$ odd.

If one of the parameters, $m$ and $s$, is even and the other one odd, three massive light neutrinos lead to $m_1$ and $m_2$ depending on both the couplings $y_1$ and $y_2$ as well as the angle $\theta_R$ and to an additional rotation in the $(12)$-plane through an angle, determined by these three quantities, similar to Case 1) and Case 2). This additional rotation acts in the same plane as $R_{12} (\theta_L)$ which appears in the PMNS mixing matrix. Thus, taking into account this additional angle only results in the replacement of $\theta_L$ by an effective angle $\widetilde{\theta}_L$ that is the sum of $\theta_L$ and the additional angle.
The results presented so far lead to lepton mixing corresponding to Case 3 a), i.e.
\begin{equation}
\label{eq:PMNSmixingCase3a}
U_{\mbox{\scriptsize{PMNS}}} = \Omega (s, m) ({\bf 3}) \, R_{12} (\theta_L) \, K_\nu
\end{equation}
and where necessary with $\theta_L$ being read as $\widetilde{\theta}_L$.

\subsubsection{Light neutrino mass spectrum and PMNS mixing matrix for Case 3 b.1)}
\label{subsec:case3numass2}

Results for Case 3 b.1) can be achieved in the simplest way by a different assignment
of the light neutrino masses to the couplings $y_f$, $f=1,2,3$, as the form of the PMNS mixing matrix for Case 3 b.1) is
\begin{equation}
\label{eq:PMNSmixingCase3b1}
U_{\mbox{\scriptsize{PMNS}}} = \Omega (s, m) ({\bf 3}) \, R_{12} (\theta_L) \, P \, K_\nu
\end{equation}
where
\begin{equation}
\label{eq:permutationP}
P= \left( \begin{array}{ccc}
0 & 1 & 0\\
0 & 0 & 1\\
1 & 0 & 0
\end{array}
\right) \, ,
\end{equation}
as discussed in~\cite{Hagedorn:2014wha}.
Concretely, a mixing pattern belonging to Case 3 b.1) can be achieved by assigning
\begin{equation}
\label{eq:numassespermmsevenoroddCase3b1}
m_1 =\frac{y_3^2 \, \langle H \rangle^2}{M} \; , \;\;  m_2 = \frac{y_1^2 \, \langle H \rangle^2}{M}  \;\;\; \mbox{and} \;\;\; m_3 = \frac{y_2^2 \, \langle H \rangle^2}{M}
\end{equation}
for the choices $m$ and $s$ either both even or both odd. The matrix $K_\nu$ is then given by $P^T \, K_\nu \, P$ with $K_\nu$ as in Eq.~(\ref{eq:msevenoddCase3aKnu}).
For the choice $m$ even and $s$ odd as well as for $m$ odd and $s$ even we find
that setting the off-diagonal elements in the matrix in Eq.~(\ref{eq:mnumiddle}) to zero corresponds to strong IO
for $y_2=0$ and thus
\begin{equation}
\label{eq:strongNOmevensoddCase3b1}
m_2=\frac{y_1^2 \, \langle H \rangle^2}{M}  \, |\cos 2 \, \theta_R|   \;\; \mbox{and} \;\; m_3=0
\end{equation}
with
\begin{equation}
\label{eq:m3Case3b1}
m_1 =\frac{y_3^2 \, \langle H \rangle^2}{M} \, ,
\end{equation}
when the permutation necessary for a mixing pattern of Case 3 b.1) is taken into account.
The matrix $K_\nu$ then is
\begin{equation}
	\label{eq:strongIOmevensoddCase3b1Knu}
	K_\nu = \left( \begin{array}{ccc}
			\pm 1 & 0 & 0\\
			0 & \pm i^{c_R+1} & 0\\
			0 & 0 & 1
		\end{array}
	\right)
	\;\; \mbox{(strong IO)}
\end{equation}
with $+1$ assigned to the vanishing light neutrino mass.
As we have already commented, we do not consider special choices of $\theta_R$ in order to obtain a diagonal matrix
from the one in Eq.~(\ref{eq:mnumiddle}) and we thus omit this possibility.

It is clear that for $m$ and $s$ both even or both odd all light neutrinos can be massive without changing the results for the PMNS
mixing matrix, shown in Eq.~(\ref{eq:PMNSmixingCase3b1}), whereas for either $m$ or $s$ being even
with the other one being odd we have to proceed as for Case 3 a) and replace $\theta_L$ by $\widetilde{\theta}_L$ that is  the sum of $\theta_L$ and the additional angle, depending on the couplings $y_1$ and $y_2$ and the angle $\theta_R$.

Since we consider an example with $m$ even and $s$ odd for Case 3 b.1) in the numerical analysis in section~\ref{sec:LowScalelepto},
we mention explicit formulae for the light neutrino masses $m_i$ and the additional angle for this case. The masses read
\begin{equation}
m_1 = \frac{y_3^2 \, \langle H \rangle^2}{M} \;\; \mbox{and} \;\; m_{2,3} = \frac{\langle H \rangle^2}{2 \, M} \, \Big| (y_1^2-y_2^2) \, \cos 2 \, \theta_R \pm \sqrt{4 \, y_1^2 \, y_2^2 + (y_1^2-y_2^2)^2 \, \cos^2 2 \, \theta_R} \Big|
\end{equation}
and the tangent of twice the additional angle (expressed in terms of $\theta_L$ and $\widetilde{\theta}_L$) is given by
\begin{equation}
\tan 2 \, (\widetilde{\theta}_L - \theta_L ) = - \frac{2 \, y_1 \, y_2}{y_1^2+y_2^2} \, \tan 2 \, \theta_R \, .
\end{equation}

\subsection{Stability of light neutrino masses and PMNS mixing matrix with respect to RH neutrino mass splitting}
\label{subsec:splittingcorr}

Throughout this work, we assume that the splitting $\kappa$ (and $\lambda$) has a negligible impact on light neutrino masses and the PMNS mixing matrix.
While this is certainly true for very small mixing between LH and RH neutrinos, i.e.~at the seesaw line when $y_f \sim \sqrt{M \, m_f}/\langle H \rangle$ with $m_f$ being the light neutrino mass, compare Eq.~(\ref{eq:mnu}) and e.g.~Eq.~(\ref{eq:m2Case1}) for Case 1), close to special values of the angle $\theta_R$, e.g.~$\cos 2 \theta_R \approx 0$ in Case 1), see Eqs.~(\ref{eq:strongNOCase1}) and (\ref{eq:strongIOCase1}), at least one of the
couplings $y_f$ may become large. The smallness of the light neutrino masses  then results from a cancellation between the couplings $y_f$ and the trigonometric functions multiplying them, e.g.~$m_3 \propto y_3^2 \, |\cos 2 \theta_R|$ in Eq.~\eqref{eq:strongNOCase1}.
This cancellation can be disturbed by other small parameters, such as the splitting $\kappa$ among the RH neutrino masses, see Eq.(\ref{eq:RHmassspectrum}).
In fact, if we include the leading correction to $m_3$, which is at order $\mathcal{O}(\kappa)$, we find for strong NO in Case 1)
\begin{align}
	m_3 \approx \frac{y_3^2 \, \langle H \rangle^2}{M} |\kappa + \cos 2 \theta_R|\,.
\end{align}
To keep the corrections due to the splitting $\kappa$ small, we should therefore impose
\begin{align}
\label{eq:kappacorrectionscriterion}
	\kappa \ll | \cos 2\theta_R |\,.
\end{align}
Alternatively, we could consider an inverse seesaw-like scenario~\cite{Wyler:1982dd,Mohapatra:1986aw,Mohapatra:1986bd,Bernabeu:1987gr}, in which the corrections due to $\kappa$
give the dominant contribution to the light neutrino masses, for $|\cos 2 \theta_R|  \ll \kappa$.
This option would also lead to a (slight) modification of the PMNS mixing matrix, since $\kappa$ gives rise to off-diagonal contributions in Eq.~\eqref{eq:mnumiddle} that induce not only a rotation in the (13)-plane for Case 1). In the current analysis, we primarily focus on the case where $\kappa \ll | \cos 2\theta_R |$, and thus neglect contributions to the light neutrino masses and to the PMNS mixing matrix from the splitting $\kappa$ (and $\lambda$).

Likewise, we assume corrections, arising from renormalisation group (RG) running, to both light neutrino masses and the PMNS mixing matrix to be small.

\section{Lepton mixing data}
\label{sec50}

In the following, we present examples of group theory parameters, e.g.~the index $n$, for each of the cases, Case 1) through Case 3 b.1), that lead to a reasonable agreement with the global fit data on lepton mixing angles and light neutrino mass
splittings, provided by the NuFIT collaboration (NuFIT 5.1, October 2021, without SK atmospheric data)~\cite{Esteban:2020cvm}.\footnote{See \href{http://www.nu-fit.org/}{http://www.nu-fit.org/}.}
Note that we do not include any information from the global fit on the value of the CP phase $\delta$ in this  $\Delta\chi^2$-analysis, but instead separately comment on the impact of including this information for the different cases.
All presented examples are taken from the study performed in~\cite{Hagedorn:2014wha}.

As discussed in section~\ref{sec3}, in several occasions the angle $\theta_L$ becomes replaced by an angle $\widetilde{\theta}_L$ which is the sum of $\theta_L$ and an additional angle, depending on the couplings $y_f$, $f=1,2,3$, and the angle $\theta_R$, if the lightest neutrino mass $m_0$ is non-zero. In the following, we, nevertheless, always refer to the angle, appearing in the PMNS mixing matrix $U_{\mbox{\scriptsize{PMNS}}}$, as $\theta_L$ and comment briefly on the instances in which this has to be read as $\widetilde{\theta}_L$.

\subsection{Case 1)}
\label{sec501}

As the analysis in~\cite{Hagedorn:2014wha} has shown, the three lepton mixing angles only depend on the angle $\theta_L$ and are independent of the  index $n$ as well as of the chosen CP transformation $X (s)$.
For
\begin{equation}
\label{eq:thetaLCase1}
\theta_L \approx 0.183 ~(0.184) \,
\end{equation}
all three lepton mixing angles can be accommodated at the $3 \, \sigma$ level or better for light neutrino masses with NO (IO), i.e.~
\begin{equation}
\label{eq:mixinganglesCase1}
\sin^2 \theta_{13}\approx 0.0220 ~(0.0222) \;\; , \;\; \sin^2 \theta_{12}\approx 0.341 \;\; , \;\; \sin^2 \theta_{23} \approx 0.605 ~(0.606)
\end{equation}
which corresponds to a $\Delta\chi^2$-value of $\Delta \chi^2\approx 11.9$  for light neutrino masses with NO and of $\Delta \chi^2\approx 11.2$  for a light neutrino mass spectrum with IO.\footnote{When computing the $\Delta \chi^2$-value for IO, we subtract the value $\Delta \chi^2=\chi^2_\mathrm{IO}-\chi^2_\mathrm{NO}=2.6$ which corresponds to the overall $\Delta \chi^2$ for IO with respect to NO, that is favoured
by current global fit data~\cite{Esteban:2020cvm}.}
We note that $\sin^2 \theta_{12}$ is bounded to be larger than $1/3$ for Case 1) and  that the atmospheric mixing angle takes a value not close to maximal mixing.

Two of the three leptonic CP phases, the Dirac phase $\delta$ and the Majorana phase $\beta$, are in general trivial.
A trivial value of $\delta$, $\delta=0$ or $\delta=\pi$, is compatible at the $3 \, \sigma$ level and at the $1 \, \sigma$ level with the global fit data~\cite{Esteban:2020cvm}, respectively, for light neutrino masses with NO, while for IO only $\delta=0$ remains compatible with the data at the $3 \, \sigma$ level. The only non-trivial CP phase is the Majorana phase $\alpha$ that is fixed by $X (s)$.\footnote{When commenting on Majorana phases, we always refer to the general case in which  none of the light neutrinos is massless and thus both Majorana phases, $\alpha$ and $\beta$, are physical, as done in the analysis in~\cite{Hagedorn:2014wha}.} The magnitude of its sine is given by the magnitude of $\sin 6 \, \phi_s$.
For analytic formulae of the lepton mixing parameters we refer to~\cite{Hagedorn:2014wha}.

We remind that the numerical values of the angle $\theta_L$ given in Eq.~(\ref{eq:thetaLCase1})
are obtained under the assumption of strong NO and strong IO, respectively. If the lightest neutrino mass $m_0$ is non-zero, this angle has to be read as
$\widetilde{\theta}_L$. From the latter, we can then determine $\theta_L$ for fixed values of the couplings $y_1$ and $y_3$ and of the angle $\theta_R$, see also comments in section~\ref{subsec:case1mixing}.

The smallest value of the index $n$ fulfilling the constraints $n > 2$ even, $3 \nmid n$ and $4 \nmid n$ is $n=10$. For this choice, the parameter $s$ varies as $0 \leq s \leq n-1=9$. All numerical results for low-scale leptogenesis are presented for this value of $n$, see section~\ref{sec53}.

\subsection{Case 2)}
\label{sec502}

As discussed in~\cite{Hagedorn:2014wha}, the lepton mixing angles  depend on the ratio $u/n$ (the parameter $\phi_u$) and the angle $\theta_L$. It turns out that small values of $u/n$ ($\phi_u$) are required, up to symmetry transformations shown in~\cite{Hagedorn:2014wha},\footnote{In the present analysis we only refer to situations in which there is no contribution from the charged lepton sector to the PMNS mixing matrix, i.e.~in which the charged
lepton masses are already canonically ordered.} in order to accommodate the smallness of the reactor mixing angle $\theta_{13}$
\begin{equation}
\label{eq:constraintunCase2}
-0.1 \lesssim u/n \lesssim 0.12 \;\; \mbox{corresponding to} \;\; -0.31 \lesssim \phi_u \lesssim 0.37 \, .
\end{equation}
The value of $\theta_L$ should be close to $\theta_L \approx 0$ or $\theta_L \approx \pi$. As explained in sections~\ref{subsec:case2numass} and~\ref{subsec:case2mixing}, for $t$ odd and the light neutrino mass $m_0$ being non-zero we have to replace $\theta_L$ by $\widetilde{\theta}_L$ and have to compute from the latter $\theta_L$ for fixed values of the couplings $y_1$ and $y_3$ and of the angle $\theta_R$. Like for Case 1), $\sin^2 \theta_{12}$ is always bounded from below by $1/3$.
The explicit formulae for the lepton mixing angles and CP invariants can be found in~\cite{Hagedorn:2014wha}. Here we only recall that the Dirac phase $\delta$ and the Majorana phase $\beta$
depend on $\phi_u$, while the Majorana phase $\alpha$ is (mainly) determined by $\phi_v$ (and thus the choice of $t$, see Eqs.~(\ref{defuv}) and (\ref{defphiuphiv})).
Indeed, it turns out that the magnitude of the sine of $\alpha$ is given by the magnitude of $\sin \phi_v$.
Detailed numerical results, i.e.~tables with examples of $n$ and $u$ as well as $v$ and $\theta_L$ along with further explanations, can be
found in~\cite{Hagedorn:2014wha}.

\begin{table}
\begin{center}
\begin{tabular}{|c||c|c|c|}
\hline
$u$ & $u=-1$ & $u=0$ & $u=+1$\\
\hline
$\theta_L$ & $\ba 0.146\\(0.148)\ea$   & $0.184$ & $\ba 0.146\\ (0.148) \ea$ \\[0.01in]
\hline
$\sin^2 \theta_{12}$  & $0.341$ & $0.341$ & $0.341$\\[0.01in]
$\sin^2 \theta_{13}$ & $\ba 0.0222\\ (0.0224) \ea$ & $\ba 0.0222 \\ (0.0224) \ea$ & $\ba 0.0222\\ (0.0224) \ea$  \\[0.01in]
$\sin^2 \theta_{23}$ & $0.437$ & $0.5$ & $0.563$\\[0.01in]
\hline
$\Delta \chi^2$  & $\ba 9.25\\ (11.2)\ea$ & $\ba 10.8 \\(12.5) \ea$ & $\ba 8.27 \\(8.62) \ea$ \\[0.01in]
\hline
\end{tabular}
\caption{{\small {\bf Case 2)} Results of $\theta_L$ and lepton mixing angles
for $n=14$ and the admitted values of the parameter $u$. Note that other combinations of $u$ and $\theta_L$ generated with symmetry transformations, discussed in~\cite{Hagedorn:2014wha}, as well as
considering $\pi-\theta_L$ also lead to a viable description of the experimental data on lepton mixing angles. We additionally show the value of $\Delta \chi^2$, arising from a fit to the global fit data, provided by the NuFIT collaboration~\cite{Esteban:2020cvm}. Results in brackets refer to light neutrino masses with IO.
}}
\label{tab:Case2n14}
\end{center}
\end{table}

In the following, we focus on the choice $n=14$ for the index, fulfilling all constraints on $n$
($n$ even, $3 \nmid n$ and $4 \nmid n$). For this value of $n$, Eq.~(\ref{eq:constraintunCase2}) is fulfilled by $u=0$ and $u=\pm 1$.
Since we distinguish the different sub-cases according to whether $s$ and $t$ are even or odd, we list the combinations of $s$ and $t$ which
correspond to the values $u=0$ and $u=\pm 1$ for $n=14$. For $u=0$ all admitted values of $t$ are even, i.e.
\begin{equation}
(s, t) = (0,0), \, (1,2), \, (2,4), \, (3,6), \, (4,8), \, (5,10), \, (6,12) \, ,
\end{equation}
while for $u=\pm 1$ we only have odd values of $t$, i.e.
\begin{eqnarray}
\nonumber
u=-1 && (s, t) = (0,1), \, (1,3), \, (2,5), \, (3,7), \, (4,9), \,  (5,11), \, (6,13)\\
\nonumber
u=+1  &&  (s, t) = (1,1), \, (2,3), \, (3,5), \, (4,7), \, (5,9) , \, (6,11), \, (7,13) \, .
\end{eqnarray}
The values of $\theta_L$ (or $\widetilde{\theta}_L$ for $t$ odd and $m_0$ non-zero), leading to results of the lepton mixing angles in agreement with the global fit data~\cite{Esteban:2020cvm} at the $3 \, \sigma$ level or better, are given in Tab.~\ref{tab:Case2n14}.
These choices of $n$ and $u$ as well as combinations of $s$ and $t$ are mainly used in the numerical analysis of low-scale leptogenesis, see section~\ref{sec54}.

\begin{figure}
\centering
\includegraphics[width=0.9\textwidth]{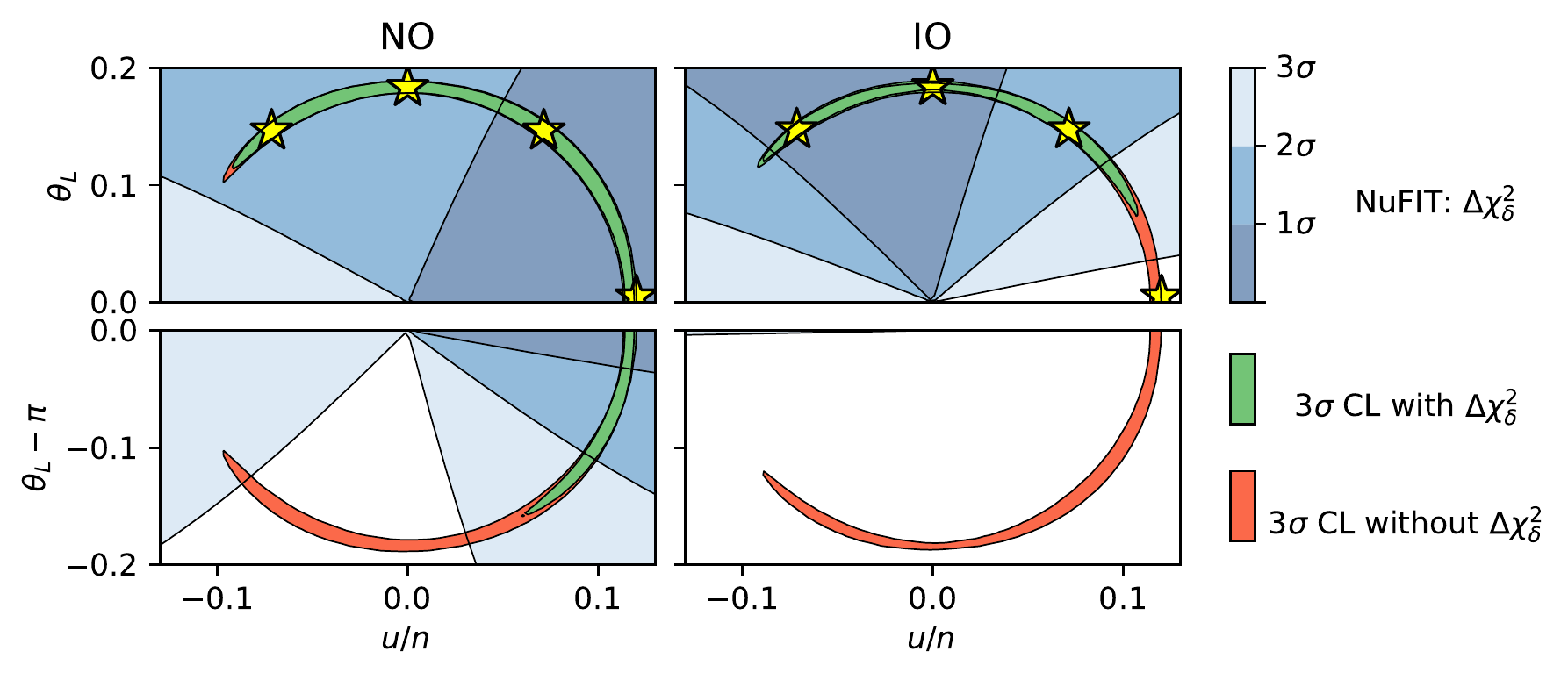}
\caption{{\small {\bf Case 2)} Impact of constraints on the value of the CP phase $\delta$ for light neutrino masses with NO (left plot) and IO (right plot), respectively. We show the areas, leading to an agreement with the global fit data at the $3 \, \sigma$ level or better,
in the $u/n-\theta_L$-plane.
The orange area corresponds to the case in which only the information on the three lepton mixing angles is taken into account.
In different shades of blue we display the range of the CP phase $\delta$ at the $1\, \sigma$ (dark blue), $2 \, \sigma$ (blue) and $3 \, \sigma$ (light blue) level, as preferred by the global fit data. As one can see,
this constraint disfavours certain areas in the $u/n-\theta_L$-plane.
When combining it with the information on the lepton mixing angles,
the green area results as favoured at the $3 \, \sigma$ level. The constraints on the CP phase $\delta$ have a moderate impact for light neutrino masses with NO, but substantially reduce the favoured areas in the case of IO.
The specific choices $n=14$, $u=0$ and $u=\pm 1$ and their corresponding values of $\theta_L$ for $\theta_L$ small are indicated by yellow stars. Furthermore, we highlight the choice $u/n=0.12$ and $\theta_L=0$ with a yellow star.
}}
\label{fig:case2NOIOchi2}
\end{figure}

As commented, the Dirac phase $\delta$ only depends on $\phi_u$ (and $\theta_L$) like the lepton mixing angles. We can thus visualise the parameter space, preferred at the $3 \, \sigma$ level by the global fit data on the lepton mixing angles, in the $u/n-\theta_L$-plane and impose the additional constraints, arising from the CP phase $\delta$, in this plane. This is shown in Fig.~\ref{fig:case2NOIOchi2}.
In this plot the choices $n=14$, $u=0$ and $u=\pm 1$ for $\theta_L$ being small
are highlighted by yellow stars.
They correspond to the following values of the CP phase $\delta$: $\sin\delta=-1$ for $u=0$ and $\theta_L \approx 0.184$ and $\sin\delta\approx -0.811 \, (-0.813)$ for $u=\pm 1$ and $\theta_L \approx 0.146 \, (0.148)$. The values in brackets are for light neutrino masses with IO.

\subsection{Case 3 a) and Case 3 b.1)}
\label{sec503}

We present numerical examples for both cases, Case 3 a) and Case 3 b.1), which are taken from
the analysis in \cite{Hagedorn:2014wha}. Like for Case 2), we choose numerical examples, where the masses of the charged leptons are not permuted. Compared to Case 1) and Case 2), the dependence of the PMNS mixing matrix on the flavour and CP symmetries as well as on the residual symmetry $G_\nu$ is more complicated in Case 3 a) and Case 3 b.1). For this reason, more combinations of group theory parameters are analysed in order to explore the features of Case 3 a) and Case 3 b.1). At the same time, we mainly focus on light neutrino masses with NO, since this mass ordering is (slightly) favoured by current global fit data~\cite{Esteban:2020cvm}. Furthermore, examples for the case of a light neutrino mass spectrum with IO are already studied for Case 1) and Case 2).

\subsubsection{Case 3 a)}

As shown in~\cite{Hagedorn:2014wha}, the smallness of the reactor mixing angle $\theta_{13}$ is related to small $m/n$ ($1-m/n$). Also the atmospheric mixing angle $\theta_{23}$
only depends on this parameter combination. The smallest viable choice of the index $n$ is $n=16$.
For this value of $n$, the two choices of the parameter $m$, $m=1$ and $m=15$,
lead to
\begin{equation}
\sin^2 \theta_{13} \approx 0.0254 \; , \;\;
\sin^2 \theta_{23} \approx 0.613 \;\, (m=1) \;\;\; \mbox{and} \;\;\; \sin^2 \theta_{23} \approx 0.387 \;\, (m=15) \, .
\end{equation}
The results for $\theta_{13}$ and $\theta_{23}$ are  outside the $3 \, \sigma$ ranges of the global fit~\cite{Esteban:2020cvm}, $\Delta \chi^2_{13}\approx 21.8$ for $\sin^2\theta_{13}$ and $\Delta \chi^2_{23}\approx 6.25 \, (19)$  for $\sin^2\theta_{23}$ and $m=1$ ($m=15$), assuming light neutrino masses with NO.
However, they can be brought into accordance with the experimental data, if corrections in an explicit model are taken into account, see e.g.~\cite{Lin:2009bw} for a model with the flavour symmetry $A_4$ in which the reactor mixing angle is generated with the help of corrections only. 
It is worth noting that $n$ has to be even and thus both values of $m$ turn out to be odd.
According to the different choices of $s$, $0 \leq s \leq n-1=15$, there are 16 possible CP transformations $X (s)$. All of them
give rise to a viable fit to the solar mixing angle $\theta_{12}$. Indeed, for most of them there are two values of $\theta_L$ (or $\widetilde{\theta}_L$ depending on the combination of $m$ and $s$, see section~\ref{subsec:case3numass}) which allow
for such a fit. One of these values is typically $\theta_L \approx 0$ or $\theta_L \approx \pi$. Hence, both $s$ even and $s$ odd can be studied with this example numerically, whereas $m$ has to be necessarily odd.
In Tab.~\ref{tab:Case3an16} we list the results for $m=1$ and for all possible values of $s$. Those for $m=15$ can be obtained with the help of the symmetry transformations, discussed in~\cite{Hagedorn:2014wha}.

In order to also capture the case $m$ even, we use the choice $n=34$ and $m=2$ which gives rise to the same values for the reactor and atmospheric mixing angles as $n=17$ and $m=1$, studied in~\cite{Hagedorn:2014wha},
since the ratios $m/n$ coincide and lead to
\begin{equation}
\sin^2 \theta_{13} \approx 0.0225 \;\;\; \mbox{and} \;\;\; \sin^2 \theta_{23} \approx 0.607 \,.
\end{equation}
For the value of $\Delta \chi^2_{13 \, (23)}$, the contribution to $\Delta \chi^2$ due to $\sin^2 \theta_{13}$ ($\sin^2 \theta_{23}$), we find for this choice $\Delta \chi^2_{13 \, (23)}\approx 0.243 \, (4.16)$ for light neutrino masses with NO.
Numerical results for the different values of $s$, $0 \leq s \leq n-1=33$, can be found in~\cite{Hagedorn:2014wha} for $s$ even (using the analysis for $n=17$ and $m=1$) and are repeated for convenience in Tab.~\ref{tab:Case3an34seven}. For $s$ odd, they are
mentioned in Tab.~\ref{tab:Case3an34sodd}. Both these tables can be found in appendix~\ref{appE}. Like for the choice $n=16$ and $m=1$, a viable fit to the solar mixing angle is possible for all choices of the CP transformation $X (s)$ and for most of them for two different values of $\theta_L$ (or $\widetilde{\theta}_L$).

\begin{figure}
	\centering
	\includegraphics[width=\textwidth]{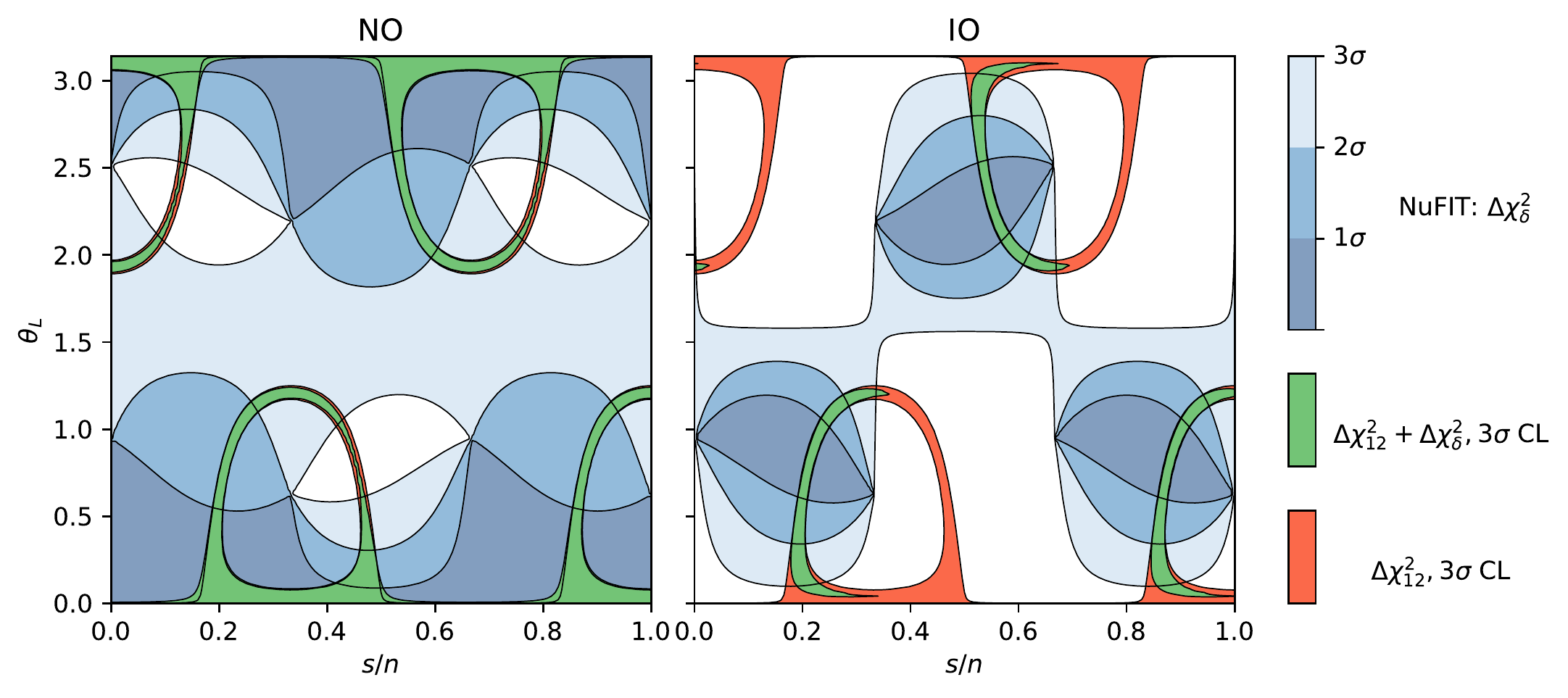}
	\caption{{\small {\bf Case 3 a)} Impact of constraints on the value of the CP phase $\delta$ for light neutrino masses with NO (left plot) and IO (right plot), respectively. We show the areas, leading to an agreement with the global fit data on the solar mixing angle $\theta_{12}$ and CP phase $\delta$ at the $3 \, \sigma$ level or better,
in the $s/n-\theta_L$-plane. The conventions and colour coding are the same as in Fig.~\ref{fig:case2NOIOchi2}. The impact of imposing constraints on the CP phase $\delta$ is only very mild for light neutrino masses with NO, while adding information on the value of $\delta$ considerably affects the allowed areas in the $s/n-\theta_L$-plane in the case of IO.}}
\label{fig:case3aNOIOchi2}
\end{figure}

The CP phases $\delta$, $\alpha$ and $\beta$ all have in general a non-trivial dependence on the parameters $n$, $m$, $s$ and $\theta_L$ and formulae for these together with a detailed numerical discussion can be found in~\cite{Hagedorn:2014wha}. Here we only highlight a few interesting features. The impact of imposing constraints on the CP phase $\delta$ from the global fit data~\cite{Esteban:2020cvm} can be seen in the $s/n-\theta_L$-plane in Fig.~\ref{fig:case3aNOIOchi2}. This figure is adapted from~\cite{Hagedorn:2014wha} (see figure 6). Furthermore, we note that the sine of the Majorana phase $\alpha$ approximately equals in magnitude $\sin 6 \, \phi_s$, while the sine of the Majorana phase $\beta$ is suppressed for $\theta_L \approx 0, \pi$ and has a rather complicated dependence on $\phi_s$ otherwise.

\begin{table}
\begin{center}
\begin{tabular}{c}
$
\begin{array}{|l||c|c|c|c|c|c|c|c|}
\hline
\;\;\;\; s & 0 & 1 & 2 & 3 & 4 & 5 & 6 & 7 \\
\hline
\;\;\;\; \theta_L & \ba1.93\\ \left[3.10\right]\ea & \ba 2.00\\\left[3.09\right]\ea & \ba2.40\\ \left[3.03\right]\ea & 0.265 & \ba0.0568\\ \left[1.07\right]\ea & \ba0.0404\\ \left[1.20\right]\ea & \ba0.0429\\ \left[1.18\right]\ea & \ba0.0737\\  \left[0.957\right]\ea  \\
\hline
\;  \sin^{2}\theta_{12} & 0.305 & 0.305 & 0.305 & 0.317 & 0.305 & 0.305 & 0.305 & 0.305  \\
\hline
\;  \Delta\chi^2_{12} &  \sim 0 &  \sim 0   &  \sim 0 & 1.12 &  \sim 0  &  \sim 0  & \sim 0 &  \sim 0   \\
\hline
\end{array}
$
\end{tabular}
\\[0.1in]
\begin{tabular}{c}
$
\begin{array}{|l||c|c|c|c|c|c|c|c|}
\hline
\;\;\;\; s & 8 &  9 & 10 & 11 & 12 & 13 & 14 & 15 \\
\hline
\;\;\;\; \theta_L  & 0  & \ba2.18\\ \left[3.07\right]\ea  & \ba1.96\\ \left[3.10\right]\ea  & \ba1.94\\ \left[3.10\right]\ea  & \ba2.07\\ \left[3.08\right]\ea   & 2.88 & \ba0.114\\ \left[0.740\right]\ea  & \ba0.0479\\ \left[1.14\right]\ea \\
\hline
\;  \sin^{2}\theta_{12} &  0.342 & 0.305 &  0.305 & 0.305 &  0.305 & 0.317 &  0.305 & 0.305 \\
\hline
\;  \Delta\chi^2_{12} & 8.51 & \sim 0 & \sim 0 & \sim 0 & \sim 0 & 1.12 & \sim 0 & \sim 0 \\
\hline
\end{array}
$
\end{tabular}
\caption{{\small  \textbf{Case 3 a)} Results for the smallest value of the index $n$, $n=16$, and $m=1$ together with $\sin^2 \theta_{13} \approx 0.0254$ and $\sin^2 \theta_{23} \approx 0.613$.
For $m=15$ the same results are found up to
$\sin^2 \theta_{23} \approx 0.387$ and $\theta_L$ being replaced by $\pi-\theta_L$ (leaving aside very minor differences due to the fit). We display the contribution to $\Delta \chi^2$ arising from the fitting of the solar mixing angle $\theta_{12}$, $\Delta \chi^2_{12}$, for light neutrino masses with NO. We write $\sim 0$ when $\Delta \chi^2_{12}<10^{-3}$. The presence of a second best fit value for $\theta_L$ is indicated in square brackets.}
\label{tab:Case3an16}
}
\end{center}
\end{table}

\newpage
\subsubsection{Case 3 b.1)}

We mention two numerical examples in order to cover all possible combinations of $m$ and $s$ being even and odd for Case 3 b.1).
As long as we do not consider any permutations arising from the charged lepton sector, the measured value of the solar mixing angle $\theta_{12}$ constrains $m$ to be
\begin{equation}
\label{eq:Case3b1mcon}
m \approx \frac{n}{2} \, ,
\end{equation}
as shown in~\cite{Hagedorn:2014wha}. Furthermore, the value of $\theta_L$ (or $\widetilde{\theta}_L$ depending on the combination of $m$ and $s$, see section~\ref{subsec:case3numass2}) has to be close to $\pi/2$ in order to achieve the correct size of $\theta_{13}$. The atmospheric mixing angle $\theta_{23}$ depends on the choice of the CP transformation $X (s)$.
The CP phases $\delta$, $\alpha$ and $\beta$ are in general all non-trivial and depend on the actual values of the parameters $n$, $m$, $s$ and $\theta_L$.
They have been studied analytically and numerically in detail in~\cite{Hagedorn:2014wha}. Like for the other cases, we focus here on the main features of the CP phases: we explore the impact of the constraints on $\delta$ from the global fit data~\cite{Esteban:2020cvm}, see below, and we note that the sines of both Majorana phases $\alpha$ and $\beta$ are the same in magnitude as $\sin 6 \, \phi_s$, if $m=\frac{n}{2}$.

For a small value of $n$ we use $n=8$. Then, the parameter $m$ is fixed to be $m=4$.
Regarding the choice of the CP symmetry, i.e.~the parameter $s$, we observe that not all of them can lead to a viable fit to the experimental
data on the lepton mixing angles, see Tab.~\ref{tab:Case3b1n8}.
Note that for this choice of $n$ and $m$ not only the PMNS mixing matrix has been discussed in~\cite{Hagedorn:2014wha}, but also the results for neutrinoless double beta
decay and for unflavoured leptogenesis in a high-scale type-I seesaw scenario can be found in~\cite{Hagedorn:2016lva}.

\begin{table}
\begin{center}
\begin{tabular}{c}
$
\begin{array}{|l||c|c|c|c|c|c|c|c|c|c|c|c|}
\hline
\;\;\;\; s & 1  & 4  & 7  \\
\hline
\;\;\;\; \theta_L & 1.31~\left[1.83\right]  & 1.31~\left[1.83\right]  &  1.31~\left[1.83\right]  \\
\hline
\; \sin^{2}\theta_{23} &  0.580~\left[0.420\right]  & 0.5 & 0.420~\left[0.580\right]\\[0.01in]
\; \sin^{2}\theta_{12} & 0.318  &  0.318  & 0.318 \\[0.01in]
\; \sin^{2}\theta_{13} & 0.0222  & 0.0222 & 0.0221\\
\hline
\;  \Delta\chi^2  & 1.48~\left[5.29\right]  & 4.12 & 5.29~ \left[1.48\right] \\
\hline
\end{array}
$
\end{tabular}
\caption{{\small  \textbf{Case 3 b.1)} Results for a small value of the index $n$, $n=8$, together with $m=4$. Note that not for all admitted values
of the parameter $s$ a viable fit to the experimental data on lepton mixing angles~\cite{Esteban:2020cvm} can be achieved. The goodness of the fit can be read off from the value of $\Delta \chi^2$ for light neutrino masses with NO. The presence of a second best fit value for $\theta_L$ is indicated in square brackets.}
\label{tab:Case3b1n8}
}
\end{center}
\end{table}

\begin{figure}
	\centering
	\includegraphics[width=0.99\textwidth]{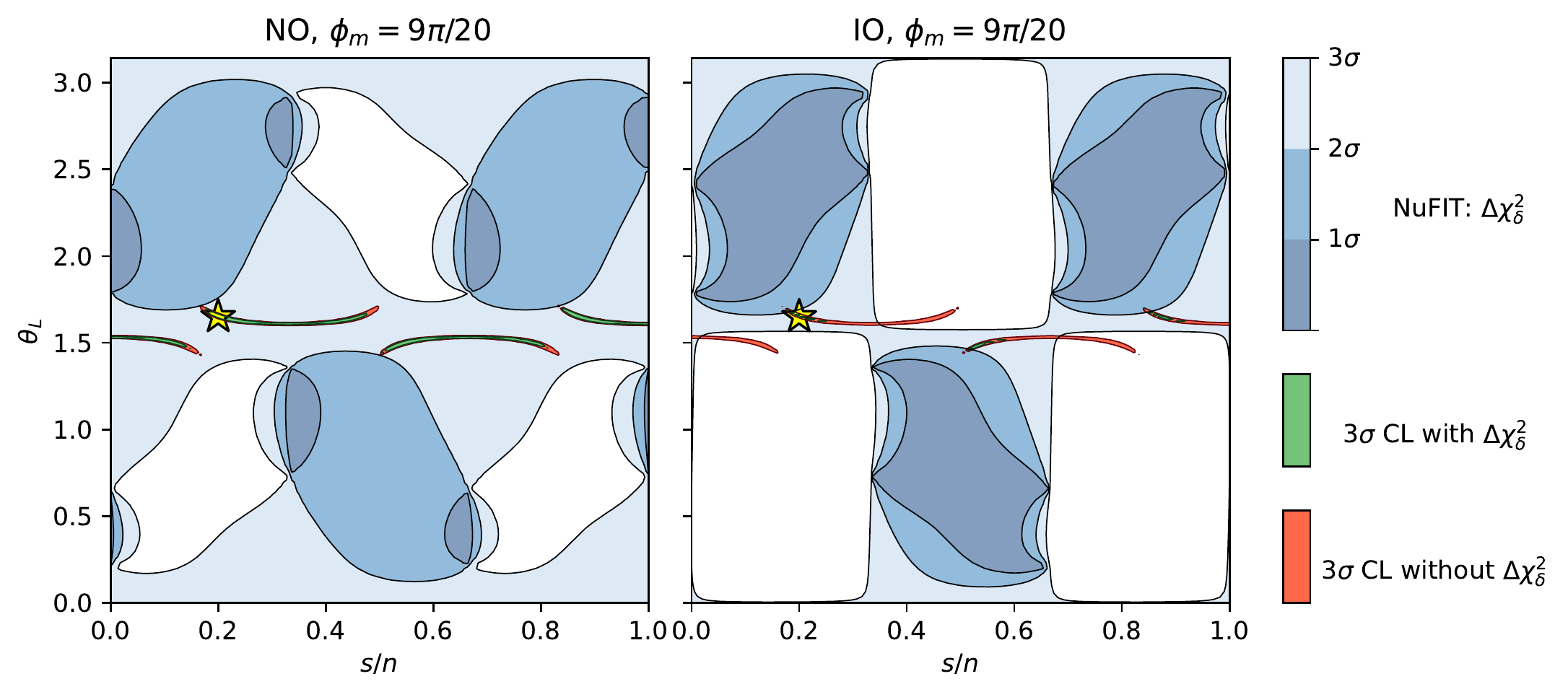}\\
	\includegraphics[width=0.99\textwidth]{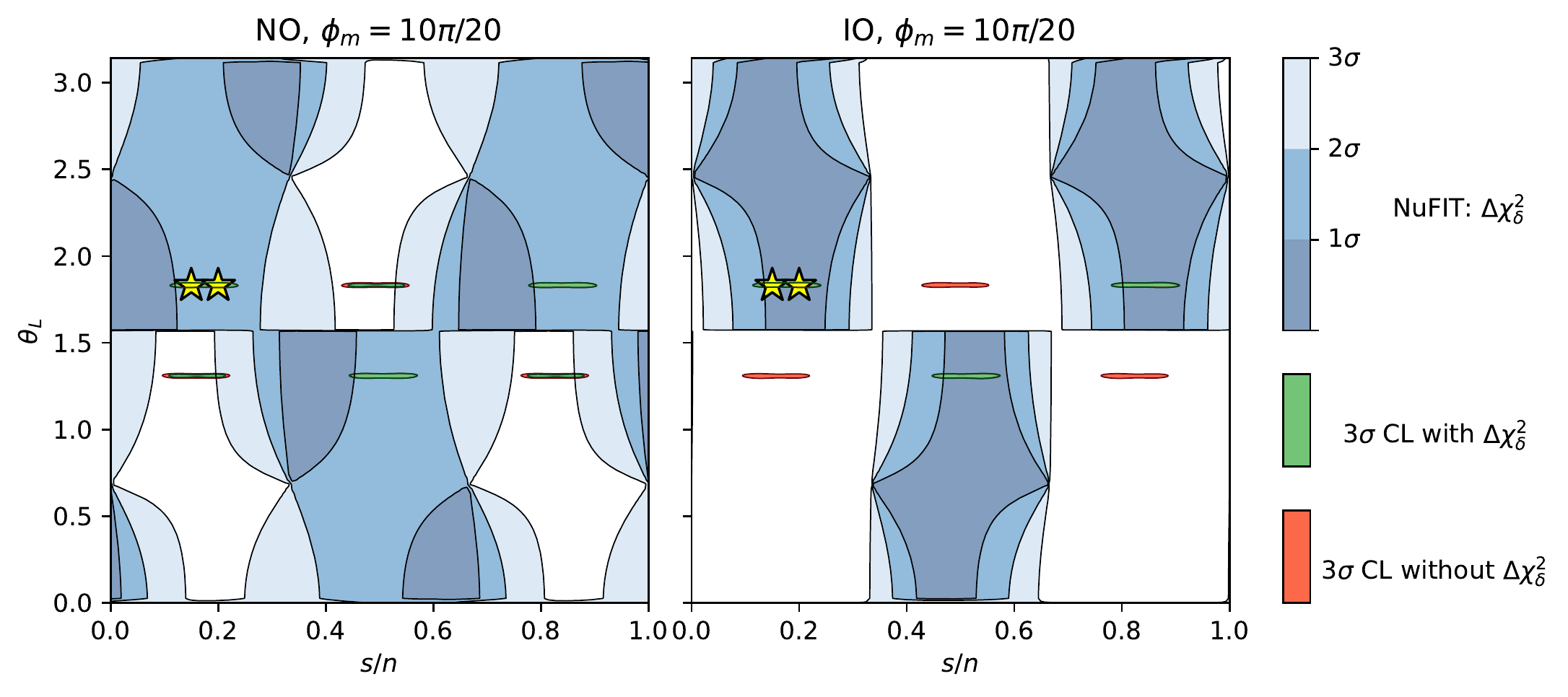}
	\caption{{\small
		{\bf Case 3 b.1)} Impact of the constraints on the value of the CP phase $\delta$ for light neutrino masses with NO (left column) and IO (right column) for different values of $\phi_m$, $\phi_m=\frac{9 \, \pi}{20}$ ($n=20$ and $m=9$, plots in the upper row) and $\phi_m=\frac{10 \,  \pi}{20}$ ($n=20$ and $m=10$, plots in the lower row).
		We show the areas, leading to an agreement with the global fit data on the lepton mixing angles and the CP phase $\delta$ at the $3 \, \sigma$ level or better, in the $s/n-\theta_L$-plane.
		The conventions and colour coding are the same as in Fig.~\ref{fig:case2NOIOchi2}. The impact of imposing constraints on the CP phase $\delta$ is only mild for light neutrino masses with NO, while adding information on the value of $\delta$ considerably affects the allowed areas in the $s/n-\theta_L$-plane in the case of IO.
		The choice $s=4$ corresponding to $s/n=4/20$ is represented by a yellow star in all plots. For $\phi_m=\frac{10 \,  \pi}{20}$ we additionally highlight $s=3$ meaning $s/n=3/20$ with a yellow star.
}}
\label{fig:case3bNOIOchi2}
\end{figure}

A value of $n$, fulfilling all constraints and leading to more than one admissible value of $m$, is $n=20$. According to Eq.~\eqref{eq:Case3b1mcon} three values of $m$ are admitted, namely $m=9$, $m=10$ and $m=11$, see
also~\cite{Hagedorn:2014wha}.
In Tab.~\ref{tab:Case3b1n20m9} we display the results for $m=9$ and $0 \leq s \leq n-1=19$.
The results for $m=11$ can be obtained from those shown in Tab.~\ref{tab:Case3b1n20m9}, using symmetry transformations of the parameters $m$, $s$ and $\theta_L$ as explained in~\cite{Hagedorn:2014wha}.
The results, obtained for $m=10$, are found in Tab.~\ref{tab:Case3b1n20m10} and share some features with those shown for $n=8$ (and $m=4$). These tables can be found in appendix~\ref{appE}. For more details  see~\cite{Hagedorn:2014wha}.
In Fig.~\ref{fig:case3bNOIOchi2} we show the impact of the global fit data for the CP phase $\delta$ on these results in the $s/n-\theta_L$-plane for both $m=9$ (plots in the upper row) and $m=10$ (plots in the lower row) for $n=20$ for light neutrino masses with NO and IO. We clearly see that adding the constraints on the CP phase $\delta$ only mildly affects the results for light neutrino masses with NO, while it reduces by at least a factor of two the areas, favoured at the $3 \, \sigma$ level or better, in the $s/n-\theta_L$-plane for light neutrino masses with IO.
These plots should be compared to similar plots shown in~\cite{Hagedorn:2014wha} (see figure 8; note that in~\cite{Hagedorn:2014wha} $m=11$ has been chosen instead of $m=9$). We note that we focus on Case 3 b.1) in the discussion of low-scale leptogenesis, see section~\ref{sec55}.

\section{Low-scale leptogenesis}
\label{sec:LowScalelepto}

This section is dedicated to the analysis of low-scale leptogenesis in the outlined scenario. The parameter space which is scanned
comprises the different cases, Case 1) through
Case 3 b.1), the choices of group
theory parameters and the angle $\theta_L$ as well as the parameters related to the light neutrino mass spectrum and to the RH neutrinos such as their mass scale $M$, the splittings $\kappa$ and $\lambda$ and the angle $\theta_R$. We take a representative subset of examples for the group theory parameters and the angle $\theta_L$ for Case 1) through Case 3 b.1) from the preceding section.

\subsection{Preliminaries}
\label{sec51}

The common thermal leptogenesis scenario
(``vanilla leptogenesis")
requires that $M_i > 10^9$ GeV~\cite{Davidson:2002qv}. Flavour effects can slightly lower this bound~\cite{Dev:2017trv}, but the heavy neutrinos $N_i$ remain out of reach for direct searches.
However, there are several options  of making leptogenesis for $M_i$ below the  TeV  scale feasible~\cite{Hambye:2001eu}, including
$i)$ a degenerate particle spectrum~\cite{Flanz:1994yx,Covi:1996wh}, 
$ii)$ an approximate conservation of charges or
$iii)$ a hierarchy among coupling constants.
Within the low-scale type-I seesaw framework, see Eq.~\eqref{L}, all three of them can play a role.
One possibility to classify different leptogenesis scenarios is the way in which the deviation from equilibrium required for baryogenesis~\cite{Sakharov:1967dj} is achieved, namely either in particle freeze-out and decays (\emph{freeze-out scenario})
or during their approach to thermal equilibrium (\emph{freeze-in  scenario}).
\emph{Resonant leptogenesis} during heavy neutrino freeze-out~\cite{Pilaftsis:2003gt,Pilaftsis:2005rv} is commonly associated with a degeneracy in the eigenvalues of the Majorana mass matrix $M_R$, i.e.~option $i)$. Option 
$ii)$ is realised in the so-called Akhmedov-Rubakov-Smirnov (ARS) mechanism~\cite{Akhmedov:1998qx,Asaka:2005pn} during heavy neutrino freeze-in,
when the approximate conservation of a generalised lepton number  permits the generation of a sizeable asymmetry in the LH fields by hiding an asymmetry of opposite sign  from the sphalerons in the RH neutrinos.\footnote{
The ARS mechanism requires in addition a mass degeneracy when one considers two generations of RH neutrinos, but this is not needed for more than two, see~\cite{Asaka:2005pn}.}
 Option $iii)$ can play a role during both freeze-in and freeze-out. Flavour-hierarchical Yukawa couplings lead to hierarchical equilibration rates, which can  delay the equilibration of one heavy neutrino or lead to flavour-hierarchical washout~\cite{Canetti:2014dka,Garbrecht:2014bfa}.

The freeze-out scenario is usually associated with $M_i$ above the electroweak scale, while the freeze-in mechanism has originally been proposed for $M_i$ in the GeV range.
However, detailed investigations of the parameter space have shown that the ranges of $M_i$ and $U_{\alpha i}^2$, in which both mechanisms operate, widely overlap~\cite{Klaric:2020phc,Klaric:2021cpi,Drewes:2021nqr}, cf.~\cite{Klaric:2021cpi} and references therein for a detailed discussion.
They can be described by the same set of quantum kinetic equations which are reminiscent of the density matrix equations for light neutrinos \cite{Sigl:1993ctk} and can be derived from non-equilibrium quantum field theory, cf.~\cite{Garbrecht:2018mrp} for a recent review,
\begin{subequations}\label{QKE}
	\begin{align}
		i \frac{d n_{\Delta_\alpha}}{dt}
		&= -2 i \frac{\mu_{\alpha}}{T} \int \frac{d^{3} k}{(2 \pi)^{3}} \operatorname{Tr}\left[\Gamma_{\alpha}\right] f_{N}\left(1-f_{N}\right) 
		\quad +i \int \frac{d^{3} k}{(2 \pi)^{3}} \operatorname{Tr}\left[\tilde{\Gamma}_{\alpha}\left(\bar{\rho}_{N}-\rho_{N}\right)\right],
		\label{kin_eq_a}
		\\
		i \, \frac{d\rho_{N}}{dt}
		&= \left[H_{N}, \rho_{N}\right]-\frac{i}{2}\left\{\Gamma, \rho_{N}-\rho_{N}^{eq} \right\} 
		\quad-\frac{i}{2} \sum_{\alpha} \tilde{\Gamma}_{\alpha}\left[2 \frac{\mu_{\alpha}}{T} f_{N}\left(1-f_{N}\right)\right] ,
		\label{kin_eq_b}
		\\
		i \, \frac{d \bar{\rho}_{N}}{d t}
		&= -\left[H_{N}, \bar{\rho}_{N}\right]-\frac{i}{2}\left\{\Gamma, \bar{\rho}_{N}-\rho_{N}^{eq} \right\}
		\quad+\frac{i}{2} \sum_{\alpha} \tilde{\Gamma}_{\alpha}\left[2 \frac{\mu_{\alpha}}{T} f_{N}\left(1-f_{N}\right)\right] \; .
		\label{kin_eq_c}
\end{align}\label{kin_eq}\end{subequations}
Here $\rho_N$ and $\bar{\rho}_{N}$ are the momentum averaged density matrices for the two helicities of the heavy neutrinos,
$H_N$ is an effective Hamiltonian,
$f_N$ is the Fermi-Dirac distribution for heavy neutrinos, $\mu_{\alpha}$ are flavoured lepton chemical potentials,
and  $\Gamma$, $\Gamma_\alpha$ and $\tilde{\Gamma}_\alpha$ are different thermal interaction rates.
The computation of these rates has been an active field of research~\cite{Biondini:2017rpb,Laine:2022pgk}. In the present analysis we use the results of~\cite{Ghiglieri:2017gjz} combined with the extrapolation to the non-relativistic regime from~\cite{Klaric:2021cpi}.
The flavoured lepton chemical potentials $\mu_{\alpha}$ are related to the comoving lepton number densities $n_{\Delta_\alpha}$ by a susceptibility matrix~\cite{Buchmuller:2005eh,Garbrecht:2019zaa}.

We solve the set of equations, given in Eq.~\eqref{QKE}, for two types of initial conditions for the heavy neutrinos $N_i$,  vanishing and thermal initial abundances.
Vanishing initial conditions apply in scenarios, in which the reheating temperature is considerably lower than the scale at which new particles other than the RH neutrinos appear, so that the Lagrangian in Eq.~\eqref{L} describes a valid effective field theory at all relevant energies~\cite{Bezrukov:2008ut}.
A prominent example of this kind is the $\nu$MSM~\cite{Asaka:2005pn,Asaka:2005an},
which, in principle, could be a valid effective field theory up to the Planck scale~\cite{Bezrukov:2012sa}.
Thermal initial conditions apply in scenarios, in which the RH neutrinos have additional interactions at energies below the reheating temperature.
In both cases, we assume that all asymmetries vanish initially, cf.~\cite{Domcke:2020quw} for a related discussion.

\paragraph{Importance of heavy neutrino mass scale.}
The overall mass scale of the heavy neutrinos is one of the most important parameters determining the BAU.
There are several ways in which this mass scale plays a role.
\begin{itemize}
	\item \textbf{Interaction rates} of the heavy neutrinos in general depend on both the RH neutrino mass $M$ and the temperature $T$.
		For relativistic neutrinos with $M \ll T$, there are two types of processes that reach equilibrium at different temperatures, which depend on the helicity of the produced RH neutrino~\cite{Anisimov:2010gy,Ghiglieri:2017gjz, Eijima:2017anv}.
		If we associate a \emph{fermion number}  with helicity, these processes can be either fermion number conserving (FNC)  or fermion number violating (FNV).
		The FNC processes entering the matrices $\Gamma$ and $H_N$ only depend on the RH neutrino masses indirectly, through the Yukawa couplings. This
		 leads to a typical equilibration temperature 
		\begin{align}
			\label{eq:equilibrationFNC}
			T^\mathrm{eq}_\mathrm{FNC} \sim  \gamma_+ \, \Tr (Y_D Y_D^\dagger) \, T_0  \gtrsim \gamma_+ \, \frac{M \, T_0 \, (\sum_i m_i)}{\langle H \rangle^2} \sim T_\mathrm{sph} \times \frac{M}{10 \, \text{GeV}}\,,
		\end{align}
		where $T_\mathrm{sph} \approx 131.7$ GeV is the sphaleron freeze-out temperature, $T_0 = m_\mathrm{Pl} \sqrt{\frac{45}{4 \pi^3 g_\star}} = T^2/H$ is the comoving temperature in an expanding Universe (with $g_\star \approx 106.75$ and $m_\mathrm{Pl} \approx 1.22 \times 10^{19}$~GeV), and $\gamma_+ \sim 10^{-2}$ is a numerical coefficient associated with such processes.
		Note that the inequality in Eq.~\eqref{eq:equilibrationFNC} comes from the fact that the size of the Yukawa couplings can exceed the naive seesaw limit by several orders of magnitude for special choices of parameters.
		For sub-GeV RH neutrino masses, these FNC processes do not necessarily reach equilibrium before the sphaleron freeze-out, which could prevent successful baryogenesis.
		Such processes only lead to lepton flavour violation (LFV), and rely on washout to convert the lepton flavour asymmetries into a lepton number asymmetry.
		In contrast, the FNV processes can directly lead to a lepton number asymmetry.
		Their rate carries an additional suppression factor of $M^2/T^2$. This  leads to the equilibration temperature 
		\begin{align}
			\label{eq:equilibrationFNV}
			T^\mathrm{eq}_\mathrm{FNV} \sim \sqrt[3]{ \left(\gamma_- \frac{T^2}{M^2} \right) \Tr (Y_D Y_D^\dagger) M^2 T_0}
			\
			\gtrsim
			\
			T_\mathrm{sph} \times \frac{M}{50\, \text{GeV}}\,,
		\end{align}
		where we have approximated $\left(\gamma_- \frac{T^2}{M^2} \right) \sim 10^{-2}$ in the relativistic limit.
While they only equilibrate for $M \gtrsim 50$ GeV, they can play an important role for even lighter RH neutrinos,
as they allow for different CP-violating combinations, see section~\ref{sec6}.
In the intermediate regime $M \sim T_\mathrm{sph} \sim 100$ GeV, both FNC and FNV processes have similar interaction rates, while for $T \ll M$ we enter the non-relativistic regime, with RH neutrino decays as the dominant process -- with equal FNC and FNV rates.
On the other hand, for $T \lesssim M$, the lepton asymmetry washout rates become Boltzmann-suppressed, causing a gradual freeze-out of lepton number.
This freeze-out dominates over the sphaleron freeze-out when $M \gtrsim 10 \, T_\mathrm{sph}$.
\item \textbf{The expansion and cooling of the Universe} causes a change to the equilibrium distribution of the RH neutrinos.
For relativistic RH neutrinos, $T \gg M$, this change does not cause a sizeable deviation from equilibrium, as the RH neutrino number density to entropy ratio remains constant.
However, once the temperature becomes comparable to the RH neutrino masses, this ratio decreases, which corresponds to a deviation from equilibrium.
While this deviation is maximal around $M\sim T$, already the leading order $\mathcal{O}(M^2/T^2)$ deviation from equilibrium can be sufficient to produce the observed BAU -- allowing leptogenesis with thermal initial conditions for RH neutrinos with masses as low as a few GeV~\cite{Hambye:2016sby, Klaric:2020phc}.
This late-time contribution to the BAU can completely dominate over the asymmetry produced during freeze-in, especially if the prior asymmetries are erased by the lepton number washout.
For RH neutrinos with masses above the electroweak scale we can, therefore, expect that the BAU becomes independent of the initial conditions, unless the lepton number washout is highly flavour-hierarchical~\cite{Garbrecht:2014bfa}.
\item \textbf{The scaling of the parameter space for $M>2$ TeV} allows us to relate results of parameter scans for different RH neutrino masses.
	In~\cite{Klaric:2021cpi} it has been found that keeping the ratio
	\begin{align}
		M : Y_D Y_D^\dagger : \Delta M_{ij} /M
		\label{eq:ratio}
	\end{align}
	fixed, where $\Delta M_{ij} = M_i-M_j$,  leads to the same final BAU.
	To achieve this, we need to re-scale the mass splittings $\Delta M_{ij}/M$ for different RH neutrino masses, while
	the ratios of these masses and the Yukawa couplings
	are kept fixed by the seesaw relation, see Eq.~\eqref{eq:mnu}.\footnote{We neglect  corrections to the seesaw formula, arising from the mass splittings.} 
	This scaling only holds between the electroweak scale $T_\mathrm{EW}$~\cite{DOnofrio:2015gop} 
	and the scale $T^\mathrm{eq}_{e \mathrm{R}}$ at which the RH electron Yukawa interactions come into equilibrium~\cite{Bodeker:2019ajh}, $T_\mathrm{EW} \approx 160$ GeV $ \ll M \ll T^\mathrm{eq}_{e \mathrm{R}} \approx 85$ TeV.
\item \textbf{Initial conditions.}
	Besides washout effects discussed below, the size of the RH neutrino masses determines how sensitive the BAU is to the initial RH neutrino abundance.
	Barring flavour-hierarchical Yukawa couplings, the washout of the lepton asymmetries is generically strong, erasing any asymmetry that might be generated before the RH neutrino decays.
	For masses $M\gtrsim 2$ TeV there is, therefore, little to no dependence on the initial abundance of the RH neutrinos.
	In contrast, for lighter RH neutrinos the deviation from equilibrium, caused by the expansion of the Universe, is often subdominant to the contribution from the RH neutrino equilibration.
	This is particularly important for RH neutrino masses $M\lesssim 2$ GeV, where leptogenesis is only possible, if we assume vanishing initial abundances.
\end{itemize}

\paragraph{Impact of RH neutrino mass splittings.}
The mass splittings among the RH neutrinos are among the most important parameters determining the overall size of the BAU.
Both baryogenesis via RH neutrino oscillations and resonant leptogenesis require a degeneracy in energies that is comparable to the interaction widths. This gives the resonance condition
\begin{align}
	|\Delta M_{ij}| \sim \frac{\Gamma_N E_N}{M} \approx
	\begin{cases}
		\frac{T \, \Gamma_N}{M}  & \text{for}\quad \quad T \gg M\,,\\
		\Gamma_N  & \text{for}\quad \quad T \lesssim M\,,
	\end{cases}
	\label{eq:resCondition}
\end{align}
for heavy neutrinos $N_i$ and $N_j$ and with  $E_N$ and $\Gamma_N$ being the energy and thermal width of the heavy neutrinos, respectively.
This condition alone is not sufficient to guarantee successful baryogenesis.
 In addition, the oscillation temperature~\cite{Asaka:2005pn} has to be higher than the sphaleron freeze-out temperature
\begin{align}
	T_\mathrm{sph} \lesssim T_\mathrm{osc} =
	\begin{cases}
		\left( M \, |\Delta M_{ij}| \, T_0 \right)^{1/3} & 
		\quad \quad T \gg M \,,\\
		\left( |\Delta M_{ij}| \, T_0 \right)^{1/2} & 
		\quad \quad T \lesssim M\,,
	\end{cases}
	\label{eq:Tosc}
\end{align}
as otherwise the sphalerons freeze out before a single RH neutrino oscillation has happened.

Given the delicate dependence on the mass splittings $\Delta M_{ij}$,
one may wonder whether the splittings induced by radiative corrections to the heavy neutrino masses could have a disrupting effect on leptogenesis.
The radiative corrections to the heavy neutrino masses have previously been studied in~\cite{Roy:2010xq,Antusch:2002rr}, and can be estimated as
\begin{align}
    (4 \pi)^2 \frac{d}{d t} \hat{M}_R =
    (\hat{Y}_D^\dagger \hat{Y}_D) \, \hat{M}_R + \hat{M}_R \, (\hat{Y}_D^\dagger \hat{Y}_D)^T\,,
\end{align}
where $t=\ln \mu / \mu_0$.
If we neglect the running of the neutrino Yukawa coupling matrix $Y_D$ and solve the RG equations perturbatively, we find as correction
\begin{align}
    \delta \hat{M}_R^{RG} = \hat{M}_R (t) - \hat{M}_R (0) = \frac{t}{(4 \pi)^2} \, M \,
    \left[(\hat{Y}_D^\dagger \hat{Y}_D) + (\hat{Y}_D^\dagger \hat{Y}_D)^T \right]
    + \mathcal{O}(Y_D^2 \, \kappa) + \mathcal{O}(Y_D^2 \, \lambda)\,.
\end{align}
These corrections to the heavy neutrino mass spectrum generically have the same structure as the products of Yukawa couplings contributing to the thermal masses which enter the effective Hamiltonian in Eq.~\eqref{QKE}, i.e.~\footnote{In fact, in the closed-time-path formalism both contributions are described by the same diagrams.}
\begin{align}
    H_N = \frac{\hat{M}_R^2}{2 E_N} +
    h_+ \, (\hat{Y}_D^\dagger \hat{Y}_D) +
    h_- \, (\hat{Y}_D^T \hat{Y}_D^*)\,.
\end{align}
They can, thus, be simply absorbed through a redefinition of the numerical coefficients
\begin{align}
    h_\pm \rightarrow h_\pm +
    \frac{M^2}{E_N}\frac{t}{(4 \pi)^2}\, .
\end{align}
They are typically subdominant to $h_+$ in the relativistic regime, whereas they can give up to an $\mathcal{O}(1)$ correction to $h_-$.
Any additional CP violation introduced by these radiative corrections is, therefore, included when considering the effects of the thermal masses on the CP-violating combinations, see  section~\ref{sec6}.

\paragraph{The overall scale of the Yukawa couplings} is the main parameter determining the production and decay rates of the RH neutrinos in the early Universe.
Tiny Yukawa couplings lead to low RH neutrino  production and decay rates -- and can, therefore,  lead to a value of the BAU below the observed one. 
This is particularly important for sub-GeV RH neutrino masses, since the Yukawa couplings are expected to be too small for efficient baryogenesis, as suggested by Eq.~\eqref{eq:equilibrationFNC}.
Fortunately, this limit can be avoided for special choices of parameters, 
for which the size of the Yukawa couplings can exceed the naive seesaw limit by several orders of magnitude.
For sub-GeV RH neutrino masses, baryogenesis can, therefore, impose a lower bound on the Yukawa couplings~\cite{Canetti:2012zc}.
Besides an increased production rate, large values of the Yukawa couplings also correspond to large values of the mass splittings $\Delta M_{ij}$, as implied by the resonance condition, see  Eq.~\eqref{eq:resCondition}.

\paragraph{Impact of washout effects.}
The amount of BAU produced via low-scale leptogenesis does not only depend on the mass splittings and widths of the RH neutrinos, but also on how quickly any asymmetry is erased through washout processes, i.e.~how fast  interactions of the RH neutrinos erase the asymmetries in the different lepton flavours.

This is especially important for $M \lesssim 50$ GeV, a regime in which RH neutrino interactions do not directly lead to an overall lepton number asymmetry, but only to a lepton flavour asymmetry~\cite{Akhmedov:1998qx}.
Because the RH neutrino couplings to the three flavours of the LH lepton doublets are in general different, these asymmetries can be washed out at different rates, and convert the lepton flavour asymmetry into an overall lepton asymmetry~\cite{Asaka:2005pn}.

For larger RH neutrino masses, $M\gtrsim 50$ GeV, interactions that directly violate lepton number are no longer suppressed, and the flavoured washout becomes less important.
Nonetheless, it can still play an important role, if the lepton number asymmetry is suppressed due to a special choice of parameters, e.g. $\Delta M_{ij}=0$,
or if the washout is negligible in one of the lepton flavours~\cite{Garbrecht:2014bfa} which leads to a preservation of the asymmetry produced at high temperatures.

\subsection{Prerequisites of parameter scan}

{\renewcommand{\arraystretch}{1.2}
\begin{table}
\begin{center}
\begin{tabular}{|c|c|c|}
\hline
Parameter & Range of values & Prior \\
\hline
$M$ & $\big[50 \mbox{ MeV},~70 \mbox{ TeV}\big]$ &  Log  
\\
\hline
$\ba \kappa \\ \lambda \ea$ &  $\big[10^{-20},10^{-1}\big]$ & Log\\
\hline
$\theta_R$ &  $\big[0,2\pi\big]$ & $\ba \mbox{Linear} \\ \mbox{Log on } \theta_R-k\frac{\pi}{4}\ea$ \\
\hline
$m_0$ &  $\big[10^{-15},1\big]\cdot 0.03 \mbox{ eV}$ & Log\\
\hline
$\ba \frac{s}{n} \\ \frac{v}{n} \ea$ &  $\ba \big[0,1\big] \\ \big[0,3\big]\ea$ & Linear\\
\hline
\end{tabular}
\end{center}
\caption{{\small {\bf Range of values
for the
free parameters used in this analysis.}
``Prior'' refers -- in a slight abuse of terminology -- to the measure in parameter space taken for the randomisation and is not meant to indicate a Bayesian parameter fit. Note $k$ is an integer and its actual value depends on the case, Case 1) through Case 3 b.1).
The ratio $s/n$ is relevant for Case 1), Case 3 a) and Case 3 b.1), while $v/n$ is relevant for Case 2). 
The rest of the parameters, group theory parameters
and the angle $\theta_L$, is determined by one of the representative examples, given in section~\ref{sec50}.
}}
\label{range of values parameters}
\end{table}}

In the numerical scan we solve the quantum kinetic equations in Eq.~\eqref{QKE} for different choices of parameters as well as for 
two types of initial conditions: vanishing and thermal initial heavy neutrino abundances. We explore the viable parameter space for both of them, showing that these in general differ. Since the CP-violating combinations, derived in section~\ref{sec6}, do not depend on the initial conditions, we mostly display results for vanishing initial conditions when testing their validity.

We consider heavy neutrino masses ranging from 50 MeV to 70 TeV, see Tab.~\ref{range of values parameters}, and thus cover the entire experimentally accessible mass range.
Lower masses are strongly constrained by cosmological considerations~\cite{Hernandez:2014fha,Vincent:2014rja,Sabti:2020yrt,Boyarsky:2020dzc,Domcke:2020ety,Mastrototaro:2021wzl}, supernovae~\cite{Mastrototaro:2019vug} and
direct searches~\cite{Arguelles:2021dqn,Kelly:2021xbv}, in particular when combined with neutrino oscillation data~\cite{Drewes:2016jae,Bondarenko:2021cpc,Chrzaszcz:2019inj}. For larger masses, on the other hand, the set of equations in Eq.~\eqref{QKE} would have to be modified to treat the RH charged lepton chemical potentials dynamically.
 The results of scanning the entire range of RH neutrino masses for regions of successful generation of the BAU in the $M - U^2$-plane can be found in sections~\ref{sec53}-\ref{sec55}. For each of the different cases, Case 1) through Case 3 b.1), representative examples for the group theory parameters and the angle $\theta_L$ as well as different values of the splitting $\kappa$ are studied.
In addition to the scan of the viable parameter space, we consider four benchmark values of the RH neutrino mass scale $M$, $M \in \{1,10,100,1000\}$ GeV, cf.~Tab.~\ref{benchmark choice_alt}, in order to illustrate the parametric dependence of the BAU and the validity of the analytical results, discussed in section~\ref{sec6}.
These benchmark values for $M$ are chosen to cover the non-relativistic, intermediate and relativistic regimes of leptogenesis. Moreover, they are typical for the sensitivity reach of different experiments, see e.g.~\cite{Agrawal:2021dbo,Abdullahi:2022jlv,FCC:2018evy,CEPCStudyGroup:2018ghi,Alimena:2019zri}.

The size of the RH neutrino mass splittings is determined by the two splittings $\kappa$ and $\lambda$, introduced in section~\ref{sec2}, see Eqs.~\eqref{dMRtilde} and~\eqref{lambda definition}, respectively. They can, in principle, both vary between $0$ and $10^{-1}$, with the upper limit being given by the requirement that these splittings correspond to a small breaking of the flavour and CP symmetry  of the scenario.\footnote{At a more practical level, the
computation in the density matrix equations in Eq.~\eqref{QKE} would have to be refined to apply them to an arbitrary heavy neutrino mass spectrum.}
They govern the heavy neutrino mass spectrum in different ways. While the splitting $\kappa$ keeps two of the heavy neutrinos degenerate in mass and separates the third one, compare Eq.~(\ref{eq:RHmassspectrum}), the splitting  $\lambda$ causes a mass difference between all three of them, see Eq.~(\ref{lambda mass spectrum}).
Therefore, a non-zero value of $\lambda$ can lead to more sources of CP violation than the splitting $\kappa$ alone.
As a conservative choice, we thus focus on the effects of $\kappa$ on the generation of the BAU, and set $\lambda=0$, unless otherwise stated.
 This in turn always gives us one pair of RH neutrinos that are approximately degenerate in mass.
This degeneracy can be broken by the thermal masses, which do not always lead to additional CP violation.
Another reason to choose $\lambda =0$ comes from the observation that, as shown in section~\ref{sec:symmetries}, the correction $\delta M_R$ to the RH neutrino mass matrix $M_R$, which is proportional to the splitting $\kappa$, is invariant under the residual symmetry $G_l$, present among charged leptons, and thus can be thought of to arise from this sector in a concrete model.
In contrast, the splitting $\lambda$ is a priori not related to any residual symmetry, but rather just splits all three RH neutrino masses.
As benchmark values for $\kappa$, we  use $\kappa \in \{10^{-11},10^{-6},10^{-3}\}$.

As already mentioned in section~\ref{sec:symmetries}, the angle $\theta_R$ should range from $0$ to $2\, \pi$. As generic benchmark values, we consider $\theta_R$ being $\frac{\pi}{11}\{1,2,3,4,5\}$. This allows to span the entire range between $0$ and $\frac{\pi}{2}$. At the same time, we avoid  special values of $\theta_R$, i.e.~multiples of $\frac{\pi}{4}$, since the 
total mixing angle $U^2$ would diverge for a subset of the considered cases at such points.
We discuss results for values of $\theta_R$ close to these special values, which can reflect an enhanced residual symmetry, see section~\ref{sec4}, separately, because they are very interesting from a phenomenological viewpoint, as they are the only choices that can lead to values of the active-sterile mixing angles $U_{\alpha i}^2$ that greatly exceed the naive seesaw estimate, shown in Eq.~\eqref{eq:naiveseesawformula}.

Regarding light neutrino masses, both their mass ordering and the mass $m_0$ of the lightest neutrino are currently unknown. In the present  analysis, we, therefore, consider light neutrino masses with NO as well as IO. However, in the main text we focus on light neutrino masses with NO and show plots for IO only in case of a qualitative difference between the results for the two mass orderings. Otherwise, further results for light neutrino masses with IO can be found in appendix~\ref{appF}.
 The lightest neutrino mass $m_0$ is varied between zero and 
a mass close to its upper limit, allowed by cosmological observations~\cite{Planck:2018vyg},
\begin{equation}
\label{eq:m0max}
	m_0 = 0.03 ~~\mathrm{ eV} ~~ \mathrm{(NO)}\mbox{ and }m_0 = 0.015~ \rm{ eV}~~ (IO),
\end{equation}
see also Eq.~\eqref{eq:m0bound} in appendix~\ref{appD3}. Most of the plots are, however, presented for a benchmark value of $m_0$, $m_0$ being zero or one of the values in Eq.~(\ref{eq:m0max}).

The signs of the couplings $y_f$ can be positive or negative, see Eq.~(\ref{eq:YDgen}). For concreteness, we only display results for positive $y_f$. However, results for other choices of the signs of these couplings can be  obtained numerically in a similar way. In certain instances, it is also possible to derive them with the help of the analytical expressions, given in section~\ref{sec6}. 

Lastly, we observe that accommodating the global fit data on lepton mixing angles and potentially the current indication of the CP phase $\delta$~\cite{Esteban:2020cvm} well usually does not completely fix all group theory parameters 
for a certain case, Case 1) through Case 3 b.1). For example, for Case 1) the parameter $s$ corresponding to the choice of the CP symmetry does not impact the accommodation of the lepton mixing angles and the CP phase $\delta$, but is related to the value of the Majorana phase $\alpha$, see section~\ref{sec501}. This is similarly true for the parameter $v$ in Case 2) and $s$ in Case 3 a) and Case 3 b.1). Consequently, we do not only discuss results for a fixed value of this parameter, but also vary it in its entire allowed range, i.e.~$s/n$ ($v/n$) is treated as continuous parameter between $0$ and $1~(3)$.

\begin{table}
\begin{center}
\begin{tabular}{|c|c|}
\hline
Parameter & Benchmark values \\
\hline
$M$ &  $\{1,10,100,1000\}$ GeV\\
\hline
$\kappa$ &  $\{10^{-11},10^{-6},10^{-3}\}$\\
$\lambda$ &  $0$\\
\hline
$\theta_R$ &  $\frac{\pi}{11}\{1,2,3,4,5\}$\\
\hline
\end{tabular}
\hspace{0.3in}
\begin{tabular}{|c|c|}
\hline
Parameter & Benchmark values \\
\hline
$m_0$  [NO (IO)] &  $\ba 0.03 \mbox{ eV} ~(0.015 \mbox{ eV}) \\[0.01in] 0 \mbox{ eV} \ea $\\
\hline
$y_f$, $f=1,2,3$ &  $\geq 0$\\
\hline
\end{tabular}
\end{center}
\caption{{\small {\bf Benchmark values for the
free parameters used in this analysis.} The rest of the parameters, group theory parameters
and the angle $\theta_L$, is determined as much as possible by one of the representative examples, found in section~\ref{sec50}. The fixed values of the parameters $s$ and $v$, respectively, are given for each single example in sections~\ref{sec53}-\ref{sec55}.   
}}
\label{benchmark choice_alt}
\end{table}

\subsection{Case 1)}
\label{sec53}

In the following, we discuss the results for low-scale leptogenesis for Case 1), $n=10$ and different values of $s$. Apart from scanning the parameter space spanned by the mass scale $M$ of the RH neutrinos and the angle $\theta_R$ for different values of the splitting $\kappa$, we also explore the impact of different initial conditions, of the light neutrino mass spectrum as well as of the splitting $\lambda$.

\begin{figure}
\begin{subfigure}{.5\textwidth}
	\centering
	\includegraphics[width = \textwidth]{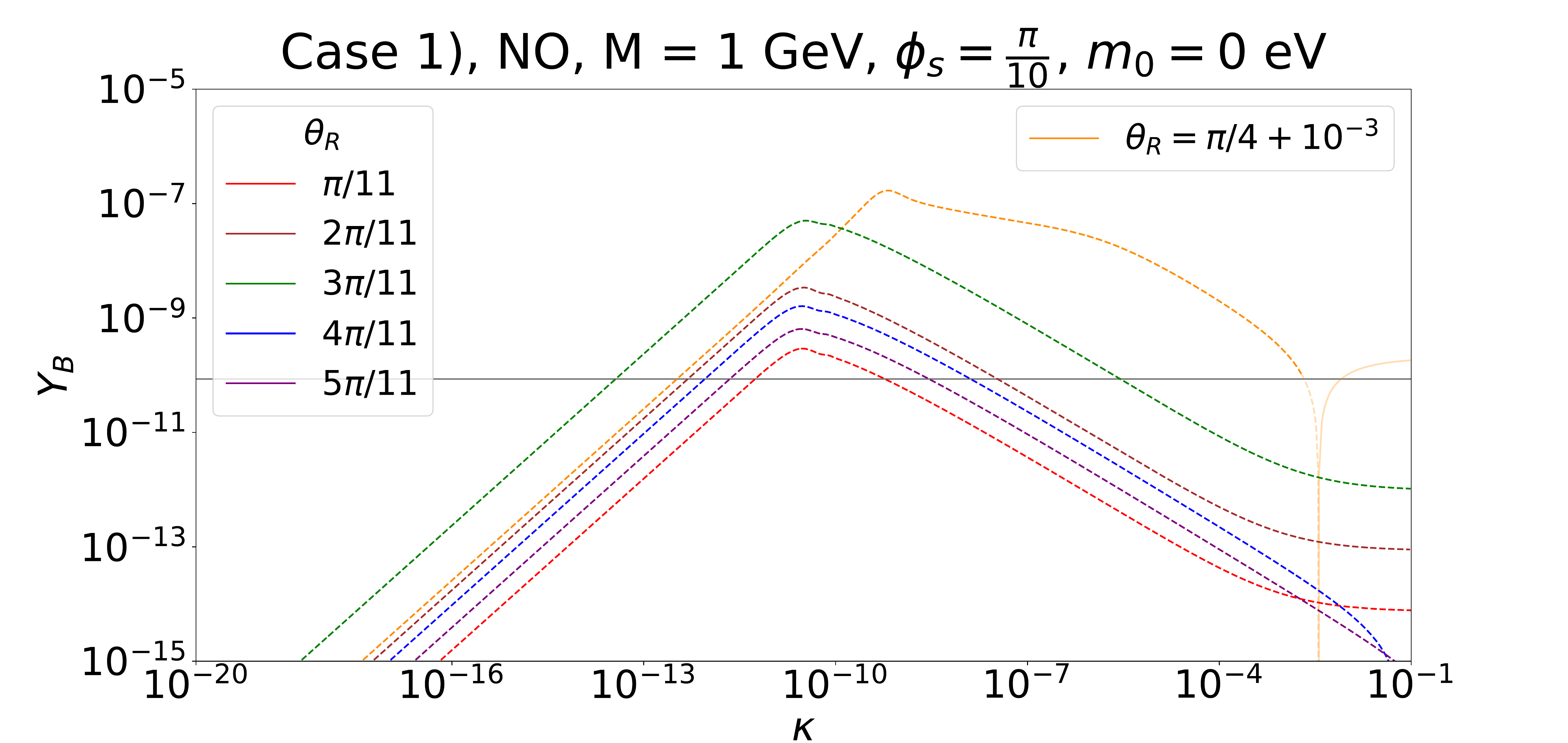}
	\caption{Vanishing initial conditions.}
\end{subfigure}
\begin{subfigure}{.5\textwidth}
	\centering
	\includegraphics[width = \textwidth]{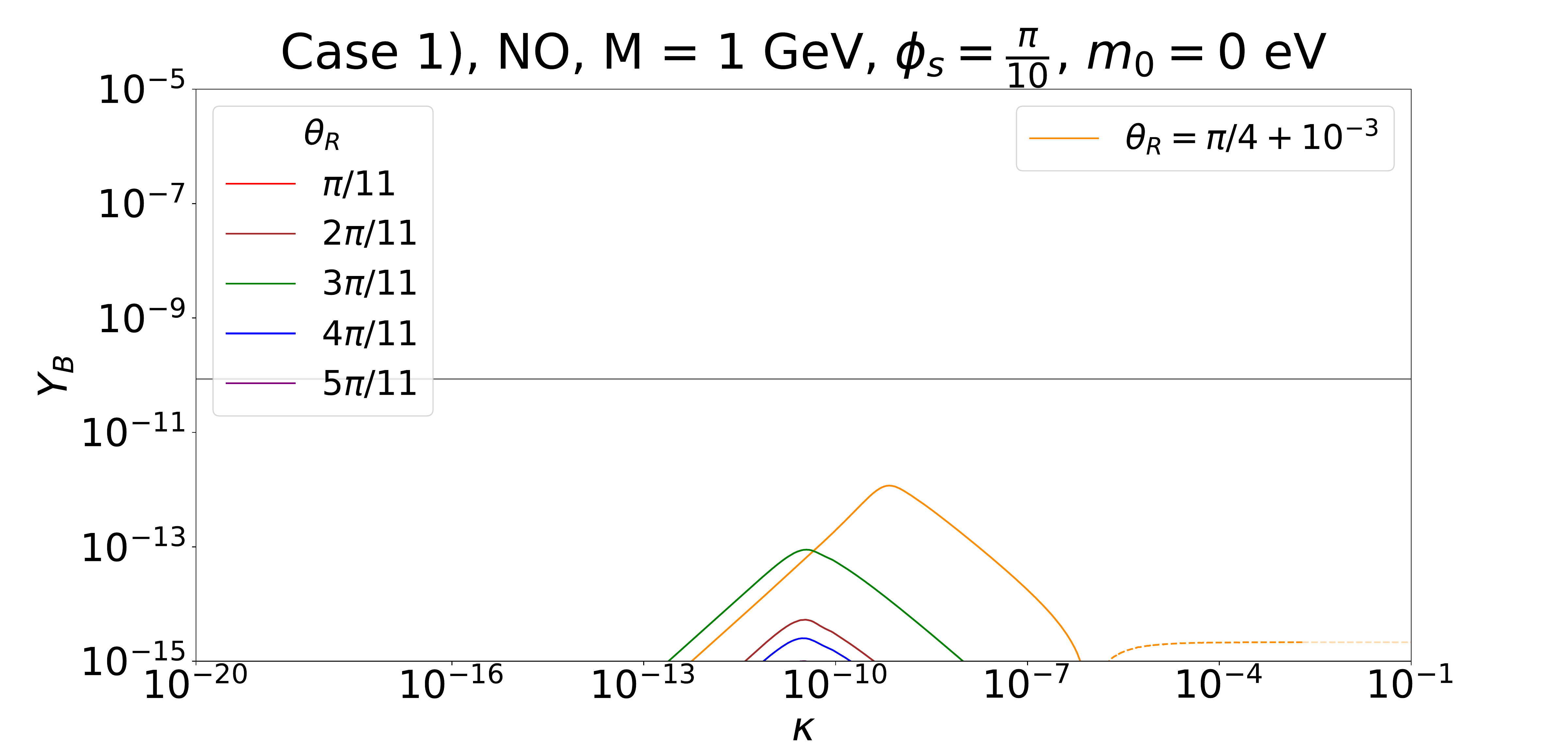}
	\caption{Thermal initial conditions.}
\end{subfigure}
\begin{subfigure}{.5\textwidth}
	\centering
	\includegraphics[width = \textwidth]{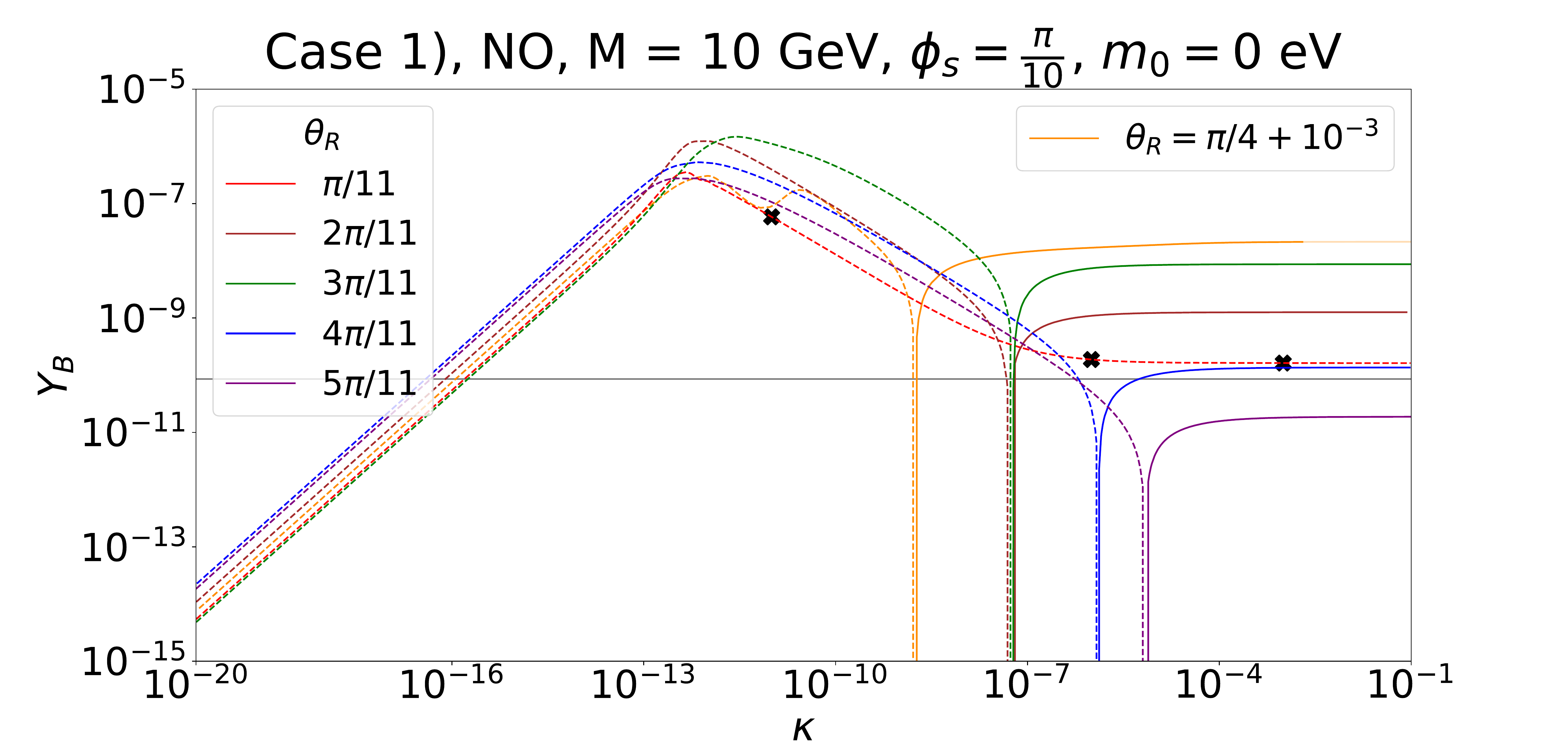}
	\caption{Vanishing initial conditions.}
	\label{NO kappa BAU 10 GeV}
\end{subfigure}
\begin{subfigure}{.5\textwidth}
	\centering
	\includegraphics[width = \textwidth]{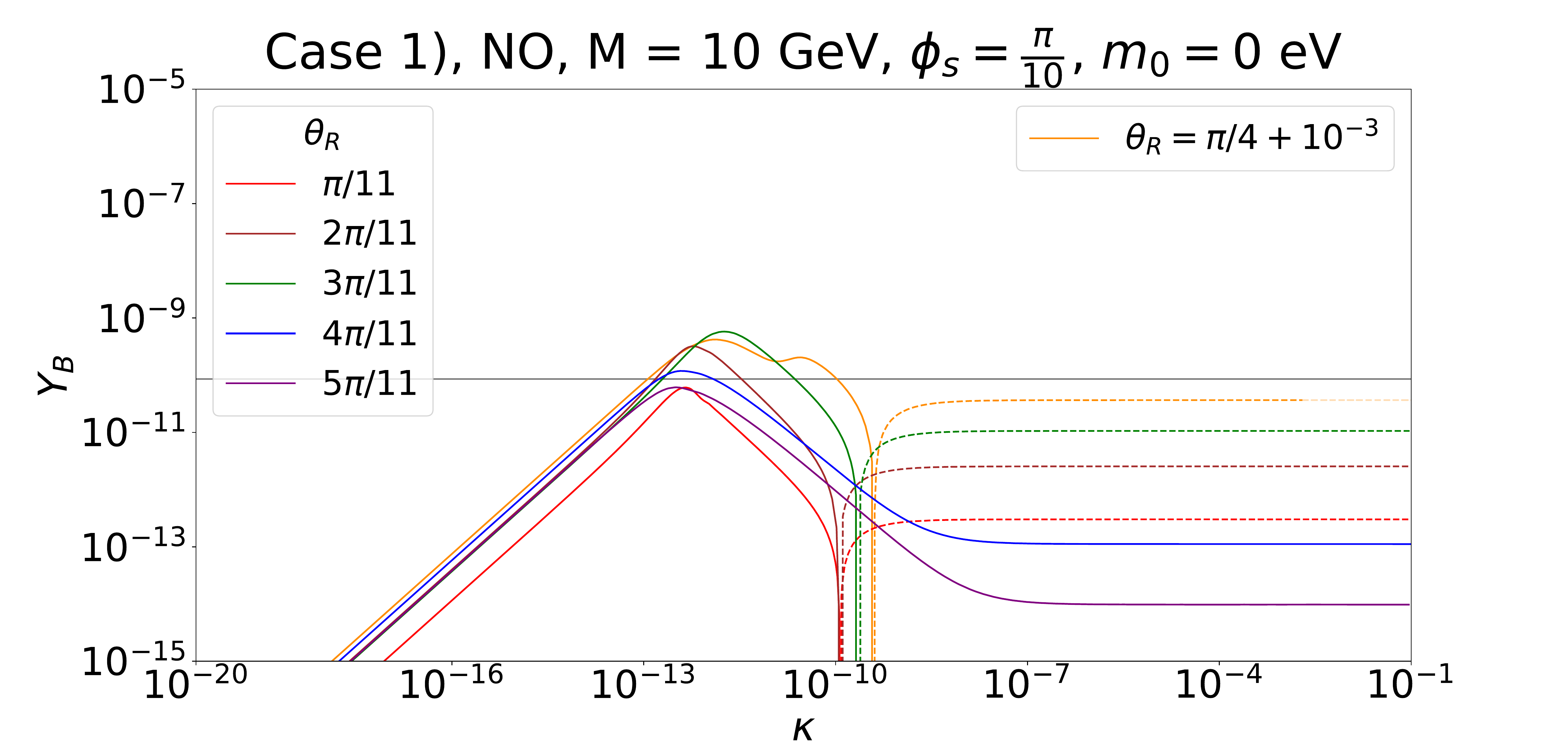}
	\caption{Thermal initial conditions.}
\end{subfigure}
\begin{subfigure}{.5\textwidth}
	\centering
	\includegraphics[width = \textwidth]{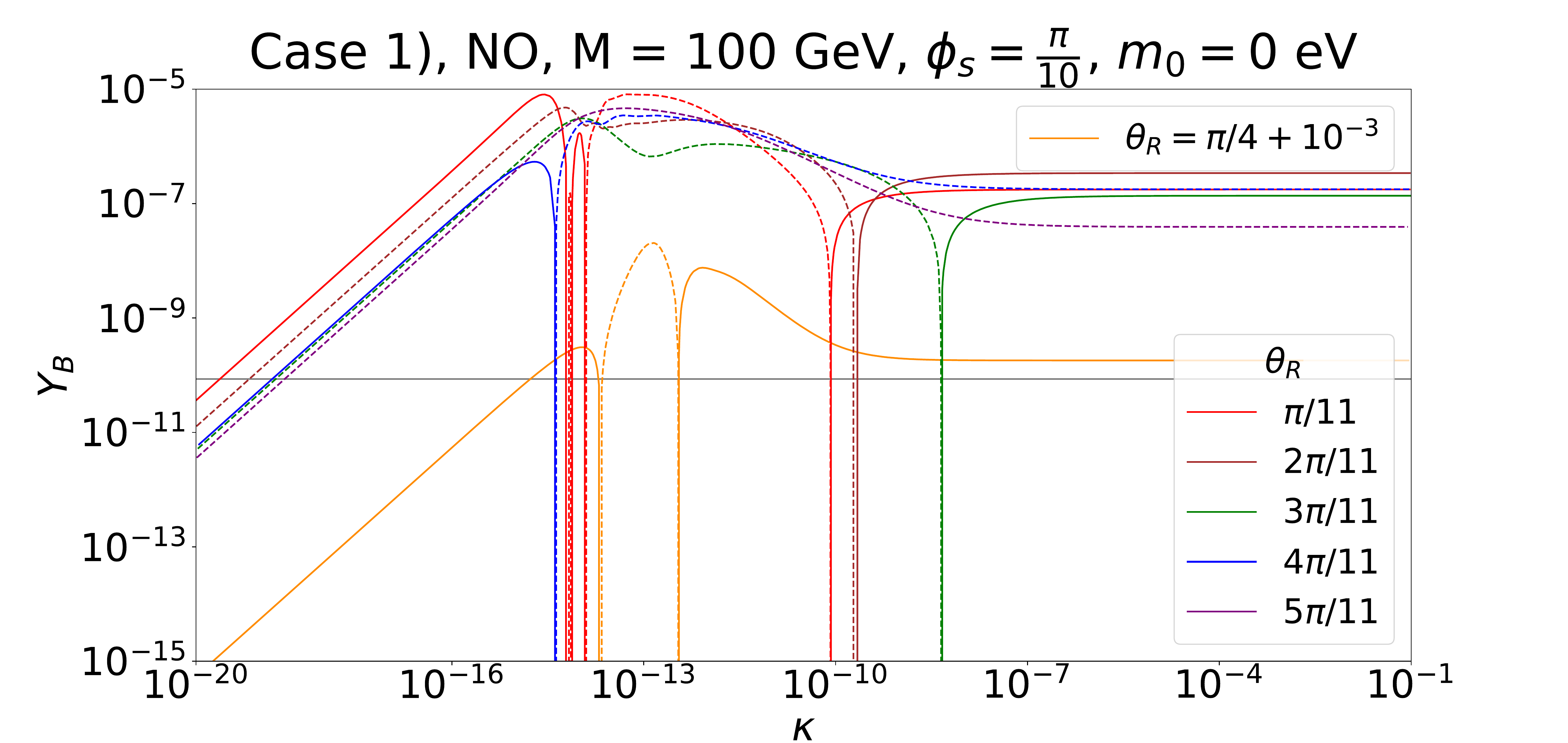}
	\caption{Vanishing initial conditions.}
\end{subfigure}
\begin{subfigure}{.5\textwidth}
	\centering
	\includegraphics[width = \textwidth]{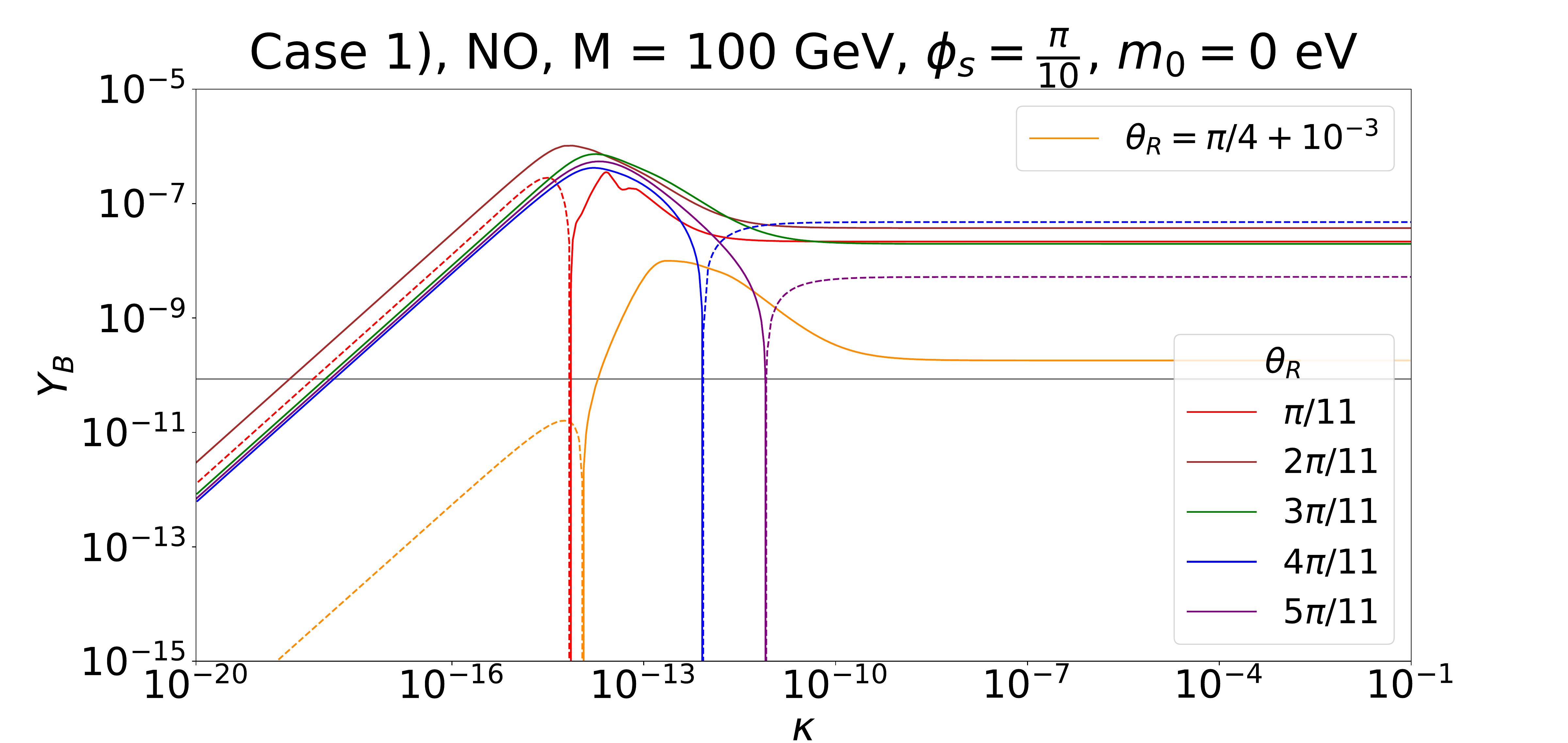}
	\caption{Thermal initial conditions.}
\end{subfigure}
\begin{subfigure}{.5\textwidth}
	\centering
	\includegraphics[width = \textwidth]{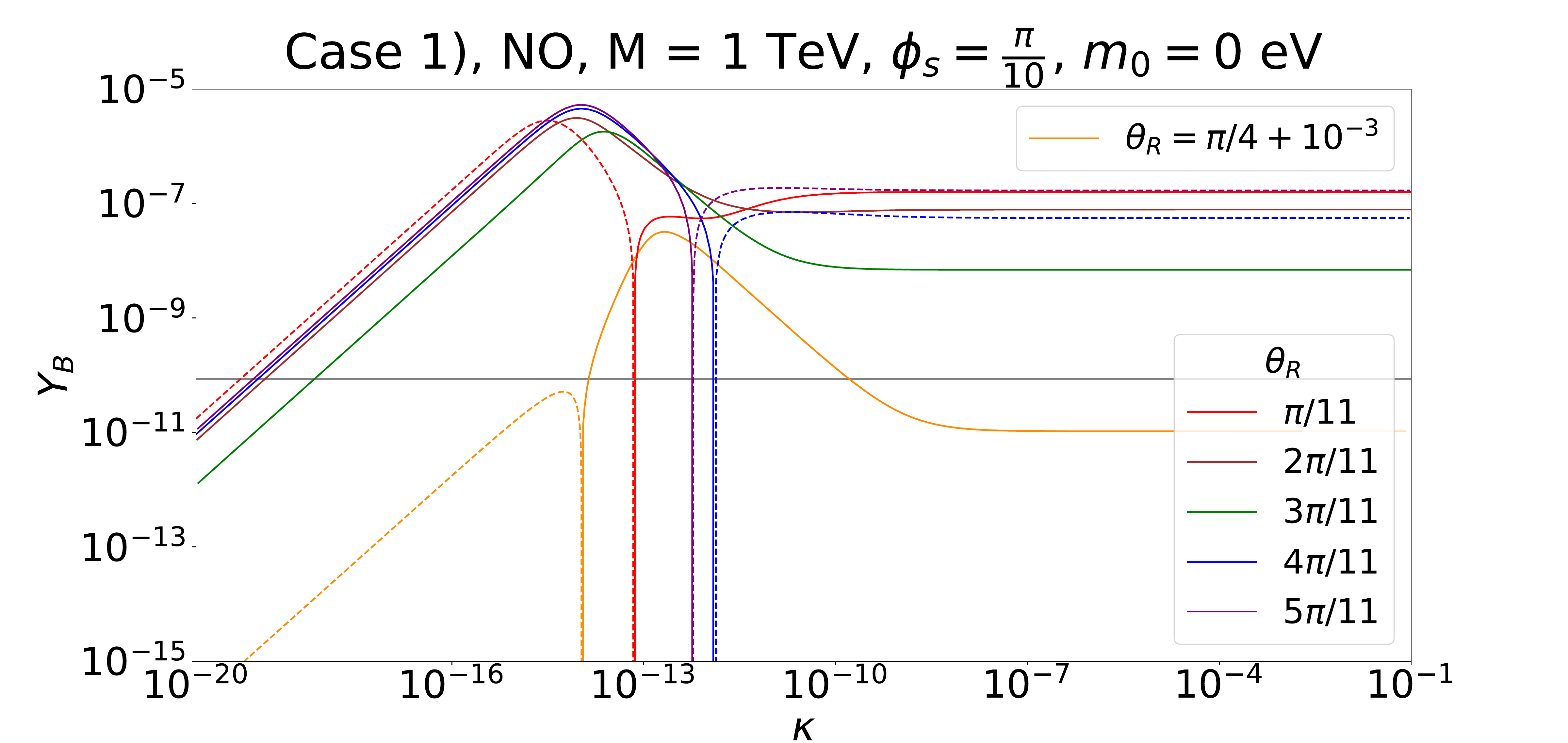}
	\caption{Vanishing initial conditions.}
\end{subfigure}
\begin{subfigure}{.5\textwidth}
	\centering
	\includegraphics[width = \textwidth]{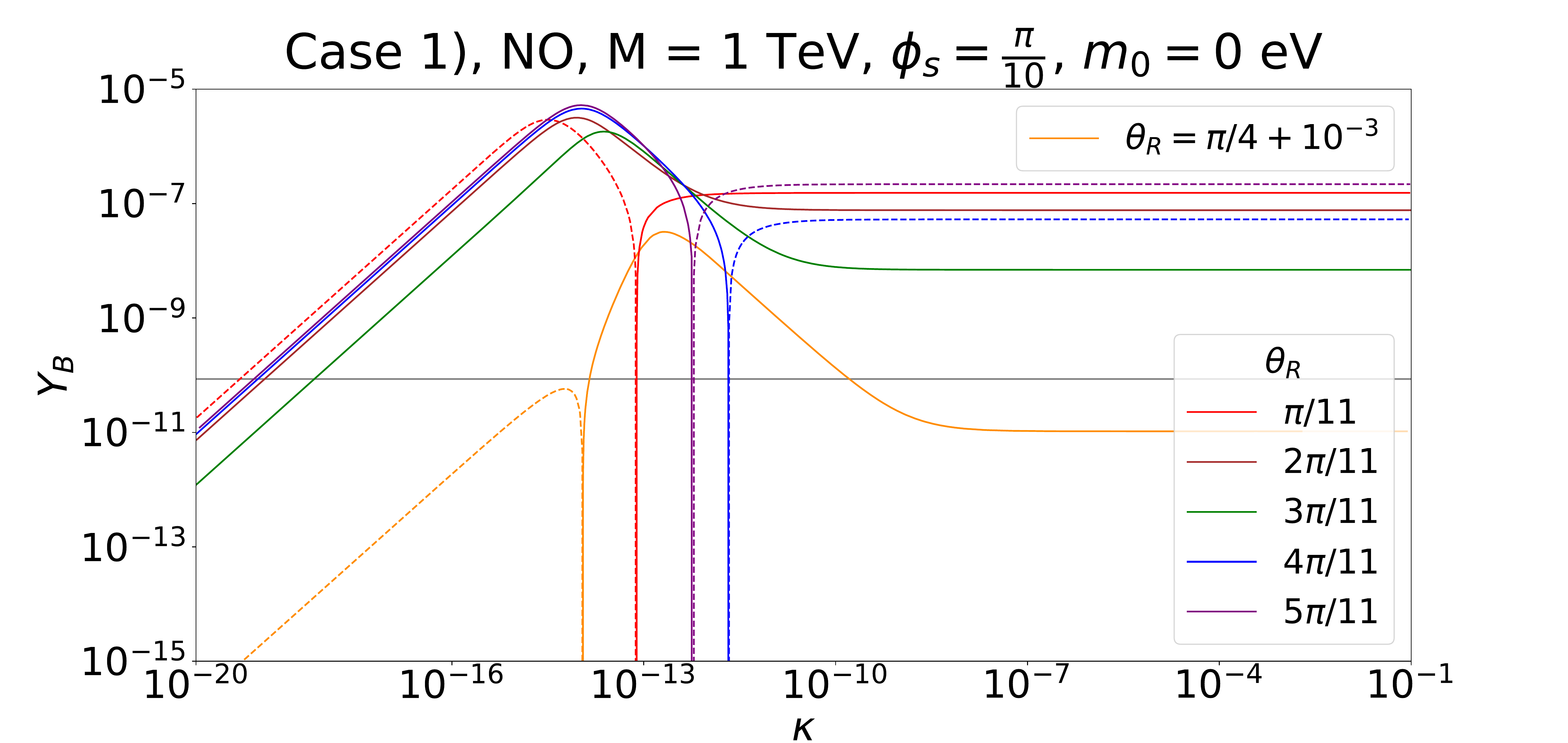}
	\caption{Thermal initial conditions.}
\end{subfigure}
\caption{{\small {\bf Case 1)} BAU ($Y_B$) as function of the splitting $\kappa$ for a Majorana mass $M=1$ GeV, $M=10$ GeV, $M=100$ GeV and $M=1$ TeV for both vanishing (left plots) and thermal initial conditions (right plots). For each of these choices, different values of the angle $\theta_R$ have been studied. Light neutrino masses have strong NO. The index $n$ of the flavour group is $n=10$. The parameter $s$ is fixed to $s=1$ corresponding to $\phi_s=\frac{\pi}{10}$.
Both negative (dashed lines) as well as positive (continuous lines) values of the BAU are represented. The grey line indicates the observed value of the BAU, $Y_B\approx 8.6 \cdot 10^{-11}$. The black crosses in plot (c) refer to particular choices of $\theta_R$ and $\kappa$ that are used as reference values in some of the following plots, e.g. in Fig.~\ref{NO phi BAU 10 GeV combined}. The choice $\theta_R=\frac{\pi}{4}+10^{-3}$ (orange curve) corresponds to a value of $\theta_R$ close to a special value, for details see text. Faint lines indicate that the condition in Eq.~\eqref{eq:kappacorrectionscriterion} is not fulfilled.}}
\label{NO kappa BAU Case I different masses}
\end{figure}

\paragraph{Importance of heavy neutrino mass scale.}
In Fig.~\ref{NO kappa BAU Case I different masses} we illustrate the dependence of the BAU on the splitting $\kappa$ for the four  benchmark values of the heavy neutrino mass scale $M$ and for both vanishing and thermal initial conditions.
When comparing the plots for $M=1$ GeV and $M=10$ GeV, we can see a clear difference in the overall scale of the BAU, irrespective of the splitting $\kappa$, caused by the smaller overall scale of the Yukawa couplings. This is confirmed, when comparing to the result for a value of $\theta_R$ close to a special value, $\theta_R=\pi/4+10^{-3}$, which leads to overall larger Yukawa couplings, and hence larger values of the BAU.
For intermediate masses, $1$ GeV $\lesssim M \lesssim 1$ TeV, we find a rather complicated dependence on the splitting $\kappa$, due to  the competing LNV and LFV processes.
Similar conclusions hold, when comparing with Fig.~\ref{NO Mass U2 Case1}, in which we show the dependence on both the RH neutrino mass scale $M$ and the total mixing angle $U^2$.
For masses $M>2$ TeV, all the dependence on the RH neutrino mass scale can be absorbed into a re-scaling of the mass splittings.
We can observe the approximate scaling of the BAU for $M\gtrsim 2$ TeV and $\kappa \gtrsim 10^{-6}$, which corresponds to a plateau in $\kappa$, and where a re-scaling of the mass splitting does not affect the BAU.

\paragraph{Impact of heavy neutrino mass splittings.}

In contrast to the remaining cases,
in Case 1) setting both $\kappa=0$ and $\lambda=0$ can never lead to successful baryogenesis.
We can verify this by evaluating the CP-violating combination $C_{\mathrm{DEG},\alpha}$, see section~\ref{sec62}.
In Fig.~\ref{NO kappa BAU Case I different masses} we can see the dependence of the BAU on the splitting $\kappa$.
As expected from the perturbative expressions in section~\ref{sec62},
for small values of $\kappa$, the BAU grows linearly with it,
until it reaches the resonance peak, after which it begins to decrease and reaches a plateau, when the resonance condition, see   Eq.~\eqref{eq:resCondition}, is no longer satisfied for $\kappa \gtrsim \Gamma_N \,  E_N/M^2$.
While one might expect that a large value of $\kappa$ removes the resonant enhancement of the BAU, we find that this is in general not the case, as two of the heavy neutrinos remain degenerate in mass irrespective of the value of $\kappa$. This plateau can also be observed in Figs.~\ref{NO Mass U2 Case1} and \ref{kappa lambda grid}. As described in section~\ref{sec6}, a useful criterion to understand whether the BAU reaches a plateau for large $\kappa$ (and $\lambda=0$) or not is to evaluate the reduced mass-degenerate CP-violating combination $C^{(23)}_{\mathrm{DEG},\alpha}$, defined in Eq.~\eqref{eq:reducedCPcombinations}. If  this combination is non-zero, there remains CP violation within the mass-degenerate two-by-two subsystem and one can expect a plateau for large $\kappa$. For Case 1), indeed, $C^{(23)}_{\mathrm{DEG},\alpha}$ do not vanish, see Eq.~\eqref{eq:reducedCPVcombinationCase1}.

\begin{figure}
	\centering
	\begin{subfigure}{.496\textwidth}
		\centering
		\includegraphics[width = \textwidth]{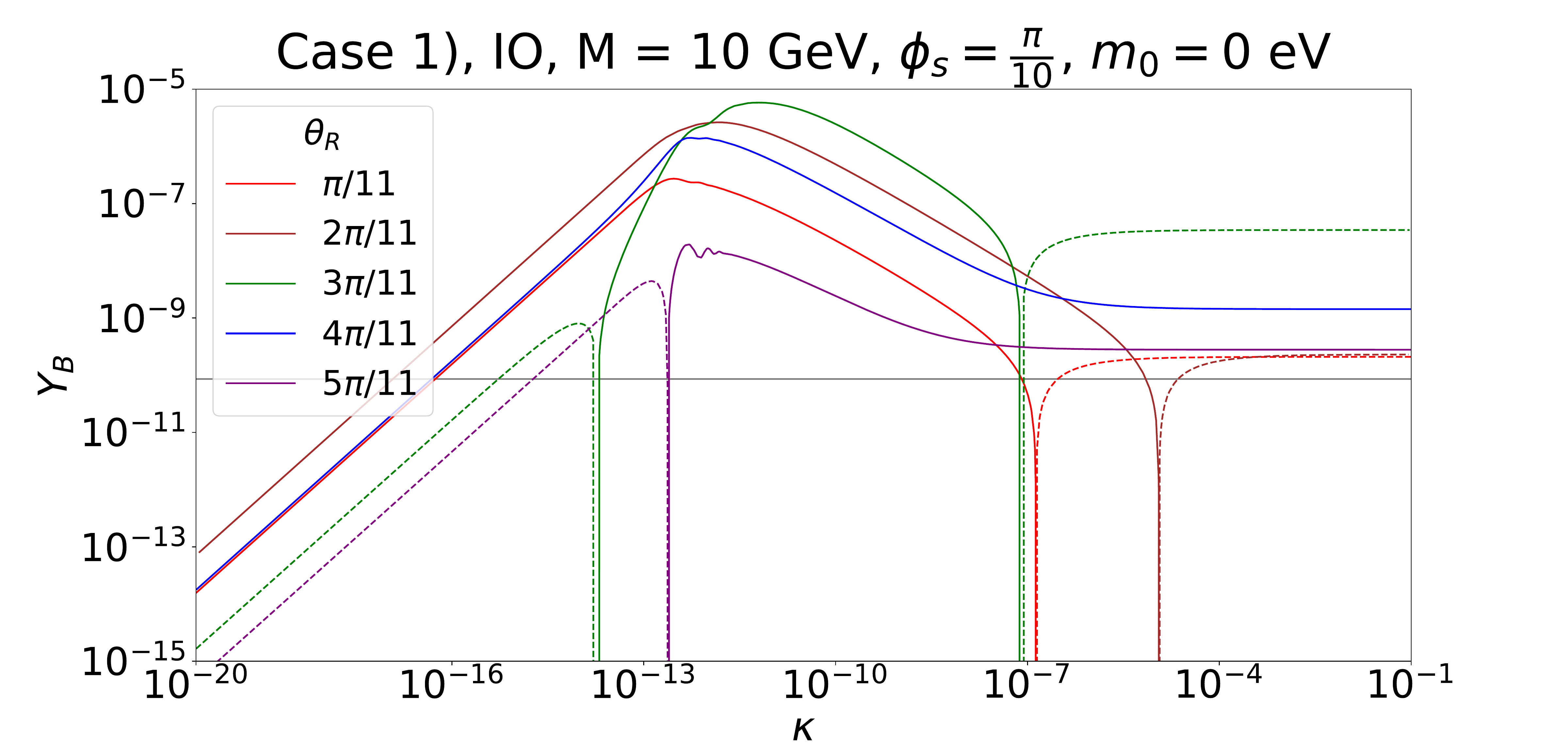}
		\caption{Vanishing initial conditions.}
		\label{kappa BAU 10 GeV IO caseI}
	\end{subfigure}
	\begin{subfigure}{.496\textwidth}
		\centering
		\includegraphics[width = \textwidth]{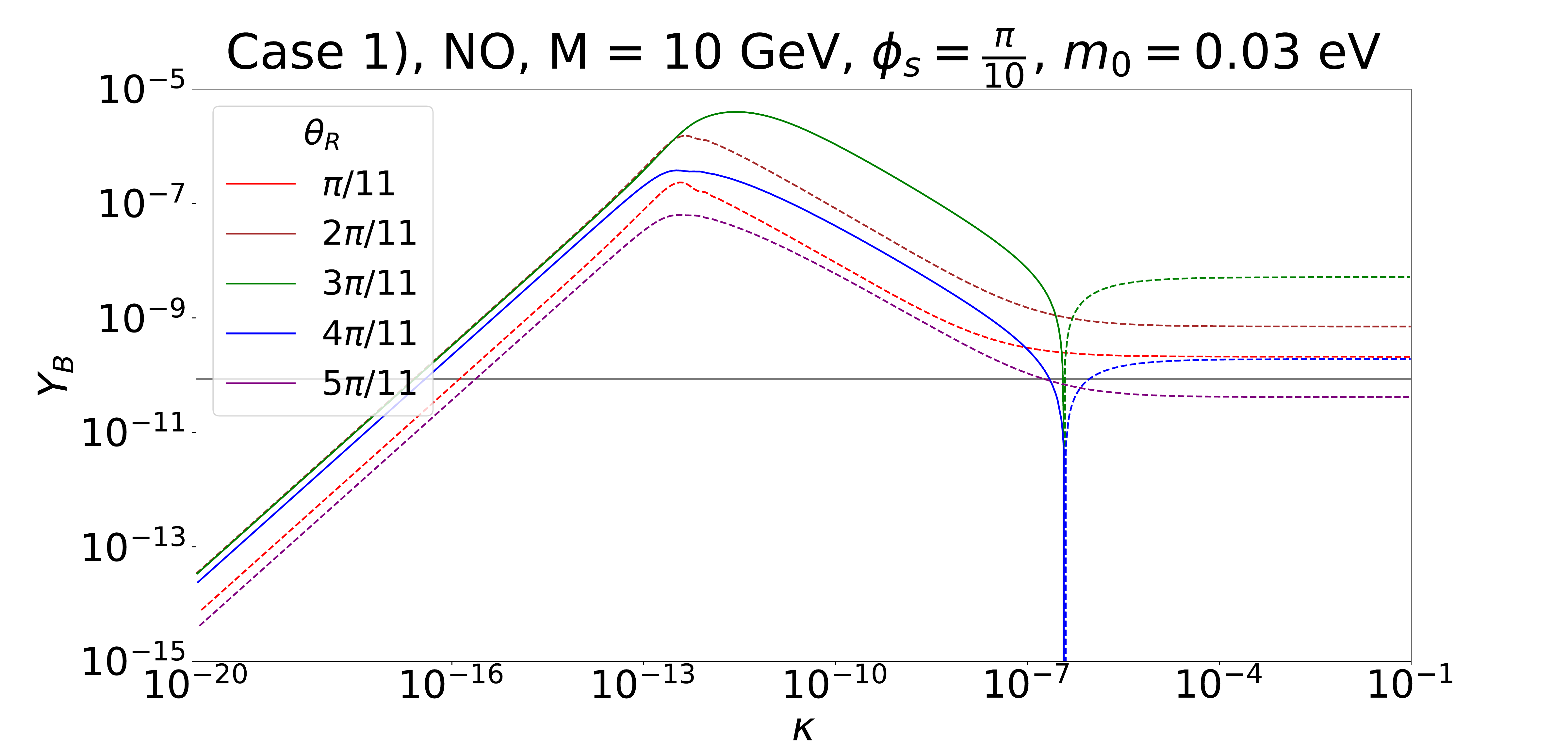}
		\caption{Vanishing initial conditions.}
		\label{kappa BAU 10 GeV NO massive case I}
	\end{subfigure}
\caption{{\small {\bf Case 1)} $Y_B$ as function of $\kappa$ for $M=10$ GeV for light neutrino masses with strong IO (left plot) and for light neutrino masses with NO, but non-zero $m_0$, $m_0=0.03$ eV (right plot). For the remaining choices see Fig.~\ref{NO kappa BAU Case I different masses}.
\label{kappa BAU 10 GeV IO and NO massive case I}
}}
\end{figure}

\paragraph{Dependence on initial conditions.}
The deviation from equilibrium required for successful leptogenesis can be realised in two ways: either during the freeze-in and production of the RH neutrinos or during their freeze-out and decay.
To distinguish these two regimes, we study two types of initial conditions, thermal and vanishing initial RH neutrino abundances.
If we assume a vanishing initial abundance of RH neutrinos, both freeze-in and freeze-out contribute to leptogenesis,
while for thermal initial abundances all of the BAU is generated during freeze-out.
Leptogenesis from freeze-out is most efficient for $M\gtrsim T$, and is therefore typically associated with RH neutrino masses above the electroweak scale.
Nonetheless, the expansion of the Universe can cause the necessary deviation from equilibrium even for GeV-scale RH neutrino masses~\cite{Hambye:2016sby,Klaric:2020phc,Drewes:2021nqr}.
Similarly, for RH neutrino masses above the electroweak scale washout is strong in general, and any asymmetries generated during freeze-in are typically erased. This can be avoided if the RH neutrinos couple hierarchically to the different lepton flavours~\cite{Garbrecht:2014bfa}.

In Fig.~\ref{NO kappa BAU Case I different masses} we show the dependence of the BAU on the initial conditions (vanishing and thermal RH neutrino abundances).
We find that baryogenesis with thermal initial conditions is possible from around $M\sim 5$ GeV for $\kappa \sim 10^{-11}$, and around $M \gtrsim 10$ GeV for larger values of  the splitting $\kappa$, $\kappa \gtrsim 10^{-6}$, as shown by the faint lines in Fig.~\ref{NO Mass U2 Case1}.
Finally, in Fig.~\ref{NO kappa BAU Case I different masses},
for RH neutrino masses above the electroweak scale we find that both thermal and vanishing initial conditions can lead to identical BAU, since our choice of  benchmark values  generically leads to strong washout of the asymmetries generated during freeze-in.

\paragraph{Role of light neutrino mass spectrum.} For the different choices of parameters, we also study the role of the light neutrino mass spectrum, i.e.~the dependence of the results for the BAU on the mass ordering of the light neutrinos, NO or IO, as well as on the lightest neutrino mass $m_0$, $m_0=0$ or $m_0$ being close to its upper bound, see Eq.~(\ref{eq:m0max}), indicated by cosmological observations~\cite{Planck:2018vyg}. As one can see, the results for light neutrino masses with strong IO show qualitatively the same features as those for light neutrino masses with NO, compare plot (a) in Fig.~\ref{kappa BAU 10 GeV IO and NO massive case I} with plot (c) in Fig.~\ref{NO kappa BAU Case I different masses}. Similarly, a non-zero value of the lightest neutrino mass $m_0$, see Fig.~\ref{kappa BAU 10 GeV IO and NO massive case I} (plot (b)), does not change the qualitative behaviour of the BAU. We, however, observe that for (strong) IO, more changes of the sign of the BAU are expected than for light neutrino masses with strong NO. A possible explanation is that the couplings $y_f$,
$f=1,2,3$, are in general larger and less hierarchical in the case of (strong) IO than for strong NO.
For completeness, we show the corresponding plots for $M=1$ GeV, $M=100$ GeV and $M=1$ TeV for light neutrino masses with strong IO
and for light neutrino masses with NO and $m_0=0.03$ eV in Fig.~\ref{kappa BAU Case I different masses IO/Massive m0} in appendix~\ref{appF1}.

\begin{figure}[t!]
	\begin{subfigure}{.5\textwidth}
		\includegraphics[width = 1.05\textwidth]{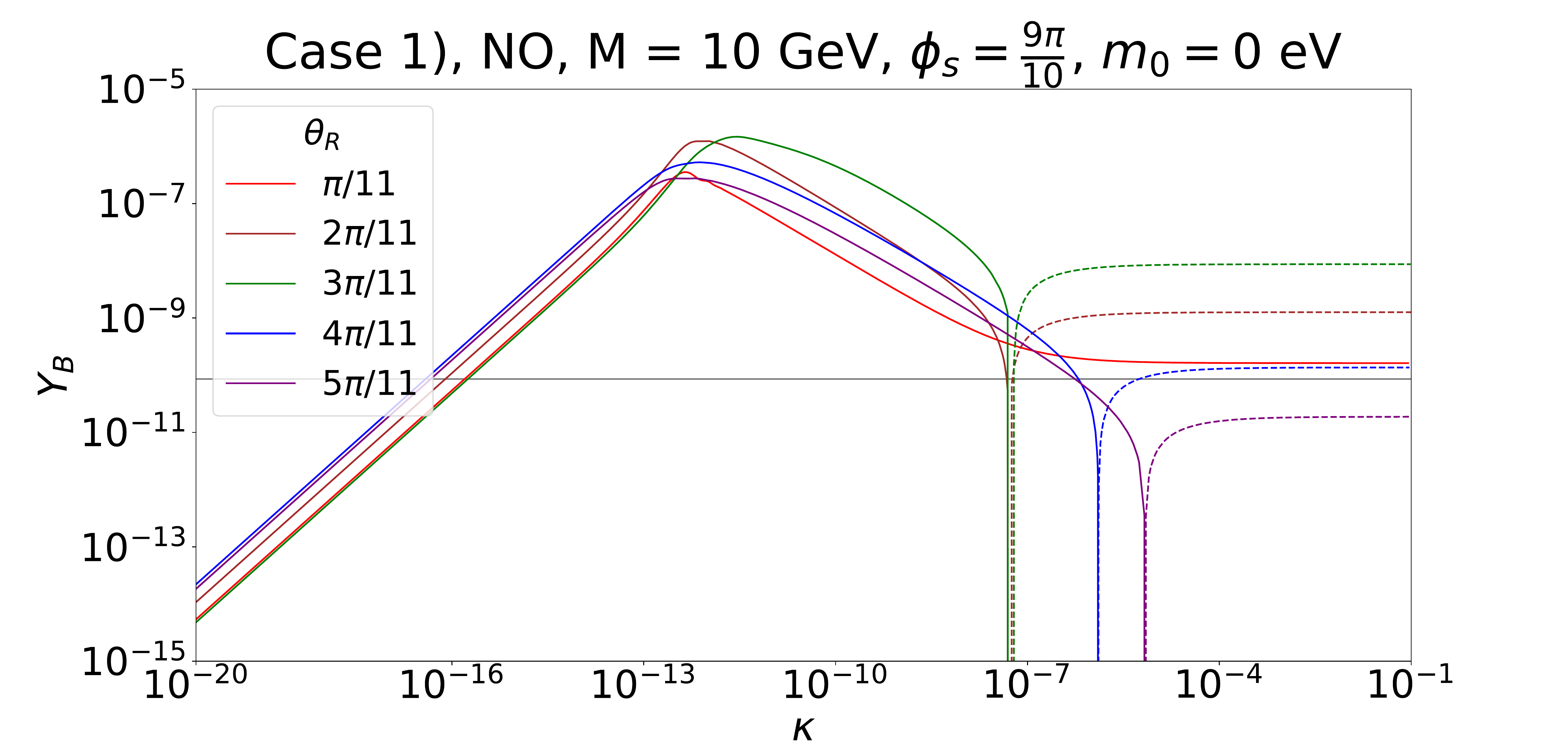}
		\caption{Vanishing initial conditions.}
		\vspace{5cm}
	\end{subfigure}
	\begin{subfigure}{.5\textwidth}
		\centering
		\includegraphics[width = 1.05\textwidth]{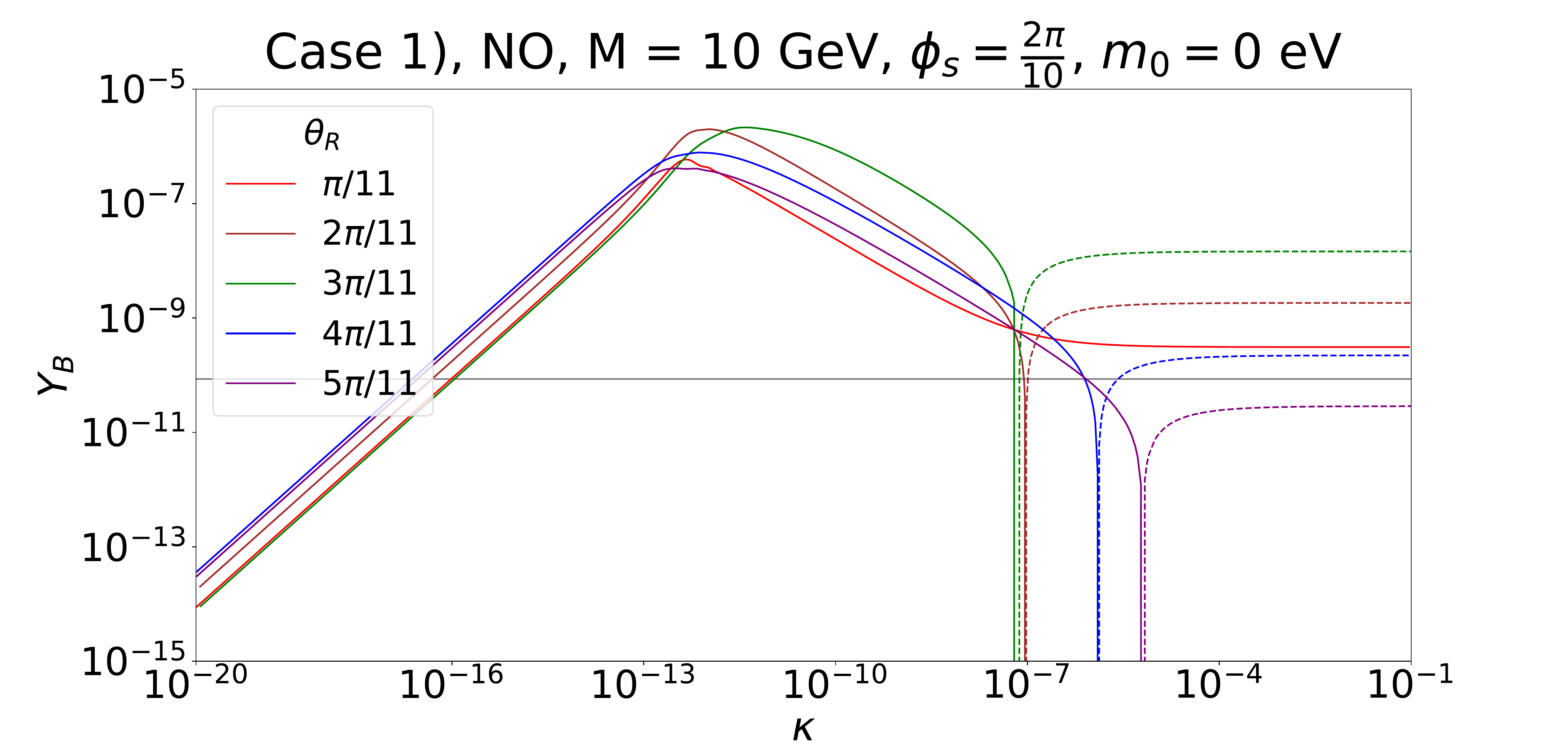}
		\caption{Vanishing initial conditions.}
		\includegraphics[width = 1.05\textwidth]{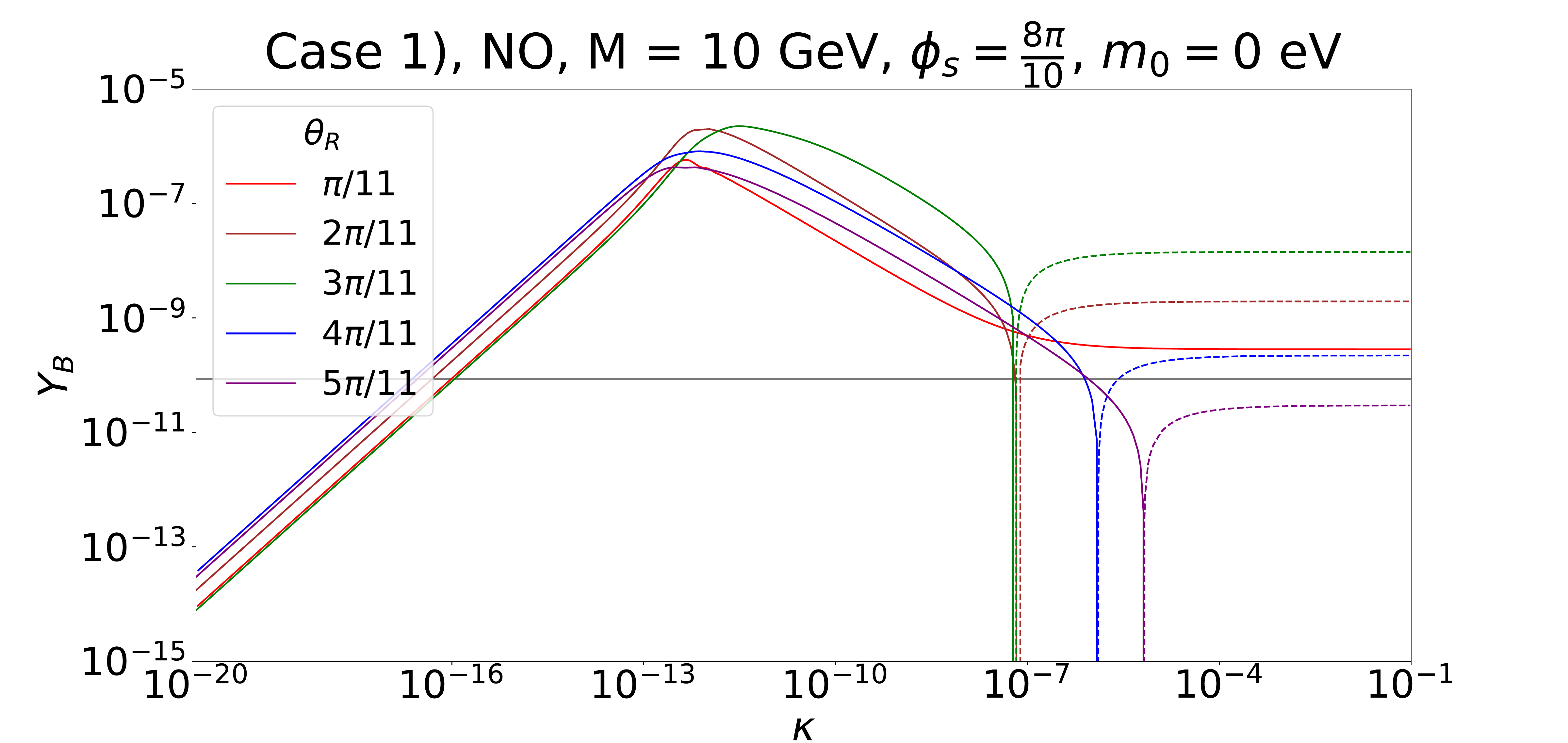}
		\caption{Vanishing initial conditions.}
	\end{subfigure}
	\caption{{\small {\bf Case 1)} $Y_B$ as function of $\kappa$ for $M=10$ GeV and different values of $\phi_s$, $\phi_s \in \frac{\pi}{10}\cdot \{2,8,9\}$. For the remaining choices see Fig.~\ref{NO kappa BAU Case I different masses}.
}}
\label{NO kappa BAU Case I different phis}
\end{figure}

\begin{figure}[t!]
	\centering
	\begin{subfigure}{.49\textwidth}
		\includegraphics[width = 1.05\textwidth]{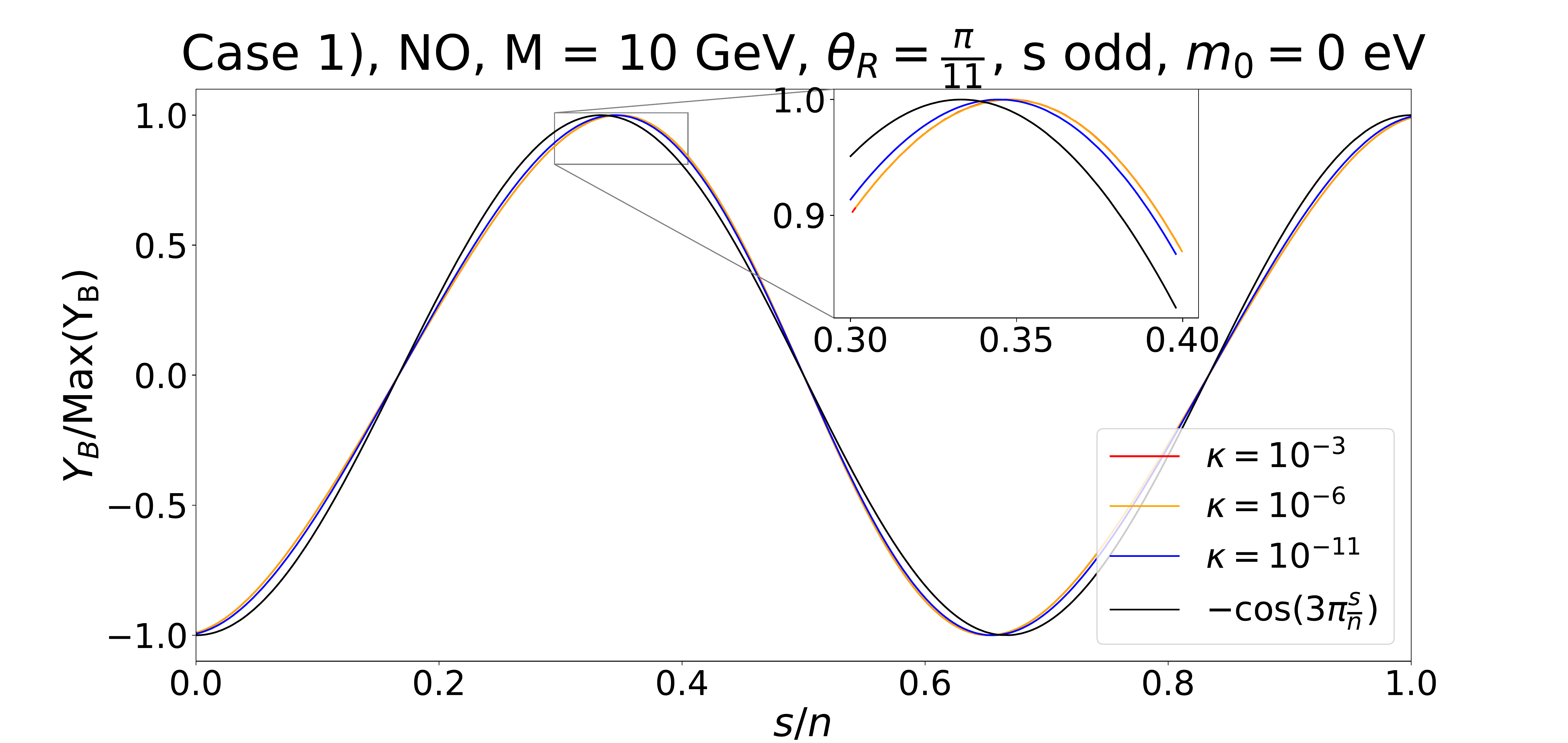}
		\caption{Vanishing initial conditions.}
	\end{subfigure}
	\begin{subfigure}{.49\textwidth}
		\includegraphics[width = 1.05\textwidth]{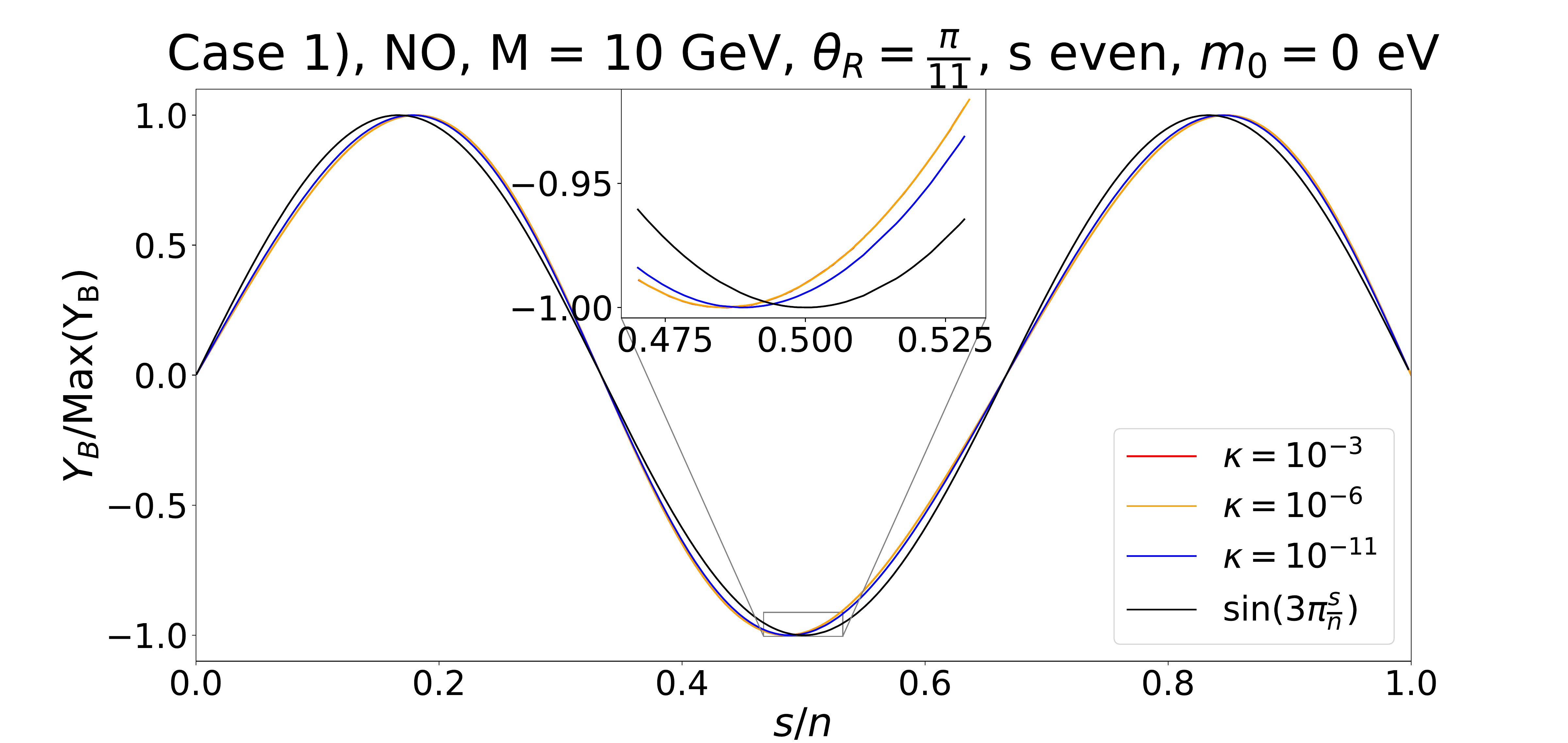}
		\caption{Vanishing initial conditions.}
	\end{subfigure}
\caption{{\small {\bf Case 1)} $Y_B$ as function of $\frac{s}{n}$, treated as continuous parameter, for odd (left plot) and even (right plot) values of $s$, respectively. The Majorana mass $M$ is chosen as $M=10$ GeV and three different values of $\kappa$, $\kappa \in \{10^{-3},10^{-6},10^{-11}\}$, are used, see the red, orange and blue curve, respectively. Additionally, we display the dependence on $\phi_s$ ($\frac{s}{n}$) as expected from the CP-violating combinations, compare Eqs.~(\ref{eq:CLFValphaCase1seven}) and (\ref{eq:CLNValphaCase1seven}), in black. The angle $\theta_R$ is fixed to $\theta_R=\frac{\pi}{11}$ and light neutrino masses are assumed to have strong NO.}}
\label{NO phi BAU 10 GeV combined}
\end{figure}

\paragraph{Dependence on CP transformation $X (s)$.}
We compare the dependence of the BAU  on the splitting $\kappa$ for different values of $\phi_s$ in Figs.~\ref{NO kappa BAU Case I different masses}, plot (c), and \ref{NO kappa BAU Case I different phis}.
In doing so, we fix $M=10 \, \mathrm{GeV}$, $m_0=0$ and consider light neutrino masses with NO and different values of $\theta_R$.
For $\phi_s=\frac{\pi}{10}$ ($n=10$ and $s=1$) and $\phi_s=\frac{9 \, \pi}{10}$ ($s=9$), compare Fig.~\ref{NO kappa BAU Case I different masses}  (plot (c)) and Fig.~\ref{NO kappa BAU Case I different phis} (plot (a)),
we note that they lead to a very similar functional dependence of the BAU on $\kappa$, whereas the sign of the BAU is opposite (in the plots the style of the curves changes from continuous to dashed lines and vice versa). This observation is in agreement with the dependence of the CP-violating combinations $C_{\mathrm{LFV},\alpha}$ and $C_{\mathrm{LNV},\alpha}$ on $\phi_s$, see Eqs.~(\ref{eq:CLFValphaCase1seven}) and (\ref{eq:CLNValphaCase1seven}) with $\sin 3 \, \phi_s$ being replaced by $-\cos 3 \, \phi_s$, since $-\cos 3 \, \phi_s\approx -0.588$ for $\phi_s=\frac{\pi}{10}$ and $-\cos 3 \, \phi_s\approx +0.588$ for $\phi_s=\frac{9 \,\pi}{10}$. This is true for all odd values of $s$ whose sum equals the index $n$ of the flavour symmetry.
In addition, we observe that for $s=5$, $s=\frac{n}{2}$ in general (recall the index $n$ is always even and not divisible by four),
the CP-violating combinations vanish, $-\cos 3 \, \phi_s=-\cos \frac{3 \, \pi}{2}=0$, and thus the BAU does so as well, as we have checked explicitly.
Furthermore, we can compare the plots of the BAU with respect to $\kappa$ for two even values of $s$ (and otherwise the same parameters as chosen before), e.g.~$\phi_s=\frac{2 \, \pi}{10}$ ($s=2$) and  $\phi_s=\frac{8 \, \pi}{10}$ ($s=8$), see the right plots (plot (b) and (c)) in Fig.~\ref{NO kappa BAU Case I different phis}. As one can see, these plots are (nearly) identical, both in size and sign of the BAU. This can also be understood with the help of the CP-violating combinations $C_{\mathrm{LFV},\alpha}$ and $C_{\mathrm{LNV},\alpha}$ which depend on $\sin 3 \, \phi_s$ for $s$ even, i.e.~$\sin 3 \, \phi_s \approx +0.951$ for both $\phi_s=\frac{2 \, \pi}{10}$ and $\phi_s=\frac{8 \, \pi}{10}$. This happens, like for all odd $s$, for all even $s$ whose sum equals the index $n$. Knowing the dependence $\sin 3 \, \phi_s$ for $s$ even, we additionally confirm the expectation that for $s=0$ the generated BAU vanishes (within numerical precision).
With Fig.~\ref{NO kappa BAU Case I different phis}, we can eventually compare the results between odd and even values of $s$ and find for $\phi_s=\frac{9 \, \pi}{10}$ ($s=9$) and $\phi_s=\frac{2 \, \pi}{10}$ ($s=2$) (plot (a) and (b))
that the BAU should have the same sign, but be slightly larger for $\phi_s=\frac{2 \, \pi}{10}$, since $-\cos 3 \, \phi_s\approx +0.588$ for $\phi_s=\frac{9 \,\pi}{10}$ and $\sin 3 \, \phi_s \approx +0.951$ for $\phi_s=\frac{2 \,\pi}{10}$ with the dependence on the rest of the parameters being the same.

In Fig.~\ref{NO phi BAU 10 GeV combined}, we check the dependence of the BAU on $\phi_s$ (treating $s/n$ as continuous parameter) for $s$ odd (plot (a)) and for $s$ even (plot (b)) for a fixed value of the Majorana mass $M$, $M=10$ GeV, a fixed value of $\theta_R$, $\theta_R=\frac{\pi}{11}$, and different sizes of the splitting $\kappa$, ranging between $10^{-11}$ and $10^{-3}$.
As one can see, the dependence $-\cos 3 \, \phi_s$ for $s$ odd and $\sin 3 \, \phi_s$ for $s$ even, expected from the CP-violating combinations, is very well reproduced and only small deviations arise, compare the black curve with the coloured curves in Fig.~\ref{NO phi BAU 10 GeV combined}.
These deviations cannot be understood in a simple way analytically, since they result from the dynamics of the system, in particular from washout effects, see section~\ref{sec51}, and are sensitive to the chosen Majorana mass $M$, the value of $\theta_R$ and the splitting $\kappa$.
Indeed, in Fig.~\ref{NO phi BAU 10 GeV combined} one sees that increasing $\kappa$ leads to larger deviations.
All observations, made here for light neutrino masses with strong NO, are qualitatively confirmed for the case of light neutrino masses with strong IO. However, we emphasise that the described deviations are in general larger, as can be seen in Fig.~\ref{IO phi BAU 10 GeV Case 1} in appendix~\ref{appF1}. In this figure, we show the BAU as function of $\phi_s$ ($s/n$) for $s$ odd, three values of $\kappa$ and two different values of $\theta_R$ in order to underline that the size of the deviations noticeably depends on the splitting $\kappa$ as well as the angle $\theta_R$.

As shown in~\cite{Hagedorn:2014wha}, only the Majorana phase $\alpha$, appearing in the PMNS mixing matrix, is non-trivial and its sine depends on $\phi_s$ through $\sin 6 \, \phi_s$ for all values of $s$ such that no CP violation arises for $s=0$ and $s=\frac{n}{2}$ ($s=5$ for our choice of the index $n$, $n=10$). This is also true for the BAU, as we have discussed.
\begin{figure}
	\begin{subfigure}{.5\textwidth}
		\centering
		\includegraphics[width = \textwidth]{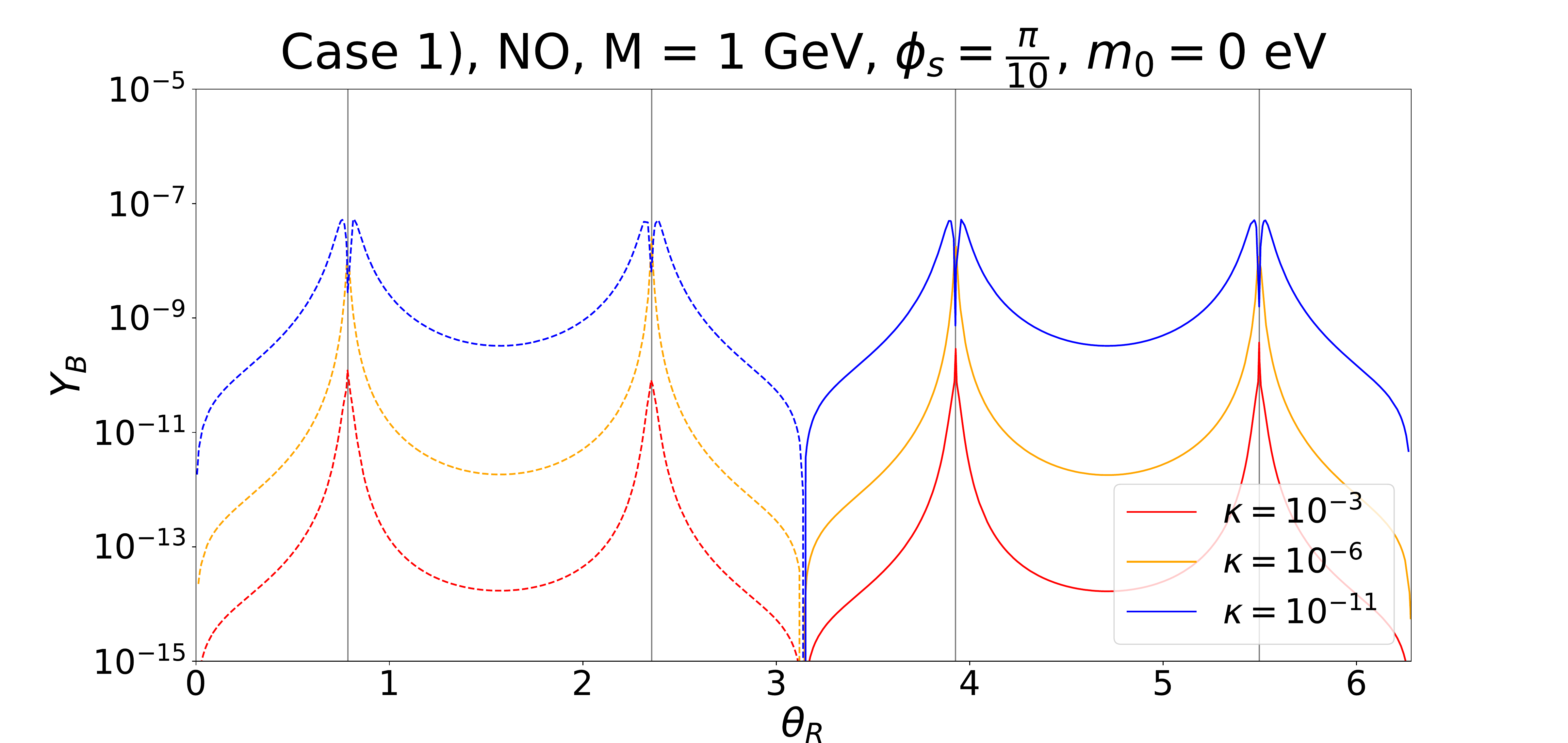}
		\caption{Vanishing initial conditions.}
	\end{subfigure}
	\begin{subfigure}{.5\textwidth}
		\centering
		\includegraphics[width = \textwidth]{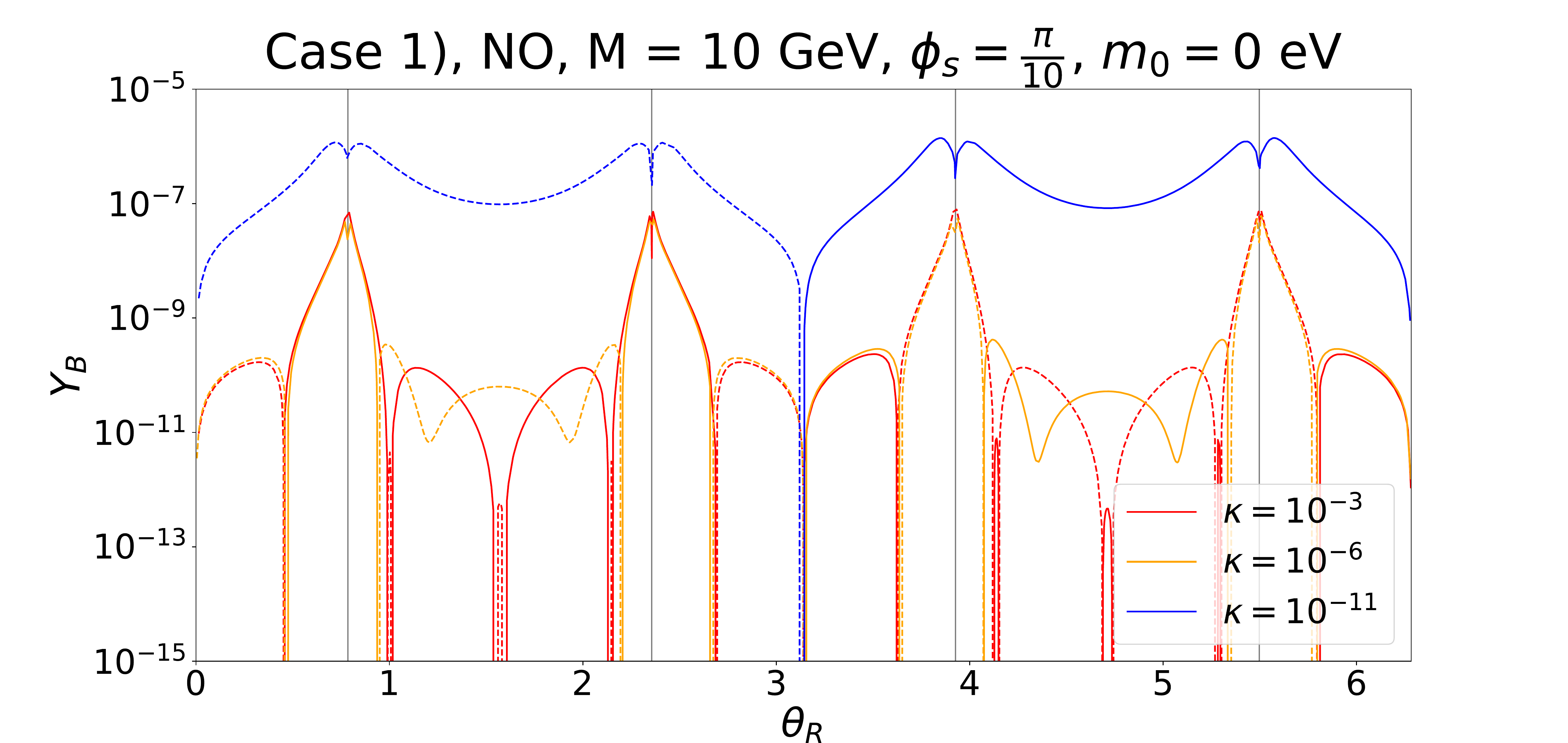}
		\caption{Vanishing initial conditions.}
	\end{subfigure}
	\begin{subfigure}{.5\textwidth}
		\centering
		\includegraphics[width = \textwidth]{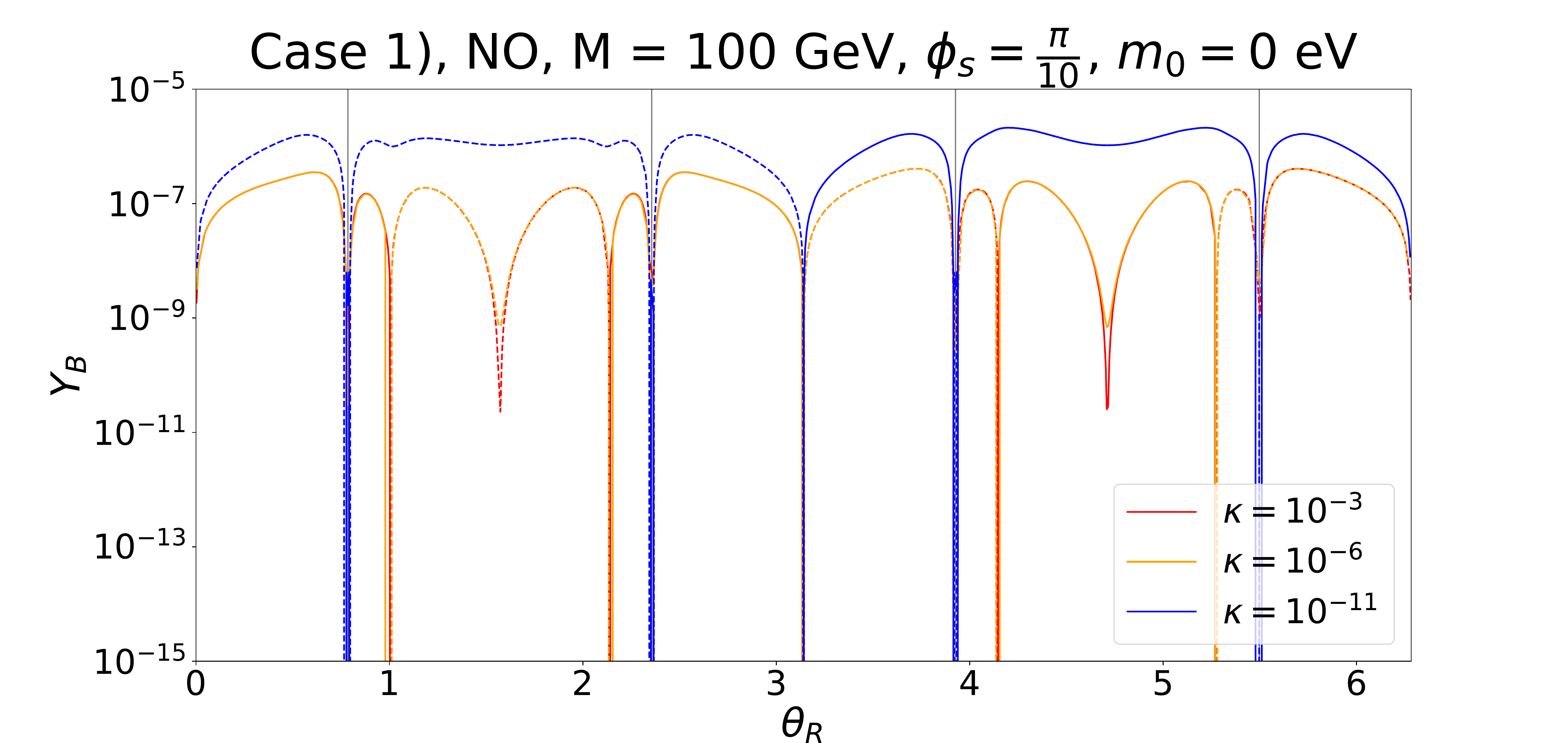}
		\caption{Vanishing initial conditions.}
	\end{subfigure}
	\begin{subfigure}{.5\textwidth}
		\centering
		\includegraphics[width = \textwidth]{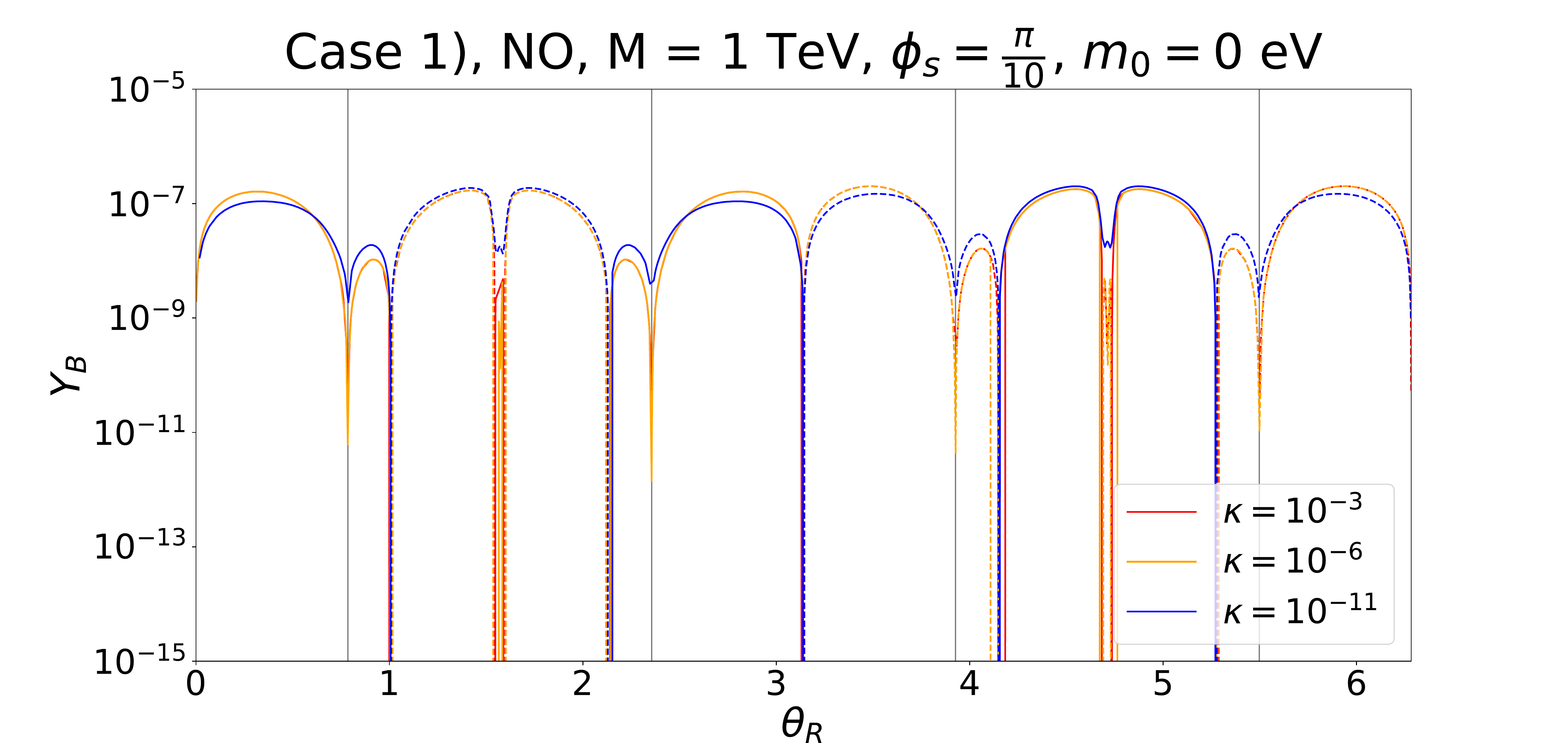}
		\caption{Vanishing initial conditions.}
	\end{subfigure}
\caption{{\small {\bf Case 1)} $Y_B$ as function of the angle $\theta_R$ for four different values of the Majorana mass, $M=1$ GeV (upper left plot), $M=10$ GeV (upper right plot), $M=100$ GeV (lower left plot) and $M=1$ TeV (lower right plot), and different values of $\kappa$, $\kappa \in \{10^{-3},10^{-6},10^{-11}\}$. Both negative (dashed lines) as well as positive (continuous lines) values of the BAU are represented. The value of $\phi_s$ is fixed to $\phi_s=\frac{\pi}{10}$ and light neutrino masses follow strong NO. The vertical grey lines correspond to regions, in which the criterion, given in Eq.~\eqref{eq:kappacorrectionscriterion}, is not fulfilled.}}
\label{NO thetaR BAU 1-10 GeV}
\end{figure}

\paragraph{Dependence on angle $\theta_R$.} We can, furthermore, investigate the dependence of the BAU on the angle $\theta_R$ for different values of $\kappa$. The result is shown in Fig.~\ref{NO thetaR BAU 1-10 GeV} for fixed values of $M$ and $\phi_s$ and three different values of $\kappa$. We observe that the dependence on $\theta_R$ is in general quite complicated and does not only follow $\sin \theta_R$ for strong NO ($y_1=0$), see Eqs.~\eqref{eq:CLFValphaCase1seven} and~\eqref{eq:CLNValphaCase1seven}, but can reveal additional enhancements of the BAU as well as additional changes of the sign of the BAU. In particular, the  Majorana mass $M$ and the size of the splitting $\kappa$ play a crucial role, compare plots (a)-(d) in Fig.~\ref{NO thetaR BAU 1-10 GeV}. Nevertheless, we recognise that at least for $\theta_R=0,\pi, 2 \, \pi$ the BAU vanishes which is in agreement with the $\sin \theta_R$-dependence of the CP-violating combinations for strong NO. The enhancement of the BAU for values of $\theta_R$ around $\theta_R= \frac{\pi}{4}$ and its odd multiples is related to the coupling $y_3$ being proportional to the inverse of $\sqrt{|\cos 2 \, \theta_R|}$ for a fixed light neutrino mass $m_3$, compare Eq.~(\ref{eq:strongNOCase1}). Moreover, the more complicated dependence on $\theta_R$ for $M\gtrsim 10$ GeV arises due to the dynamical interplay of the LNV and LFV CP-violating sources, whereas, to good approximation, only the LFV source does contribute to the generation of the BAU for $M=1$ GeV.
We have also checked that these statements are qualitatively the same for light neutrino masses with strong IO ($y_3=0$), up to the fact that in this case the BAU vanishes for $\theta_R=\frac{\pi}{2}, \frac{3 \, \pi}{2}$, as expected from the CP-violating combinations $C_{\mathrm{LFV},\alpha}$ and $C_{\mathrm{LNV},\alpha}$, which depend on $\cos \theta_R$ for $y_3=0$. Clearly, in the case of strong IO the coupling $y_1$ scales as the
inverse of $\sqrt{|\cos 2 \, \theta_R|}$ for a fixed light neutrino mass $m_1$, see Eq.~(\ref{eq:strongIOCase1}).

\begin{figure}
	\centering
	\includegraphics[width = 0.49\textwidth]{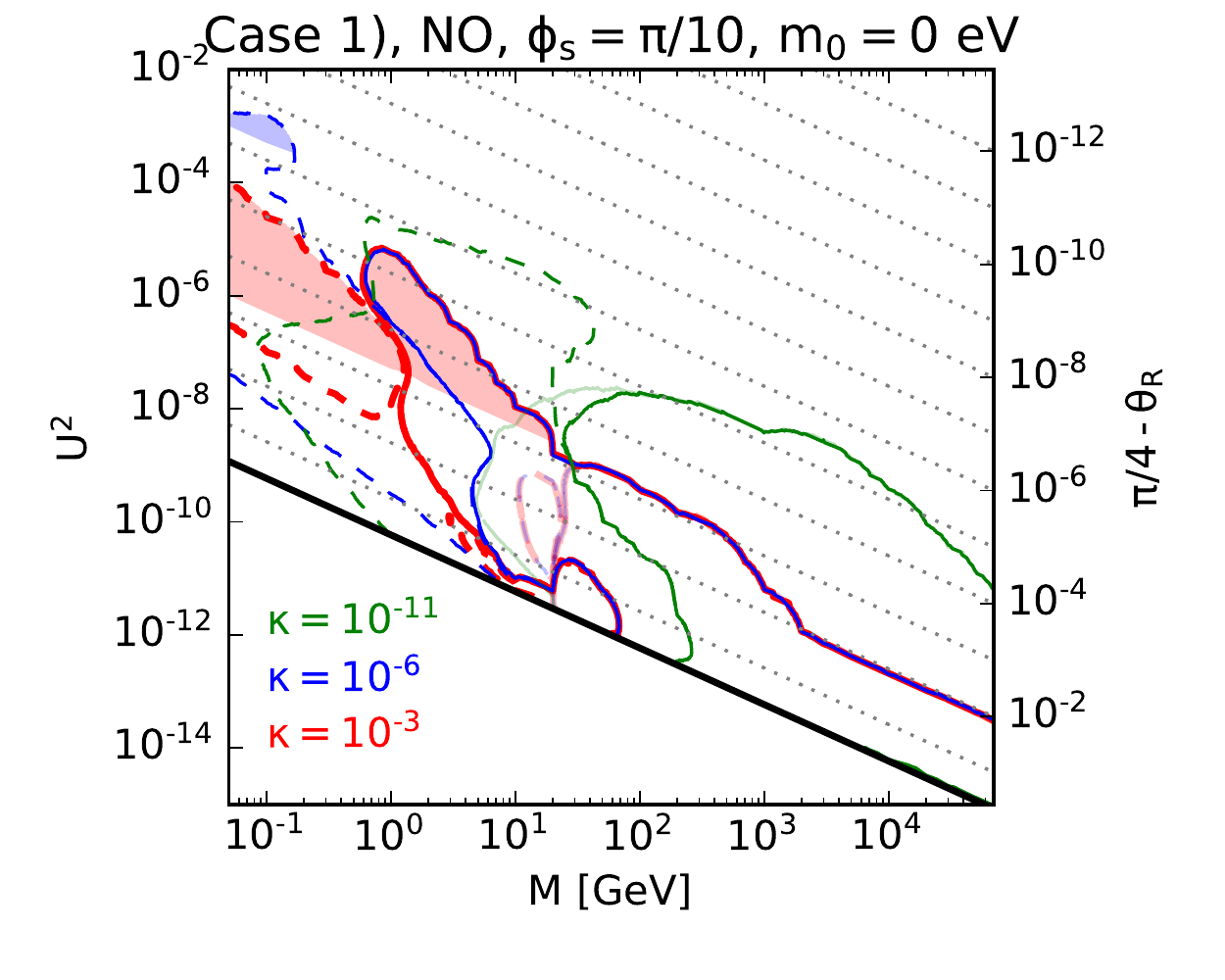}
	\includegraphics[width = 0.49\textwidth]{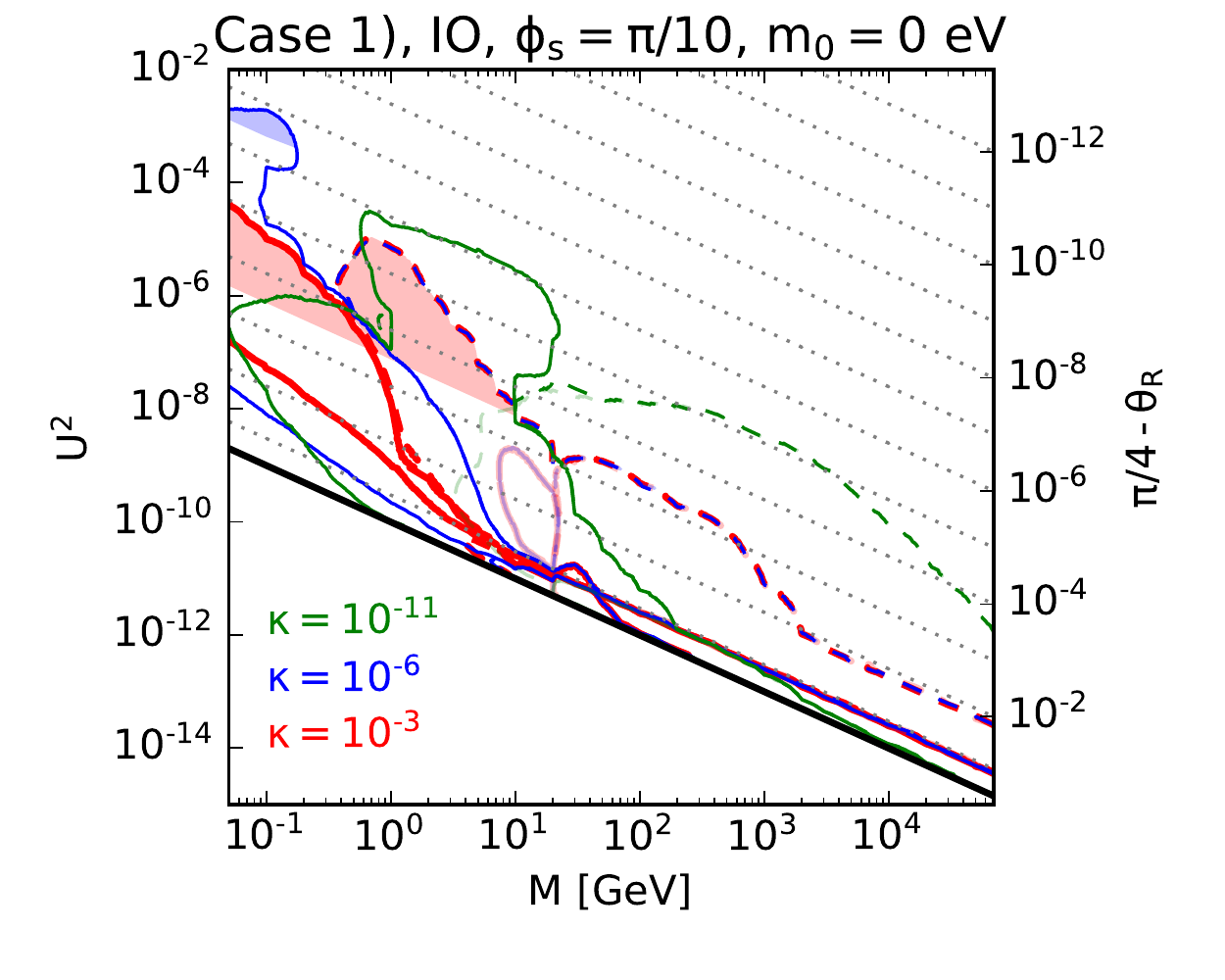}
	\caption{{\small {\bf Case 1)} Range of total mixing angle $U^2$ consistent with leptogenesis for the Majorana mass varying between $50$ MeV and $70$ TeV. The splitting $\kappa$ is chosen among the three values $\kappa \in \{10^{-3}, 10^{-6}, 10^{-11}\}$. The angle $\theta_R$ can be read off. The value of $\phi_s$ is fixed to $\phi_s=\frac{\pi}{10}$. Light neutrino masses follow strong NO (left plot) and strong IO (right plot), respectively.
		The contours corresponding to positive (negative) BAU are shown in continuous (dashed) lines. Note that these refer to vanishing initial conditions except for the lines shown in fainter colours.
		The sign of the BAU can be changed either by changing the signs of the couplings $y_f$ or the sign of
		the splitting $\kappa$ or by replacing the parameter $s$
		by $n-s$ for $s$ being odd, compare the CP-violating combinations in Eqs.~(\ref{eq:CLFValphaCase1seven}) and (\ref{eq:CLNValphaCase1seven}).
		For masses larger than $M\sim 10$ GeV, we see that the lines for $\kappa = 10^{-6}$ and $\kappa = 10^{-3}$ coincide.
		This can be verified in Fig.~\ref{NO kappa BAU Case I different masses}, where $Y_B$ becomes independent of the splitting $\kappa$ for $\kappa \gtrsim 10^{-6}$. In contrast, for $\kappa=10^{-11}$ both $Y_B$ and the allowed total mixing angle $U^2$ are generically larger for these larger masses.
		The blue and red shaded areas show the regions in which the condition in Eq.~\eqref{eq:kappacorrectionscriterion} and alike is not satisfied.
		The black lines indicate the seesaw line, defined in Eq.~\eqref{eq:naiveseesawformula}.
}}
\label{NO Mass U2 Case1}
\end{figure}

\begin{figure}[t!]
	\centering
	\includegraphics[width = \textwidth]{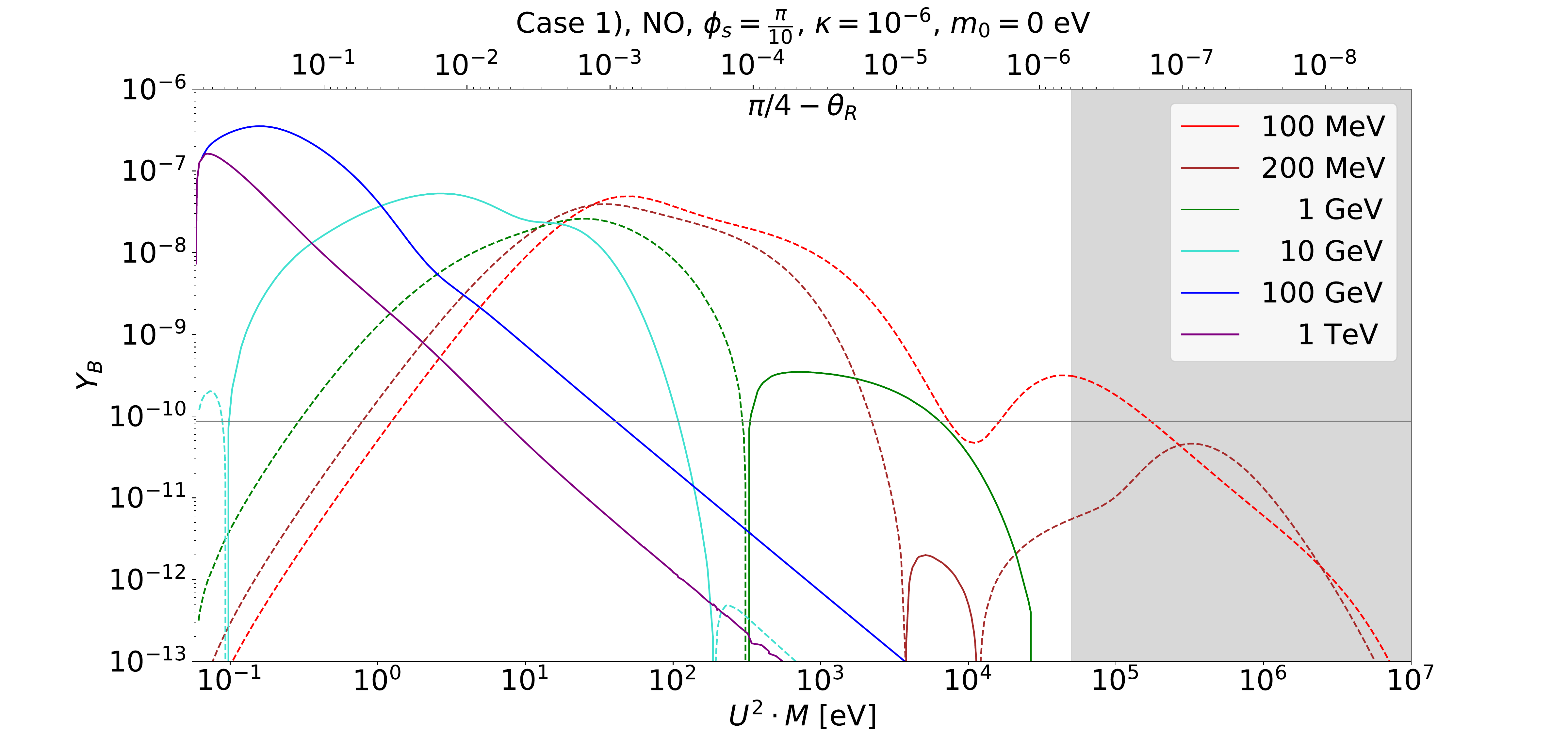}
\caption{{\small {\bf Case 1)} $Y_B$ as function of $U^2\cdot M$ for different values of the Majorana mass $M$ and assuming vanishing initial conditions. The product $U^2\cdot M$ is introduced, as it is independent of $M$ contrary to $U^2$. The splitting $\kappa$ is fixed to $10^{-6}$. Light neutrino masses follow strong NO and $\phi_s$ is chosen as $\frac{\pi}{10}$. Both negative (dashed lines) as well as positive (continuous lines) values of the BAU are represented. The grey line indicates the observed value of the BAU, $Y_B\approx 8.6 \cdot 10^{-11}$. The grey shaded area represents the region in which the criterion, given in Eq.~\eqref{eq:kappacorrectionscriterion},  is not fulfilled.}}
\label{NO BAU vs U2 10 GeV e-6 multiple curves}
\end{figure}

\begin{figure}[t!]
\begin{subfigure}{.5\textwidth}
\hspace*{-1cm}\includegraphics[trim={7.2cm 0 7.2cm 0},clip,width = 1.2\textwidth]{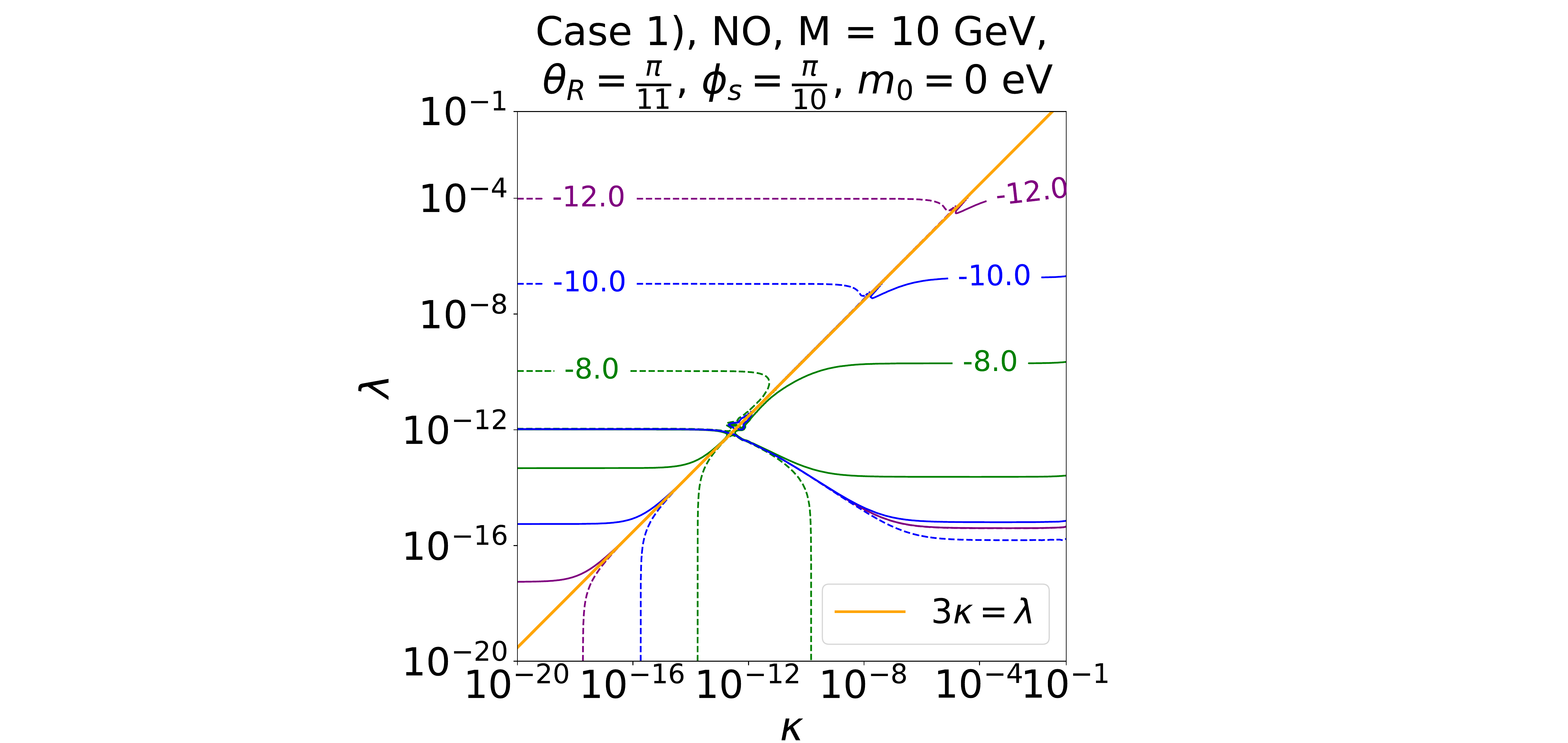}
\caption{Vanishing initial conditions.~~~~~~~}
	\end{subfigure}
	\begin{subfigure}{.5\textwidth}
		\hspace*{-0.7cm}\includegraphics[trim={7.2cm 0 7.2cm 0},clip,width = 1.2\textwidth]{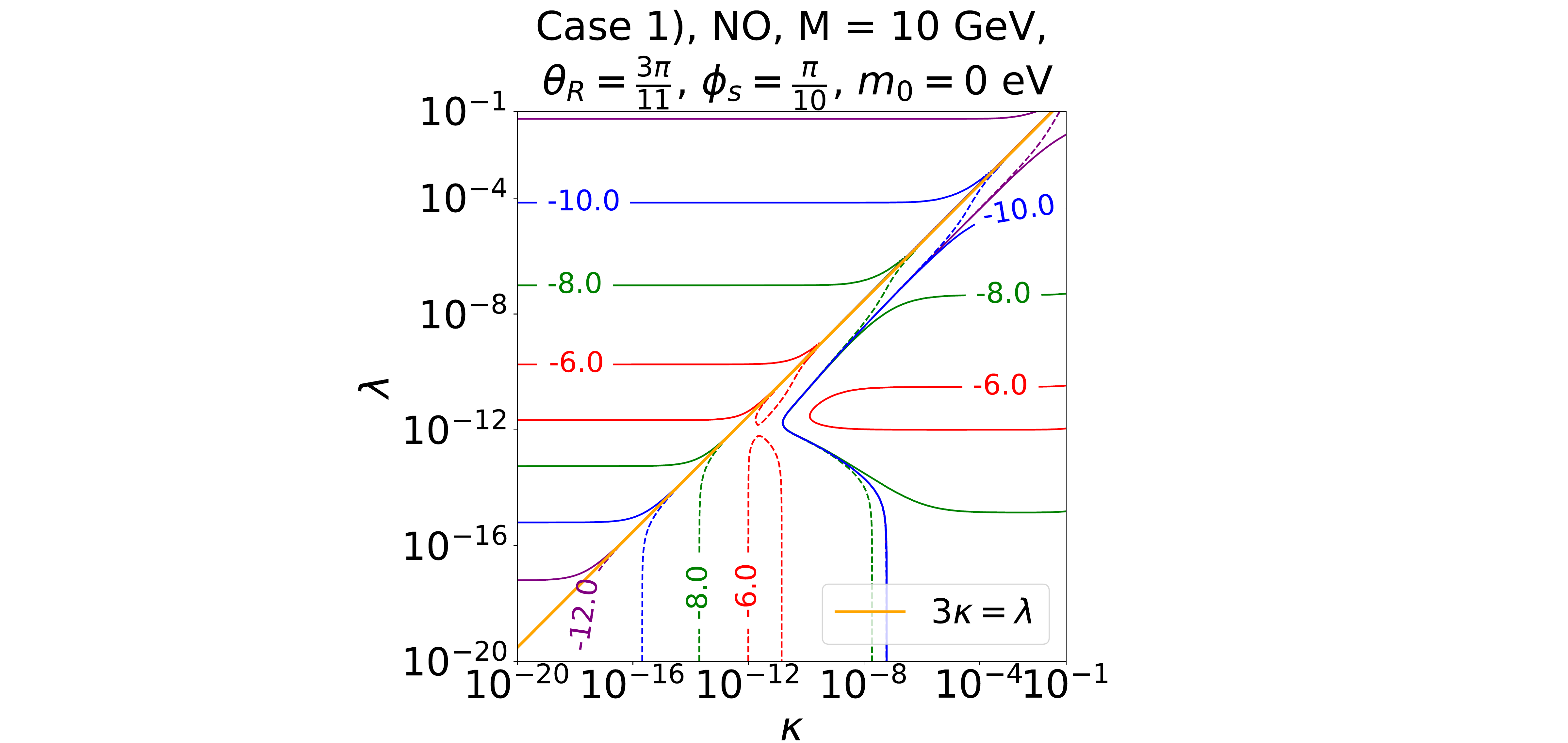}
		\caption{Vanishing initial conditions.}
	\end{subfigure}
	\caption{{\small {\bf Case 1)} Contour plots of $Y_B$ as function of both splittings $\kappa$ and $\lambda$ for two different values of the angle $\theta_R$, $\theta_R=\frac{\pi}{11}$ (left plot) and $\theta_R=\frac{3 \, \pi}{11}$ (right plot). Iso-BAU lines are represented in the $\kappa-\lambda$-plane; note $-12.0$ has to be read as $10^{-12}$ and alike.
		We, furthermore, show the line corresponding to $3 \, \kappa=\lambda$ in orange for which two of the heavy neutrinos become degenerate in mass, compare Eq.~(\ref{eq:Deltasigmas}), and the BAU vanishes, see Eqs.~(\ref{eq:CLFValphaCase1seven}) and (\ref{eq:CLNValphaCase1seven}).
		The Majorana mass is fixed to $M=10$ GeV, light neutrino masses follow strong NO and $\phi_s$ is fixed to $\phi_s=\frac{\pi}{10}$.
}}
\label{kappa lambda grid}
\end{figure}

\paragraph{Special values of $\theta_R$.} As already briefly mentioned in the previous paragraph,
we observe that for $\theta_R$ being close to an odd multiple of $\frac{\pi}{4}$ the coupling $y_3$ ($y_1$) becomes large for strong NO (IO) and a fixed value of the light neutrino mass $m_3$ ($m_1$) and thus also large values of the total mixing angle $U^2$ can be reached, much larger than expected from the naive seesaw estimate, see Eq.~\eqref{eq:naiveseesawformula}.
At the same time, we observe an enhancement of the BAU in the vicinity of these special values for $M=1$ GeV and $M=10$ GeV.
As discussed in section~\ref{sec51}, this can be explained by the increased overall scale of the Yukawa couplings which, for $M\lesssim 10$ GeV, are expected to be too small for an efficient production of the BAU, compare Eq.~\eqref{eq:equilibrationFNC}, due to slow heavy neutrino interactions. 
Simultaneously, increasing the Yukawa couplings allows for larger mass splittings consistent with the resonance  condition, see Eq.~\eqref{eq:resCondition}.
Such an enhancement around special values of $\theta_R$ has not been observed for larger masses, e.g. $M=100$ GeV and $1$ TeV, see plots (c) and (d) in Fig.~\ref{NO thetaR BAU 1-10 GeV}.
Note that, for $\theta_R$ being  chosen as one of the special values, the BAU is always strongly suppressed due to strong washout of the lepton asymmetry.
As demonstrated in Figs.~\ref{NO Mass U2 Case1} and \ref{NO BAU vs U2 10 GeV e-6 multiple curves}, the correct amount of BAU can be produced through low-scale leptogenesis in the vicinity of these special values (see Fig.~\ref{fig:Case1_MU2IC} in appendix~\ref{appF1} for dedicated plots for each value of the splitting $\kappa$).
In Fig.~\ref{NO Mass U2 Case1}, we show the contours in the $M-U^2$-plane that correspond to the generation of the correct amount of BAU, with positive and negative sign in the case of continuous and dashed lines, respectively. We do so for three different values of the splitting $\kappa$. We can see that for small values of the Majorana mass $M$ large values of the total mixing angle $U^2$ can be attained which allow for testing this scenario at different current and future experiments.
 Note that the blue and red shaded areas in Fig.~\ref{NO Mass U2 Case1} show the regions in which the condition in Eq.~\eqref{eq:kappacorrectionscriterion} and alike is not satisfied.

As we detail in section~\ref{sec41}, at these special values of $\theta_R$ the residual symmetry of the combination $Y_D^\dagger Y_D$ is enhanced.
The smallness of the deviation of $\theta_R$ from such a special value can be thus protected by this enhanced symmetry and should not be considered as a tuning of the angle $\theta_R$.

\paragraph{Effect of splitting $\lambda$.}
The dependence of the BAU on both splittings $\kappa$ and $\lambda$ is shown in Fig.~\ref{kappa lambda grid} for a fixed value of $M$ and of $\phi_s$, the light neutrino mass spectrum following strong NO and two different values of $\theta_R$.
For $\lambda \lesssim 10^{-17}$, the dependence on $\lambda$ is rather weak, unless $3 \, \kappa \lesssim \lambda$. In this case, one of the mass splittings among the RH neutrinos, and therefore the asymmetry, exactly vanishes, compare Eq.~(\ref{eq:Deltasigmas}).
For larger values of $\lambda$, $\lambda \gtrsim 10^{-15}$, with the exact value depending on $\theta_R$,
$\lambda$ almost completely dominates the generation of the BAU.
This is not unexpected, as the splitting $\lambda$ induces a mass difference between all three RH neutrinos simultaneously, see Eq.~(\ref{lambda mass spectrum}).
This also explains why the BAU decreases with $\lambda$ increasing beyond $10^{-9}$, whereas
for large values of the splitting $\kappa$ the BAU reaches a plateau, see e.g.~also the plots in Fig.~\ref{NO kappa BAU Case I different masses}. Nonetheless, we observe that even values of $\lambda \sim 10^{-4}$ can lead to successful baryogenesis.

\subsection{Case 2)}
\label{sec54}

For Case 2), we discuss the results for low-scale leptogenesis for the representative choice of the group theory parameters $n$, $u$ and $v$ (or equivalently $s$ and $t$), $n=14$, $u=0$ or $u=\pm 1$, chosen in
section~\ref{sec502}. These lead to a good fit of the experimental data on lepton mixing. Additionally, we study the choice $n=50$, $u=6$ ($t$ is  always even) corresponding to $u/n=0.12$ and hence $\theta_L=0$, compare
Eq.~(\ref{eq:constraintunCase2}) and Fig.~\ref{fig:case2NOIOchi2}. We consider several values of $v$ in order to elucidate the dependence of the generated BAU on $v$ ($\phi_v$) and the agreement between numerical results and expectations from the  CP-violating combinations.

We split this discussion into two parts according to whether $t$ is even or $t$ is odd. For the latter choice, we encounter the interesting possibility to generate the correct amount of  BAU without the splitting $\kappa$ (and $\lambda$), as indicated by the CP-violating combination $C_{\mathrm{DEG},\alpha}$ in Eq.~\eqref{eq:CDEGalpha_seventodd_Case2}.

\mathversion{bold}
\subsubsection{Choice \texorpdfstring{$t$}{t} even}
\mathversion{normal}
\label{subsec:5.4.1}

We first comment on the effect of the splitting $\kappa$, then briefly on the impact of the light neutrino mass spectrum, on the dependence of our results on the choice of the CP transformation $X (s,t)= X (u,v)$ and eventually on the
impact of the angle $\theta_R$. Note that for this choice of $t$ a non-vanishing lightest neutrino mass, $m_0 \neq 0$, does not lead to a further rotation and to the replacement of $\theta_L$
with $\widetilde{\theta}_L$, compare Eqs.~(\ref{eq:combmnuCase2steven}) and (\ref{eq:combmnuCase2tevensodd}). We thus present results for $m_0=0.03$ eV and only resort to a light neutrino mass spectrum with strong NO (IO), meaning $m_0=0$, in the cases in which we study how well the CP-violating combinations $C_{\mathrm{LFV},\alpha}$ and $C_{\mathrm{LNV},\alpha}$ can reproduce the numerical results, since the former considerably simplify for vanishing lightest neutrino mass, see Eqs.~(\ref{eq:CLFValphaCase2stevenNO}) and (\ref{eq:CLFValphaCase2stevenIO}).

\begin{figure}
	\begin{subfigure}{.5\textwidth}
		\centering
		\includegraphics[width = \textwidth]{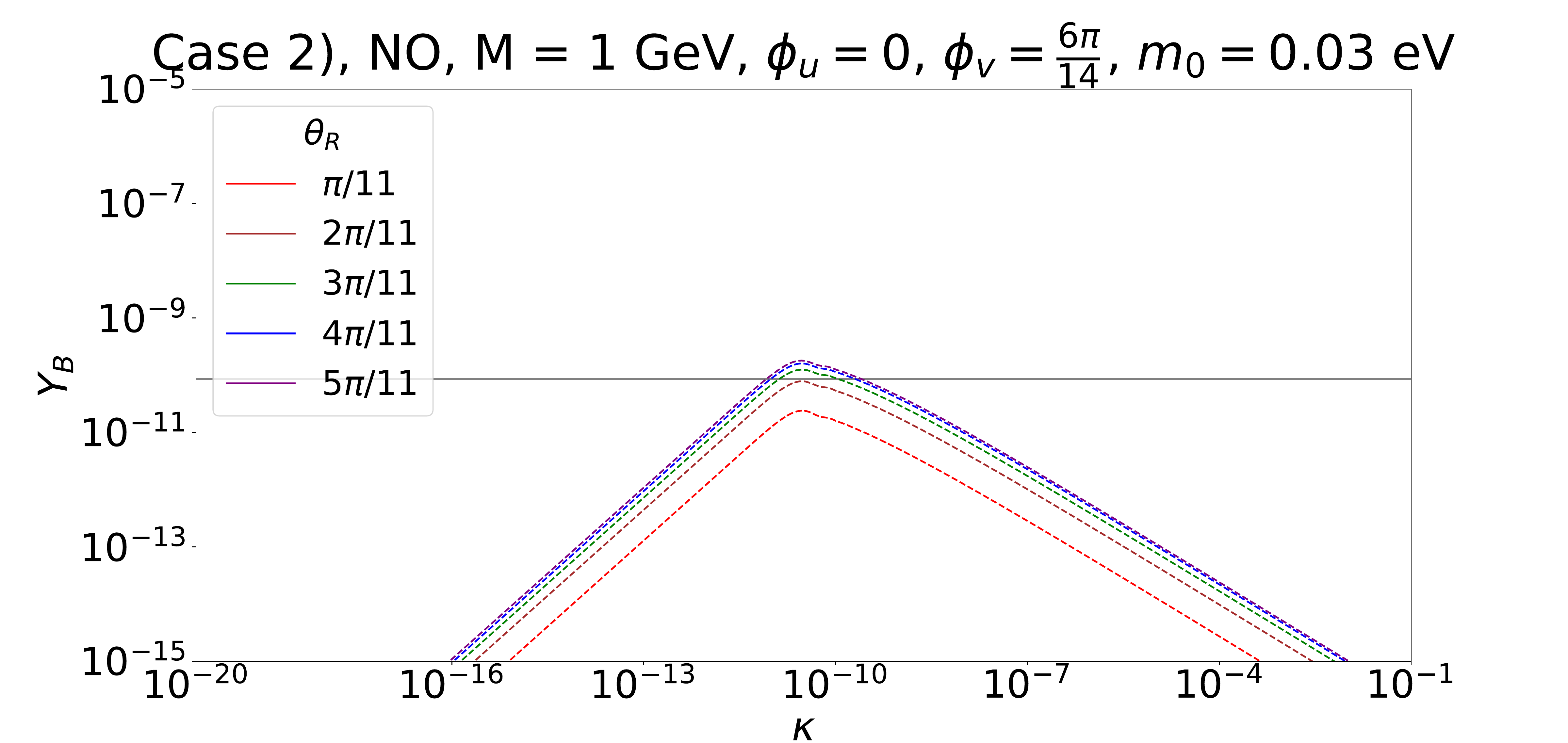}
		\caption{Vanishing initial conditions.}
	\end{subfigure}
	\begin{subfigure}{.5\textwidth}
		\centering
		\includegraphics[width = \textwidth]{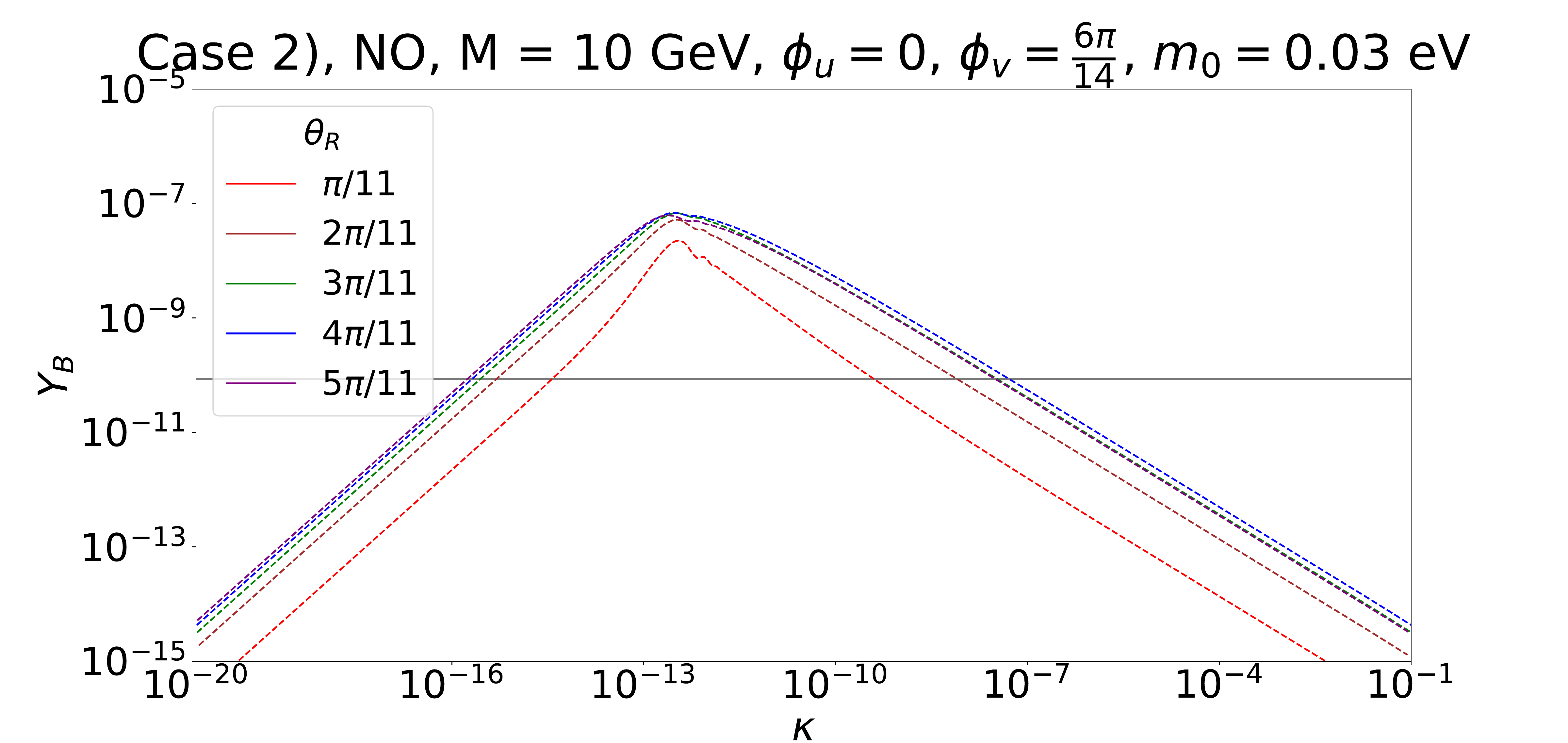}
		\caption{Vanishing initial conditions.}
		\label{NO kappa BAU 10 GeV Case II}
	\end{subfigure}
	\begin{subfigure}{.5\textwidth}
		\includegraphics[width = \textwidth]{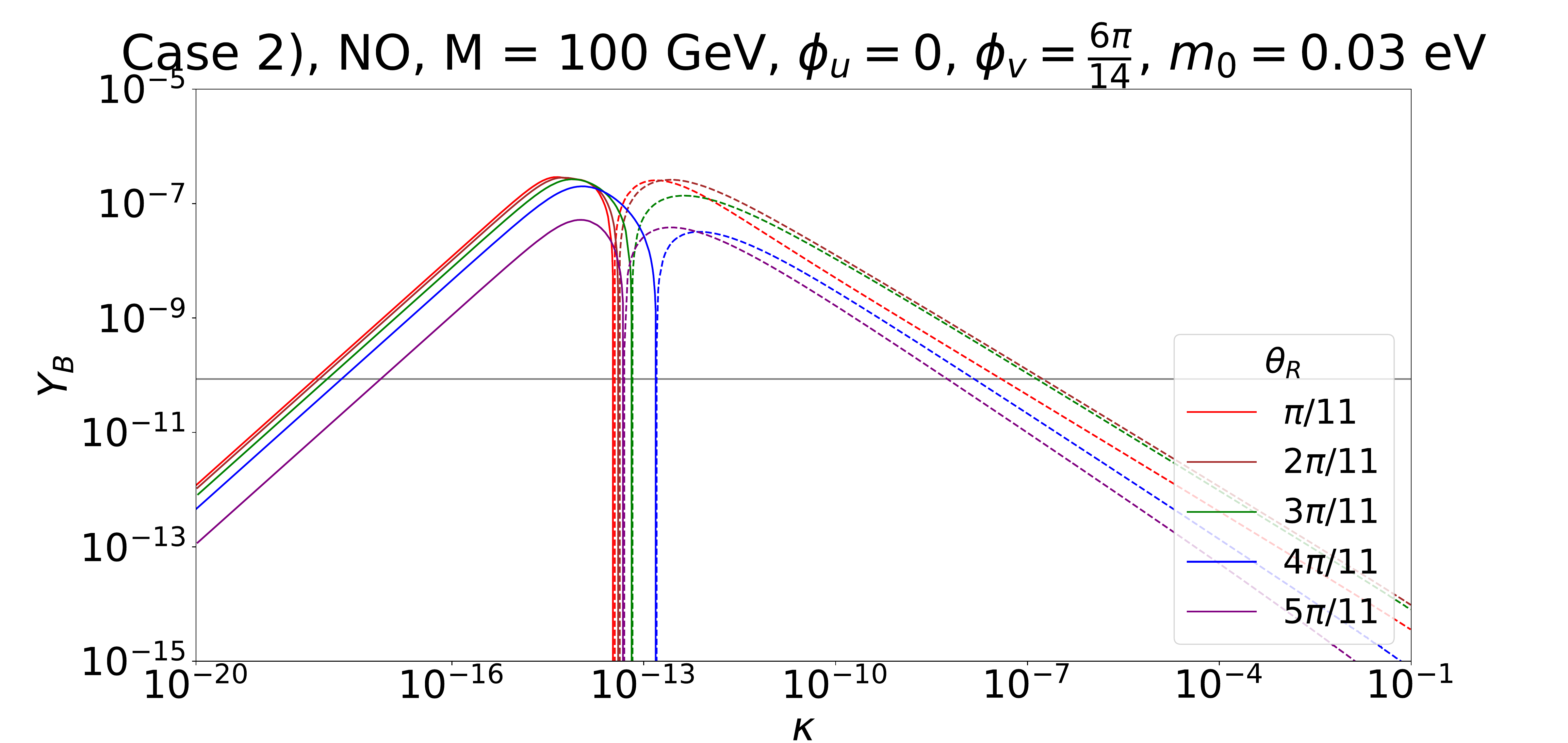}
		\caption{Vanishing initial conditions.}
	\end{subfigure}
\begin{subfigure}{.5\textwidth}
	\includegraphics[width = \textwidth]{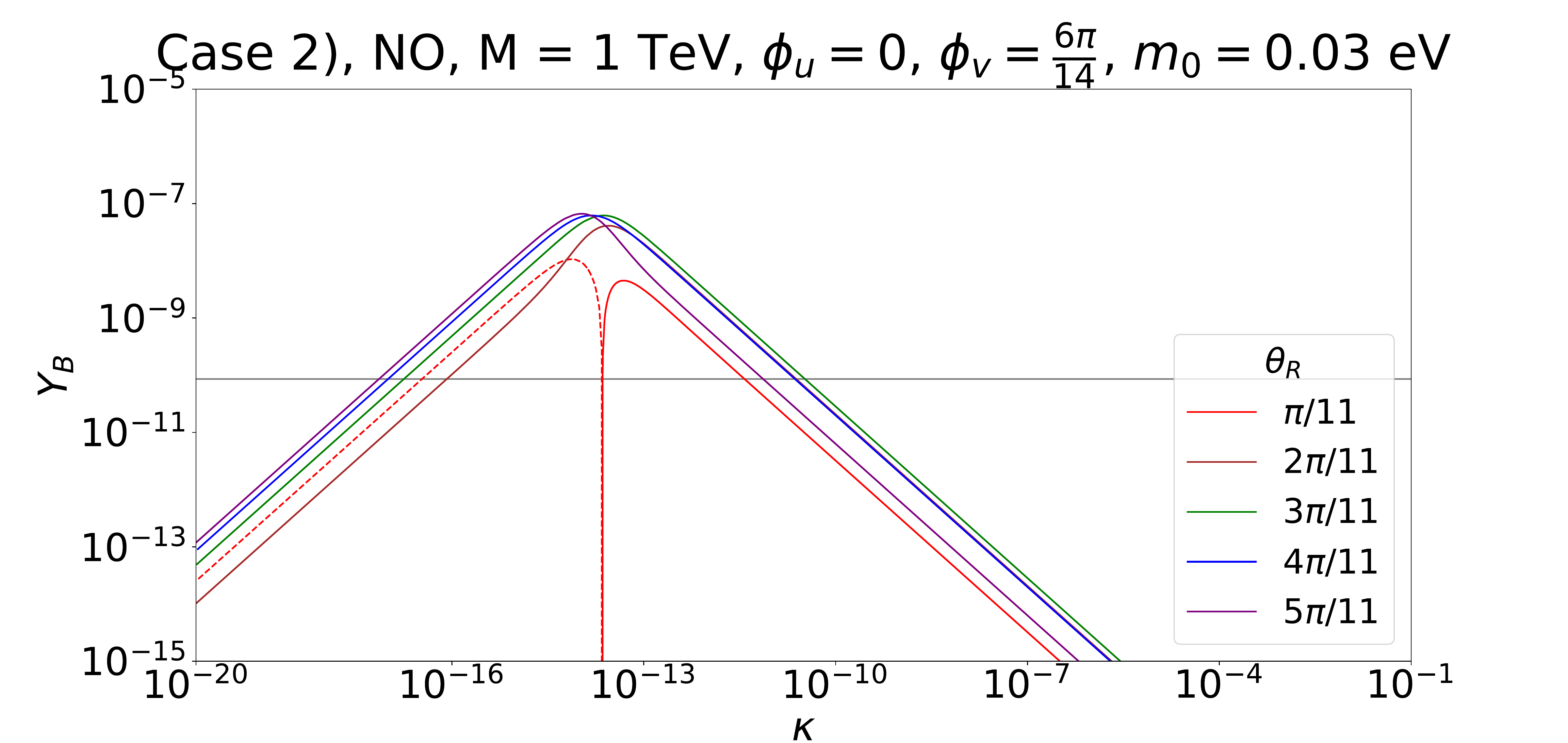}
	\caption{Vanishing initial conditions.}
\end{subfigure}
\caption{{\small {\bf Case 2)} $Y_B$ as function of the splitting $\kappa$ for a Majorana mass $M=1$ GeV (upper left plot), $M=10$ GeV (upper right plot), $M=100$ GeV (lower left plot) and $M=1$ TeV (lower right plot). For each of these choices, different values of $\theta_R$ have been studied. The parameters $s$ and $t$ are fixed to $s=1$ and $t=2$, respectively, corresponding to $\phi_u=0$ and $\phi_v=\frac{6\pi}{14}$. Light neutrino masses have NO with $m_0 = 0.03$ eV.
Both negative (dashed lines) as well as positive (continuous lines) values of the BAU are represented.  The grey line indicates the observed value of the BAU, $Y_B\approx 8.6 \cdot 10^{-11}$.}}
\label{NO kappa BAU Case II different masses}
\end{figure}

\paragraph{Effect of splitting $\kappa$.} Similar to Case 1), we observe that, for small $\kappa$, the BAU grows linearly with the splitting $\kappa$ before reaching the resonance peak. This is consistent with the observation that, for $\lambda=0$, all CP-violating combinations, derived in section~\ref{sec63}, are proportional to $\kappa$, if $t$ is even. The main difference, however, lies in the absence of a plateau for large $\kappa$, see e.g.~Fig.~\ref{NO kappa BAU Case II different masses}. The BAU monotonically decreases for $\kappa$ larger than the resonance peak, which is expected, as the reduced mass-degenerate CP-violating combination $C^{(23)}_{\mathrm{DEG},\alpha}$ is zero for $t$ even.

An additional interesting feature is the power law with which the BAU decreases. For $M=1,10,100$ GeV, the BAU decreases as $\kappa^{-\frac{2}{3}}$ for $\kappa$ larger than the resonance peak.
This observation is consistent with the analytical approximations, derived for the relativistic regime in~\cite{Drewes:2016gmt}. For larger masses, i.e. $M=1$ TeV, the BAU decreases linearly with $\kappa$, which can be expected from the behaviour of the decay asymmetries, see e.g.~equation (2) in~\cite{Dev:2017wwc}, for $\kappa$ larger than the resonance peak. 

\paragraph{Impact of light neutrino mass spectrum.} While focusing on light neutrino masses with NO and $m_0=0.03$ eV in Fig.~\ref{NO kappa BAU Case II different masses}, we also display two plots for the Majorana mass $M=10$ GeV and for light neutrino masses with strong NO and strong IO, respectively, in Fig.~\ref{NOIO kappa BAU 10 GeV Case II massless} in appendix~\ref{appF2}. As one can see, the qualitative and quantitative
behaviour of the BAU with respect to $\kappa$ is similar to what is observed in Fig.~\ref{NO kappa BAU Case II different masses}, plot (b).

\begin{figure}
	\begin{subfigure}{.5\textwidth}
		\centering
		\includegraphics[width = \textwidth]{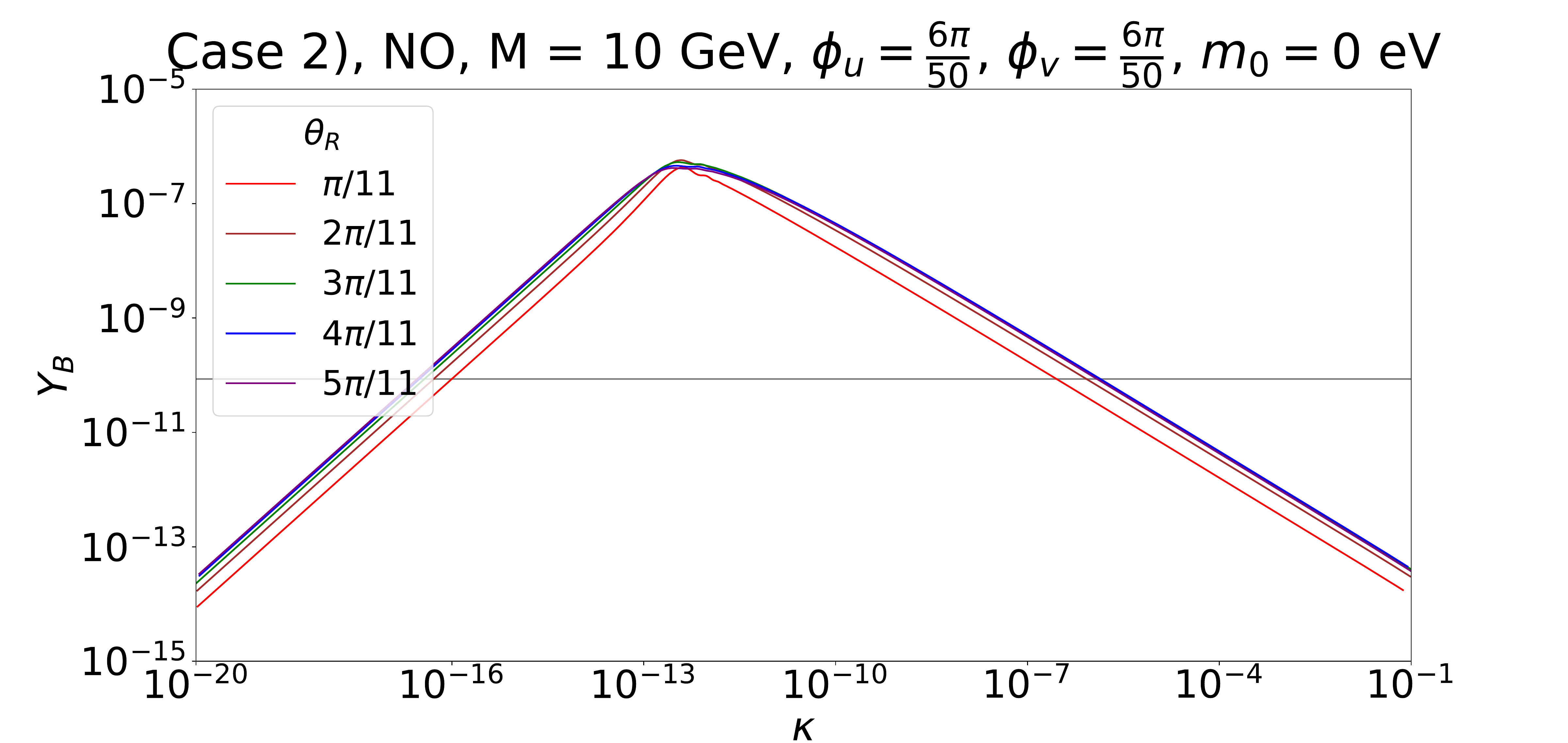}
		\caption{Vanishing initial conditions.}
	\end{subfigure}
	\begin{subfigure}{.5\textwidth}
		\includegraphics[width = \textwidth]{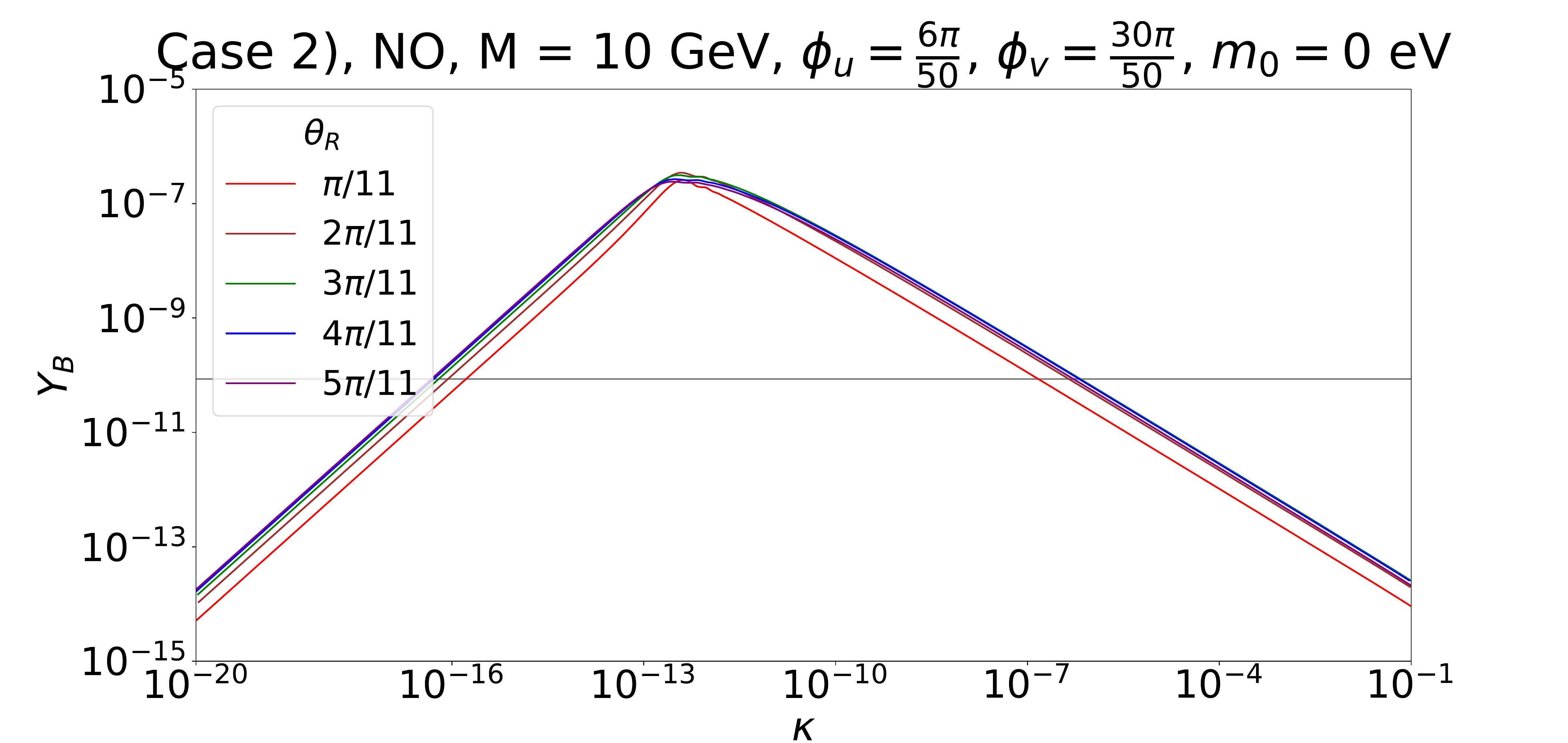}
		\caption{Vanishing initial conditions.}
	\end{subfigure}
	\begin{subfigure}{.5\textwidth}
		\includegraphics[width = \textwidth]{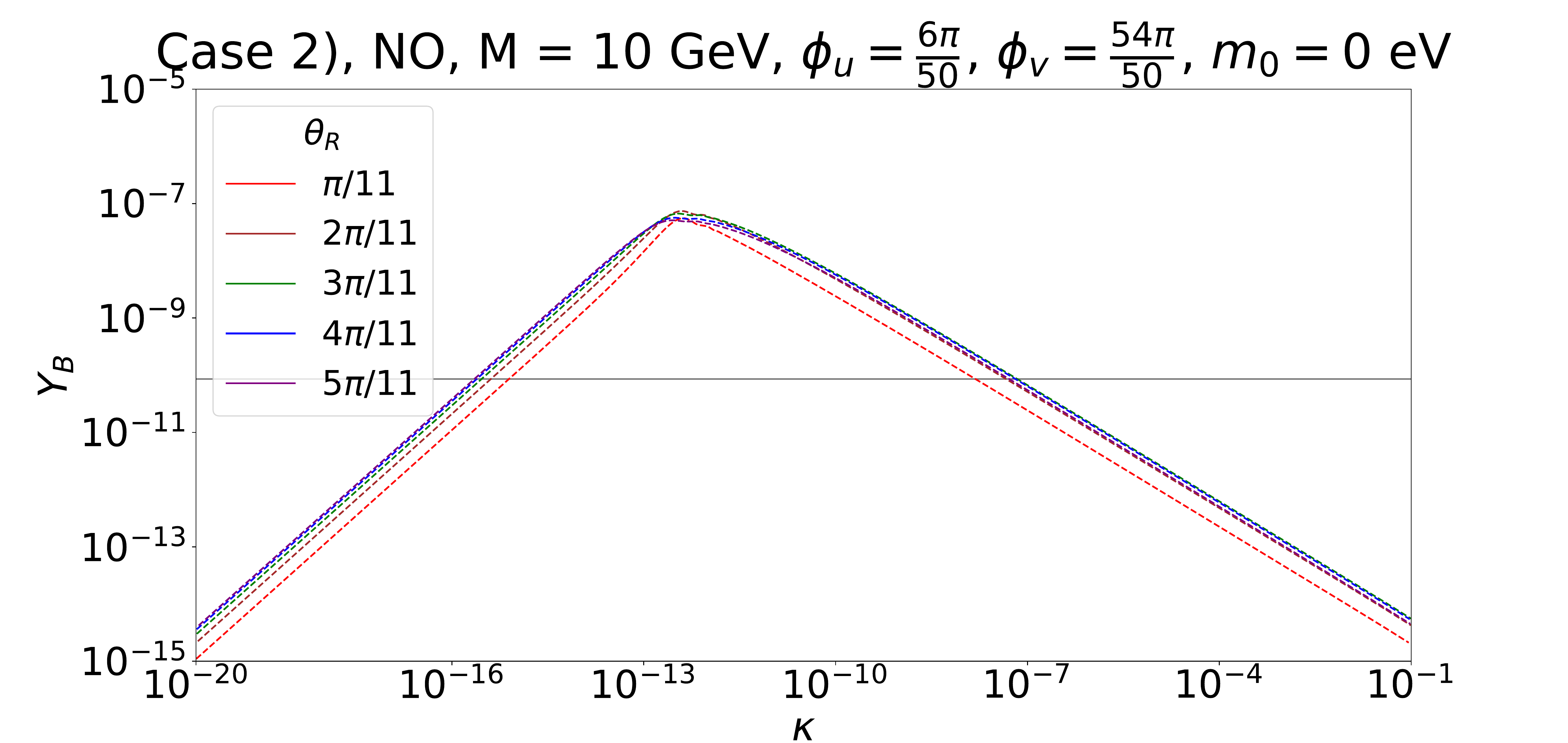}
		\caption{Vanishing initial conditions.}
	\end{subfigure}
	\begin{subfigure}{.5\textwidth}
		\includegraphics[width = \textwidth]{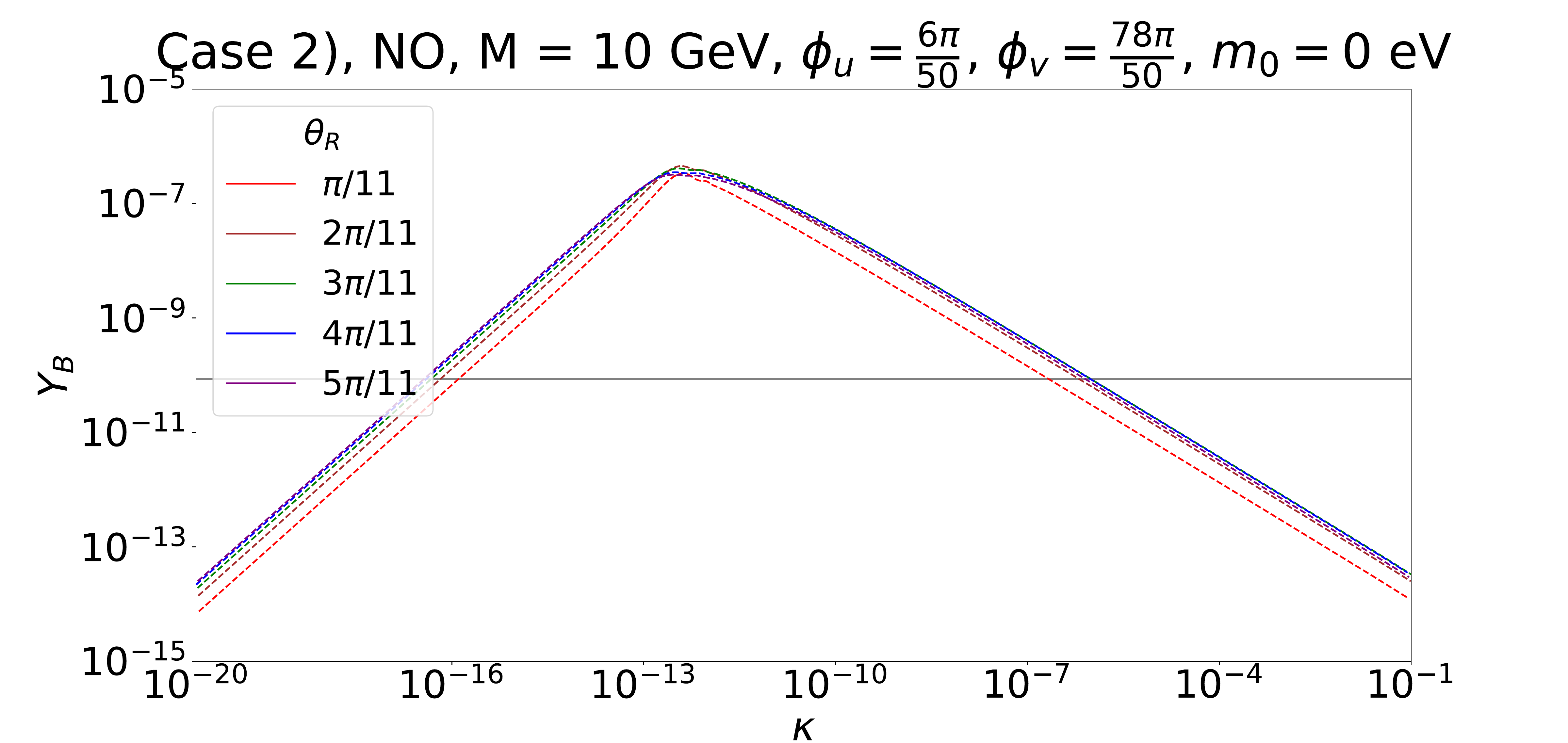}
		\caption{Vanishing initial conditions.}
	\end{subfigure}
	\caption{{\small {\bf Case 2)} $Y_B$ as function of $\kappa$ for a Majorana mass of $M=10$ GeV and for $n=50$ and the parameters $(s,t)$ equal to $(4,2)$ (upper left plot), $(8,10)$ (upper right plot), $(12,18)$ (lower left plot) and $(16,26)$ (lower right plot). Light neutrino masses have strong NO.
		For the remaining choices see Fig.~\ref{NO kappa BAU Case II different masses}.
}}
\label{NO kappa BAU 10 GeV Case II massless different phis u=0.12}
\end{figure}

\begin{figure}[t!]
	\begin{subfigure}{.5\textwidth}
		\centering
		\includegraphics[width = \textwidth]{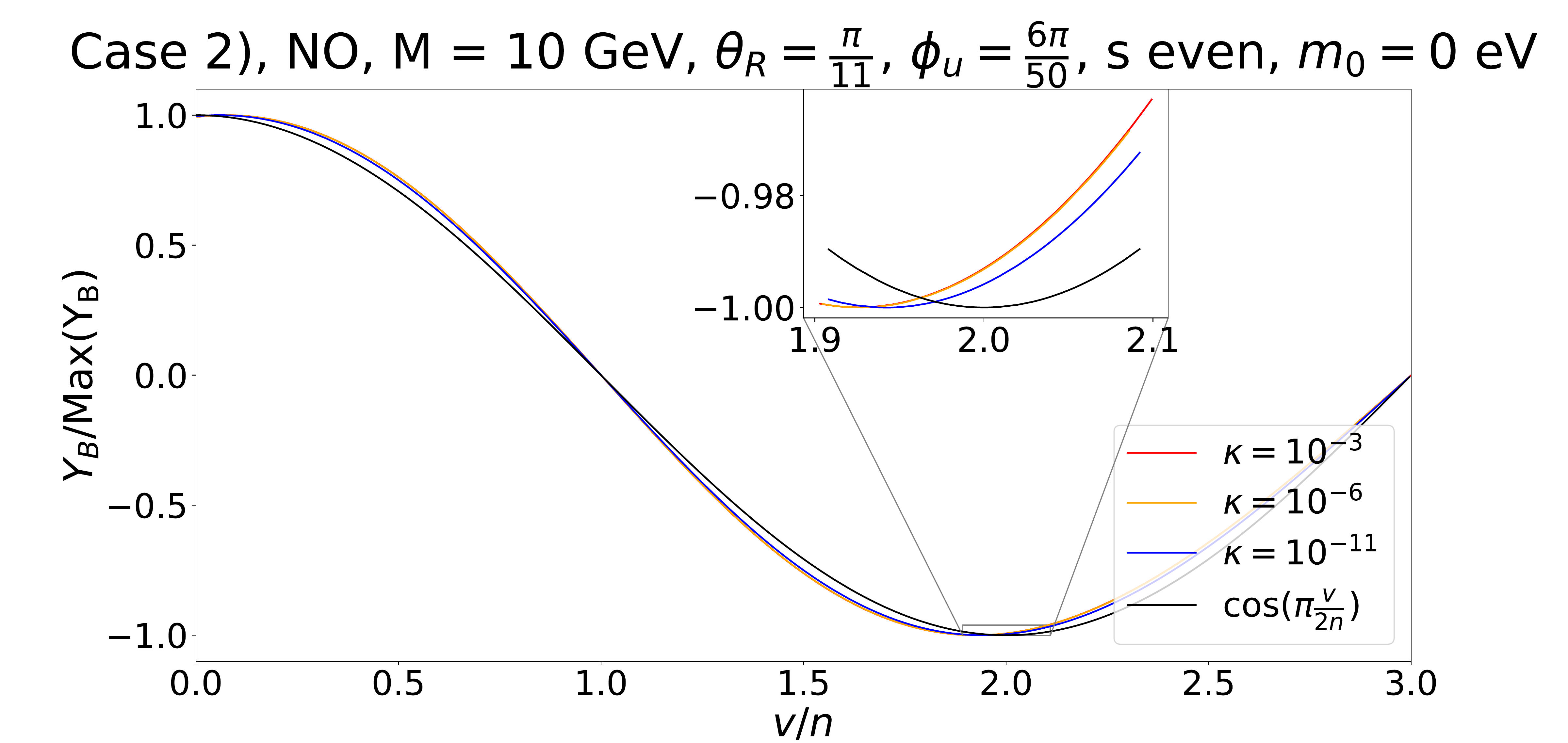}
		\caption{Vanishing initial conditions.}
	\end{subfigure}
	\begin{subfigure}{.5\textwidth}
		\centering
		\includegraphics[width = \textwidth]{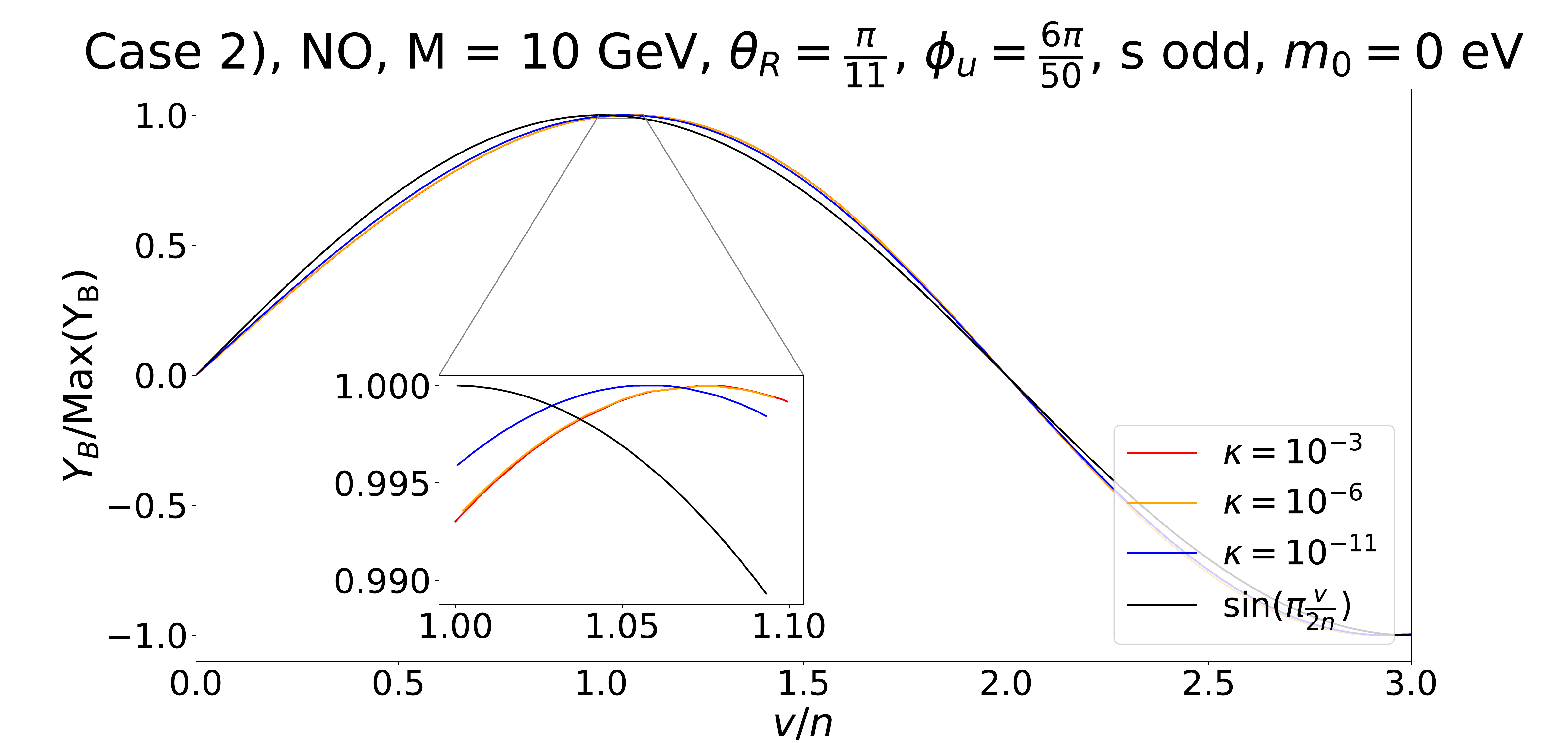}
		\caption{Vanishing initial conditions.}
	\end{subfigure}
	\caption{{\small {\bf Case 2)} $Y_B$ as function of $\frac{v}{n}$, treated as continuous parameter, for a Majorana mass $M=10$ GeV and different values of $\kappa$, $\kappa \in \{10^{-3},10^{-6},10^{-11}\}$. The group theory parameters $n$ and $u$ are chosen as $n=50$ and $u=6$
		so that $\phi_u=\frac{6 \, \pi}{50}$. This choice of $u$ corresponds to $t$ even and $s$ being even (left plot) or odd (right plot).
		Light neutrino masses have strong NO.
		We compare the numerical results (coloured curves) with the analytical expectation $\cos(\frac{\pi v}{2n})$ (for $s$  even) and $\sin(\frac{\pi v}{2n})$ (for $s$ odd) arising from the CP-violating combinations, see Eq.~\eqref{eq:CLFValphaCase2stevenNO}, shown in black.
}}
\label{NO BAU phiv e-3e-6e-11 caseII massless}
\end{figure}

\paragraph{Dependence on CP transformation $X (s,t)= X (u,v)$.} We observe that the angle $\theta_L$, which leads to a good agreement with the experimental data on lepton mixing, is non-zero for the choice of the parameters $n$ and $u$, $n=14$ and $u=0$ and $u=\pm 1$, see Tab.~\ref{tab:Case2n14}. This complicates the study of the dependence of the BAU on the CP transformation $X (s,t)= X (u,v)$ and the comparison between the numerical results and the expectations from the CP-violating combinations $C_{\mathrm{LFV},\alpha}$ and $C_{\mathrm{LNV},\alpha}$ for light neutrino masses with strong NO and IO, see Eqs.~(\ref{eq:CLFValphaCase2stevenNO})
and (\ref{eq:CLFValphaCase2stevenIO}), respectively. We thus use $n=50$ and $u=6$ as choices, since then $\theta_L=0$ leads to an acceptable fit of the lepton mixing angles, compare Eq.~(\ref{eq:constraintunCase2}) and Fig.~\ref{fig:case2NOIOchi2}. In this case,
the CP-violating combinations $C_{\mathrm{LFV},\alpha}$ and $C_{\mathrm{LNV},\alpha}$ follow $\cos \frac{\phi_v}{2}$ and $\sin \frac{\phi_v}{2}$ for strong NO and strong IO, respectively. The numerical results are shown in Figs.~\ref{NO kappa BAU 10 GeV Case II massless different phis u=0.12} and~\ref{NO BAU phiv e-3e-6e-11 caseII massless}.

In Fig.~\ref{NO kappa BAU 10 GeV Case II massless different phis u=0.12} we display the BAU with respect to the splitting $\kappa$ for a fixed value of $M$, $M=10$ GeV, and for different values of $\theta_R$. Furthermore, we choose
four different combinations of $(s,t)$ which are both even and always lead to $\phi_u=\frac{6 \, \pi}{50}$, but to different values of $v$ ($\phi_v$), i.e.~for $(s,t)=(4,2)$ (plot (a))
we have $\phi_v=\frac{6 \, \pi}{50}$ and $\cos \frac{\phi_v}{2} \approx +0.982$, for $(s,t)=(8,10)$ (plot (b))
it is $\phi_v=\frac{30 \, \pi}{50}$ and $\cos \frac{\phi_v}{2} \approx +0.588$, for $(s,t)=(12,18)$ (plot (c))
we have $\phi_v=\frac{54 \, \pi}{50}$ and $\cos \frac{\phi_v}{2} \approx -0.125$ and for $(s,t)=(16,26)$ (plot (d)), eventually,
it is $\phi_v=\frac{78 \, \pi}{50}$ and $\cos \frac{\phi_v}{2} \approx -0.771$. These differences of $\cos \frac{\phi_v}{2}$ in magnitude and, in particular, in sign are clearly visible in the four different plots (compare the exchange or not of continuous and dashed lines in order to track the possible change in the sign of the BAU due to $\cos \frac{\phi_v}{2}$).

In Fig.~\ref{NO BAU phiv e-3e-6e-11 caseII massless} we show the BAU as well, normalised to its maximal attainable value, with respect to $v/n$, treated as continuous parameter between $0$ and $3$, in order to emphasise the dependence on this combination. Again, we fix the Majorana mass to $M=10$ GeV and use $\phi_u=\frac{6 \, \pi}{50}$. Additionally, we focus on three different values of $\kappa$ only, varying between $10^{-11}$ and $10^{-3}$, and on the angle $\theta_R$ being $\theta_R=\frac{\pi}{11}$. In plot (a) of Fig.~\ref{NO BAU phiv e-3e-6e-11 caseII massless} we set $s$ to be even, while $s$ being odd is assumed in plot (b). The CP-violating combinations $C_{\mathrm{LFV},\alpha}$ and $C_{\mathrm{LNV},\alpha}$ are proportional to $\cos \frac{\phi_v}{2}$ for $s$ even and to $\sin \frac{\phi_v}{2}$ for $s$ odd for light neutrino masses with strong NO, compare Eq.~(\ref{eq:CLFValphaCase2stevenNO}) and the text below Eq.~(\ref{eq:CLFValphaCase2stevenelectron}).
In Fig.~\ref{NO BAU phiv e-3e-6e-11 caseII massless} we verify this behaviour to a very
high degree (compare the coloured curves for different values of $\kappa$ with the black curve which shows the analytic expectation). As already seen for Case 1), the deviation from the expectation is larger for larger $\kappa$. In appendix~\ref{appF2}, see Fig.~\ref{IO BAU phiv e-3e-6e-11 caseII massless}, we display for completeness the corresponding plots for light neutrino masses with strong IO. Again, the expectations from the CP-violating combinations are confirmed.

Finally, we remind that the magnitude of the sine of the Majorana phase $\alpha$, appearing in the PMNS mixing matrix, is to very good degree
determined by $\sin \phi_v= 2 \, \sin \frac{\phi_v}{2} \, \cos \frac{\phi_v}{2}$~\cite{Hagedorn:2014wha}.
Thus, at least for light neutrino masses with $m_0=0$ we explicitly see that the vanishing of the BAU implies $\sin \alpha=0$ as well.

\begin{figure}
	\begin{subfigure}{.5\textwidth}
		\centering
		\includegraphics[width = \textwidth]{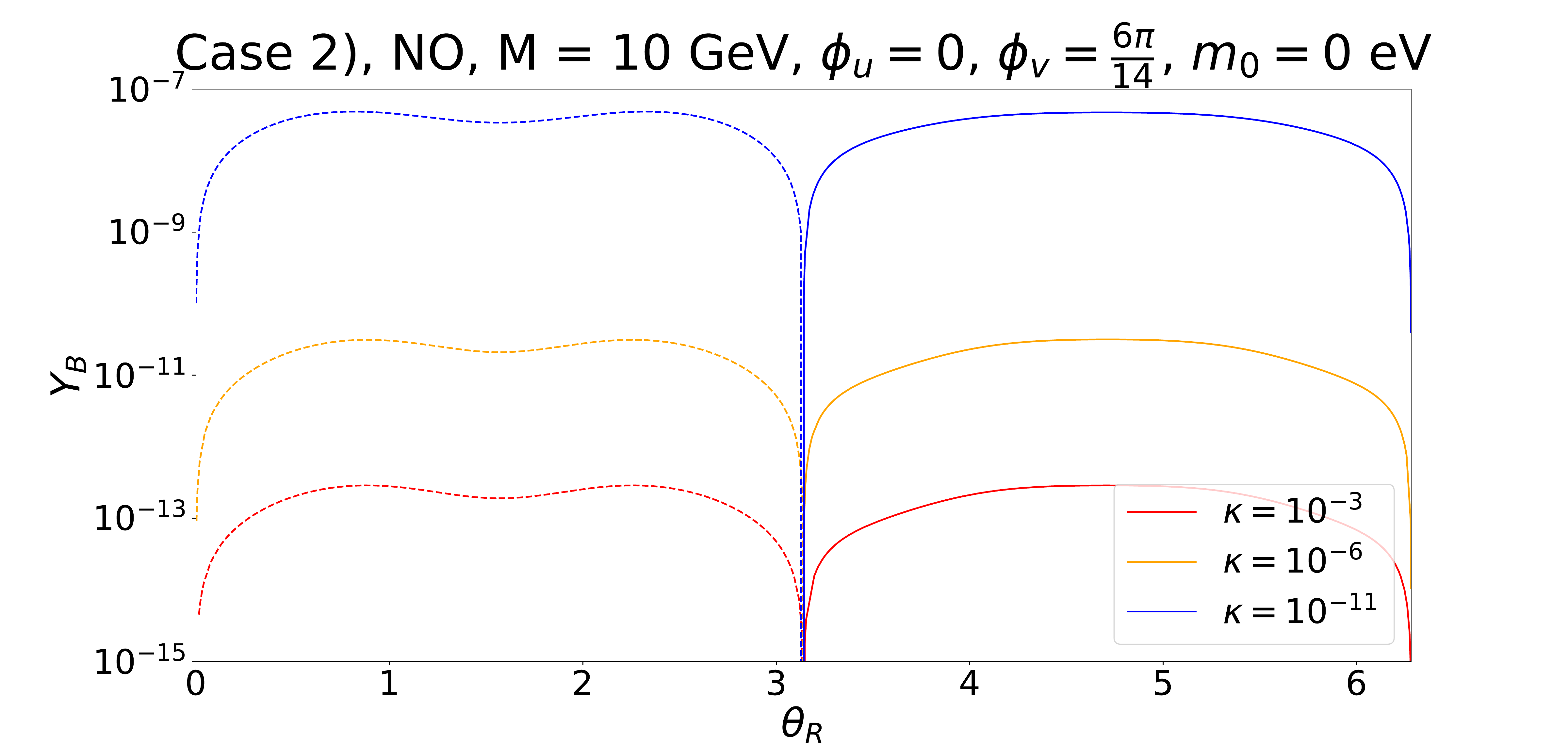}
		\caption{Vanishing initial conditions.}
	\end{subfigure}
	\begin{subfigure}{.5\textwidth}
		\centering
		\includegraphics[width = \textwidth]{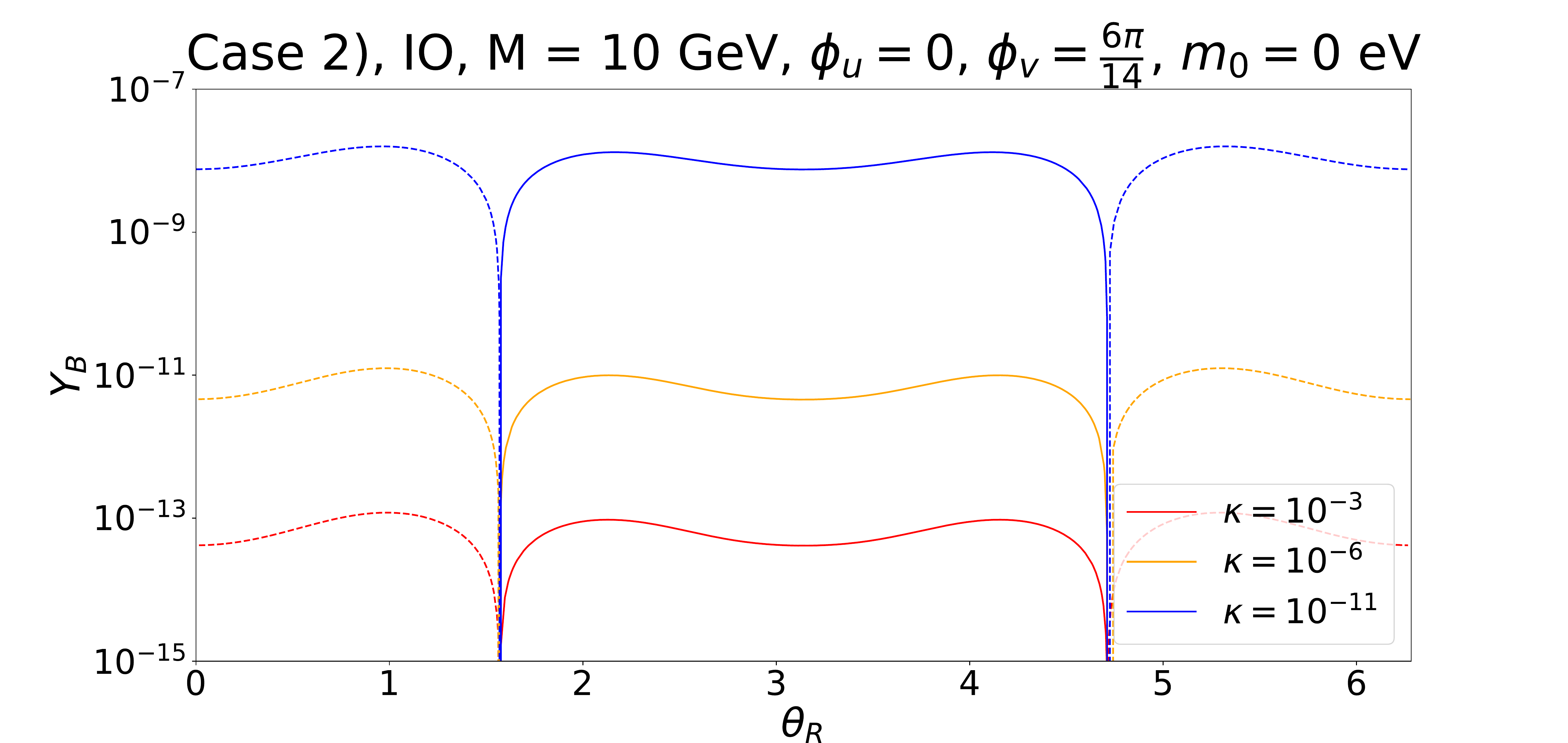}
		\caption{Vanishing initial conditions.}
	\end{subfigure}
\caption{{\small {\bf Case 2)} $Y_B$ as function of the angle $\theta_R$ for a Majorana mass $M=10$ GeV and different values of $\kappa$, $\kappa \in \{10^{-3},10^{-6},10^{-11}\}$. Results for both light neutrino masses with strong NO (left plot) and with strong IO (right plot) are displayed. Negative (dashed lines) as well as positive (continuous lines) values of the BAU are represented. The parameters $(s,t)$ have been fixed to $(1,2)$ which corresponds to $\phi_u=0$ and $\phi_v=\frac{6\pi}{14}$.}}
\label{NOIO BAU vs thetaR e-3e-6e-11 caseII massless}
\end{figure}

\paragraph{Dependence on angle $\theta_R$.} In Fig.~\ref{NOIO BAU vs thetaR e-3e-6e-11 caseII massless}, we fix the value of the Majorana mass $M$, $M=10$ GeV, as well as the values of $n$, $u$ and $v$ ($s$ and $t$) such that $\phi_u=0$ and $\phi_v=\frac{6 \, \pi}{14}$ and plot the BAU with respect to the angle $\theta_R$.  As is shown in Eqs.~(\ref{eq:CLFValphaCase2stevenNO}) and (\ref{eq:CLFValphaCase2stevenIO}), for a light neutrino mass spectrum with strong NO and strong IO, the CP-violating combinations $C_{\mathrm{LFV},\alpha}$ and $C_{\mathrm{LNV},\alpha}$ reveal a simple dependence on the angle $\theta_R$, i.e.~they are proportional to $\sin\theta_R$ and $\cos\theta_R$, respectively.
We thus expect vanishing BAU for
$\theta_R=0, \pi, 2 \, \pi$ for strong NO and $\theta_R=\frac{\pi}{2}, \frac{3 \, \pi}{2}$ for strong IO. This expectation is confirmed very well, as we can see in Fig.~\ref{NOIO BAU vs thetaR e-3e-6e-11 caseII massless}. Beyond this, the dependence of the BAU on the angle $\theta_R$ is also determined by the washout and, thus, is more complicated. We note that, contrary to Case 1), the masses $m_i$ of the light neutrinos, and consequently also the couplings $y_f$, $f=1,2,3$, do not depend on the angle $\theta_R$, see Eq.~(\ref{eq:tevenCase2numasses}). For this reason, the total mixing angle $U^2$ is always close to the seesaw line and special values of
$\theta_R$, at which the BAU is strongly enhanced, are absent in this case, compare Fig.~\ref{NOIO BAU vs thetaR e-3e-6e-11 caseII massless} and Fig.~\ref{NO thetaR BAU 1-10 GeV} for Case 1).

\mathversion{bold}
\subsubsection{Choice \texorpdfstring{$t$}{t} odd}
\mathversion{normal}
\label{sec:choicetodd}

For this choice of $t$, we first study the dependence of the BAU on the size of the splitting $\kappa$ for $m_0=0$.
Then, we analyse the interesting case in which the correct amount of BAU can be generated without the splitting $\kappa$ (and $\lambda$),
if the lightest neutrino mass $m_0$ is non-zero. Furthermore, we scrutinise the interplay between the different contributions to the generation of the BAU, encoded in the
CP-violating combinations $C_{\mathrm{LFV},\alpha}$ and $C_{\mathrm{LNV},\alpha}$ as well as $C_{\mathrm{DEG},\alpha}$, respectively, for non-zero $\kappa$ and $m_0$. Eventually, we comment on special values of $\theta_R$ which render
one of the couplings $y_f$, $f=1,2,3$, large, compare Eqs.~(\ref{eq:strongNOCase2}) and (\ref{eq:strongIOCase2}), and thus can lead to large total mixing angle $U^2$, while achieving the correct amount
of BAU.

\begin{figure}
	\begin{subfigure}{.5\textwidth}
		\centering
		\includegraphics[width = \textwidth]{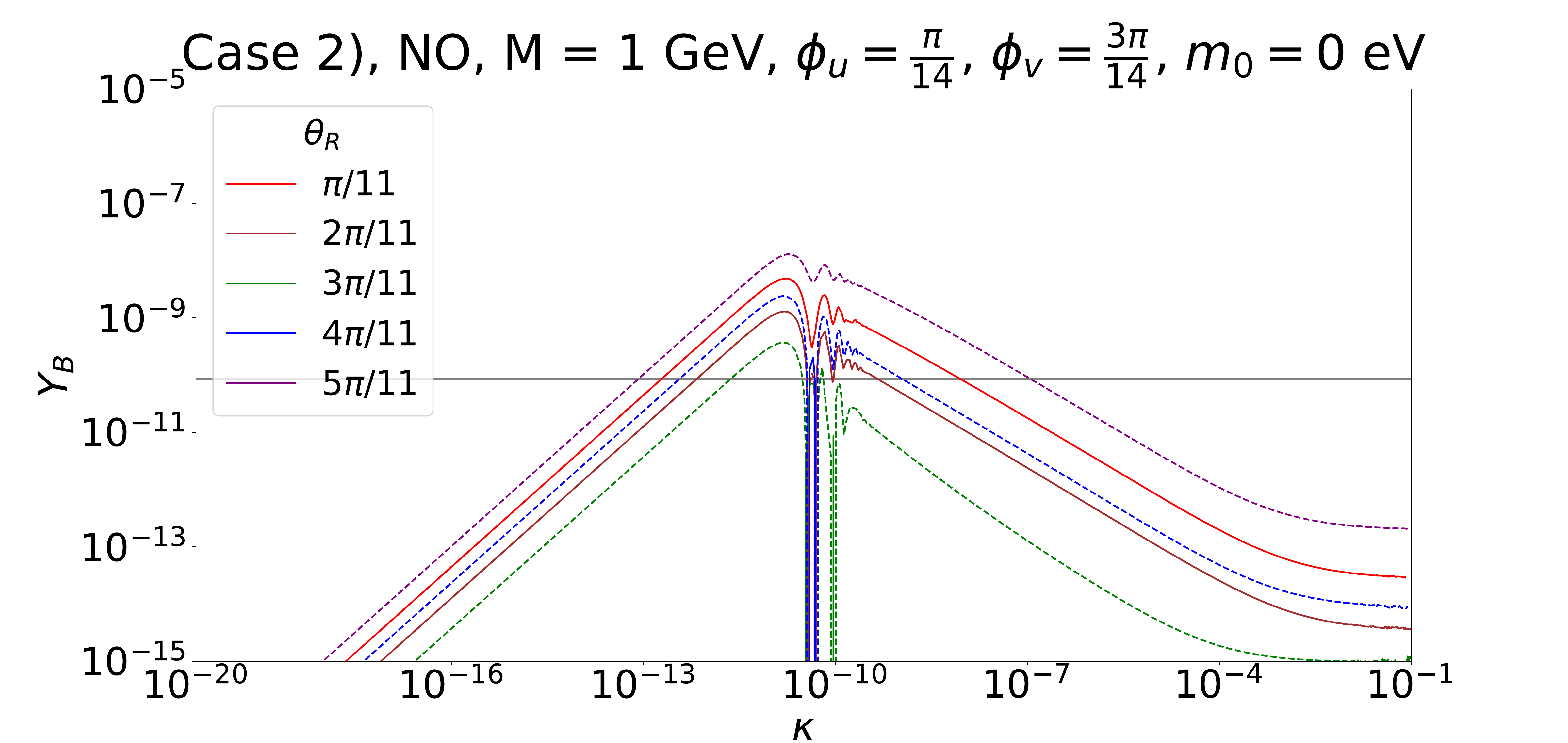}
		\caption{Vanishing initial conditions.}
	\end{subfigure}
	\begin{subfigure}{.5\textwidth}
		\includegraphics[width = \textwidth]{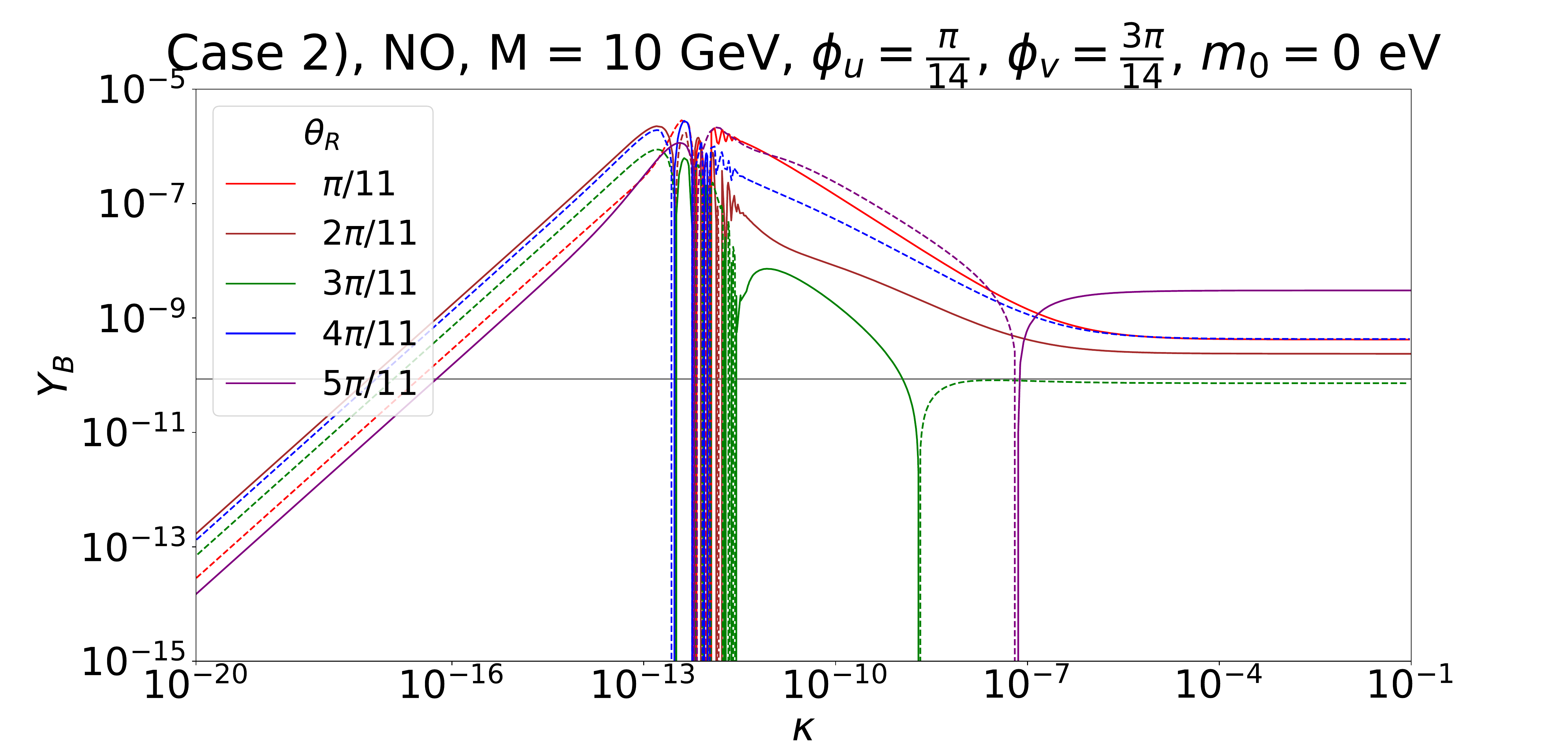}
		\caption{Vanishing initial conditions.}
		\label{NO kappa BAU 10 GeV Case II todd}
	\end{subfigure}
	\begin{subfigure}{.5\textwidth}
		\centering
		\includegraphics[width = \textwidth]{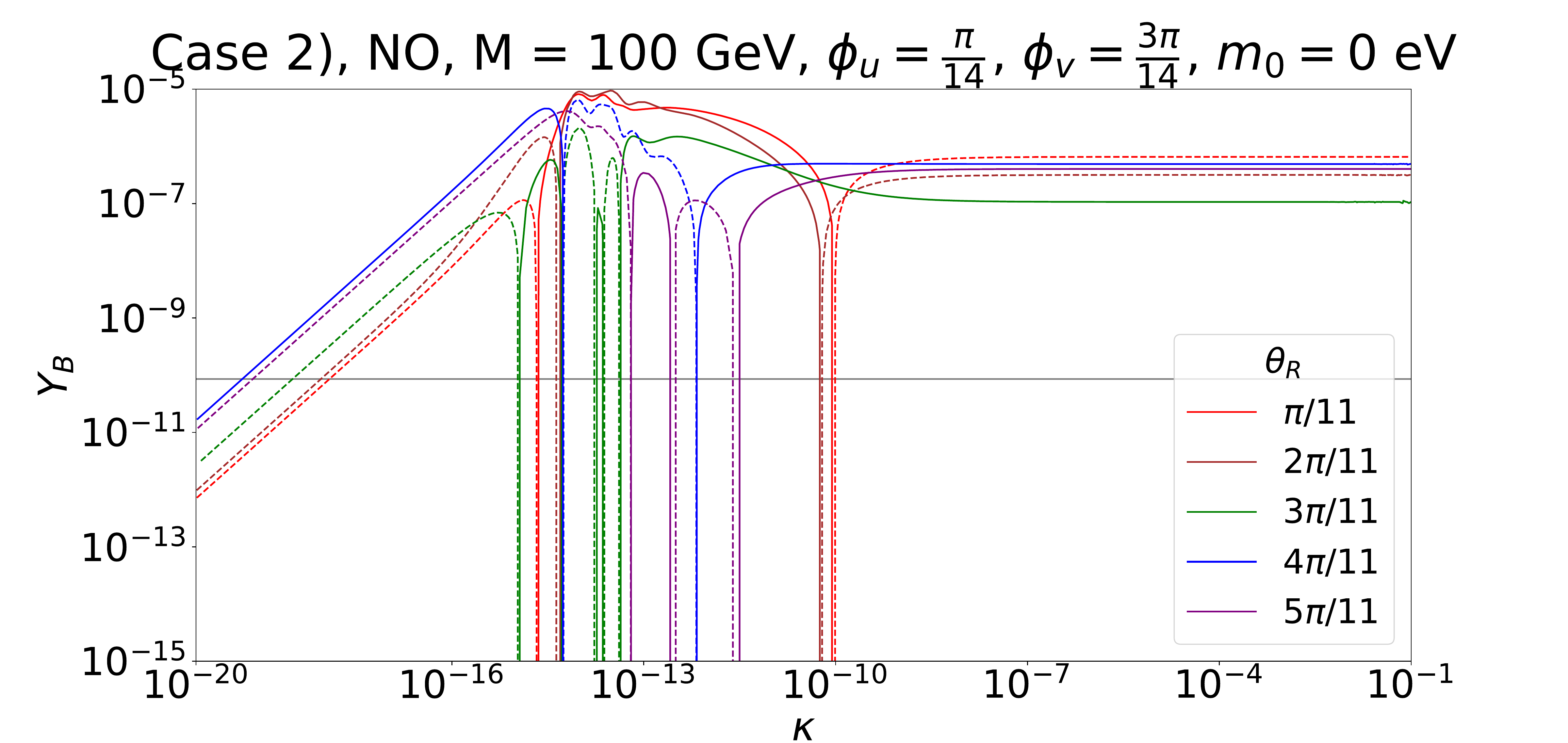}
		\caption{Vanishing initial conditions.}
		\label{NO kappa BAU 100 GeV Case II todd}
	\end{subfigure}
	\begin{subfigure}{.5\textwidth}
		\includegraphics[width = \textwidth]{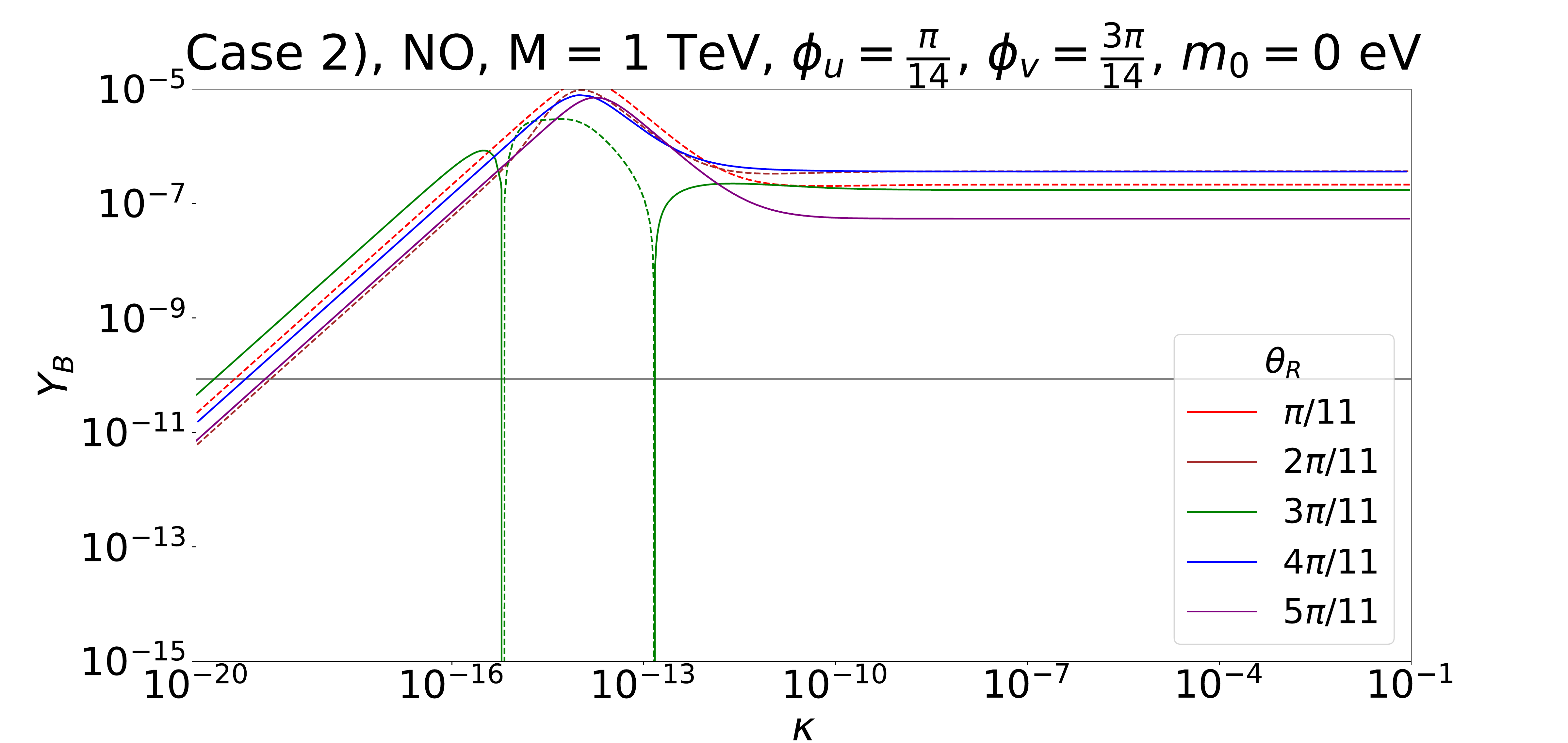}
		\caption{Vanishing initial conditions.}
	\end{subfigure}
	\caption{{\small {\bf Case 2)} $Y_B$ as function of $\kappa$ for a Majorana mass $M=1$ GeV (upper left plot), $M=10$ GeV (upper right plot), $M=100$ GeV (lower left plot) and $M=1$ TeV (lower right plot). For each of these choices, different values of $\theta_R$ have been studied. The parameters $(s,t)$ have been fixed to $(1,1)$ which corresponds to $\phi_u=\frac{\pi}{14}$ and $\phi_v=\frac{3\pi}{14}$. Light neutrino masses
		have strong NO.
Both negative (dashed lines) as well as positive (continuous lines) values of the BAU are represented. The grey line indicates the observed value of the BAU, $Y_B\approx 8.6 \cdot 10^{-11}$. The large number of oscillations around $\kappa \approx 10^{-12}$ to $10^{-11}$ in plot (b) is an artefact of the momentum averaged set of equations, see Eq.~\eqref{kin_eq}, used in this work. Explicitly including the momentum dependence would remove these, see~\cite{Ghiglieri:2017csp}.}}
\label{NO kappa BAU Case II todd different masses}
\end{figure}

\paragraph{Effect of splitting $\kappa$.}
In Fig.~\ref{NO kappa BAU Case II todd different masses} we can see the dependence of the BAU on the splitting $\kappa$. The overall behaviour is quite similar to Case 1), and we can once again observe a plateau for $\kappa \gg \Gamma_N \, E_N/M^2$. 
The presence of this plateau for $t$ being odd  is implied by the non-vanishing of the reduced mass-degenerate CP-violating combination $C^{(23)}_{\mathrm{DEG},\alpha}$, see Eq.~\eqref{eq:CDEG23_seventodd_Case2}.
For  GeV-scale RH neutrino masses, see plot (b) in Fig.~\ref{NO kappa BAU Case II todd different masses}, starting with $\kappa \gtrsim 10^{-12}$, the BAU becomes highly oscillatory.
These oscillations continue until $\kappa \sim 10^{-10}$, where they are damped by an effective ``decoherence term'' that has been introduced in~\cite{Klaric:2021cpi} instead of explicitly switching off the fast oscillations, as has been done in~\cite{Canetti:2012kh,Canetti:2010aw} -- which would lead to a vanishing source term at late times.
This procedure ensures that they approach the correct time average~\cite{Garbrecht:2011aw}, and, at the same time, keeps the oscillations without averaging when they are sufficiently slow.
We note that this is not a physical phenomenon, but only introduced to reduce the numerical stiffness of the problem.
The results in this region require a more thorough investigation, since this damping is not physical and should be decreased until the BAU is independent of it.
Finally, we note that these oscillations are not entirely physical in the relativistic regime, but are an artefact of neglecting the full momentum dependence of the system, c.f.~\cite{Ghiglieri:2017csp}, as each momentum mode $k$ has a different oscillation frequency $\sim M \Delta M_{ij}/k$, which erases the effect of oscillations in the BAU.
It is interesting to note that we observe these oscillations exactly when transitioning between the relativistic and non-relativistic regimes,
where all momentum modes oscillate with the same frequency $\sim \Delta M_{ij}$,
and the oscillations become coherent, which is also reflected in the BAU.
Nonetheless, the envelope of the curves can serve as a rough order-of-magnitude estimate for the BAU in this regime.

In most of the analysis, we focus on the choice $u=1$, see e.g.~Fig.~\ref{NO kappa BAU Case II todd different masses}. However, we have also studied the behaviour of the BAU with respect to the splitting $\kappa$ for $u=-1$ (corresponding to $s=0$ and $t=1$) and otherwise
the same choice of parameters as in Fig.~\ref{NO kappa BAU Case II todd different masses}.
For the Majorana mass $M$ being $M=10$ GeV, the results are displayed in plots (a) and (b) in Fig.~\ref{NOIO kappa BAU 10 GeV Case II todd u-1} in appendix~\ref{appF2} for light neutrino masses with strong NO and with strong IO, respectively. We can, in particular, compare plot (a) in Fig.~\ref{NOIO kappa BAU 10 GeV Case II todd u-1} with plot (b) in Fig.~\ref{NO kappa BAU Case II todd different masses} and see that the qualitative behaviour
of the BAU with $\kappa$ is very similar, while the value of the BAU for large values of $\kappa$ turns out to be somewhat lower for $u=-1$ than
for $u=1$.

\begin{figure}
	\begin{subfigure}{.5\textwidth}
		\centering
		\includegraphics[width = \textwidth]{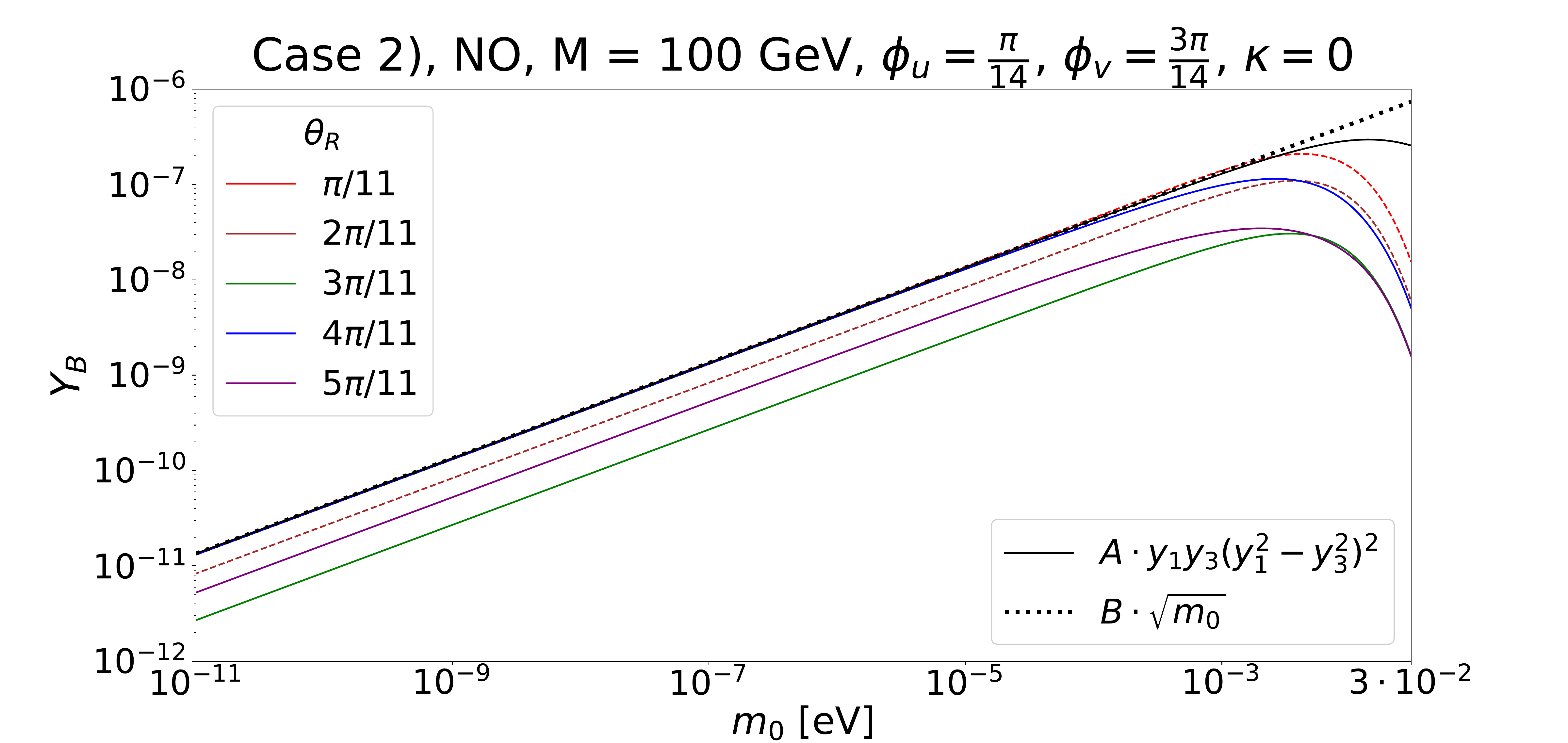}
		\caption{Vanishing initial conditions.}
	\end{subfigure}
	\begin{subfigure}{.5\textwidth}
		\centering
		\includegraphics[width = \textwidth]{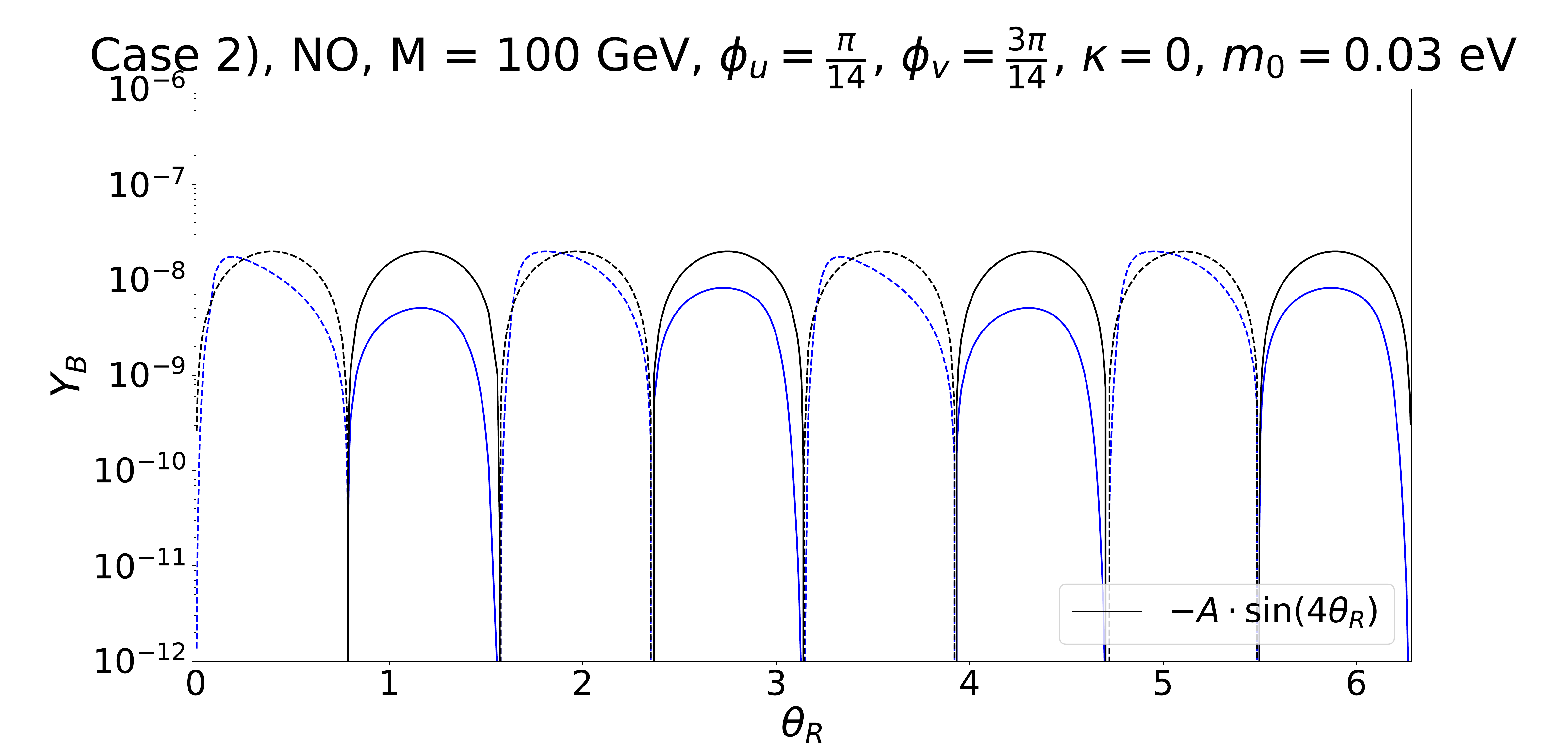}
		\caption{Vanishing initial conditions.}
	\end{subfigure}
	\caption{{\small {\bf Case 2)} Results for $Y_B$ in the absence of splittings, $\kappa=0$ (and $\lambda=0$). We set the Majorana mass $M$ to $M=100$ GeV and fix the group theory parameters
		$n$, $s$ and $t$ to $n=14$, $s=1$ and $t=1$ such that $\phi_u=\frac{\pi}{14}$ and $\phi_v=\frac{3 \, \pi}{14}$.
		Light neutrino masses follow NO.
The left plot displays $Y_B$ as function of the lightest neutrino mass $m_0$ for different values of $\theta_R$. In addition, we compare, for $\theta_R=\frac{\pi}{11}$, the numerical results (red curve) with the following two analytical expressions: $A \cdot y_1 \, y_3 \, (y_1^2-y_3^2)^2$ with $A \approx -6.14 \cdot 10^{31}$ (continuous black line) and $B \cdot \sqrt{m_0}$ with $B \approx -4.26 \cdot 10^{-6} \mbox{eV}^{-1/2}$ (dotted black line). The first expression reflects the dependence of the CP-violating combination $C_{\mathrm{DEG},\alpha}$ on the couplings $y_f$, compare Eq.~(\ref{eq:CDEGalpha_seventodd_Case2}), while the second one gives the overall dependence of the BAU on $m_0$. The right plot shows $Y_B$ as function of the angle $\theta_R$. We compare the numerical result (blue curve) to the analytical expectation from the CP-violating combination $C_{\mathrm{DEG},\alpha}$ which is proportional to $\sin 4\,  \theta_R$ (black curve). The coefficient $A$ has been chosen to be Max($|Y_B|$). Both negative (dashed lines) as well as positive (continuous lines) values of the BAU are represented.}}
\label{NO thetaR/m0 BAU 100 GeV caseII odd no mass splittings}
\end{figure}

\begin{figure}
\begin{subfigure}{.5\textwidth}
	\centering
	\includegraphics[width = \textwidth]{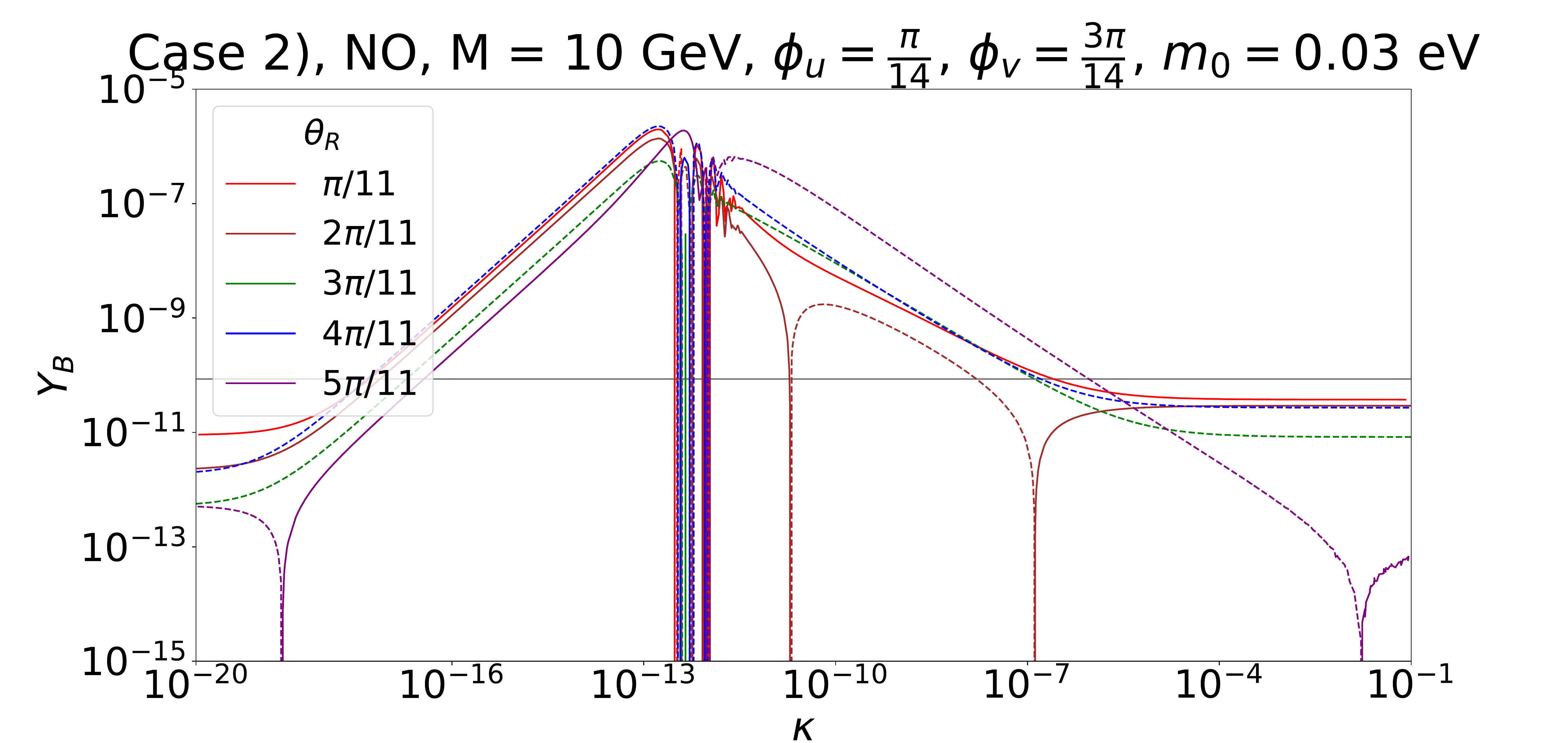}
	\caption{Vanishing initial conditions.}
\end{subfigure}
\begin{subfigure}{.5\textwidth}
	\centering
	\includegraphics[width = \textwidth]{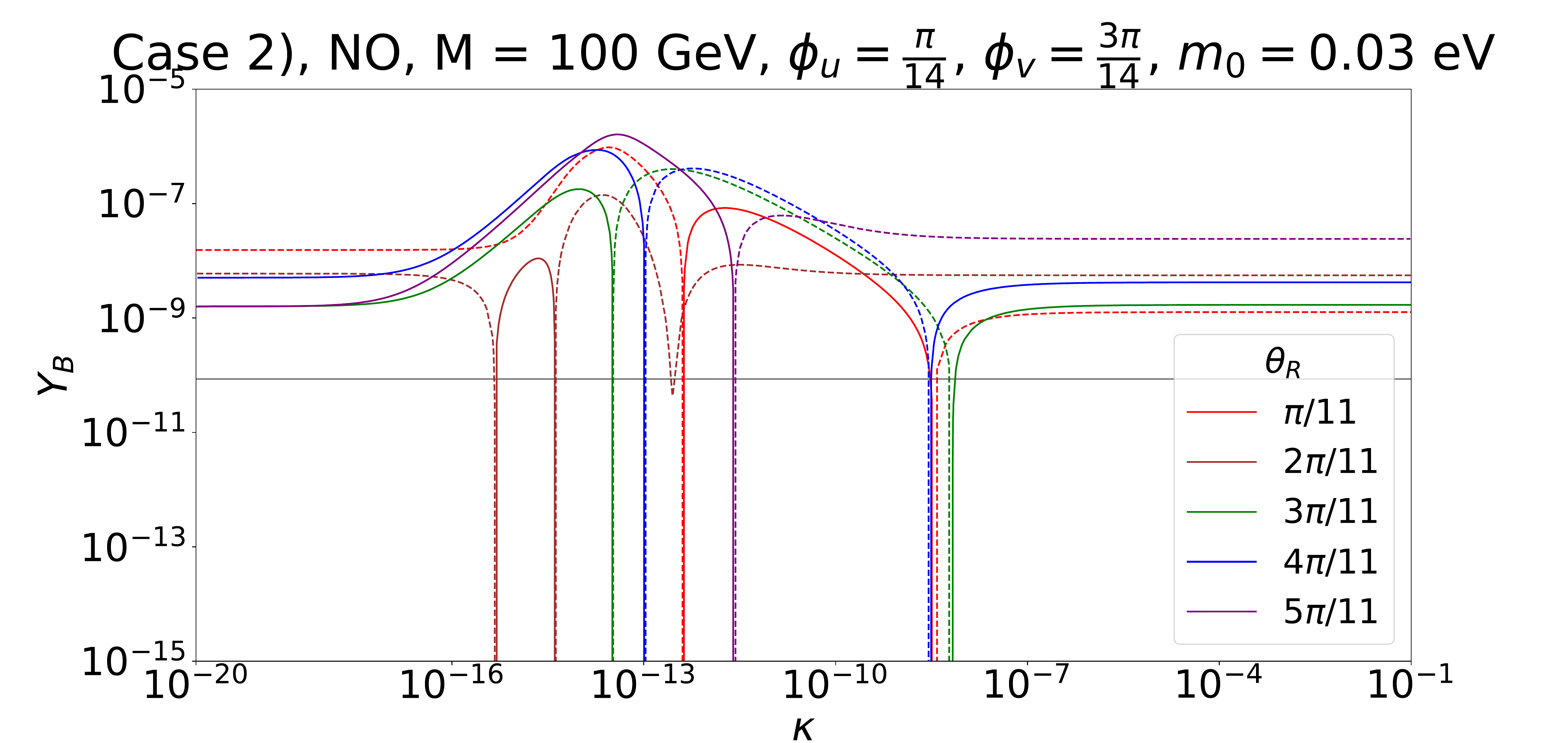}
	\caption{Vanishing initial conditions.}
\end{subfigure}
\caption{{\small {\bf Case 2)} $Y_B$ as function of the splitting $\kappa$ for different values of $\theta_R$ and for $M=10$ GeV (left plot) and $M=100$ GeV (right plot). The group theory parameters $n$, $u$ and $v$ are taken to be $n=14$, $u=1$ and $v=3$ (equivalent to $s=1$ and $t=1$).
Light neutrino masses follow NO and $m_{0}$ has been set to $0.03$ eV. Since both the splitting $\kappa$ and $m_0$ do not vanish, we can study the interplay between different contributions to the BAU, associated with the three different CP-violating combinations $C_{\mathrm{LFV},\alpha}$,  $C_{\mathrm{LNV},\alpha}$ and $C_{\mathrm{DEG},\alpha}$.
For the rest of the choices see Fig.~\ref{NO kappa BAU Case II todd different masses}.
}
}
\label{NO BAU vs kappa caseII odd massive}
\end{figure}

\paragraph{Results for vanishing splittings.} Leptogenesis is possible,
even if $\hat{M}_R$, see Eq.~(\ref{eq:MRUR}), is proportional to the identity  matrix. As pointed out in~\cite{Antusch:2017pkq}, this is caused by thermal effects, more precisely by the mismatch between the heavy neutrino mass basis and interaction basis at finite temperature, i.e., by the different flavour structures in $H_N$ and the FNC and FNV contributions to the thermal rate $\Gamma$ in Eq.~\eqref{QKE}.
This general behaviour can be verified for the scenario at hand by noting that, for $\kappa=0$ and $\lambda=0$, CP violation can only appear through thermal effects. The CP-violating combination $C_{\mathrm{DEG},\alpha}$, indeed, appears due to the mismatch between the heavy neutrino thermal masses and the decay rate $\Gamma$. This has as consequence that the BAU is very sensitive to the mass of the heavy neutrinos in this regime. We observe that it is enhanced by approximately four orders of magnitude between $M=10$ GeV and $M=100$ GeV, compare plot (a) in Fig.~\ref{NO thetaR/m0 BAU 100 GeV caseII odd no mass splittings} for $M=100$ GeV with plot (a) in Fig.~\ref{NO m0 BAU 10 GeV/1 TeV caseII odd no mass splittings} in appendix~\ref{appF2} for $M=10$ GeV, something not observed for $\kappa\neq 0$, compare e.g.~plots (b) and (c) in Fig.~\ref{NO kappa BAU Case II todd different masses}. Such an enhancement is also visible in Fig.~\ref{NO BAU vs kappa caseII odd massive} for very small values of $\kappa$.

For Case 2), leptogenesis with $\kappa=0$ and $\lambda=0$ requires
the lightest neutrino mass $m_0$ to be non-zero, as can be checked numerically as well as analytically with the help of the CP-violating
combination $C_{\mathrm{DEG},\alpha}$. This combination is proportional to both couplings $y_1$ and $y_3$, see Eq.~(\ref{eq:CDEGalpha_seventodd_Case2}) and the text below Eq.~(\ref{eq:CLNVsummed_seventodd_Case2}). The exact dependence of $C_{\mathrm{DEG},\alpha}$ on these couplings
is $y_1 \, y_3 \, (y_1^2-y_3^2)^2$. This dependence is confirmed in Fig.~\ref{NO thetaR/m0 BAU 100 GeV caseII odd no mass splittings}, plot (a), where we plot the value of the BAU with respect to the lightest neutrino mass $m_0$.
We have fixed not only $\kappa=0$ and $\lambda=0$, but also the Majorana mass $M$ to $M=100$ GeV. The choices of $n$, $u$ and $v$ are $n=14$, $u=1$ and $v=3$ (corresponding to $s$ and $t$ being both one) such that $\phi_u=\frac{\pi}{14}$ and $\phi_v=\frac{3 \, \pi}{14}$. For all five different values of $\theta_R$ that are shown, we clearly see that the dependence of the BAU on the couplings is very well reproduced by the analytic result (compare the coloured curves with the black continuous curve), if $m_0$ is smaller than $10^{-2}$ eV. Since we show the BAU with respect to $m_0$, we note in addition that the resulting curves for all studied values of $\theta_R$ are proportional to $\sqrt{m_0}$ for $m_0 \lesssim 10^{-3}$ eV.
We note that, for completeness, the same type of plots can be found for $M=10$ GeV and $M=1$ TeV in Fig.~\ref{NO m0 BAU 10 GeV/1 TeV caseII odd no mass splittings} in appendix~\ref{appF2}.

In a similar way, we can verify the expected dependence of the BAU on the angle $\theta_R$ which should be of the form $\sin 4 \, \theta_R$ according to the expression for the CP-violating combination $C_{\mathrm{DEG},\alpha}$ in Eq.~(\ref{eq:CDEGalpha_seventodd_Case2}). The numerical results are shown in plot (b) in Fig.~\ref{NO thetaR/m0 BAU 100 GeV caseII odd no mass splittings}. We have used the same values for $M$, $\kappa$, $n$, $u$ and $v$ as in plot (a) of this figure, but now fixed the lightest neutrino mass to its largest possible value for NO, $m_0=0.03$ eV. We can see that the values of $\theta_R$ that should lead to vanishing BAU coincide with the expectation that $\theta_R$ should be a multiple of $\frac{\pi}{4}$. Nevertheless, the functional dependence on the angle $\theta_R$ is also determined by the washout, as can be inferred from the
visible differences between the numerical (blue curve) and the analytical result (black curve) in plot (b) in Fig.~\ref{NO thetaR/m0 BAU 100 GeV caseII odd no mass splittings}.

\paragraph{Results for $\kappa$ and $m_0$ non-zero.}  The interplay between the different contributions associated with the CP-violating combinations
$C_{\mathrm{LFV},\alpha}$ and $C_{\mathrm{LNV},\alpha}$ as well as $C_{\mathrm{DEG},\alpha}$ is studied for two different values of $M$, $M=10$ GeV and $M=100$ GeV, and several values
of $\theta_R$. The light neutrino mass spectrum has NO and the lightest neutrino mass is $m_0=0.03$ eV. Furthermore, we use $n=14$, $u=1$ and $v=3$ (corresponding to $s=1$ and $t=1$), that give rise to $\phi_u=\frac{\pi}{14}$ and $\phi_v=\frac{3 \, \pi}{14}$. As we can see in Fig.~\ref{NO BAU vs kappa caseII odd massive}, for very small values of $\kappa$ a value of the BAU which is independent of $\kappa$ is observed,
corresponding to the value of the BAU for vanishing splitting $\kappa$. The enhancement of the BAU for $\kappa$ in the range $10^{-16} \lesssim \kappa \lesssim 10^{-7}$ is due to the contributions related to the CP-violating combinations $C_{\mathrm{LFV},\alpha}$ and $C_{\mathrm{LNV},\alpha}$. For even larger values of $\kappa$, the value of the BAU is mostly determined
by the reduced mass-degenerate CP-violating combination $C^{(23)}_{\mathrm{DEG},\alpha}$ and attains a plateau.

\begin{figure}[t!]
	\centering
	\includegraphics[width = 0.49\textwidth]{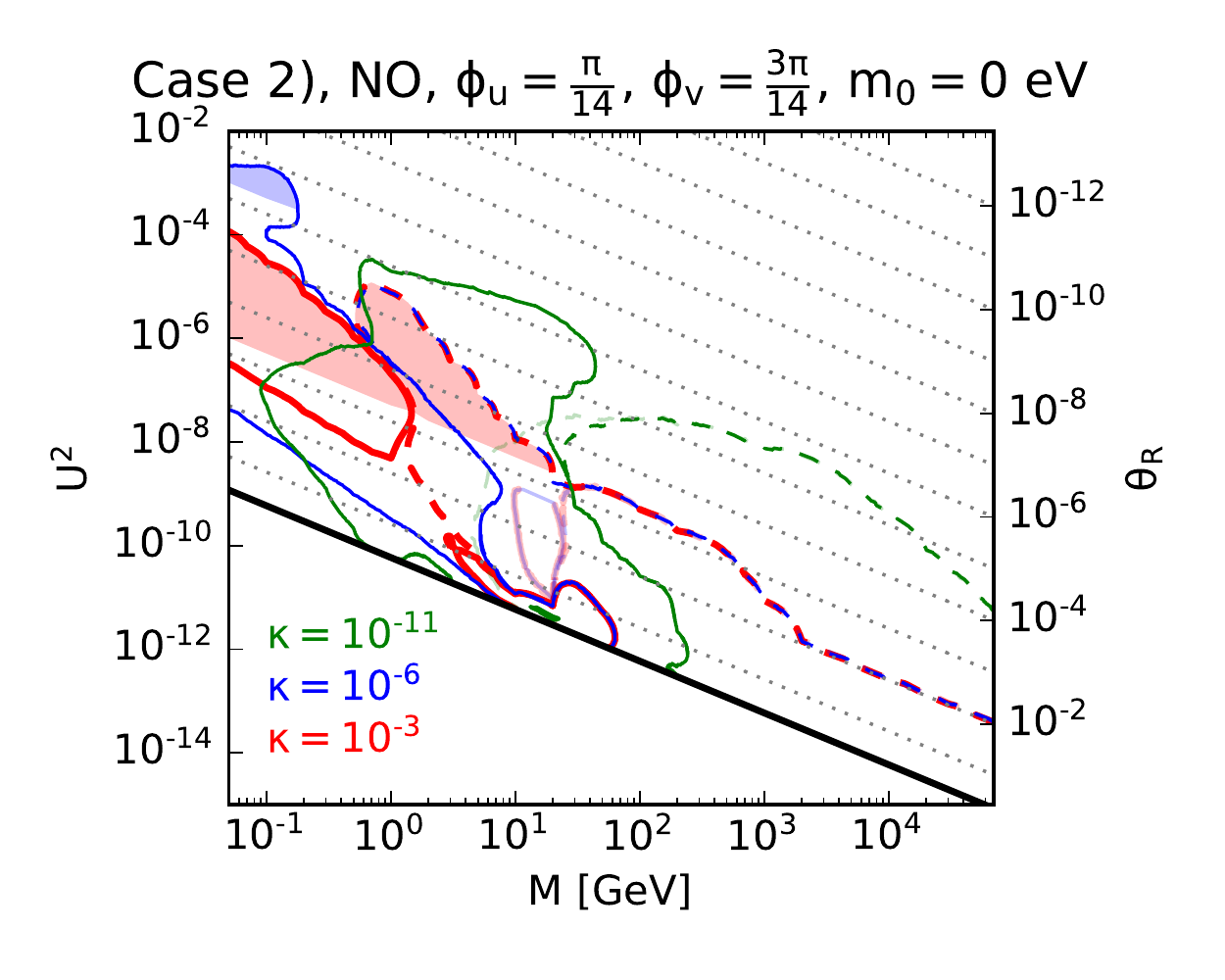}
	\includegraphics[width = 0.49\textwidth]{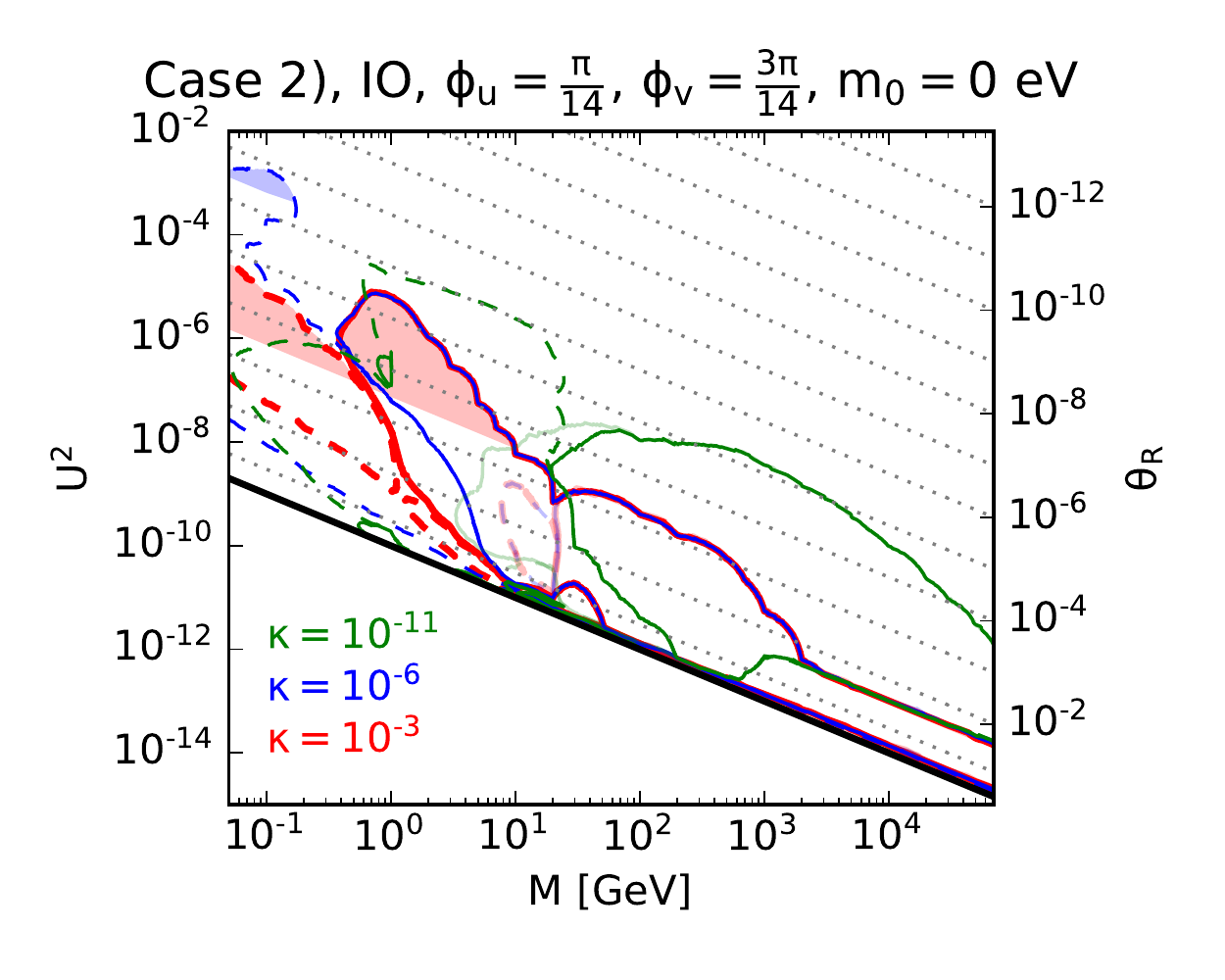}
	\caption{{\small {\bf Case 2)} Range of total mixing angle $U^2$ consistent with leptogenesis for the Majorana mass $M$ varying between $50$ MeV and $70$ TeV. The splitting $\kappa$ is chosen among the three values $\kappa \in \{10^{-3}, 10^{-6}, 10^{-11}\}$. The angle $\theta_R$ can be read off. The values of $\phi_u$ and $\phi_v$ are fixed to $\phi_u=\frac{\pi}{14}$ and $\phi_v=\frac{3 \, \pi}{14}$ corresponding to $t$ odd. Light neutrino masses follow strong NO (left plot) and strong IO (right plot), respectively.
		The contours corresponding to positive (negative) value of the BAU are shown in continuous (dashed) lines. Note that these refer to vanishing initial conditions except for the lines shown in fainter colours.
		For masses larger than $M\sim 10$ GeV, we see that the lines for $\kappa = 10^{-6}$ and $\kappa = 10^{-3}$ coincide.
		This can be verified in Fig.~\ref{NO kappa BAU Case II todd different masses}, where $Y_B$ becomes independent of the splitting $\kappa$ for $\kappa \gtrsim 10^{-6}$. In contrast, for $\kappa=10^{-11}$ both $Y_B$ and the allowed total mixing angle $U^2$ are generically larger for larger masses $M$. The blue and red shaded areas show the regions in which a condition like the one in Eq.~(\ref{eq:kappacorrectionscriterion}) is not fulfilled. The black lines indicate the seesaw line.
	Note the striking similarity between these results and those obtained in Case 1), see Fig.~\ref{NO Mass U2 Case1}.
	}}
	\label{NO Mass U2 Case2}
\end{figure}

\paragraph{Special values of $\theta_R$.} For $t$ being odd, the light neutrino masses $m_i$ depend, like for Case 1), on the angle $\theta_R$. In particular, we observe that for strong NO (IO) the coupling $y_3$ ($y_1$) has to be proportional to $1/\sqrt{|\sin 2 \, \theta_R|}$ for a fixed value
of $m_3$ ($m_1$). So, for values of $\theta_R$ close to multiples of $\pi/2$, one of the couplings can be very large and in turn also the
total mixing angle $U^2$. Similar to Case 1), in the vicinity of such special values a sufficient amount of BAU can also be generated.
This is shown in Fig.~\ref{NO Mass U2 Case2}, where we plot in the $M-U^2$-plane the contours (continuous lines for positive BAU and dashed ones for negative BAU) corresponding to the observed amount of BAU for different values of the splitting $\kappa$ (see Fig.~\ref{fig:Case2_MU2IC} in appendix~\ref{appF2} for dedicated plots for each value of the splitting $\kappa$). The choices of the group theory parameters are $n=14$, $u=1$ and $v=3$ (corresponding to $s=1$ and $t=1$),
while all further choices and conventions for colour coding are the same as in Fig.~\ref{NO Mass U2 Case1} for Case 1). We note the striking similarity between Figs.~\ref{NO Mass U2 Case1} and \ref{NO Mass U2 Case2}, up to the overall sign of the BAU (exchange of continuous and dashed lines). In appendix~\ref{appF2}, see Fig.~\ref{NO BAU vs U2 10 GeV e-6 multiple curves caseII}, we also show the dependence
of the BAU on the combination $U^2 \cdot M$ for one value of $\kappa$, $\kappa=10^{-6}$, and several different values of the Majorana
mass $M$, ranging between $M=100$ MeV and $M=1$ TeV.

In section~\ref{sec41}, we show that for the special values of $\theta_R$ associated with $1/\sqrt{|\sin 2 \, \theta_R|}$ becoming large, i.e.~$\theta_R$ being a multiple of $\pi/2$, the residual symmetry of the combination $Y_D^\dagger Y_D$ is enhanced. Consequently, the size of the deviation of $\theta_R$
from one such value is controlled by the amount of breaking of this enhanced symmetry.

\subsection{Case 3 a) and Case 3 b.1)}
\label{sec55}

In this section, we exemplify the results for the BAU, obtained from low-scale leptogenesis, for Case 3 a) and Case 3 b.1).
We focus the discussion on Case 3 b.1), since for Case 3 a) the accommodation of the lepton mixing angles
requires rather large values of the index $n$ of the flavour group, $n \geq 16$, compare section~\ref{sec503}.

As shown in section~\ref{sec33}, we have to distinguish two different sub-cases for Case 3 a) and Case 3 b.1), either $m$ and $s$ are both even (both odd) or $m$ is even and $s$ odd (or vice versa). In order to cover these different sub-cases, we mainly fix $m$ to be even
and split the discussion into two parts, $s$ even and $s$ odd. For the latter choice, it is possible to generate the correct amount of BAU
without the splitting $\kappa$ (and $\lambda$) and for a light neutrino mass spectrum with strong NO. While the possibility to achieve
the BAU without splittings is also given for the choice $t$ odd for Case 2), see section~\ref{sec:choicetodd}, the even more minimal option in which we
also have strong NO, i.e.~$m_0=0$, can instead only be realised for Case 3 b.1).
To cover both sub-cases, we choose the index $n$ to be $n=20$ and set $m=10$. The latter fixes the choice of the generator of the residual $Z_2$ symmetry in the neutrino sector, compare section~\ref{subsec:case3ressymm}.
As shown in section~\ref{sec503} and in Tab.~\ref{tab:Case3b1n20m10} in appendix~\ref{appE}, for this choice several values of $s$, even as well as odd, allow to accommodate the data on lepton mixing well. We note that for $s$ odd we also consider other choices of $n$ and $m$ in order to elucidate the dependence of the BAU on $s/n$, treated as continuous parameter. At the end of this section, we briefly discuss the choice $m$ odd and $s$ even, as it is the only combination of $m$ and $s$ that leads to a non-zero value of the BAU for large $\kappa$.

\mathversion{bold}
\subsubsection{Choice \texorpdfstring{$m$}{m} even and \texorpdfstring{$s$}{s} even}
\mathversion{normal}

For this choice of $m$ and $s$, we observe that the PMNS mixing matrix always only depends on the angle $\theta_L$
independently of the value of the lightest neutrino mass, compare Eq.~(\ref{eq:combmnuCase3amsevenodd}). Given this, we set $m_0$ to its largest possible value
for light neutrino masses with NO and only choose $m_0=0$ in order to confront the numerical results with simple formulae for the CP-violating
combinations $C_{\mathrm{LFV},\alpha}$
and $C_{\mathrm{LNV},\alpha}$, see Eq.~(\ref{eq:CLFValphamsevenNO}). We first discuss the dependence of the BAU on the splitting $\kappa$ and then for $m_0=0$
we scrutinise its dependence on the choice of the CP transformation $X (s)$ as well as on the angle $\theta_R$.

\begin{figure}
	\begin{subfigure}{.5\textwidth}
		\centering
		\includegraphics[width = \textwidth]{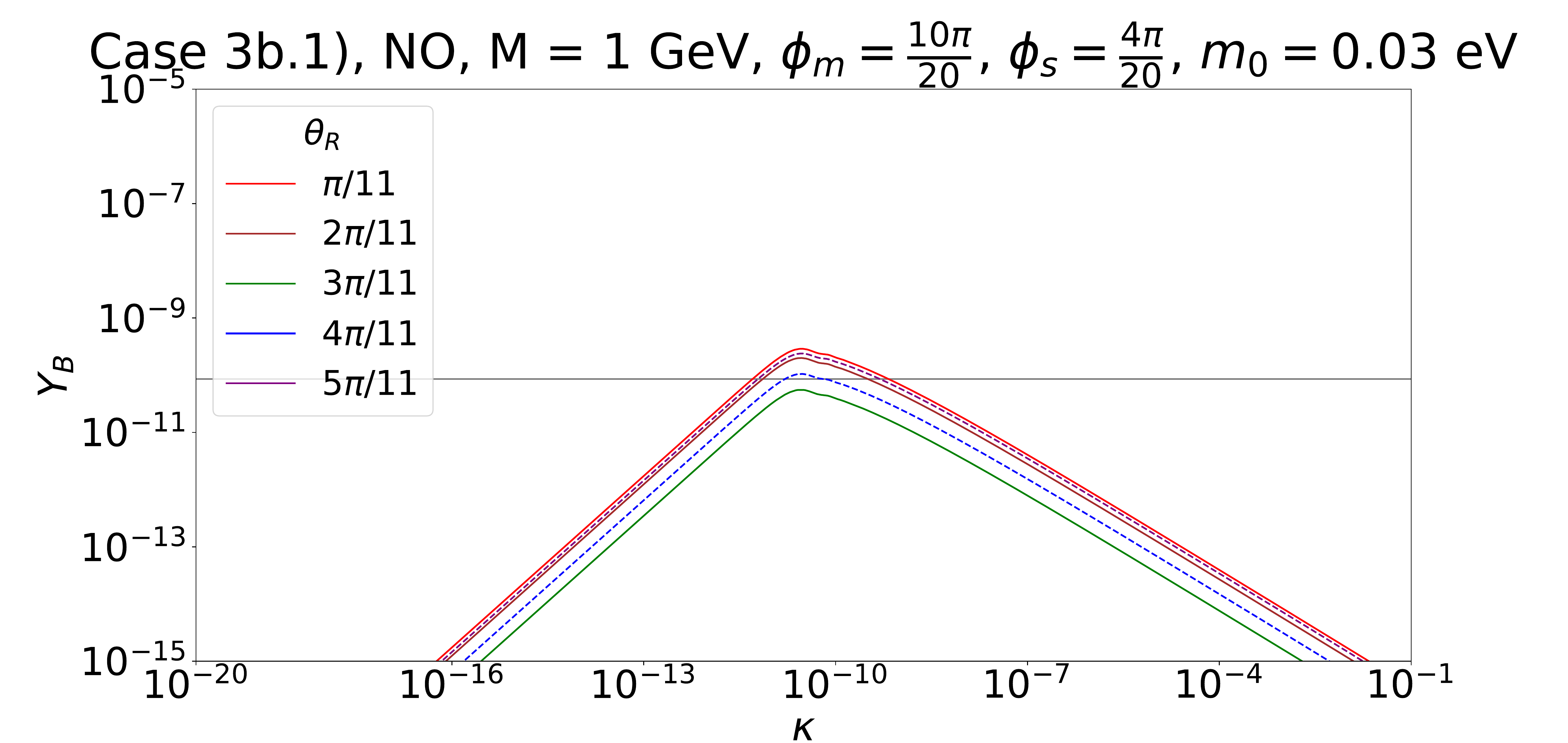}
		\caption{Vanishing initial conditions.}
	\end{subfigure}
	\begin{subfigure}{.5\textwidth}
		\centering
		\includegraphics[width = \textwidth]{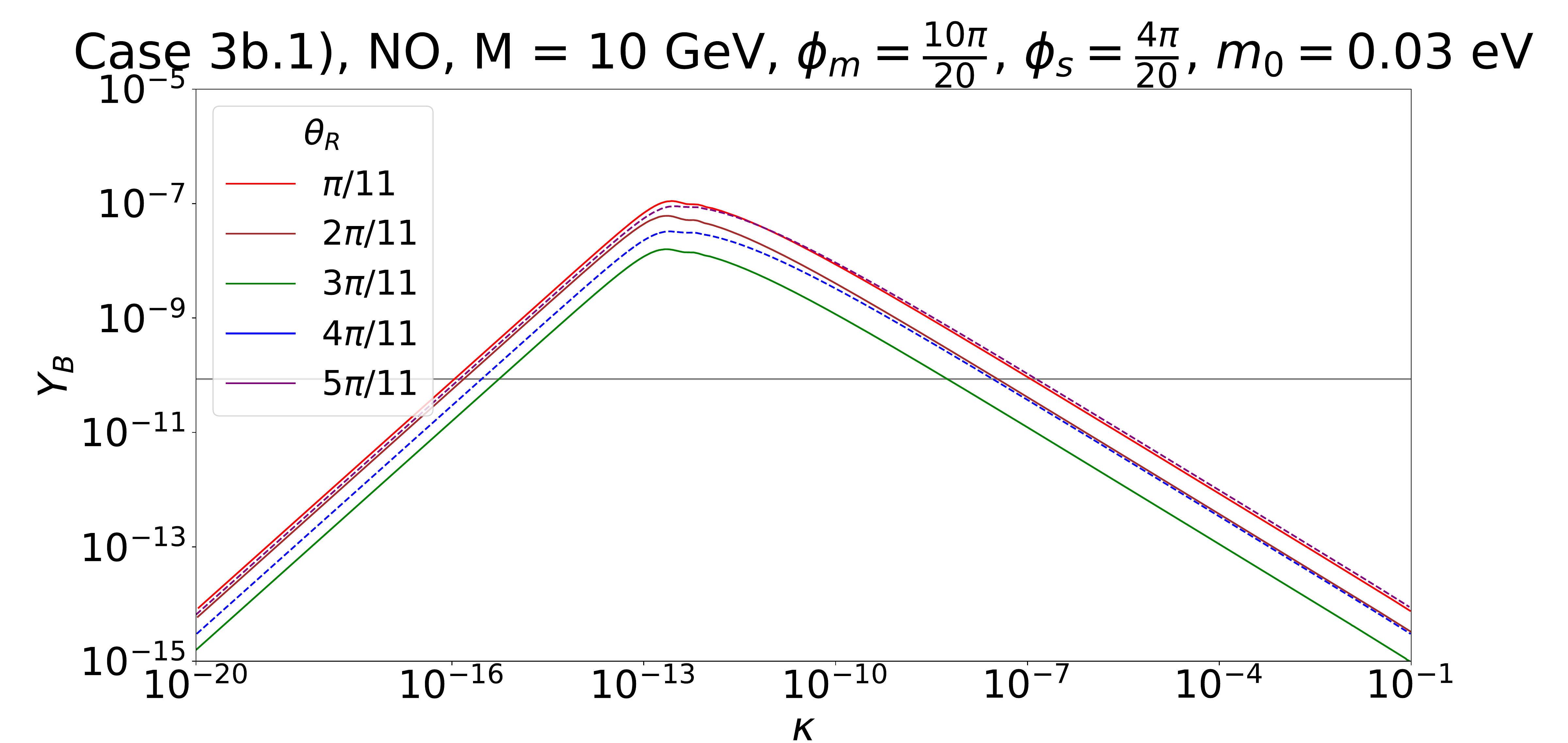}
		\caption{Vanishing initial conditions.}
		\label{NO kappa BAU 10 GeV IIIb even massive}
	\end{subfigure}
	\begin{subfigure}{.5\textwidth}
		\includegraphics[width = \textwidth]{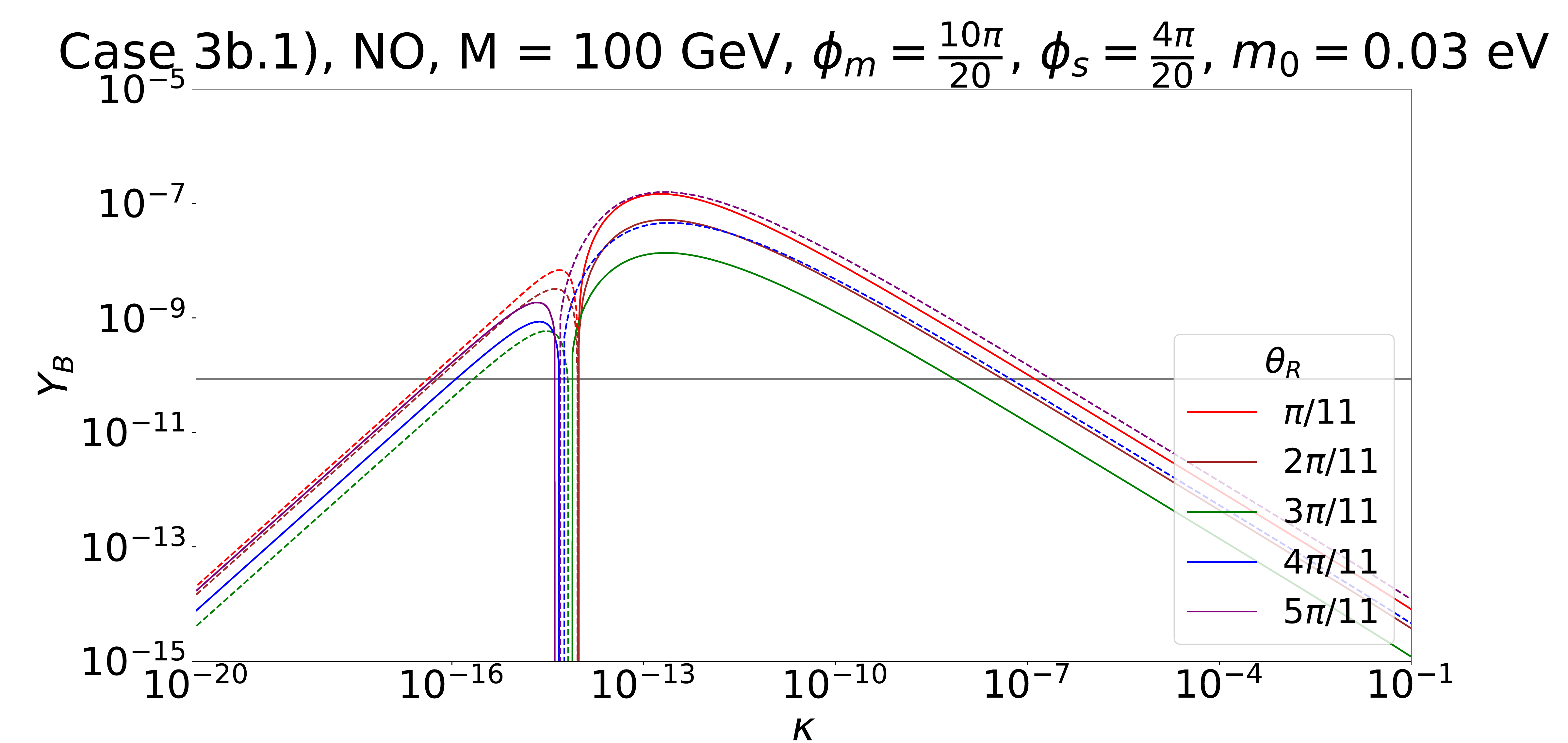}
		\caption{Vanishing initial conditions.}
	\end{subfigure}
	\begin{subfigure}{.5\textwidth}
		\includegraphics[width = \textwidth]{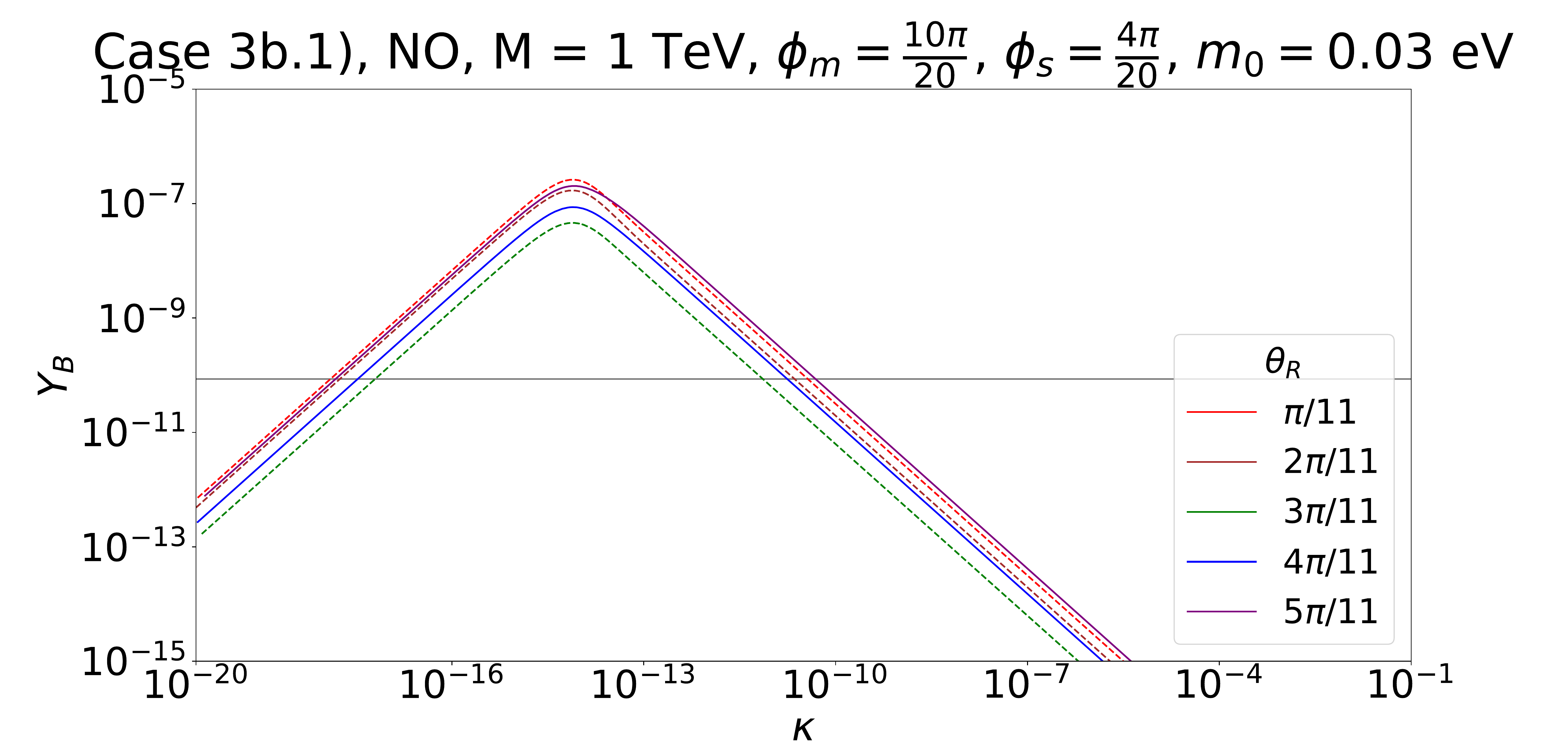}
		\caption{Vanishing initial conditions.}
	\end{subfigure}
	\caption{{\small {\bf Case 3 b.1)} $Y_B$ as function of the splitting $\kappa$ for a Majorana mass $M=1$ GeV (upper left plot), $M=10$ GeV (upper right plot), $M=100$ GeV (lower left plot) and $M=1$ TeV (lower right plot). For each of these plots, different values of $\theta_R$ have been studied. The group theory parameters $n$, $m$ and $s$ are fixed to $n=20$, $m=10$ and $s=4$ so that $\phi_m=\frac{10 \, \pi}{20}$ and $\phi_s=\frac{4 \, \pi}{20}$.
Light neutrino masses are assumed to follow NO with $m_0=0.03$ eV.
Both negative (dashed lines) as well as positive (continuous lines) values of the BAU are represented. The grey line indicates the observed value of the BAU, $Y_B\approx 8.6 \cdot 10^{-11}$.}}
\label{NO kappa BAU Case IIIb different masses massive lightest neutrino}
\end{figure}

\paragraph{Effect of splitting $\kappa$.} In Fig.~\ref{NO kappa BAU Case IIIb different masses massive lightest neutrino} we display the BAU
with respect to $\kappa$ for four different values of the Majorana mass $M$, different values of the angle $\theta_R$ and for fixed values of
the group theory parameters, $n=20$, $m=10$ and $s=4$, such that $\phi_m=\frac{10 \, \pi}{20}$ and $\phi_s=\frac{4 \, \pi}{20}$. Light neutrino
masses have NO with $m_0=0.03$ eV. The dependence of the BAU on $\kappa$ is very similar to Case 2), $t$ even, compare Figs.~\ref{NO kappa BAU Case IIIb different masses massive lightest neutrino} and \ref{NO kappa BAU Case II different masses}. We observe a linear growth of the BAU with $\kappa$ up to the resonance peak after which the BAU decreases as a known power law of $\kappa$. We refer the reader to  section~\ref{subsec:5.4.1} for more details. While Fig.~\ref{NO kappa BAU Case IIIb different masses massive lightest neutrino} assumes $m_0\neq 0$, its equivalent for $m_0=0$ is displayed in appendix~\ref{appF3}, see Fig.~\ref{NO kappa BAU Case IIIb different masses massless lightest neutrino}. The overall behaviour is mostly similar to Fig.~\ref{NO kappa BAU Case IIIb different masses massive lightest neutrino}.

\begin{figure}
\begin{subfigure}{.5\textwidth}
	\centering
\includegraphics[width = \textwidth]{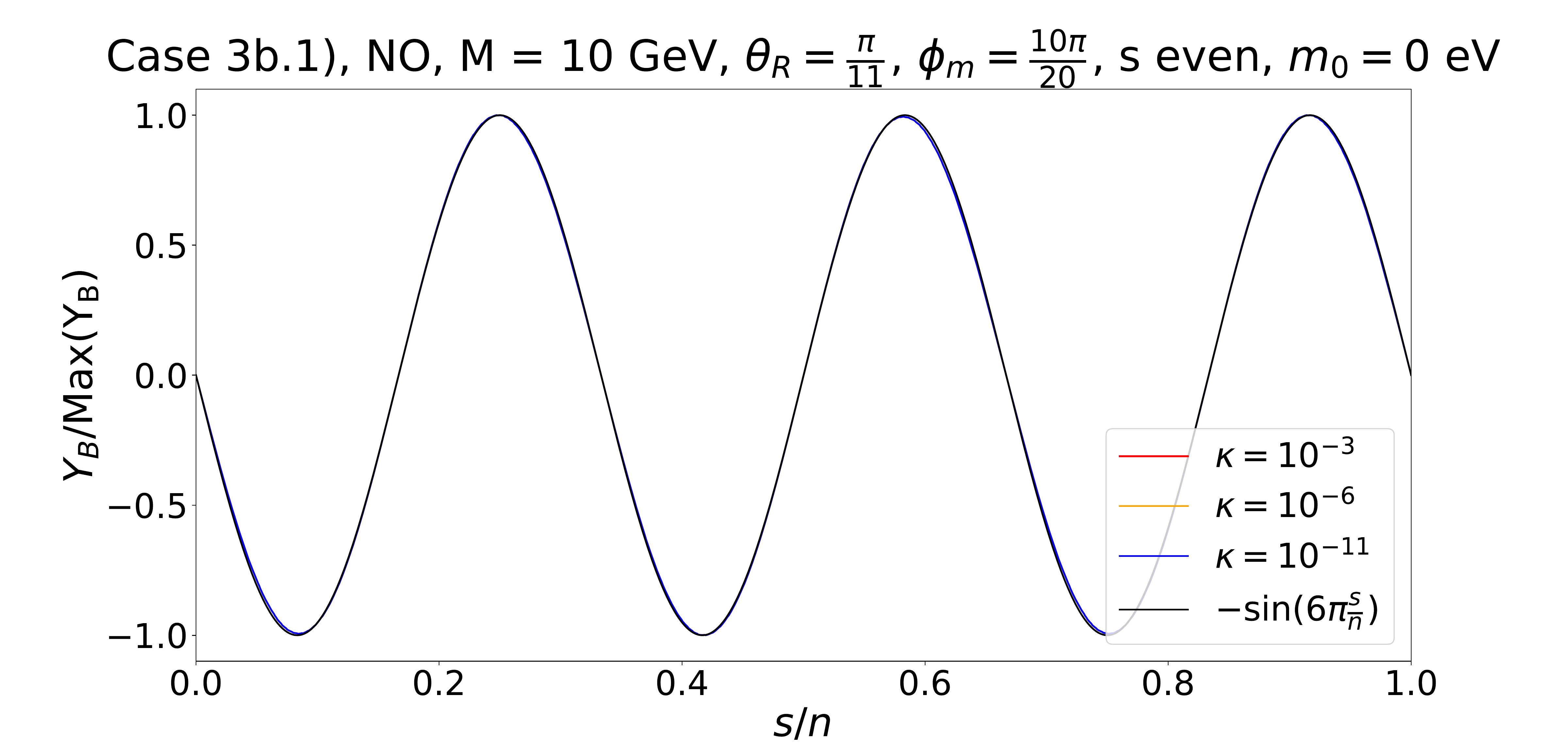}
\caption{Vanishing initial conditions}
\end{subfigure}
\begin{subfigure}{.5\textwidth}
\includegraphics[width = \textwidth]{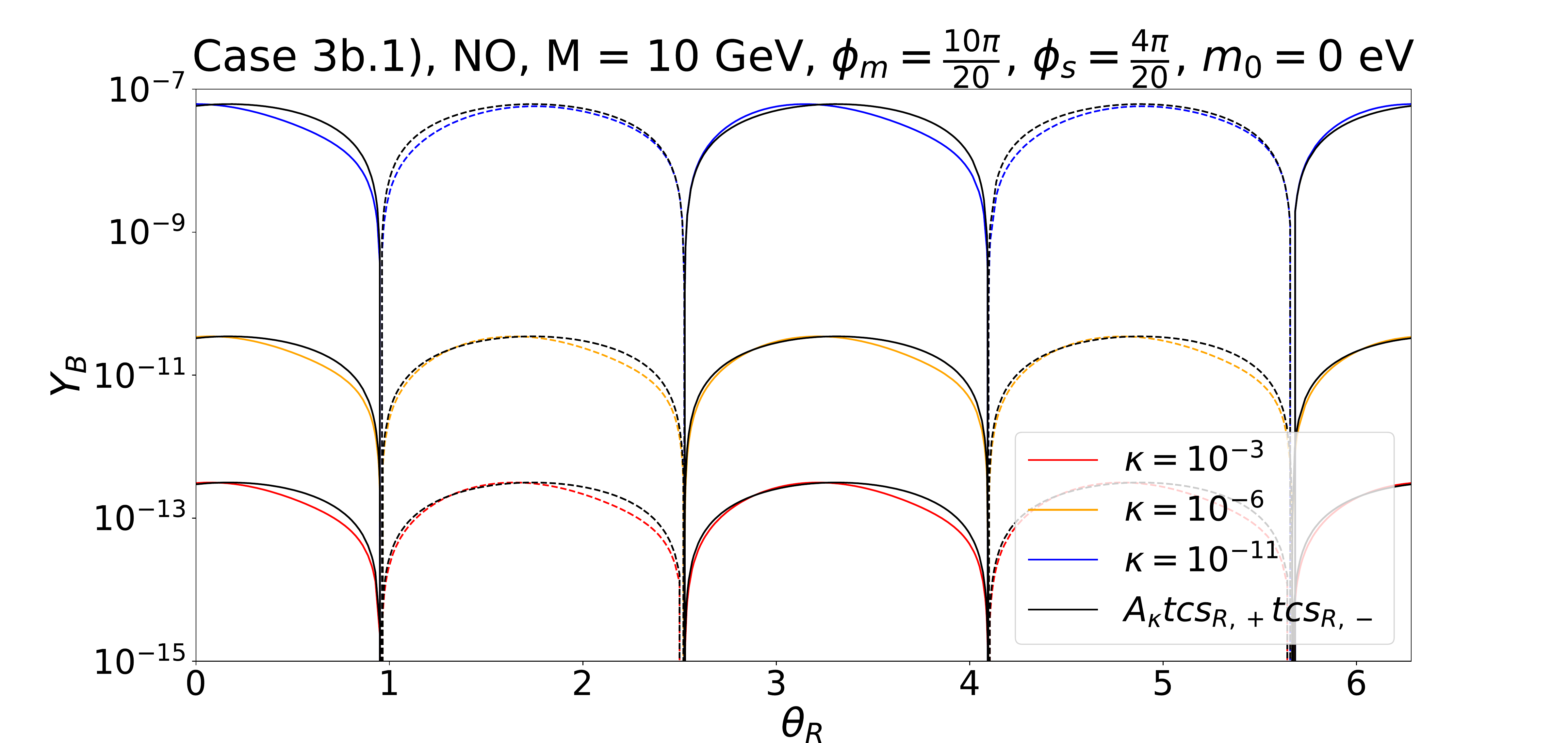}
\caption{Vanishing initial conditions}
\end{subfigure}
\caption{{\small \textbf{Case 3 b.1)} The left plot displays $Y_B$ as function of $\frac{s}{n}$, treated as continuous parameter. We assume that
$s$ is even. The Majorana mass $M$ is fixed to $M=10$ GeV, the angle $\theta_R$ has the value $\theta_R=\frac{\pi}{11}$ and three different values of $\kappa$ are used, $\kappa \in \{10^{-3},10^{-6},10^{-11}\}$. The group theory parameters $n$ and $m$ are chosen as in Fig.~\ref{NO kappa BAU Case IIIb different masses massive lightest neutrino}. Light neutrino masses have strong NO.
The right plot displays $Y_B$ as function of the angle $\theta_R$. The same values for $M$, $\kappa$, $n$ and $m$ as well as the same light
neutrino mass spectrum as in plot (a) of this figure are used. The parameter $s$ is fixed to $s=4$ and thus $\phi_s=\frac{4 \, \pi}{20}$.
The coefficient $A_{\kappa}$ is given by $\frac{\mbox{Max}\big[|Y_B|\big]}{\mbox{Max}\big[tcs_{R,+}tcs_{R,-}\big]}$, where $tcs_{R,+}$ and $tcs_{R,-}$ are defined in Eq.~\eqref{eq:definitiontcs},  for each $\kappa$. Both negative (dashed lines) as well as positive (continuous lines) values of the BAU are represented.}
}
\label{NO BAU phim/thetaR 10 GeV caseIIIb}
\end{figure}

\paragraph{Dependence on CP transformation $X (s)$.} In order to study better the dependence of the BAU on $X (s)$ or equivalently $s/n$ (taken as continuous parameter), we focus on a light neutrino mass spectrum with strong NO. Furthermore, we fix the Majorana mass $M$ to $M=10$ GeV and the angle $\theta_R$ to $\theta_R=\frac{\pi}{11}$ in addition to $n=20$ and $m=10$ and thus $\phi_m=\frac{10 \, \pi}{20}$.
In Fig.~\ref{NO BAU phim/thetaR 10 GeV caseIIIb}, plot (a), we show the generated BAU, normalised to its value that can be maximally achieved, with respect to $s/n$ for three different values of $\kappa$, ranging between $10^{-11}$ and $10^{-3}$. From the CP-violating combinations $C_{\mathrm{LFV},\alpha}$ and $C_{\mathrm{LNV},\alpha}$, see Eq.~(\ref{eq:CLFValphamsevenNO}), we expect for this choice of $m$ and $s$ that the BAU depends on $\sin 3 \, \phi_s$. However, as we clearly see in Fig.~\ref{NO BAU phim/thetaR 10 GeV caseIIIb}, plot (a), the dependence on $\phi_s$ is rather $\sin 6 \, \phi_s = 2 \, \cos 3 \, \phi_s \, \sin 3 \, \phi_s$, see black curve. This can be explained by noticing that also the washout depends on $s/n$, see Eq.~\eqref{eq:flavouredwashoutratioCase3}.
Indeed, we find that by summing the CP-violating combinations weighted by the flavoured washout parameter $f_\alpha$ over the lepton flavour $\alpha$ the dependence on $\sin 6 \, \phi_s$ is recovered for $m=\frac{n}{2}$, see Eq.~(\ref{eq:sumalphaCLFValphafalphaCase3}).

We note that for light neutrino masses with strong IO we find instead vanishing BAU in agreement with the fact that the CP-violating combinations
$C_{\mathrm{LFV},\alpha}$ and $C_{\mathrm{LNV},\alpha}$, see Eq.~(\ref{eq:CLFValphamsevenIO}), are proportional to $\Delta \sigma_{23}$
and thus to the splitting $\lambda$, compare Eq.~(\ref{eq:Deltasigmas}), which we have set to zero.

Interestingly enough, the magnitude of the sines of the Majorana phases $\alpha$ and $\beta$, appearing in the PMNS mixing matrix, equals
the magnitude of $\sin 6 \, \phi_s$ for $m=\frac{n}{2}$, as chosen here, see section~\ref{sec503}.

\paragraph{Dependence on angle $\theta_R$.} In Fig.~\ref{NO BAU phim/thetaR 10 GeV caseIIIb}, plot (b), we investigate the dependence of the BAU on $\theta_R$ for a fixed value of the Majorana mass, $M=10$ GeV, the light neutrino
mass spectrum with strong NO and $n=20$ and $m=10$ so that $\phi_m=\frac{10 \, \pi}{20}$. Furthermore, we fix $s=4$, $\phi_s=\frac{4 \, \pi}{20}$, and display the results for the same three values
of $\kappa$ as in plot (a) of this figure. Consulting the formulae of the CP-violating combinations $C_{\mathrm{LFV},\alpha}$ and $C_{\mathrm{LNV},\alpha}$, see Eq.~(\ref{eq:CLFValphamsevenNO}), we observe that these predict a dependence on $\theta_R$ which is of the form $tcs_{R,-} \, tcs_{R,+}= (\sqrt{2} \, \cos \theta_R-\sin \theta_R) \, (\cos \theta_R+ \sqrt{2} \, \sin \theta_R)$. Indeed, we find a vanishing value for the BAU, if $\theta_R \approx 0.955$ and $\theta_R \approx 4.10$, where $tcs_{R,-}$ is zero, as well as if $\theta_R \approx 2.53$ and $\theta_R \approx 5.67$, where $tcs_{R,+}$ vanishes. Beyond this, deviations of the coloured curves, that display the numerical results for the three values of $\kappa$, in Fig.~\ref{NO BAU phim/thetaR 10 GeV caseIIIb}, plot (b), from the black curves, that represent the analytical expectation from the CP-violating combinations, are clearly visible and are due to washout effects.

\mathversion{bold}
\subsubsection{Choice \texorpdfstring{$m$}{m} even and \texorpdfstring{$s$}{s} odd}
\mathversion{normal}

For $m$ even and $s$ odd, we first analyse the dependence of the BAU on the splitting $\kappa$ for different values of the Majorana mass $M$, then its dependence on the CP transformation
$X (s)$ and on the angle $\theta_R$ for vanishing splitting $\kappa$. Light neutrino masses are assumed to follow strong NO.
Furthermore, we study the amount of BAU which can be generated for $\theta_R$ close to a  special value.
We perform this analysis for the choice $n=20$, $m=10$ and $s=3$ and hence $\phi_m=\frac{10 \, \pi}{20}$ and
$\phi_s=\frac{3 \, \pi}{20}$, unless otherwise stated.

\begin{figure}
	\begin{subfigure}{.5\textwidth}
		\centering
		\includegraphics[width = \textwidth]{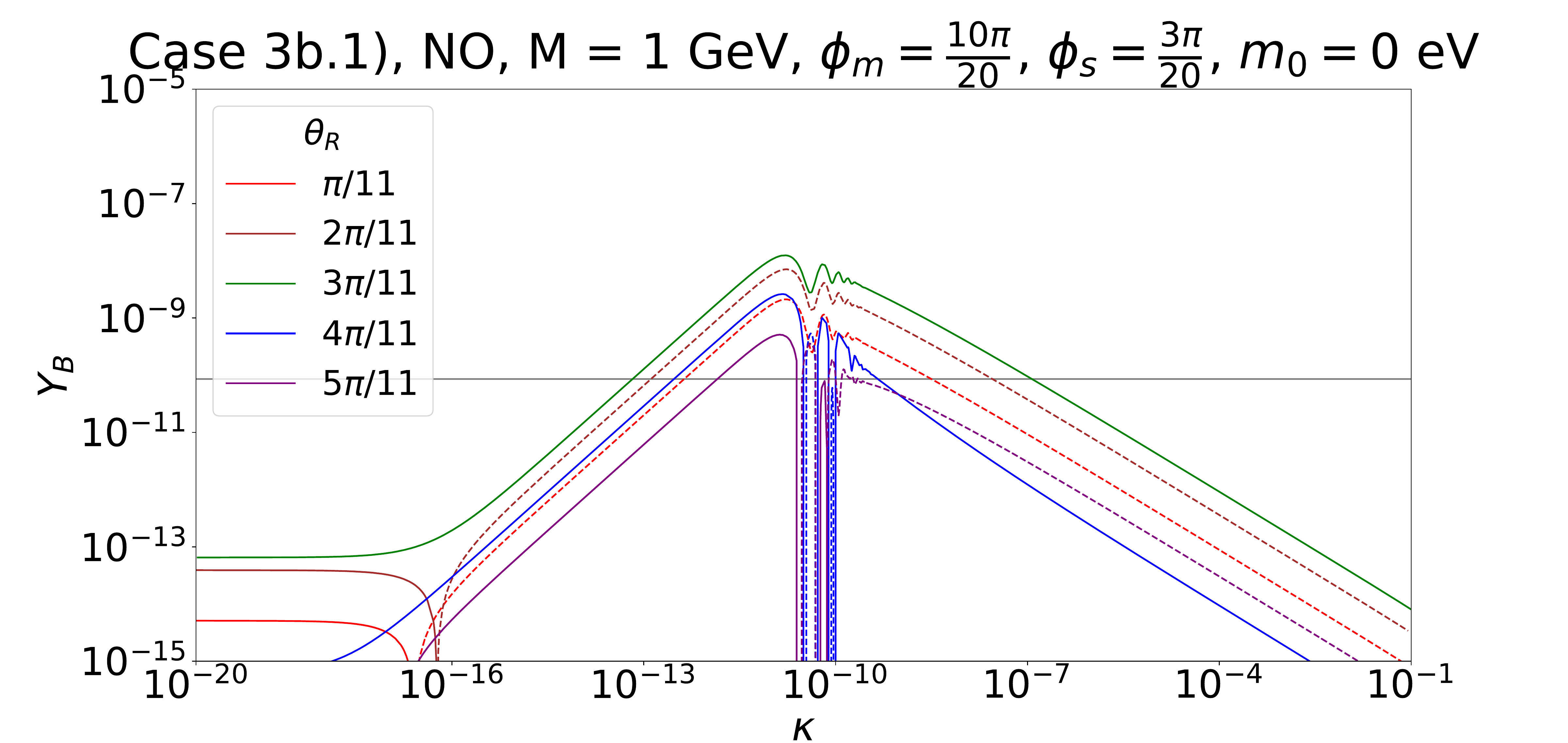}
		\caption{Vanishing initial conditions.}
	\end{subfigure}
	\begin{subfigure}{.5\textwidth}
		\centering
		\includegraphics[width = \textwidth]{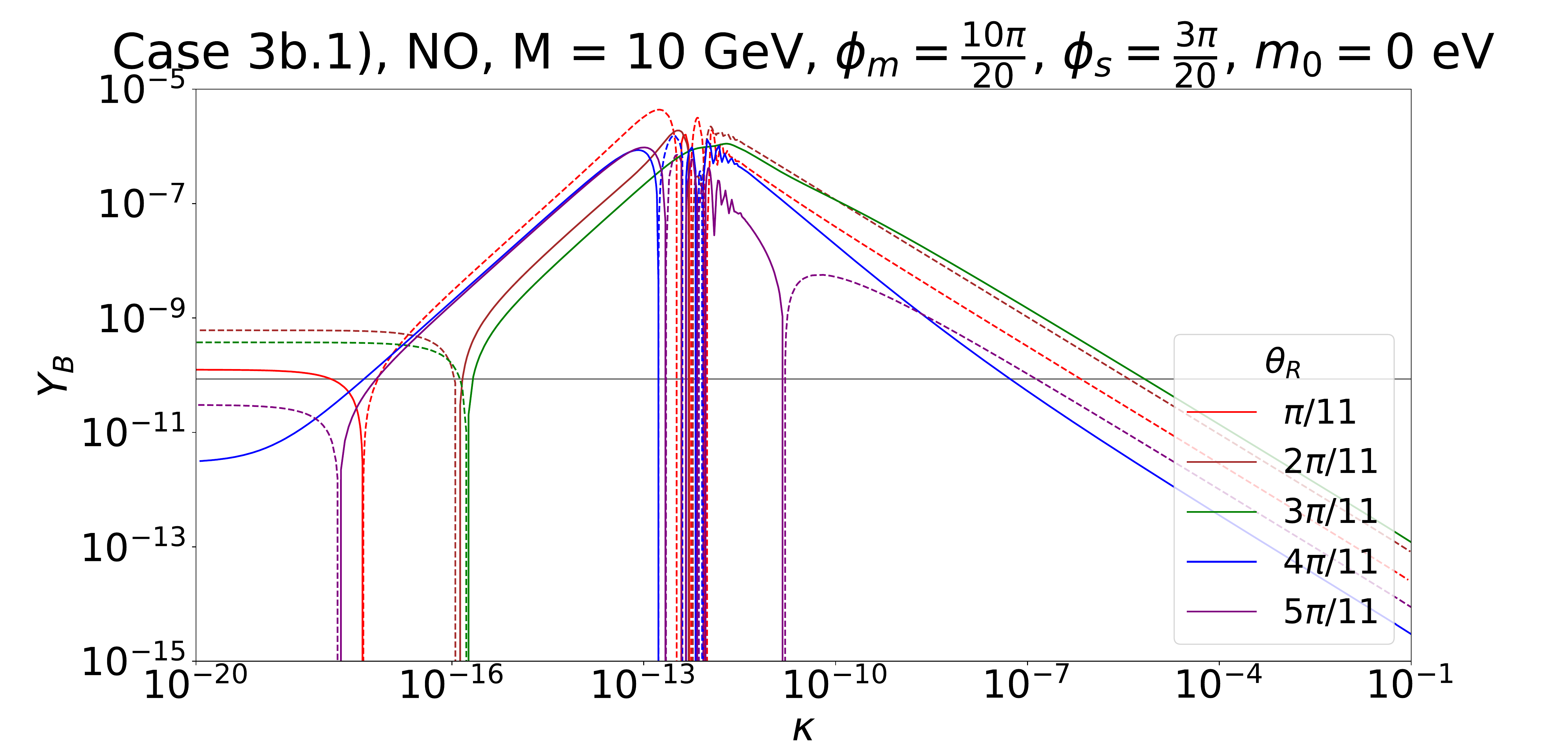}
		\caption{Vanishing initial conditions.}
		\label{NO kappa BAU 10 GeV IIIb odd massless}
	\end{subfigure}
	\begin{subfigure}{.5\textwidth}
		\includegraphics[width = \textwidth]{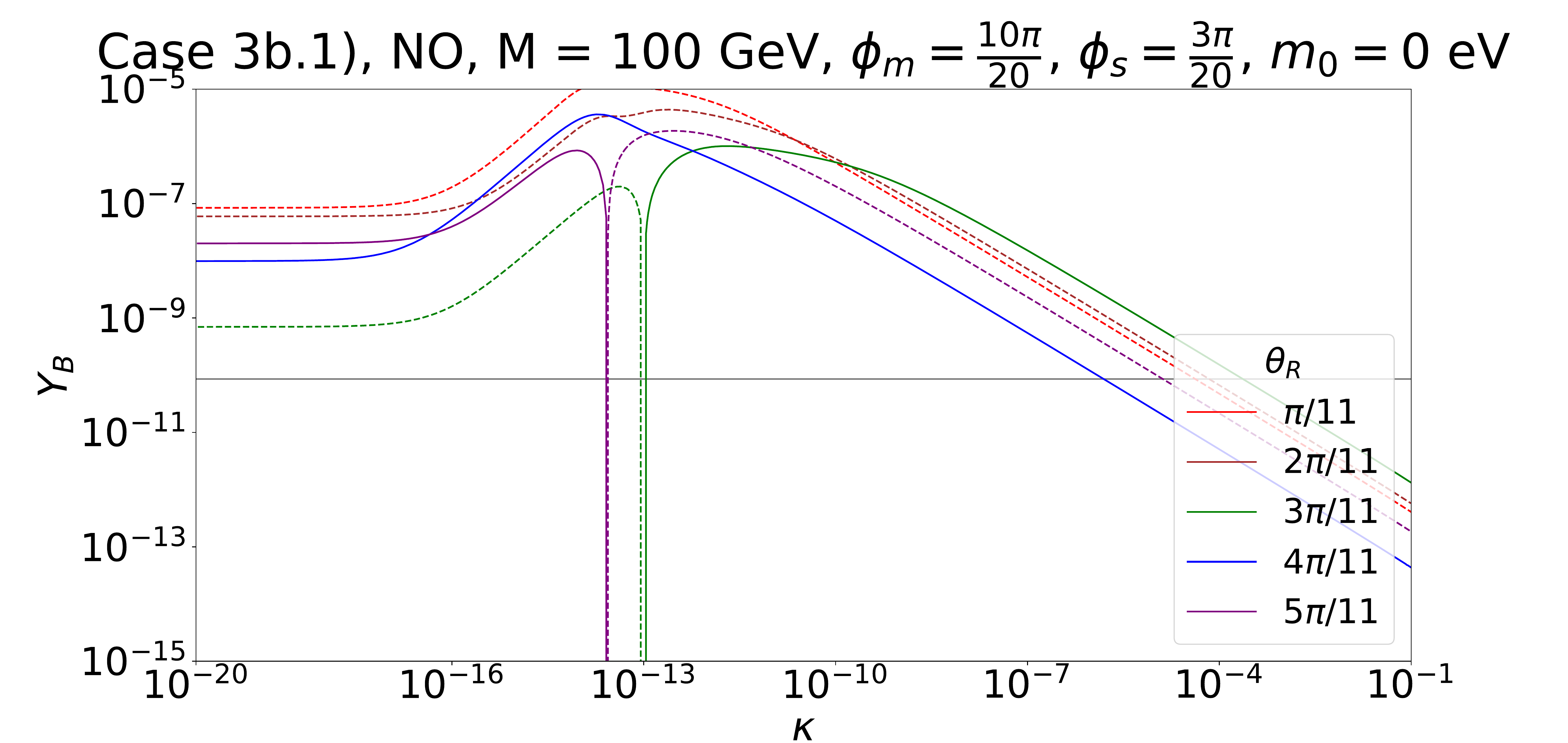}
		\caption{Vanishing initial conditions.}
	\end{subfigure}
	\begin{subfigure}{.5\textwidth}
		\includegraphics[width = \textwidth]{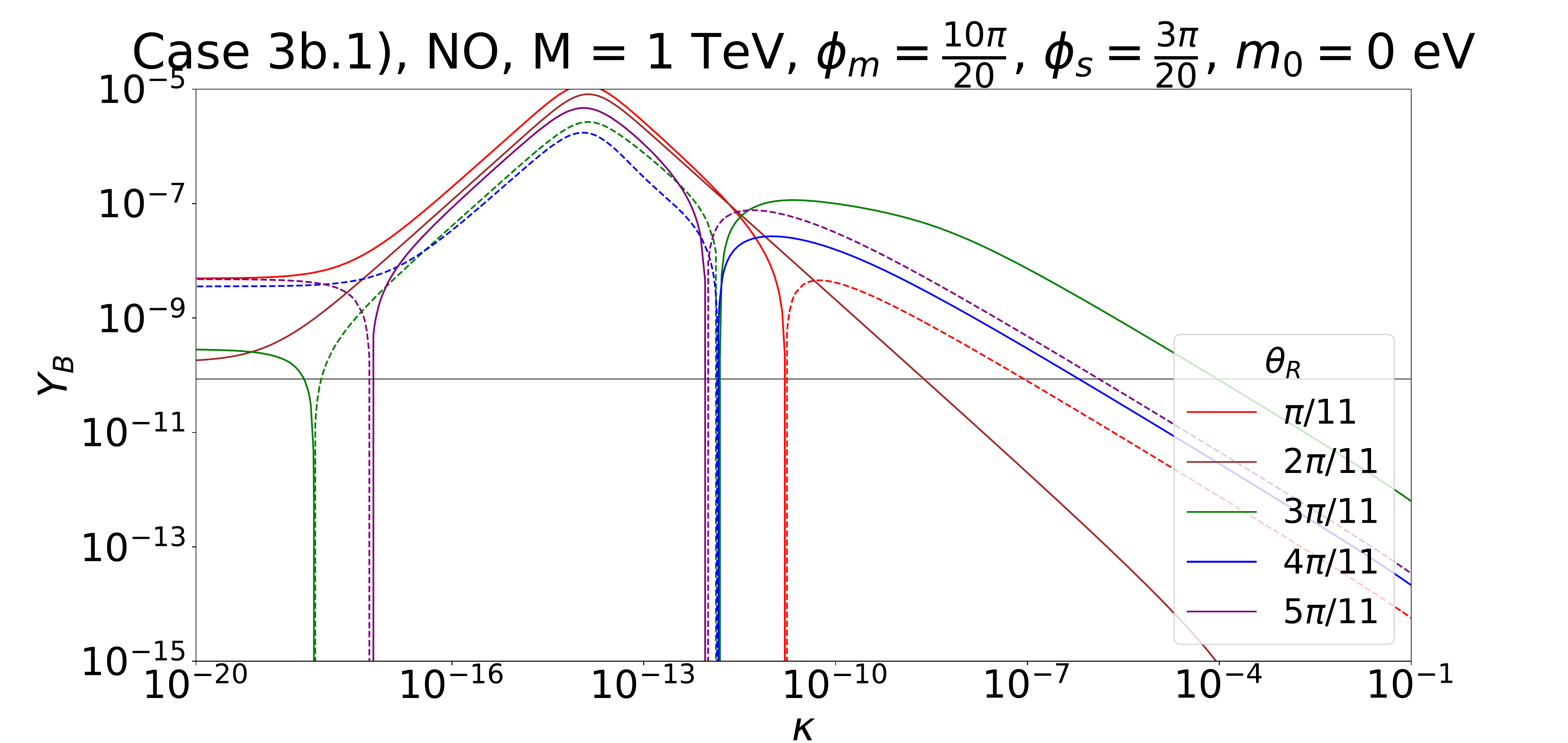}
		\caption{Vanishing initial conditions.}
	\end{subfigure}
	\caption{{\small {\bf Case 3 b.1)} $Y_B$ as function of the splitting $\kappa$ for a Majorana mass $M=1$ GeV (upper left plot), $M=10$ GeV (upper right plot), $M=100$ GeV (lower left plot) and $M=1$ TeV (lower right plot). For each of these plots, different values of $\theta_R$ have been studied. The group theory parameters $n$, $m$ and $s$ are fixed to $n=20$, $m=10$ and $s=3$. Light neutrino masses follow strong NO.
		For the rest of the choices see Fig.~\ref{NO kappa BAU Case IIIb different masses massive lightest neutrino}.
}}
\label{NO kappa BAU Case IIIb s odd different masses massless lightest neutrino}
\end{figure}

\paragraph{Effect of splitting $\kappa$.} Fig.~\ref{NO kappa BAU Case IIIb s odd different masses massless lightest neutrino} illustrates the dependence of the BAU on the splitting $\kappa$ for four different values of the Majorana mass $M$ and several values of $\theta_R$. Light neutrino masses have strong NO. One observes the same monotonic decrease of the BAU for $\kappa$ larger than values corresponding to the resonance peak as in Figs.~\ref{NO kappa BAU Case II different masses} and \ref{NO kappa BAU Case IIIb different masses massive lightest neutrino}. This is explained by the vanishing of the reduced mass-degenerate CP-violating combination $C^{(23)}_{\mathrm{DEG},\alpha}$, see Eq.~\eqref{eq:reducedCPcombinations}. For $\kappa\lesssim 10^{-16}$, we see that the BAU reaches a plateau similar to what happens for Case 2), $t$ odd and $m_0$ non-zero, see Fig.~\ref{NO BAU vs kappa caseII odd massive}. For completeness, we show the corresponding plots for light neutrino masses with NO and
$m_0=0.03$ eV in appendix~\ref{appF3}, see Fig.~\ref{NO kappa BAU Case IIIb s odd different masses massive lightest neutrino}.
As one can see, the behaviour of the BAU with $\kappa$ is qualitatively similar.

\paragraph{Results for vanishing splitting $\kappa$ and $m_0=0$.} Since the CP-violating combination $C_{\mathrm{DEG},\alpha}$ is non-zero for the choice $m$ even and $s$ odd, we expect to have a non-vanishing
value of the BAU also in the absence of the splitting $\kappa$ (and $\lambda$). Even more, we can check that also the lightest neutrino mass $m_0$ can be set to zero, if light neutrino masses follow NO,
and still a non-zero value of the BAU is achieved. We see in Fig.~\ref{NO kappa BAU Case IIIb s odd different masses massless lightest neutrino} that for very small splitting $\kappa$ and $m_0=0$ a plateau is reached for the value of the BAU. This value can be even larger than the observed value, $Y_B \approx 8.6 \cdot 10^{-11}$, depending on the choice of the Majorana mass $M$. Also for $m_0=0.03$ eV and otherwise the same choices of parameters,
we clearly see that the generated BAU reaches a plateau for very small values of $\kappa$ and that sizeable values of the BAU can be  obtained in this regime, compare Fig.~\ref{NO kappa BAU Case IIIb s odd different masses massive lightest neutrino} in appendix~\ref{appF3}.

With Fig.~\ref{NO BAU phis caseIIIb odd} we scrutinise the dependence of the BAU on the CP transformation $X (s)$. In plot (a) in this figure, we fix $M=10$ GeV, $\phi_m=\frac{10 \, \pi}{20}$ and additionally
$\theta_R=\frac{\pi}{11}$. As one can see for $s$ odd, the BAU seems to follow $\sin 6 \, \phi_s$, whereas the CP-violating combination $C_{\mathrm{DEG},\alpha}$ depends on $\sin 3 \, \phi_s$, see Eq.~(\ref{eq:CDEGalpha_soddmeven_Case3}). Like for the choice
$m$ and $s$ both even, this discrepancy can be explained by noticing that the washout depends on $s/n$, see the flavoured washout parameter $f_\alpha$ in Eq.~(\ref{eq:flavouredwashoutratioCase3}) as well as the sum of $C_{\mathrm{DEG},\alpha}$ weighted by $f_\alpha$ over the lepton flavour $\alpha$ in Eq.~(\ref{eq:sumalphaCDEGalphafalphaCase3}) and, in particular, for $m=\frac{n}{2}$ in Eq.~(\ref{eq:sumalphaCDEGalphafalphamn2Case3}). In order to better demonstrate this fact, we have also considered the choice $n=50$
and $m=24$ that leads to $\phi_m=\frac{24 \, \pi}{50} = 0.48 \, \pi$, see plot (b) in Fig.~\ref{NO BAU phis caseIIIb odd}. In this case, the effect of the washout is different,  since a further contribution to the BAU which is proportional to $\sin 3 \, \phi_s$ arises, compare Eq.~(\ref{eq:sumalphaCDEGalphafalphaCase3}). We hence observe a deviation from the proportionality to $\sin 6 \, \phi_s$ (black curve). In particular, a shift of three of the values of $s/n$ which lead to a vanishing value of the BAU is witnessed, i.e.~$s/n \approx 0.167$, $s/n=0.5$ and $s/n \approx 0.833$.
These three values of $s/n$ are associated with the zeros of $\cos 3 \, \phi_s$. On the other hand, the values of $s/n$, which are associated with the zeros of $\sin 3 \, \phi_s$, i.e.~$s/n=0$,
$s/n \approx 0.333$, $s/n\approx 0.667$ and $s/n=1$, and which also lead to a vanishing BAU, are not subject to any shift. For values of $n$ and $m$ that lead to a larger deviation of
$m/n$ from $m/n=1/2$ (and thus from $\phi_m=\frac{\pi}{2}$) this effect is much more pronounced and, indeed, for $n=20$ and $m=8$ and otherwise the same choice for $M$, $\theta_R$ and
the light neutrino mass spectrum as well as $s$ odd only the zeros of $\sin 3 \, \phi_s$ determine the values of $s/n$ that lead to a vanishing of the BAU. This is shown in Fig.~\ref{NO phis BAU 10 GeV phim820 caseIIIb}
in appendix~\ref{appF3}. Inspecting this figure, we note that not only the values of $s/n$ with vanishing BAU, but also the overall dependence of the BAU on $s/n$ is reproduced to some extent by the
function $\sin 3 \, \phi_s$ (black curve). We, however, emphasise that ratios of $m/n$ deviating this much from $m/n=1/2$ in general lead to lepton mixing angles that cannot accommodate the
experimental data at the $3 \, \sigma$ level or better.

As already mentioned for the choice $m$ and $s$ both even and also correct for $s$ odd, the magnitude of the sines of the Majorana phases $\alpha$ and $\beta$, appearing in the PMNS mixing matrix, equals the magnitude of $\sin 6 \, \phi_s$ for $m=\frac{n}{2}$, which is fulfilled for the combination $n=20$ and $m=10$.

\begin{figure}
	\begin{subfigure}{.5\textwidth}
		\centering
		\includegraphics[width = \textwidth]{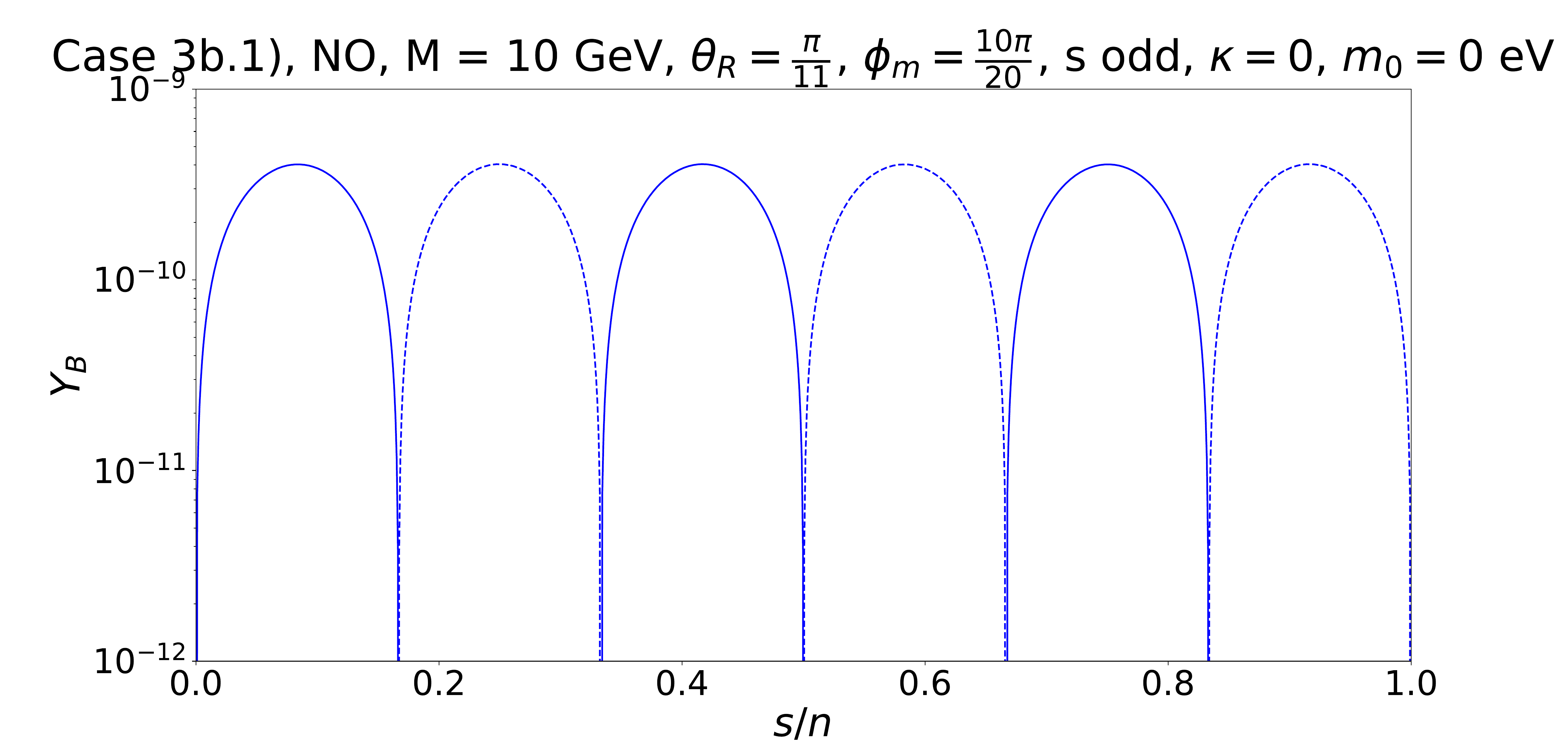}
		\caption{Vanishing initial conditions.}
	\end{subfigure}
	\begin{subfigure}{.5\textwidth}
		\includegraphics[width = \textwidth]{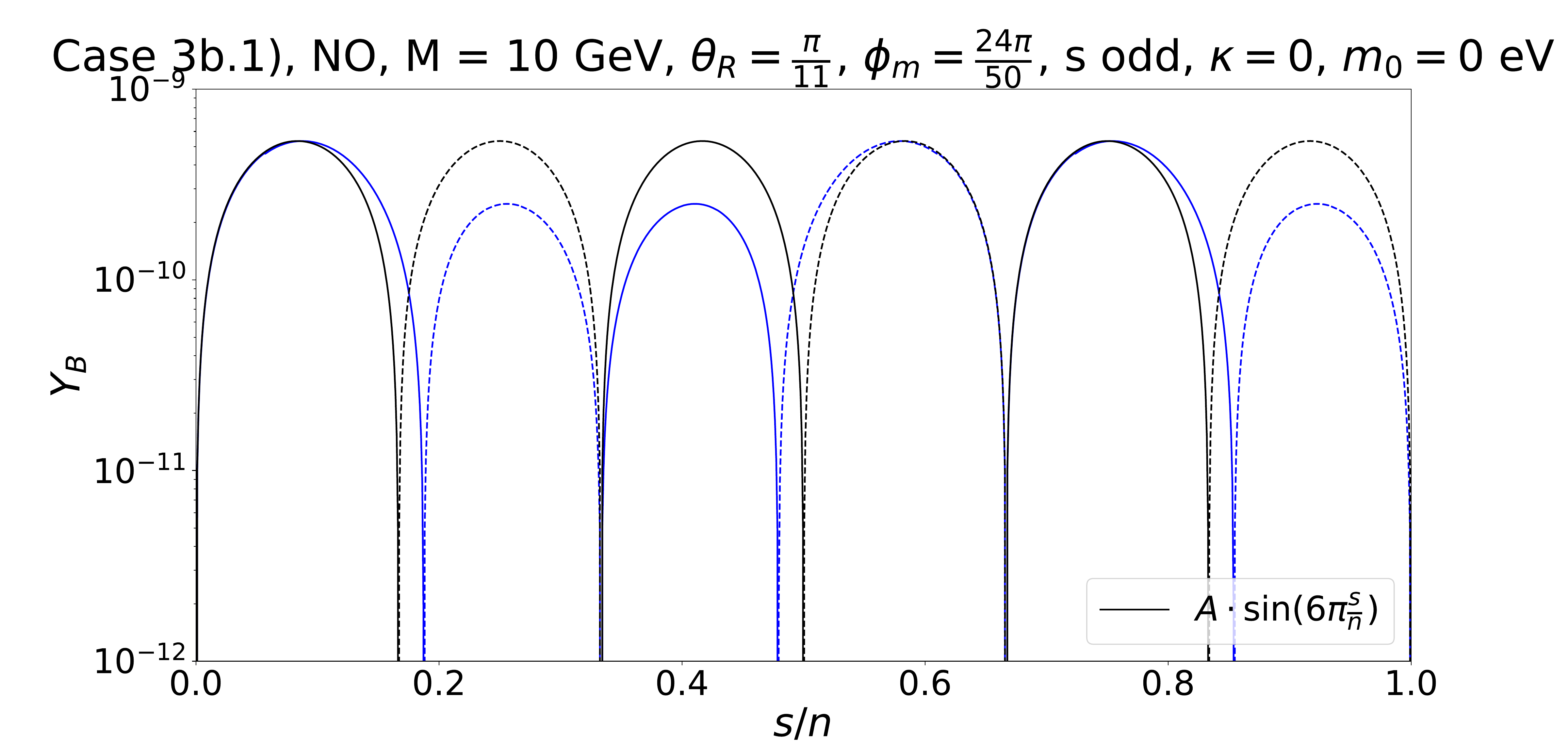}
		\caption{Vanishing initial conditions.}
	\end{subfigure}
	\caption{{\small {\bf Case 3 b.1)} $Y_B$ as function of $s/n$, treated as continuous parameter, in absence of the splitting $\kappa$
		(and $\lambda$). Light neutrino masses follow strong NO. The Majorana mass $M$ is fixed to $M=10$ GeV and the angle $\theta_R$ to $\theta_R=\frac{\pi}{11}$. The parameter $s$ is assumed to be odd. The left plot shows the result for $n$ and $m$ being $n=20$ and $m=10$
		and thus $\phi_m=\frac{10 \, \pi}{20}$, while the right plot is for $n$ and $m$ being $n=50$ and $m=24$
		and thus $\phi_m=\frac{24 \, \pi}{50}$. In the right plot, we show in black the function $\sin 6 \, \phi_s$,
		which, however, does not agree with the expectations from the CP-violating combination $C_{\mathrm{DEG},\alpha}$, see Eq.~\eqref{eq:CDEGalpha_soddmeven_Case3}.
		For more details, see text. The coefficient $A$ is given by $\mbox{Max}\big[|Y_B|\big]$. Both negative (dashed lines) as well as positive (continuous lines) values of the BAU are represented.
}}
\label{NO BAU phis caseIIIb odd}
\end{figure}

\begin{figure}
\begin{subfigure}{.5\textwidth}
	\centering
\includegraphics[width = 1.05\textwidth]{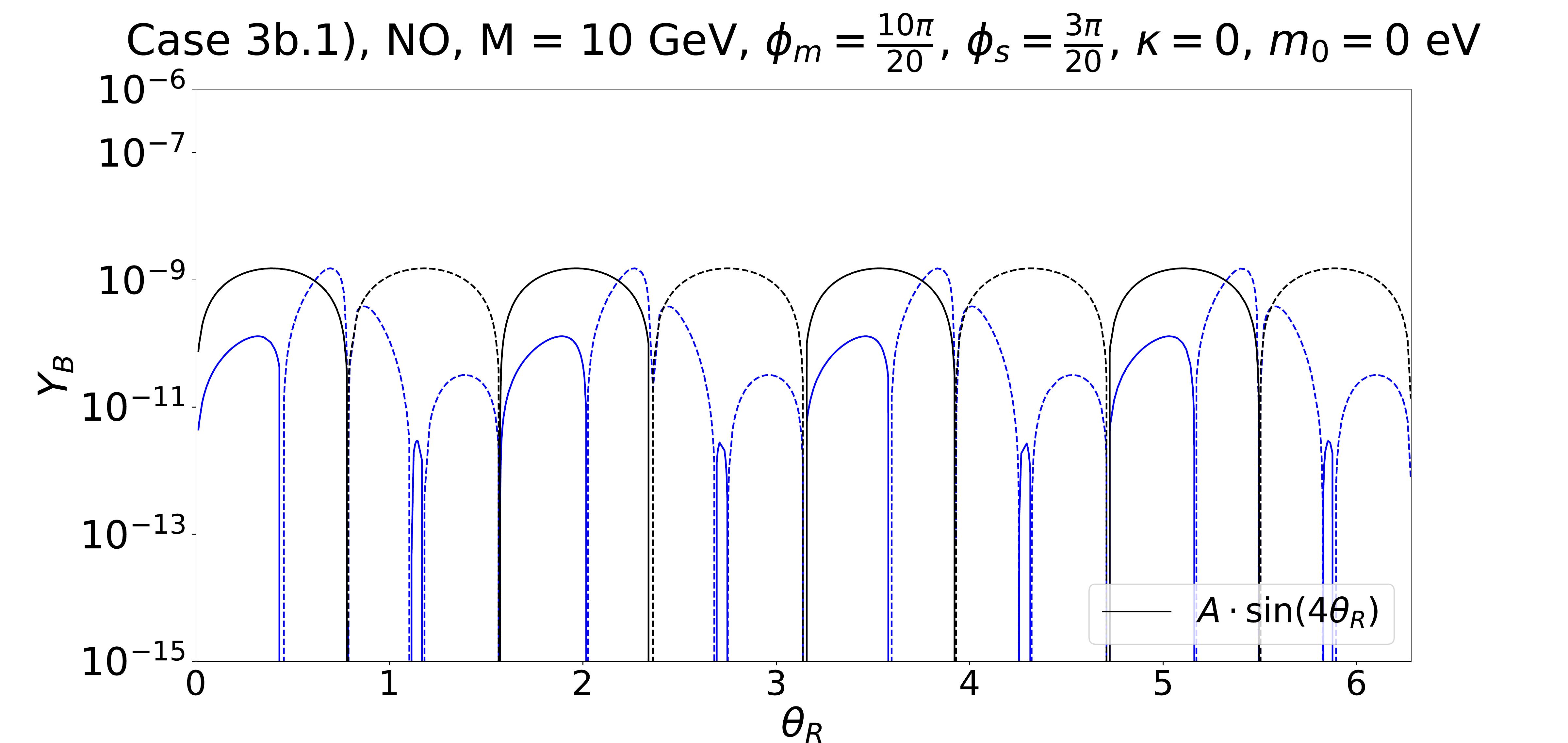}
\caption{Vanishing initial conditions.}
\end{subfigure}
\begin{subfigure}{.5\textwidth}
\includegraphics[width = 1.05\textwidth]{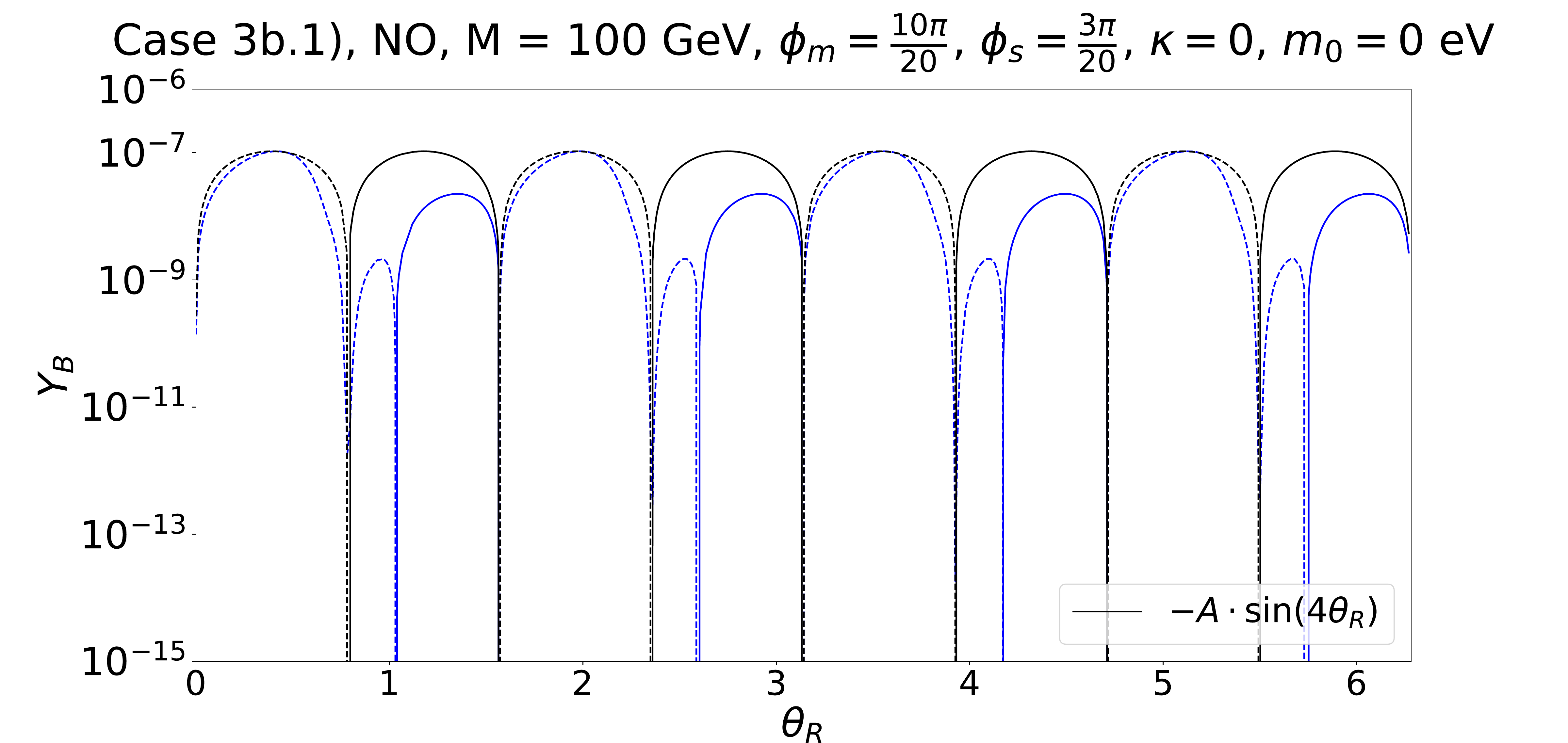}
\caption{Vanishing initial conditions.}
\end{subfigure}
\caption{{\small {\bf Case 3 b.1)} $Y_B$ as function of the angle $\theta_R$ for a Majorana mass $M=10$ GeV (left plot) and $M=100$ GeV (right plot). The splitting $\kappa$ (and $\lambda$) is set to zero. Light neutrino masses follow strong NO. The group theory parameters $n$, $m$ and $s$
are fixed to $n=20$, $m=10$ and $s=3$. We compare the numerical result (blue curve) to the analytical expectation from the CP-violating combination $C_{\mathrm{DEG},\alpha}$ which is proportional to $\sin 4\,  \theta_R$ (black curve). 
The coefficient $A$ is given by $\mbox{Max}\big[|Y_B|\big]$. Both negative (dashed lines) as well as positive (continuous lines) values of the BAU are represented.}}
\label{NO BAU thetaR caseIIIb odd}
\end{figure}

The CP-violating combination $C_{\mathrm{DEG},\alpha}$ is proportional to $\sin 4 \, \theta_R$, see Eq.~(\ref{eq:CDEGalpha_soddmeven_Case3}). We test the hypothesis of the BAU being also proportional to $\sin 4 \, \theta_R$ for two different values
of the Majorana mass $M$, $M=10$ GeV and $M=100$ GeV, $s=3$ such that $\phi_s=\frac{3 \, \pi}{20}$ and otherwise the same choice of $n$, $m$ and the light neutrino mass spectrum as
in Fig.~\ref{NO BAU phis caseIIIb odd}, plot (a). As one can see in Fig.~\ref{NO BAU thetaR caseIIIb odd}, the values of $\theta_R$ which lead to a vanishing of the BAU are only partly captured by $\sin 4 \, \theta_R$ that predicts $\theta_R$ to be a multiple
of $\pi/4$. Furthermore, we note that for the larger value of $M$ the dependence of the BAU on $\theta_R$ is better reproduced by the analytical expectation.

Eventually, we comment on the dependence of the BAU on the lightest neutrino mass $m_0$ for light neutrino masses with NO. For the same choices of the parameters $n$, $m$, $s$, $M$
and $\kappa$, as made in Fig.~\ref{NO BAU thetaR caseIIIb odd}, plot (a), we show the BAU as function of $m_0$ for five different values of $\theta_R$ in Fig.~\ref{NO m0 BAU 10 GeV caseIIIb} in appendix~\ref{appF3}.
The generated BAU is constant for $m_0$ smaller than $10^{-3}$ eV to $10^{-2}$ eV, with the exact value depending on the chosen value of $\theta_R$, as can be seen in Fig.~\ref{NO m0 BAU 10 GeV caseIIIb}.
The dependence of the BAU on $m_0$ is determined by its dependence on the couplings $y_f$, $f=1,2,3$.
From the CP-violating combination $C_{\mathrm{DEG},\alpha}$ we expect a proportionality to the combination $y_1 \, y_2 \, (y_1^2-y_2^2)^2$, see Eq.~(\ref{eq:CDEGalpha_soddmeven_Case3}),  which is approximately constant for small $m_0$. 
This is confirmed by the behaviour observed in Fig.~\ref{NO m0 BAU 10 GeV caseIIIb}.

\begin{figure}
	\centering
	\includegraphics[width = 0.49\textwidth]{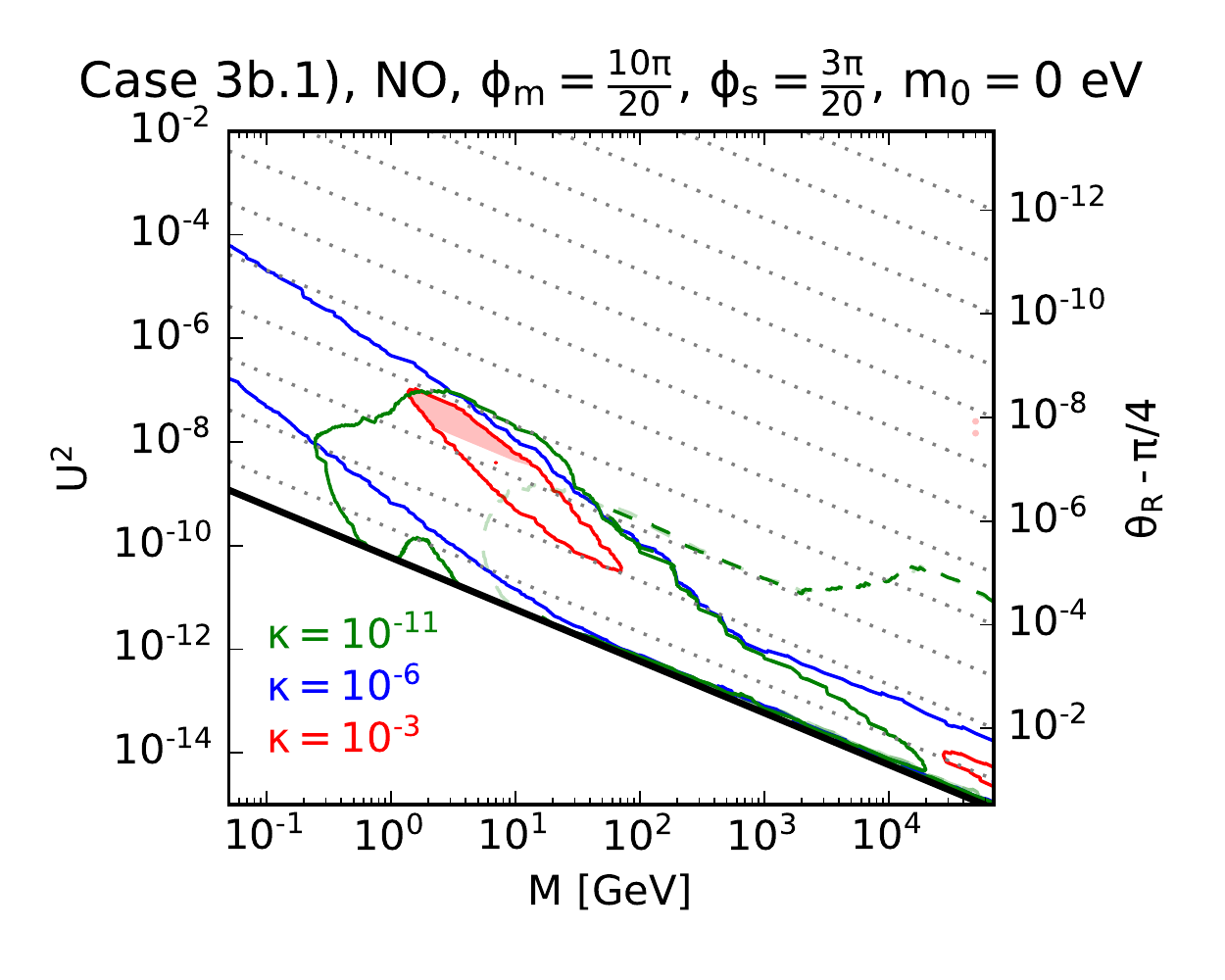}
	\includegraphics[width = 0.49\textwidth]{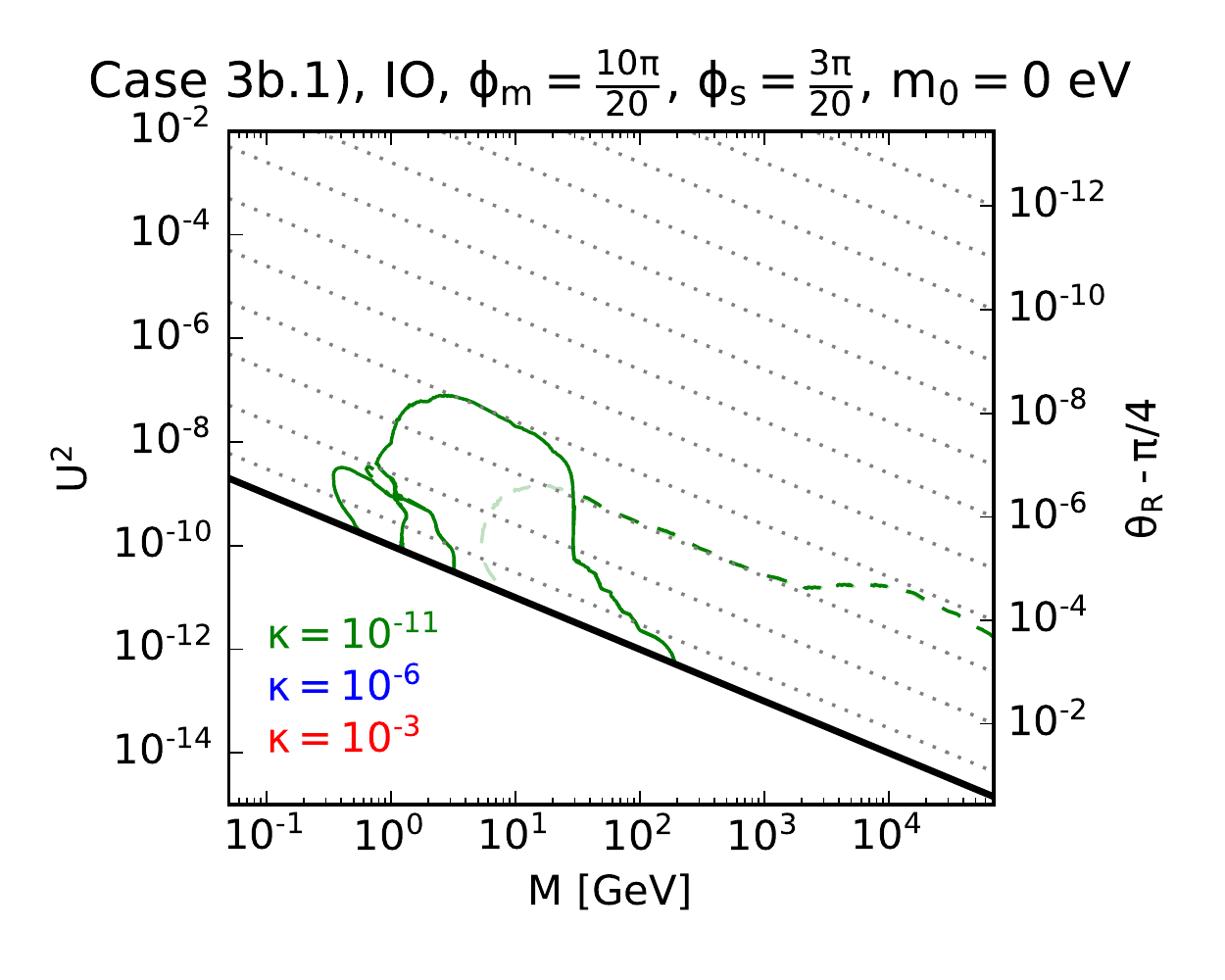}
	\caption{{\small {\bf Case 3 b.1)} Range of total mixing angle  $U^2$ consistent with leptogenesis for the Majorana mass $M$ varying between $50$ MeV and $70$ TeV. The splitting $\kappa$ is chosen among the three values $\kappa \in \{10^{-3}, 10^{-6}, 10^{-11}\}$.
		However, we note that for $\kappa=10^{-3}$ and $\kappa=10^{-6}$ the observed amount of BAU can only be generated for light neutrino masses with strong NO (left plot) and not for light neutrino masses with strong IO (right plot).
		The angle $\theta_R$ can be read off.
		The values of $\phi_m$ and $\phi_s$ are fixed to $\phi_m=\frac{10\, \pi}{20}$ and $\phi_s=\frac{3 \, \pi}{20}$ corresponding to $m$ even and $s$ odd.
		The contours corresponding to positive (negative) values of the BAU are shown in continuous (dashed) lines. Note that these refer to vanishing initial conditions except for the lines shown in fainter colours. The red shaded area shows the region in which a condition like the one in Eq.~(\ref{eq:kappacorrectionscriterion}) is not fulfilled.
		The black lines indicate the seesaw line.}
		}
\label{NO Mass U2 Case3}
\end{figure}

\paragraph{Special values of $\theta_R$.} For both $m$ even, $s$ odd and $m$ odd, $s$ even, the total mixing angle $U^2$ strongly depends on the angle $\theta_R$ and, similar to Case 1), can lie several orders of magnitude above the seesaw line for $\theta_R$ close to an odd multiple of $\frac{\pi}{4}$. In Fig.~\ref{NO Mass U2 Case3}, we display the parameter space consistent with leptogenesis in the $M-U^2$-plane for the particular choice $n=20$, $m=10$ and $s=3$ (see Fig.~\ref{fig:Case3b_MU2IC} in appendix~\ref{appF3} for dedicated plots for each value of the splitting $\kappa$). As one can see, the viable parameter space is more constrained than for the examples studied in Case 1) and Case 2), compare Figs.~\ref{NO Mass U2 Case1} and ~\ref{NO Mass U2 Case2}. This can also be observed in Fig.~\ref{NO BAU vs U2 10 GeV e-6 multiple curves caseIIIb odd} in appendix~\ref{appF3}, when comparing it to Figs.~\ref{NO BAU vs U2 10 GeV e-6 multiple curves} and \ref{NO BAU vs U2 10 GeV e-6 multiple curves caseII}. For strong IO, Fig.~\ref{NO Mass U2 Case3} implies that the observed amount of BAU cannot be generated for $\kappa \gtrsim 10^{-6}$.

\mathversion{bold}
\subsubsection{Choice \texorpdfstring{$m$}{m} odd and \texorpdfstring{$s$}{s} even}
\mathversion{normal}

\begin{figure}
	\begin{subfigure}{.5\textwidth}
		\centering
		\includegraphics[width = 1.05\textwidth]{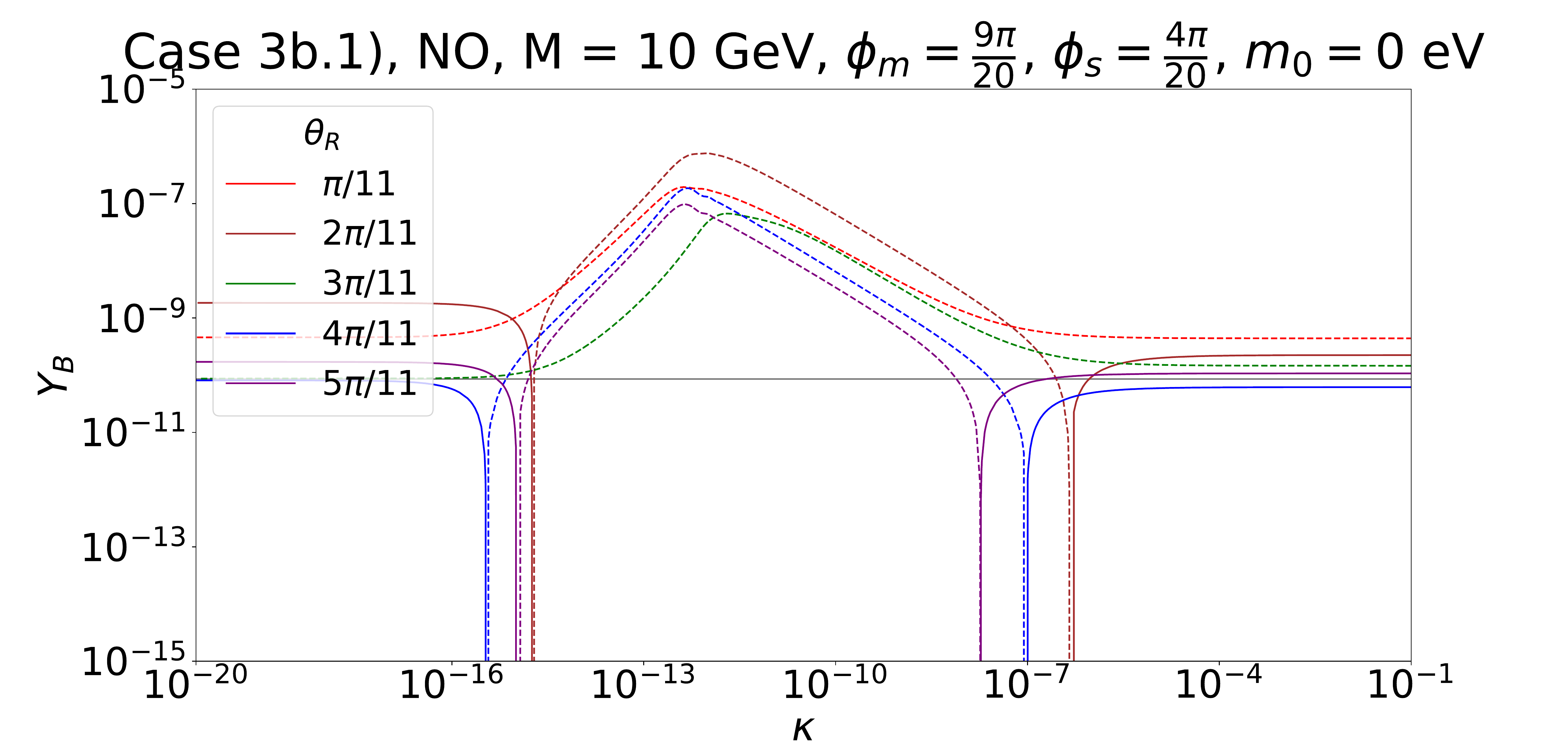}
		\caption{Vanishing initial conditions.}
	\end{subfigure}
	\begin{subfigure}{.5\textwidth}
		\includegraphics[width = 1.05\textwidth]{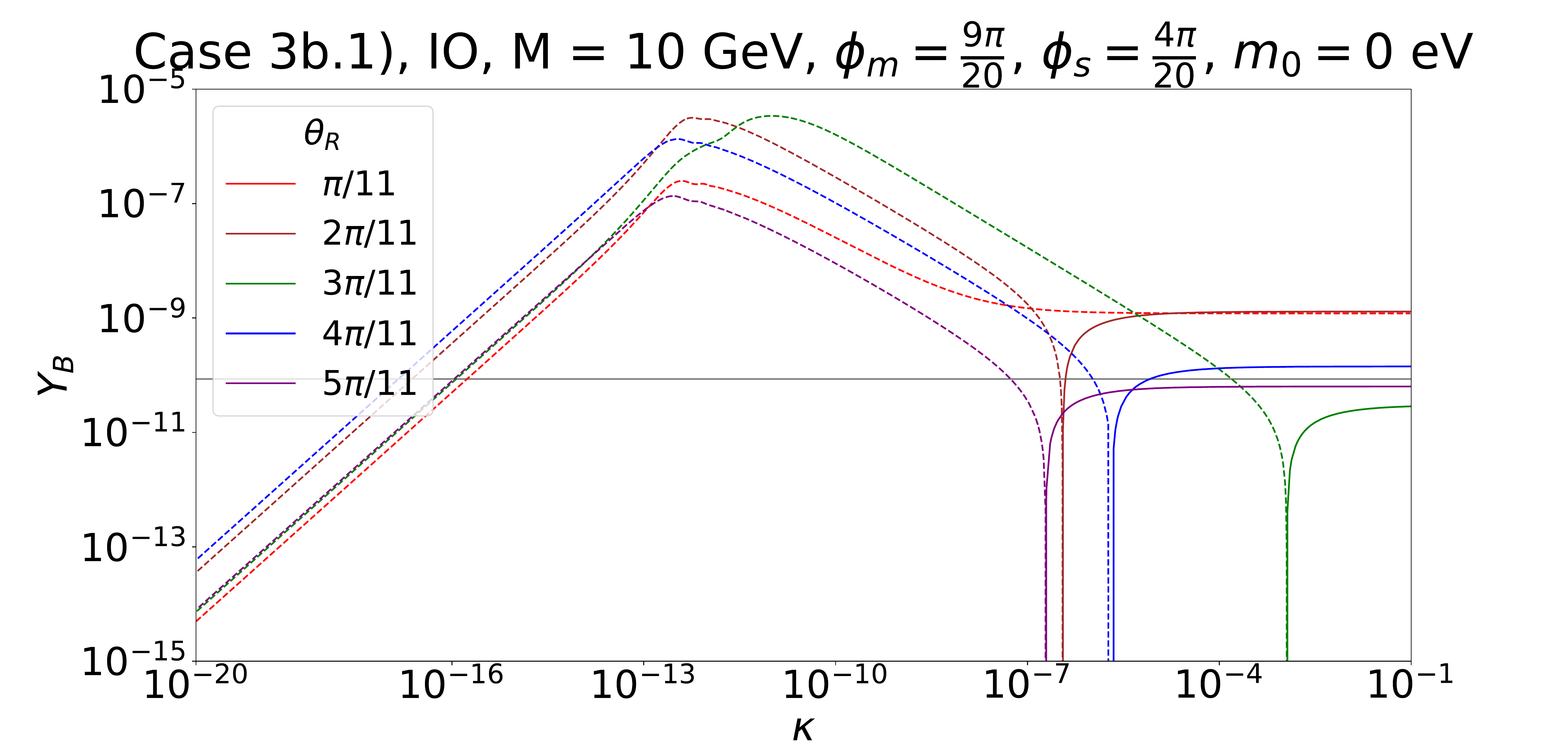}
		\caption{Vanishing initial conditions.}
	\end{subfigure}
\caption{{\small {\bf Case 3 b.1)} $Y_B$ as function of the splitting $\kappa$ for a Majorana mass $M=10$ GeV. Light neutrino masses follow strong NO (left plot) and strong IO (right plot), respectively. For each of these plots, different values of $\theta_R$ have been studied. The group theory parameters $n$, $m$ and $s$ are fixed to $n=20$, $m=9$ and $s=4$. Both negative (dashed lines) as well as positive (continuous lines) values of the BAU are represented. The grey line indicates the observed value of the BAU, $Y_B\approx 8.6 \cdot 10^{-11}$.}}
\label{NOIO BAU kappa caseIIIb odd phi4920}
\end{figure}

For $m$ odd and $s$ even, we only display the BAU as  function of the splitting $\kappa$ for both light neutrino masses with strong NO and with strong IO and a fixed value of the Majorana mass $M$, $M=10$ GeV, in order to illustrate its behaviour for large values of $\kappa$. We fix $n$, $m$ and $s$ to $n=20$, $m=9$ and $s=4$, respectively. 
Contrary to the other choices of $m$ and $s$ for Case 3 b.1), using $m$ odd and $s$ even leads to a non-zero value of the reduced mass-degenerate 
CP-violating combination $C^{(23)}_{\mathrm{DEG},\alpha}$, see Eq.~\eqref{eq:CDEGalphareduced_moddseven}. This observation is confirmed by the two plots in Fig.~\ref{NOIO BAU kappa caseIIIb odd phi4920}, in which the BAU reaches a  plateau for large $\kappa$. Notice that, for strong IO, no BAU can be generated for $\kappa=0$ in accordance with the results, shown in Eq.~\eqref{eq:CDEGalpha_soddmeven_Case3} and the text below.

\section{Analytic expressions for source and washout terms}
\label{sec6}

In this section, we first define different CP-violating combinations\footnote{ See~\cite{Hernandez:2015wna} for a treatment using CP invariants in the fully relativistic regime.} as well as the flavoured washout parameter in section~\ref{sec61}. These play a crucial role in the generation of the BAU. While the dynamics of the latter can in general be intricate, the considered scenario only contains a small number of parameters. It is, thus, possible to understand the main dependencies of the BAU on the
different parameters, e.g.~on the splitting $\kappa$ as well as on the
choice of the CP transformation $X (s)$ in Case 1). We then present expressions of the different CP-violating combinations and of the flavoured washout parameter for Case 1) through Case 3 b.1) in sections~\ref{sec62}-\ref{sec64}
and in appendix~\ref{appG}.

\subsection{Definition of CP-violating combinations and washout parameter}
\label{sec61}

The quantum kinetic equations should in principle be solved by numerically integrating Eqs.~\eqref{QKE}.
When comparing the different cases, it is nonetheless useful to compare the combinations of mass and Yukawa matrices that lead to CP violation.
We identify them by perturbatively solving Eqs.~\eqref{QKE} in both $H_N$ and $\Gamma$.
The leading term giving rise to the lepton asymmetries is given by
\begin{align}
	\mathrm{Tr} \left[ \tilde{\Gamma}_\alpha (\bar{\rho}_N - \rho_N) \right] \propto
	\mathrm{Tr} \left( \tilde{\Gamma}_\alpha \, \left[H_N, \Gamma \right] \right) \,,
\end{align}
where we recall that $\alpha$ corresponds to the lepton flavour index $\alpha=e,\mu, \tau$.
We then insert the full expressions for $\Gamma$, $\tilde{\Gamma}_\alpha$ and $H_N$ to find three independent CP-violating combinations, characterised by different combinations of Yukawa couplings
\begin{eqnarray}
C_{\mathrm{LFV},\alpha} &=& i \, \mathrm{Tr} \Big( \left[ \hat{M}_R^2, \hat{Y}_D^\dagger \,  \hat{Y}_D\right] \,  \hat{Y}_D^\dagger \, P_\alpha \,  \hat{Y}_D \Big) \; ,
\label{eq:defLFVcombination}
\\
C_{\mathrm{LNV},\alpha} &=&  i \, \mathrm{Tr} \Big( \left[ \hat{M}_R^2, \hat{Y}_D^\dagger \,  \hat{Y}_D\right] \,  \hat{Y}_D^T \, P_\alpha \,  \hat{Y}_D^* \Big) \; ,
\label{eq:defLNVcombination}
\\
C_{\mathrm{DEG},\alpha} &=& i \, \mathrm{Tr} \Big( \left[ \hat{Y}_D^T \,  \hat{Y}_D^* , \hat{Y}_D^\dagger \,  \hat{Y}_D\right] \,  \hat{Y}_D^T \, P_\alpha \,  \hat{Y}_D^* \Big) \; ,
\label{eq:defCdegalpha}
\end{eqnarray}
with $P_\alpha$ denoting the projector on the lepton flavour $\alpha$,
\begin{equation}
P_e = \left(
\begin{array}{ccc}
1 & 0 & 0\\
0 & 0 & 0\\
0 & 0 & 0
\end{array}
\right) \;\; , \;\;
P_\mu = \left(
\begin{array}{ccc}
0 & 0 & 0\\
0 & 1 & 0\\
0 & 0 & 0
\end{array}
\right) \;\; , \;\;
P_\tau = \left(
\begin{array}{ccc}
0 & 0 & 0\\
0 & 0 & 0\\
0 & 0 & 1
\end{array}
\right) \; .
\end{equation}
These combinations are defined in the mass basis of the RH neutrinos,
as indicated by the hat.
In the following, we summarise their main features.

\paragraph{The LFV CP-violating combination $C_\mathrm{LFV}$} is the dominant combination when the
heavy neutrinos are in the relativistic regime, i.e. $M \ll T$. It is the dominant contribution in the case of freeze-in leptogenesis, as the prefactors of the remaining two combinations are proportional to $(M/T)^2$.
The combination $C_\mathrm{LFV, \alpha}$, however, only leads to a lepton flavour asymmetry, since the sum over the lepton flavour index $\alpha$ vanishes,  $\sum_\alpha C_{\mathrm{LFV},\alpha} = 0$.
Any asymmetry generated in this way therefore crucially depends on a flavoured washout,
which can convert a lepton flavour asymmetry into a lepton number asymmetry.
 Consequently, the resulting BAU is of the order $\mathcal{O}(Y_D^6)$.

\paragraph{The LNV CP-violating combination $C_\mathrm{LNV}$} in general directly violates lepton number, with
\begin{equation}
\label{eq:CLNVtotal}
C_{\mathrm{LNV}} = \sum_\alpha C_{\mathrm{LNV},\alpha} \neq 0\,.
\end{equation}
This combination is typically subdominant in the relativistic regime, where $(M/T)^2 \ll 1$, but it can have a sizeable contribution to the BAU for intermediate masses $M \sim 10$ GeV, as it leads to a BAU already at the order $\mathcal{O}(Y_D^4)$.
In the non-relativistic regime, its prefactors are usually of the same size as those for $C_\mathrm{LFV}$, and both contributions have to be taken into account.\footnote{The situation can be very different at temperatures above $10^{12}$ GeV~\cite{Blanchet:2006ch}, where it is not possible to distinguish the lepton flavours and baryogenesis is only possible via the LNV CP-violating combination $C_\mathrm{LNV}$.}

\paragraph{The mass-\emph{degenerate} CP-violating combination $C_\mathrm{DEG}$} is the only of the three combinations
that can lead to a non-zero value of the BAU, if the RH neutrino masses are degenerate, since it can be non-vanishing for zero splittings $\kappa$ and  $\lambda$. This type of asymmetry is only possible at intermediate temperatures $M/T \sim 1$,
as the thermal mass and decay matrices commute in the relativistic and non-relativistic limits.
This combination only produces a lepton flavour asymmetry, since $\sum_\alpha C_{\mathrm{DEG},\alpha}=0$.
The BAU coming from this combination is therefore only  generated at the order  $\mathcal{O}(Y_D^8)$.

\paragraph{Comparison of CP-violating combinations with decay asymmetries.} A different approach to leptogenesis in the non-relativistic regime relies on standard Boltzmann equations to compute the BAU. The key quantities, governing the generation of the BAU, are in this case the decay asymmetries $\epsilon$, see e.g.~\cite{Dev:2017trv}.
We show in the following that one can relate the flavoured decay asymmetries $\epsilon_{i\alpha}$ to the CP-violating combinations $C_{\mathrm{LFV},\alpha}$ and $C_{\mathrm{LNV},\alpha}$.
On the one hand, the corresponding part of $\epsilon_{i\alpha}$ reads\footnote{We omit a possible regulator and factor $(\hat{Y}^\dagger \hat{Y})_{ii}$ that both can in general depend on the index $i$ of the RH neutrino. For the sake of readability, we here suppress the subscript $D$ of the neutrino Yukawa coupling matrix $Y_D$.}
\begin{equation}
	\label{eq:Garvdecayasymmetries}
	\sum_j \mbox{Im} \left(\hat{Y}_{i\alpha}^\dagger \hat{Y}_{\alpha j} \right) \, \mbox{Re} \left((\hat{Y}^\dagger \hat{Y})_{ij} \right) \, (M_i^2-M_j^2) \; .
\end{equation}
On the other hand, expanding the expressions for $C_{\mathrm{LFV},\alpha}$ and $C_{\mathrm{LNV},\alpha}$, found in Eqs.~\eqref{eq:defLFVcombination} and \eqref{eq:defLNVcombination}, gives
\begin{equation}
	C_{\mathrm{LFV},\alpha} = i \sum_{i,j} M_j^2 \, \{(\hat{Y}^T \hat{Y}^*)_{ij} \, (\hat{Y}^\dagger_{i \alpha}  \hat{Y}_{\alpha j})-(\hat{Y}^\dagger \hat{Y})_{ij} \, (\hat{Y}^T_{i \alpha}  \hat{Y}^*_{\alpha j}) \}
\end{equation}
and
\begin{equation}
	C_{\mathrm{LNV},\alpha} = i \sum_{i,j} M_j^2 \, \{(\hat{Y}^T \hat{Y}^*)_{ij} \, (\hat{Y}^T_{i \alpha} \hat{Y}^*_{\alpha j})-(\hat{Y}^\dagger \hat{Y})_{ij} \, (\hat{Y}^\dagger_{i \alpha} \hat{Y}_{\alpha j}) \} \; ,
\end{equation}
respectively. Combining these two results and exchanging the indices $i$ and $j$,
we obtain
\begin{equation}
	\label{eq:LFV-LNVcombination}
	\frac{1}{2}\big(C_{\mathrm{LFV},\alpha}-C_{\mathrm{LNV},\alpha}\big) = \sum_{i,j} (M_i^2-M_j^2) \, \mbox{Re} \left((\hat{Y}^\dagger \hat{Y})_{ij} \right) \, \mbox{Im} \left( \hat{Y}_{i\alpha}^\dagger \hat{Y}_{\alpha j} \right)
\end{equation}
which is very similar to the expression in Eq.~\eqref{eq:Garvdecayasymmetries}.

\paragraph{The reduced CP-violating combinations} can be defined for a particular subset of the three heavy neutrinos. For instance, in the limit of large $\kappa$, i.e.~$\lambda \ll \kappa \lesssim 1$, it is often sufficient to consider oscillations among only two of the three heavy neutrinos.
In this limit we introduce the three-by-two neutrino Yukawa coupling matrix
$\hat{Y}_{(23)}$
\begin{align}
	(\hat{Y}_{(23)})_{\alpha i} = (\hat{Y}_D)_{\alpha i}\quad \text{for}\quad i \in \{ 2, 3 \}\,,
\end{align}
and the analogous two-by-two heavy neutrino mass matrix.
We can use them to calculate the same CP-violating combinations as before.
For $\lambda = 0$, in general only the mass-\emph{degenerate} CP-violating combination is non-vanishing
\begin{align}
	\label{eq:reducedCPcombinations}
	C^{(23)}_{\mathrm{DEG},\alpha} = i \, \mathrm{Tr} \Big(
		\left[ {\hat{Y}_{(23)}}^T \,  {\hat{Y}_{(23)}}^* , {\hat{Y}_{(23)}}^\dagger \,  \hat{Y}_{(23)}\right] \,
		{\hat{Y}_{(23)}}^T \, P_\alpha \,  {\hat{Y}_{(23)}}^*
	\Big) \; .
\end{align}
It can be used to indicate the presence of a non-vanishing value of the BAU for large $\kappa$, when one of the three heavy neutrinos effectively decouples from oscillations, since $C^{(23)}_{\mathrm{DEG},\alpha}$ quantifies CP violation associated with the remaining degenerate pair of heavy neutrinos.
Clearly, the sum over the lepton flavour index $\alpha$ is zero, $\sum_\alpha C^{(23)}_{\mathrm{DEG},\alpha} = 0$.

\paragraph{Flavoured washout parameter.}
Out of the three CP-violating combinations mentioned above, two do not violate the total lepton number, but only lepton flavour. Thus,
if the CP-violating combination $C_\mathrm{LNV}$ in Eq.~(\ref{eq:CLNVtotal}) is zero, or if the heavy neutrino masses are too small for the generation of the lepton number asymmetry to be efficient, no overall lepton asymmetry will be produced initially.
Instead, the interactions of the heavy neutrinos only lead to lepton flavour asymmetries, which sum up to a vanishing lepton asymmetry. However,
the three lepton flavour asymmetries are generally washed out at different rates, which leads to an overall lepton asymmetry and thus also to a non-zero value of the BAU.
To quantify this, we introduce the \emph{flavoured washout parameter}
\begin{align}
	\label{eq:flvw}
	f_\alpha = \frac{(\hat{Y}_D \hat{Y}_D^\dagger)_{\alpha \alpha}}{ \Tr(\hat{Y}_D \hat{Y}_D^\dagger)}\; .
\end{align}
There are two cases in which it is particularly important to consider
the ratios of the lepton washouts
\begin{itemize}
	\item {\bf equal washouts} can suppress the conversion of lepton flavour asymmetries to an overall lepton asymmetry. This can happen when either
		$f_\alpha = 1/3$ for $\alpha=e, \mu, \tau$, or all lepton flavour asymmetry is generated in two flavours only, e.g.~in case the yields are $Y_e=0$ and $Y_\mu = -Y_\tau$, and the washouts in those flavours are equal, e.g.~$f_\mu = f_\tau$.
	\item {\bf flavour-hierarchical washout} can prevent an overall large washout from completely erasing early lepton asymmetries.
		This typically happens, if one of the three flavours, $\tilde{\alpha}$, is washed out at a much lower rate than the other two, i.e.~$f_{\tilde{\alpha}} \ll 1$.
\end{itemize}

\vspace{0.2in}
\noindent It is convenient to define the following quantities
\begin{equation}
\label{eq:Deltasigmas}
\Delta \sigma_{12} = (3 \, \kappa-\lambda) \, (2+\kappa+\lambda) \;\;\; \mbox{and} \;\;\; \Delta \sigma_{23} = \lambda \, (1-\kappa) \, .
\end{equation}
The vanishing of $\Delta \sigma_{ij}$ indicates that the mass  splitting between the $i^\mathrm{th}$ and $j^\mathrm{th}$ RH neutrino becomes zero. If the first factor of
$\Delta \sigma_{ij}$ is zero the two RH neutrinos have the same mass, while for the second factor being zero, they are degenerate in mass, but have opposite
CP parity. Furthermore, we note that the vanishing of the second factor of $\Delta \sigma_{ij}$ is only possible, if one of the splittings, $\kappa$ or $\lambda$,
becomes of order one. This is a situation we consider as unnatural.
In the case of vanishing splitting $\lambda$, we have $\Delta \sigma_{12}=3 \, \kappa \, (2+\kappa)$ and $\Delta \sigma_{23}=0$.
We also introduce the quantities
\begin{equation}
\Delta y^2_{ij} = y_i^2 - y_j^2
\end{equation}
defined for $i$ being smaller than $j$ as well as
\begin{equation}
\Sigma y^2 = y_1^2 + y_2^2 + y_3^2 \; .
\end{equation}

\subsection{Case 1)}
\label{sec62}

We first note that $C_{\mathrm{DEG},\alpha}$ is zero. Thus, in order to achieve non-zero BAU we need to have at least one splitting $\kappa$ or $\lambda$ non-vanishing.
In the following, we distinguish between $s$ even and $s$ odd.
Then, for $s$ being even, we find for $C_{\mathrm{LFV},\alpha}$
\begin{equation}
\label{eq:CLFValphaCase1seven}
C_{\mathrm{LFV},\alpha}= \frac{4}{9}\, M^2 \, y_2 \, \Delta \sigma_{12}  \, \Big(  y_1 \, \Delta y^2_{12}  \, \cos \theta_{L,\alpha} \, \cos \theta_R - y_3 \, \Delta y^2_{23}  \, \sin\theta_{L,\alpha}\,\sin\theta_R\Big) \, \sin 3 \, \phi_s
\end{equation}
and similarly for $C_{\mathrm{LNV},\alpha}$
\begin{eqnarray}\nonumber
C_{\mathrm{LNV},\alpha} &=& \frac{4}{9}\, M^2 \, y_2 \, \Delta \sigma_{12} \, \Big(  y_1 \, (\Delta y^2_{23}  - \Delta y^2_{13}  \, \cos 2 \, \theta_R) \, \cos \theta_{L,\alpha} \, \cos \theta_R
\\
\label{eq:CLNValphaCase1seven}
&& \;\;\;\;\;\;\;\;\;\;\;\;\;\;\;\;\;\;\;\;\;\;\;\;\;\;\;\;\;\;\;\;
- y_3 \, (\Delta y^2_{12} + \Delta y^2_{13} \, \cos 2 \, \theta_R) \, \sin\theta_{L,\alpha}\,\sin\theta_R\Big) \, \sin 3 \, \phi_s \, .
\end{eqnarray}
Note that summing $C_{\mathrm{LNV},\alpha}$ over the lepton flavour index $\alpha$ gives zero.
We have defined the quantity
\begin{equation}
\label{eq:Case1rhoalphadef}
\theta_{L,\alpha} =\theta_L + \rho_\alpha \, \frac{4 \, \pi}{3} \;  \quad {\rm with} \quad \; \rho_e=0, \ \rho_\mu=+1, \ \rho_\tau=-1 \; .
\end{equation}
For $s$ odd, we have to replace $\sin 3 \, \phi_s$ by $-\cos 3 \, \phi_s$ in the above formulae.

\vspace{0.2in}
\noindent In the limit of large splitting $\kappa$ (and $\lambda=0$),
we find for the reduced mass-degenerate CP-violating combination $C^{(23)}_{\mathrm{DEG},\alpha}$ that it is
\begin{eqnarray}\nonumber
	C^{(23)}_{\mathrm{DEG},\alpha} &=&
	\frac{4}{27} \, y_2 \, \Delta y_{13}^2 \, \left(
		2 \, \Delta y_{12}^2 - \Delta y_{13}^2 \, (1+ 2 \cos 2 \theta_R)
	\right) \,
	\left(
	y_1 \cos \theta_{L,\alpha} \sin \theta_R \right.
	\\ && \left. \;\;\;\;\;\;\;\;\;\;\;\;\;\; \;\;\;\;\;\;\;\;\;\;\;\;\;\;\;\;\;\;\;\;\;\;\;\;\;\;\;\;\;\;\;\;\;\;\;\;\;\;\;\;\;\;\;\;\;
		- y_3 \sin \theta_{L,\alpha} \cos \theta_R
	\right) \,
	\sin 2\, \theta_R \, \sin 3 \,  \phi_s
	\label{eq:reducedCPVcombinationCase1}
\end{eqnarray}
for $s$ even,
while for $s$ odd we have to replace $\sin 3 \, \phi_s$ by
$ -\cos 3 \, \phi_s$.

\vspace{0.2in}
\noindent The flavoured washout parameter $f_\alpha$, introduced  in Eq.~\eqref{eq:flvw},
is the same for $s$ even and $s$ odd, and only depends on the  couplings $y_f$ and the angle $\theta_L$
\begin{equation}
	f_\alpha =
	\frac13 \, \left( 1 + \frac{\Delta y_{13}^2}{\Sigma y^2} \cos2 \, \theta_{L, \alpha} \right) \; .
\end{equation}

\vspace{0.2in}
\noindent The recent work~\cite{Chauhan:2021xus} studies resonant leptogenesis in a low-scale type-I seesaw scenario with the same flavour and CP symmetries. For strong NO  and $s$ even, we notice that the combination $\frac{1}{2}\big(C_{\mathrm{LFV},\alpha}-C_{\mathrm{LNV},\alpha}\big)$, obtained by subtracting Eqs.~\eqref{eq:CLFValphaCase1seven} and \eqref{eq:CLNValphaCase1seven}, reads
\begin{equation}
\frac{1}{2}\big(C_{\mathrm{LFV},\alpha}-C_{\mathrm{LNV},\alpha}\big)
= -\frac{4}{9}\, M^2 \, \Delta \sigma_{12} \, y_2 \, y_3 \, \sin \theta_{L,\alpha}  \, \big(y_2^2-y_3^2 \,\sin^2 \theta_R\big) \, \sin \theta_R \, \sin 3 \, \phi_s
\end{equation}
and thus has the same dependence on the parameters $s$, $y_f$, $\theta_L$ and $\theta_R$ as the flavoured decay asymmetries in~\cite{Chauhan:2021xus}, compare equation~(4.10). This is expected from Eqs.~\eqref{eq:Garvdecayasymmetries} and \eqref{eq:LFV-LNVcombination}. We have also checked that there is agreement between their and our results for strong IO as well as for $s$ odd.

\subsection{Case 2)}
\label{sec63}

For Case 2), we can distinguish between the following combinations: $s$ and $t$ both even, $s$ even and $t$ odd, $s$ odd and $t$ even as well as $s$ and $t$ both odd.
We first discuss the CP-violating combinations for $s$ and $t$ both even. Like in Case 1), we find that for vanishing splittings $\kappa$ and $\lambda$ all CP-violating
combinations are zero. So, $C_{\mathrm{DEG},\alpha}$ is always zero. If the splittings are non-zero, we have for $C_{\mathrm{LFV},\alpha}$
\begin{eqnarray}\nonumber\label{eq:CLFValphaCase2steven}
C_{\mathrm{LFV},\alpha}&=&\frac{4}{9} \, M^2 \, \Big( -y_2 \, \Delta\sigma_{12}  \, (y_1 \, \Delta y^2_{12} \,\sin\theta_L\, \cos\theta_R + y_3 \, \Delta y^2_{23} \, \cos\theta_L\, \sin\theta_R) \, \sin \frac{\phi_{u,\alpha}}{2}  \, \cos \frac{\phi_v}{2}
\\  \nonumber
&&\;\;\;\;\;\;\;\;\;\;\;\;\;+ y_2  \, \Delta\sigma_{12} \, (y_1 \, \Delta y^2_{12} \,\cos\theta_L \, \cos\theta_R- y_3 \, \Delta y^2_{23} \, \sin\theta_L\, \sin\theta_R) \, \cos \frac{\phi_{u,\alpha}}{2} \, \sin \frac{\phi_v}{2}
\\
&&\;\;\;\;\;\;\;\;\;\;\;\;\;- \frac 12 \, y_1 \, y_3 \, \Delta y^2_{13} \, (\Delta \sigma_{12}+ 6 \, \Delta \sigma_{23})  \, \sin  2 \, \theta_R \, \sin \phi_{u,\alpha} \Big)
\end{eqnarray}
and for $C_{\mathrm{LNV},\alpha}$ that it coincides with $C_{\mathrm{LFV},\alpha}$ up to an overall sign, i.e.~$C_{\mathrm{LNV},\alpha}=-C_{\mathrm{LFV},\alpha}$. Thus, the sum of $C_{\mathrm{LNV},\alpha}$ over the lepton flavour index $\alpha$ gives zero. We have introduced the quantity
\begin{equation}
	\phi_{u,\alpha} = \frac{\pi \, u}{n} + \rho_\alpha \, \frac{4 \, \pi}{3}  = \phi_u + \rho_\alpha  \, \frac{4 \, \pi}{3} \; \quad {\rm with} \quad \; \rho_e=0 \, , \,  \rho_\mu=-1 \, , \, \rho_\tau=+1 \; .
\end{equation}
It is instructive to consider the formulae for strong NO ($y_1=0$) and strong IO ($y_3=0$) as well as the special case $u=0$ for the electron flavour, $\alpha=e$. For strong NO we have
\begin{equation}
\label{eq:CLFValphaCase2stevenNO}
C_{\mathrm{LFV},\alpha}=-\frac{4}{9} \, M^2 \, y_2\, y_3 \, \Delta y^2_{23} \, \Delta\sigma_{12} \, \Big( \cos\theta_L \, \sin \frac{\phi_{u,\alpha}}{2}  \, \cos \frac{\phi_v}{2}  + \sin\theta_L\, \cos \frac{\phi_{u,\alpha}}{2} \, \sin \frac{\phi_v}{2} \Big) \,  \sin\theta_R
\end{equation}
and for strong IO we find
\begin{equation}
\label{eq:CLFValphaCase2stevenIO}
C_{\mathrm{LFV},\alpha}=\frac{4}{9} \, M^2 \, y_1 \, y_2 \,  \Delta y^2_{12}  \, \Delta\sigma_{12} \, \Big( - \sin\theta_L \, \sin \frac{\phi_{u,\alpha}}{2}  \, \cos \frac{\phi_v}{2} + \cos\theta_L  \, \cos \frac{\phi_{u,\alpha}}{2} \, \sin \frac{\phi_v}{2}  \Big) \, \cos\theta_R \, .
\end{equation}
For $u=0$ and the electron flavour, $\alpha=e$, we get
\begin{equation}
\label{eq:CLFValphaCase2stevenelectron}
C_{\mathrm{LFV},e}=\frac{4}{9} \, M^2 \, y_2  \, \Delta\sigma_{12} \, \Big( y_1 \, \Delta y^2_{12} \,\cos\theta_L \, \cos\theta_R- y_3 \, \Delta y^2_{23} \, \sin\theta_L\, \sin\theta_R\Big) \, \sin \frac{\phi_v}{2} \; .
\end{equation}
For the combination $s$ odd and $t$ even we get the same results for the CP-violating combinations, just replacing $\phi_v$ by $\phi_v-\pi$ in the mentioned equations.

For the combination $s$ even and $t$ odd, we find instead non-zero $C_{\mathrm{DEG},\alpha}$
\begin{equation}
\label{eq:CDEGalpha_seventodd_Case2}
C_{\mathrm{DEG},\alpha} = \frac 13 \, y_1 \, y_3 \, (\Delta y_{13}^2)^2 \, \sin 4 \, \theta_R \, \sin \phi_{u,\alpha} \, .
\end{equation}
Interestingly enough, $C_{\mathrm{DEG},\alpha}$ vanish in the case of strong NO and strong IO.
If the splittings $\kappa$ and $\lambda$ are non-zero, also $C_{\mathrm{LFV},\alpha}$ and $C_{\mathrm{LNV},\alpha}$ are non-zero. We have for $C_{\mathrm{LFV},\alpha}$
\begin{eqnarray}\nonumber
C_{\mathrm{LFV},\alpha}&=&\frac{2}{9} \, M^2 \, \Big( y_2 \, \Delta \sigma_{12} \Big[(y_3 \, \Delta y^2_{23}\, \sin\theta_L \, cs_{R,-}+ y_1 \, \Delta y^2_{12}\, \cos\theta_L\, cs_{R,+})  \, \cos \frac{\phi_{u,\alpha}}{2} \, \cos \frac{\phi_v}{2}
\\  \nonumber
&&- (y_1 \, \Delta y^2_{12}\,\sin\theta_L \, cs_{R,-} + y_3 \, \Delta y^2_{23}\, \cos\theta_L\, cs_{R,+})  \, \sin \frac{\phi_{u,\alpha}}{2} \, \cos \frac{\phi_v}{2}
\\  \nonumber
&&+(y_1 \, \Delta y^2_{12}\, \cos\theta_L\, cs_{R,-}  - y_3 \, \Delta y^2_{23}\, \sin\theta_L\, cs_{R,+})  \, \cos \frac{\phi_{u,\alpha}}{2} \, \sin \frac{\phi_v}{2}
\\  \nonumber
&&- (y_3 \, \Delta y^2_{23}\, \cos\theta_L\, cs_{R,-}  - y_1 \, \Delta y^2_{12}\, \sin\theta_L\, cs_{R,+}) \, \sin \frac{\phi_{u,\alpha}}{2} \, \sin \frac{\phi_v}{2}\Big]
\\
&&+  y_1 \, y_3 \, \Delta y^2_{13} \, [\Delta\sigma_{12}+ 6 \, \Delta\sigma_{23}] \, \sin 2 \, \theta_L \, \cos \phi_{u,\alpha} \Big)
\end{eqnarray}
with $cs_{R,\pm}=\cos\theta_R\pm\sin\theta_R$. For strong NO and strong IO this result simplifies considerably. For strong NO, we have
\begin{eqnarray}\nonumber
C_{\mathrm{LFV},\alpha}&=&\frac{2}{9} \, M^2 \, y_2 \, y_3 \, \Delta y^2_{23} \, \Delta \sigma_{12} \, \Big( (\sin\theta_L  \, \cos \frac{\phi_{u,\alpha}}{2} \, \cos \frac{\phi_v}{2} -  \cos\theta_L\, \sin \frac{\phi_{u,\alpha}}{2} \, \sin \frac{\phi_v}{2}) \, cs_{R,-}
\\
&&\;\;\;\;\;\;\;\;\;\;\;\;\; \;\;\;\;\;\;\;\;\;\;\;\;\;- (\cos\theta_L \, \sin \frac{\phi_{u,\alpha}}{2} \, \cos \frac{\phi_v}{2} + \sin\theta_L \, \cos \frac{\phi_{u,\alpha}}{2} \, \sin \frac{\phi_v}{2}) \, cs_{R,+} \Big)
\end{eqnarray}
and for strong IO we get
\begin{eqnarray}\nonumber
C_{\mathrm{LFV},\alpha}&=&\frac{2}{9} \, M^2 \, y_1 \, y_2 \, \Delta y^2_{12} \, \Delta \sigma_{12} \Big( (-\sin\theta_L \, \sin \frac{\phi_{u,\alpha}}{2} \, \cos \frac{\phi_v}{2}
+\cos\theta_L\,  \cos \frac{\phi_{u,\alpha}}{2} \, \sin \frac{\phi_v}{2}) \, cs_{R,-}
\\
&&\;\;\;\;\;\;\;\;\;\;\;\;\; \;\;\;\;\;\;\;\;\;\;\;\;\;+ (\cos\theta_L \, \cos \frac{\phi_{u,\alpha}}{2} \, \cos \frac{\phi_v}{2} + \sin\theta_L \, \sin \frac{\phi_{u,\alpha}}{2} \, \sin \frac{\phi_v}{2}) \, cs_{R,+}
\Big) \, .
\end{eqnarray}
Furthermore, we can get a simple formula for the electron flavour, $\alpha=e$, taking into account that $u/n$ is small and thus also $\phi_u$,
\begin{eqnarray}
C_{\mathrm{LFV},e}&\approx&\frac{2}{9} \, M^2 \, \Big( y_2 \, \Delta \sigma_{12} \Big[(y_3 \, \Delta y^2_{23}\, \sin\theta_L \, cs_{R,-}+ y_1 \, \Delta y^2_{12}\, \cos\theta_L\, cs_{R,+})   \, \cos \frac{\phi_v}{2}
\\  \nonumber
&&\!\!\!\!\!\!\!\!\!\!\!\!\!\!\!\!\!\!+(y_1 \, \Delta y^2_{12}\, \cos\theta_L\, cs_{R,-}  - y_3 \, \Delta y^2_{23}\, \sin\theta_L\, cs_{R,+})   \, \sin \frac{\phi_v}{2} \Big]
+  y_1 \, y_3 \, \Delta y^2_{13} \, [\Delta\sigma_{12}+ 6 \, \Delta\sigma_{23}] \, \sin 2 \, \theta_L \Big)\; .
\end{eqnarray}
For $C_{\mathrm{LNV},\alpha}$ we refer to appendix~\ref{appG1}. Summing $C_{\mathrm{LNV},\alpha}$ over the lepton flavour index $\alpha$ we find
\begin{equation}
\label{eq:CLNVsummed_seventodd_Case2}
C_{\mathrm{LNV}} =  -\frac{1}{3} \, M^2 \,  (\Delta y^2_{13})^2 \, [\Delta \sigma_{12} + 6 \, \Delta\sigma_{23}] \, \sin 4 \, \theta_R \; .
\end{equation}
For the combination $s$ and $t$ both odd we get the same results for the CP-violating combinations, just replacing $\phi_v$ by $\phi_v+\pi$ in the mentioned equations.

\vspace{0.2in}
\noindent In the limit of large splitting $\kappa$ (and $\lambda=0$) we  find that the reduced mass-degenerate CP-violating combination $C^{(23)}_{\mathrm{DEG},\alpha}$ vanishes for $t$ even and all choices of $s$, while it is non-zero for $t$ odd and reads for $s$ even
\begin{eqnarray}\nonumber
C^{(23)}_{\mathrm{DEG},\alpha} &=&
\frac{2}{27} \, \Delta y_{13}^2 \, \Big( y_2 \, \Big\{\Big( \Delta y_{12}^2 -\Delta y_{23}^2 - \Delta y_{13}^2 \, \sin 2 \, \theta_R \Big)\Big[\Big(y_1 \, \cos \theta_L \, cs_{R,-} + y_3 \, \sin \theta_L \, cs_{R,+} \Big) \,
\\ \nonumber
&&\;\;\;\;\;\;\;\;\;\;\;\;\;\;\;\;\;\;\;\; 
\times \cos \frac{\phi_{u, \alpha}}{2} \, \cos \frac{\phi_v}{2} \, + \Big(y_1 \, \sin \theta_L\, cs_{R,-} - y_3 \, \cos\theta_L\, cs_{R,+} \Big) \, \sin \frac{\phi_{u, \alpha}}{2} \, \sin \frac{\phi_v}{2}\Big]
\\ \nonumber
&&+\Big( \Delta y_{12}^2 -\Delta y_{23}^2 + \Delta y_{13}^2 \, \sin 2 \, \theta_R \Big) \, \Big[\Big(y_1 \, \cos \theta_L \, cs_{R,+} - y_3 \, \sin \theta_L \, cs_{R,-} \Big) \, \cos \frac{\phi_{u, \alpha}}{2} \, \sin \frac{\phi_v}{2}
\\ \nonumber
&&\;\;\;\;\;\;\;\;\;\;\;\;\;\;\;\;\;\;\;\;\;\;\;\;\;\;\;\;\;\;\;\;\;\;\;\;\;\;\;\;\;\;\;\;\;\;\; - \Big(y_1 \, \sin \theta_L\, cs_{R,+} + y_3 \, \cos\theta_L\, cs_{R,-} \Big) \, \sin \frac{\phi_{u, \alpha}}{2} \, \cos \frac{\phi_v}{2}\Big]
\Big\}
\\ \nonumber
&&-\Big( ( y_ 1^2 \, \Delta y_{23}^2 - y_3^2 \, \Delta y_{12}^2) \, \cos 2 \, \theta_L \, \sin 2 \, \theta_R + y_1 \, y_3 \, (\Delta y_{12}^2 - \Delta y_{23}^2) \, \sin 2 \, \theta_L \, \cos 2 \, \theta_R \Big) \cos \phi_{u, \alpha}
\\
\label{eq:CDEG23_seventodd_Case2}
&&\;\;\;\;\;\;\;\;\;\;\;\;\;\;\;\;\;\;\;\;\;\;\;\;\;\;\;\;\;\;\;\;\;\;\;\;\;\;\;\;\;\;\;\;\;\;\;\;\;\;\;\;\;\;\;\;\;\;\;\;\;\;\;\; + 2 \, y_1 \, y_3 \, \Delta y_{13}^2 \, \sin 2 \, \theta_R \, \sin \phi_{u, \alpha}\Big) \, \cos 2 \, \theta_R \; .
\end{eqnarray}

For $s$ odd $\phi_v$ has to be substituted by $\phi_v+\pi$ in Eq.~(\ref{eq:CDEG23_seventodd_Case2}).

\vspace{0.2in}
\noindent The flavoured washout parameter $f_\alpha$, found in Eq.~\eqref{eq:flvw}, does not depend on the choice of $s$ and $t$ being even or odd, but only on the couplings $y_f$, the parameter $\phi_u$ and the angle $\theta_L$
\begin{align}
	f_\alpha =
	\frac13 \, \left( 1 + \frac{\Delta y_{13}^2}{\Sigma y^2} \, \cos 2 \theta_L \, \cos \phi_{u, \alpha} \right) \; .
\end{align}

\vspace{0.2in}
\noindent Similar to Case 1), we have checked explicitly that the
dependencies on the parameters $\phi_u$, $\phi_v$, $\theta_L$  and $\theta_R$, given in Eq.~\eqref{eq:CLFValphaCase2steven},  coincide with those, found in~\cite{Chauhan:2021xus}, see equation (4.13). This is expected, since for $s$ and $t$ both even $C_{\mathrm{LNV},\alpha}$ is the negative of $C_{\mathrm{LFV},\alpha}$.

\subsection{Case 3 a) and Case 3 b.1)}
\label{sec64}

Also for these cases we can distinguish two types of combinations of the parameters, which are $s$ and $m$. On the one hand, for $s$ and $m$ both even or both odd $C_{\mathrm{LFV},\alpha}$
and $C_{\mathrm{LNV},\alpha}$ are the same up to an overall sign. Consequently, summing $C_{\mathrm{LNV},\alpha}$ over the lepton flavour index $\alpha$ gives zero, i.e.~$C_{\mathrm{LNV}}=0$.
Furthermore, $C_{\mathrm{DEG},\alpha}$ vanish, meaning that non-zero CP-violating combinations for these two combinations of $s$ and $m$  require that at least one of the splittings, $\kappa$ and
$\lambda$, is non-vanishing. On the other hand, for the combinations of $s$ and $m$, where one is even and the other one odd, $C_{\mathrm{LFV},\alpha}$ and $C_{\mathrm{LNV},\alpha}$
are non-zero and different from each other, $C_{\mathrm{LNV}}$ is non-vanishing and also $C_{\mathrm{DEG},\alpha}$ are non-zero.
Since the results for Case 3 a) and Case 3 b.1) are quite lengthy, we mainly mention them for the combinations $s$ and $m$ both even as well as $s$ odd and $m$ even.

For $s$ and $m$ both even, we have for $C_{\mathrm{LFV},\alpha}$
\begin{eqnarray}\nonumber
C_{\mathrm{LFV},\alpha}&=&\frac{2}{9} \, M^2 \, \Big( \sqrt{2}\, y_1 \, y_2\, \Delta y^2_{12}\, [\Delta \sigma_{12}+ 3 \, \Delta \sigma_{23}] \, tcs_{R,-}  \, tcs_{R,+} \, \cos \phi_{m,\alpha} \, \sin 3\, \phi_s
\\  \nonumber
&&\;\;\;\;\;\;\;\;\;\;\;\;+ 3 \, \sqrt{2} \, y_3 \, \Delta \sigma_{23} \, [y_2\, \Delta y^2_{23}  \, \cos \theta_L \, tcs_{R,-} + y_1 \, \Delta y^2_{13} \sin \theta_L \, tcs_{R,+}] \, \sin \phi_{m,\alpha} \, \cos 3\, \phi_s
\\
&&\;\;\;\;\;\;\;\;\;\;\;\;- 3 \, y_3 \, \Delta \sigma_{23} \,
[y_1 \, \Delta y^2_{13}\, \cos \theta_L \, tcs_{R,+} -y_2 \, \Delta y^2_{23}\, \sin \theta_L \, tcs_{R,-}] \, \sin 2 \, \phi_{m,\alpha}
\Big) \; ,
\label{eq:CLFValphasmevenCase3}
\end{eqnarray}
where we have defined
\begin{equation}
\phi_{m,\alpha} = \frac{\pi \, m}{n} + \rho_\alpha \, \frac{4 \, \pi}{3}  = \phi_m + \rho_\alpha  \, \frac{4 \, \pi}{3} \; \quad {\rm with} \quad \; \rho_e=0 \, , \, \rho_\mu=+1 \, , \,  \rho_\tau=-1 \; ,
\end{equation}
as well as
\begin{equation}
\label{eq:definitiontcs}
tcs_{R,-}= \sqrt{2}\, \cos \theta_R -\sin\theta_R \;\;\; \mbox{and} \;\;\;  tcs_{R,+}= \cos \theta_R +\sqrt{2}\, \sin\theta_R \; .
\end{equation}
Again, we can simplify this formula by considering strong NO and strong IO. For strong NO for Case 3 a) ($y_1=0$) we find
\begin{equation}
C_{\mathrm{LFV},\alpha}=\frac{2 \, \sqrt{2}}{3} \, M^2 \, y_2\, y_3 \, \Delta y^2_{23} \, \Delta \sigma_{23} \, \Big( \cos \theta_L \, \cos 3\, \phi_s + \sqrt{2} \, \sin \theta_L  \, \cos \phi_{m,\alpha} \Big) \, tcs_{R,-} \, \sin \phi_{m,\alpha} \; ,
\end{equation}
while for strong IO for Case 3 a) and for strong NO for Case 3 b.1) ($y_3=0$) we get
\begin{equation}
\label{eq:CLFValphamsevenNO}
C_{\mathrm{LFV},\alpha}=\frac{2\, \sqrt{2}}{9} \, M^2 \, y_1 \, y_2\, \Delta y^2_{12}\, [\Delta \sigma_{12}+ 3 \, \Delta \sigma_{23}] \, tcs_{R,-}  \, tcs_{R,+} \, \cos \phi_{m,\alpha} \, \sin 3\, \phi_s  \; .
\end{equation}
For strong IO for Case 3 b.1) ($y_2=0$) we have
\begin{equation}
\label{eq:CLFValphamsevenIO}
C_{\mathrm{LFV},\alpha}=\frac{2\, \sqrt{2}}{3} \, M^2 \, y_1 \, y_3\, \Delta y^2_{13}\, \Delta \sigma_{23} \, \Big( \sin \theta_L \, \cos 3\, \phi_s - \sqrt{2} \, \cos \theta_L  \, \cos \phi_{m,\alpha} \Big)\, tcs_{R,+} \, \sin \phi_{m,\alpha} \; .
\end{equation}
We can also give approximate expressions for the electron flavour,  $\alpha=e$, for Case 3 a) and Case 3 b.1), using that the value of $m$ is strongly constrained. For Case 3 a), $m \approx 0$ ($m \approx n$), such that we have
\begin{equation}
C_{\mathrm{LFV},e}\approx (-)\frac{2\, \sqrt{2}}{9} \, M^2 \, y_1 \, y_2\, \Delta y^2_{12}\, [\Delta \sigma_{12}+ 3 \, \Delta \sigma_{23}] \, tcs_{R,-}  \, tcs_{R,+} \, \sin 3\, \phi_s \; ,
\end{equation}
while for Case 3 b.1), $m \approx \frac{n}{2}$, so that
\begin{equation}
C_{\mathrm{LFV},e}\approx\frac{2 \, \sqrt{2}}{3} \, M^2 \, y_3 \, \Delta \sigma_{23} \, \Big(y_2\, \Delta y^2_{23}  \, \cos \theta_L \, tcs_{R,-} + y_1 \, \Delta y^2_{13} \sin \theta_L \, tcs_{R,+}\Big) \, \cos 3\, \phi_s \; .
\end{equation}

For $s$ odd and $m$ even, we find
\begin{eqnarray}\nonumber
C_{\mathrm{LFV},\alpha}&=&\frac{\sqrt{2}}{9} \, M^2 \, \Big(
y_1 \, y_2\, \Delta y^2_{12} \, [\Delta \sigma_{12}+3\, \Delta \sigma_{23}] \, \sin 2 \, \theta_L \, \cos 2 \, \phi_{m,\alpha}
\\  \nonumber
&&- 3 \, \sqrt{2} \, y_3 \, \Delta \sigma_{23}  \, [y_1\, \Delta y^2_{13}\, \cos \theta_L \, \cos \theta_R+ y_2 \, \Delta y^2_{23}\, \sin \theta_L \, \sin \theta_R] \, \sin 2 \, \phi_{m,\alpha}
\\  \nonumber
&&-6 \, \sqrt{2}\, y_3 \,  \Delta \sigma_{23}\, [y_2 \, \Delta y^2_{23} \, \cos \theta_L \, \cos \theta_R +y_1 \, \Delta y^2_{13}\, \sin \theta_L \, \sin \theta_R] \, \sin \phi_{m,\alpha} \, \sin 3 \, \phi_s
\\  \nonumber
&&+6\, y_3 \, \Delta \sigma_{23} \, [y_1 \, \Delta y^2_{13} \, \sin \theta_L \, \cos \theta_R -y_2 \, \Delta y^2_{23}\, \cos \theta_L \, \sin \theta_R]\, \sin \phi_{m,\alpha} \, \cos 3 \, \phi_s
\\  \nonumber
&&+ y_1 \, y_2 \, \Delta y^2_{12} \, [\Delta \sigma_{12}+3\, \Delta \sigma_{23}]\,  \sin 2 \, \theta_R \, \cos \phi_{m,\alpha} \, \sin 3 \, \phi_s
\\
&&+ 2 \, \sqrt{2}  \, y_1 \, y_2 \, \Delta y^2_{12} \, [\Delta \sigma_{12}+3\, \Delta \sigma_{23}] \, \cos 2 \, \theta_L \, \cos \phi_{m,\alpha} \, \cos 3 \, \phi_s
\Big)
\end{eqnarray}
and for $C_{\mathrm{LNV},\alpha}$ we refer to appendix~\ref{appG2}.
Also these formulae simplify considerably for strong NO and strong IO, e.g.~for $C_{\mathrm{LFV},\alpha}$ we get for strong NO for Case 3 a) ($y_1=0$)
\begin{eqnarray}\nonumber
C_{\mathrm{LFV},\alpha}&=&-\frac{2 \, \sqrt{2}}{3} \, M^2 \, y_2 \, y_3 \, \Delta y^2_{23} \, \Delta \sigma_{23} \, \Big( \sqrt{2}\,\sin \theta_L \, \sin \theta_R \, \cos \, \phi_{m,\alpha} +\sqrt{2}\, \cos \theta_L \, \cos \theta_R  \, \sin 3 \, \phi_s
\\
&&\;\;\;\;\;\;\;\;\;\;\;\;\;\;\;\;\;\;\;\;\;\;\;\;\;\;\;\;\;\;\;\;\;\;\;\;\;\;\;\;+ \cos \theta_L \, \sin \theta_R \, \cos 3 \, \phi_s
\Big) \, \sin \phi_{m,\alpha}
\end{eqnarray}
and we have  for strong IO for Case 3 b.1) ($y_2=0$)
\begin{eqnarray}\nonumber
C_{\mathrm{LFV},\alpha}&=&-\frac{2 \, \sqrt{2}}{3} \, M^2 \,  y_1 \, y_3\, \Delta y^2_{13} \, \Delta \sigma_{23} \,  \Big( \sqrt{2} \, \cos\theta_L \, \cos\theta_R \, \cos \phi_{m,\alpha} - \sin\theta_L \, \cos\theta_R  \, \cos 3 \, \phi_s
\\
&&\;\;\;\;\;\;\;\;\;\;\;\;\;\;\;\;\;\;\;\;\;\;\;\;\;\;\;\;\;\;\;\;\;\;\;\;\;\;\;\;\;\;\;\;\;\;\;\;\;\;\;\;\;\;\;\;+ \sqrt{2} \, \sin \theta_L \, \sin \theta_R \, \sin 3 \, \phi_s
\Big) \, \sin \phi_{m,\alpha} \; .
\end{eqnarray}
Furthermore, we can look at the formulae for the electron flavour,  $\alpha=e$, and take into account the phenomenological constraints on $m$, e.g.~for $C_{\mathrm{LFV},e}$ for Case 3 a), $m\approx 0$, we find
\begin{equation}
C_{\mathrm{LFV},e}\approx\frac{\sqrt{2}}{9} \, M^2 \,  y_1 \, y_2\, \Delta y^2_{12} \, [\Delta \sigma_{12}+3\, \Delta \sigma_{23}]  \, \Big( \sin 2 \, \theta_L
+ \sin 2 \, \theta_R \,  \sin 3 \, \phi_s  + 2 \, \sqrt{2}  \, \cos 2 \, \theta_L \, \cos 3 \, \phi_s \Big)
\end{equation}
and for Case 3 b.1), $m\approx \frac{n}{2}$, we arrive at
\begin{eqnarray}\nonumber
C_{\mathrm{LFV},e}&\approx&\frac{\sqrt{2}}{9} \, M^2 \, \Big( - y_1 \, y_2\, \Delta y^2_{12} \, [\Delta \sigma_{12}+3\, \Delta \sigma_{23}] \, \sin 2 \, \theta_L
\\  \nonumber
&&-6 \, \sqrt{2}\, y_3 \,  \Delta \sigma_{23}\, [y_2 \, \Delta y^2_{23} \, \cos \theta_L \, \cos \theta_R +y_1 \, \Delta y^2_{13}\, \sin \theta_L \, \sin \theta_R] \, \sin 3 \, \phi_s
\\
&&+6\, y_3 \, \Delta \sigma_{23} \, [y_1 \, \Delta y^2_{13} \, \sin \theta_L \, \cos \theta_R -y_2 \, \Delta y^2_{23}\, \cos \theta_L \, \sin \theta_R]\, \cos 3 \, \phi_s
\Big) \; .
\end{eqnarray}
Summing $C_{\mathrm{LNV},\alpha}$ over the lepton flavour index $\alpha$ leads to
\begin{equation}
C_{\mathrm{LNV}}= \frac{\sqrt{2}}{3} \, M^2 \,  (\Delta y^2_{12})^2 \, [\Delta \sigma_{12}+ 3\, \Delta \sigma_{23}] \, \sin 4 \, \theta_R \; .
\end{equation}
Eventually, we have for $C_{\mathrm{DEG},\alpha}$
\begin{equation}
\label{eq:CDEGalpha_soddmeven_Case3}
C_{\mathrm{DEG},\alpha} = \frac{\sqrt{2}}{3} \, y_1 \, y_2 \,  (\Delta y^2_{12})^2 \,  \sin 4 \, \theta_R \, \cos \phi_{m,\alpha} \, \sin 3 \, \phi_s \; .
\end{equation}
We note that for strong NO for Case 3 a) ($y_1=0$) as well as for strong IO for Case 3 b.1) ($y_2=0$) this type of CP-violating combinations vanishes.

For $s$ even and $m$ odd, we have similar results, e.g.~for $C_{\mathrm{LNV}}$ we find
\begin{equation}
C_{\mathrm{LNV}}= \sqrt{2} \, M^2 \,  (\Delta y^2_{12})^2 \, \Delta \sigma_{23}\, \sin 4 \, \theta_R
\end{equation}
and for $C_{\mathrm{DEG},\alpha}$ the same as in Eq.~(\ref{eq:CDEGalpha_soddmeven_Case3}) up to an overall sign.

\vspace{0.2in}
\noindent In the limit of large splitting $\kappa$ (and $\lambda=0$) we find that the
reduced mass-degenerate CP-violating combination $C^{(23)}_{\mathrm{DEG},\alpha}$
is zero for three of the four possible combinations of $s$ and $m$, namely for
$s$ and $m$ both even or both odd as well as for $s$ odd and $m$ even. Only
for $s$ even and $m$ odd we have a non-vanishing result which reads
\begin{eqnarray}\nonumber
\!\!\! \!\!\! \!\!\!C^{(23)}_{\mathrm{DEG},\alpha} &=&\!\!\!
\frac{2}{27} \, \Delta y_{12}^2 \, \Big(2 \, \Delta y_{23}^2+\Delta y_{12}^2 \, (1+ 5 \, \cos 2 \, \theta_R) \Big)
\\
&&\!\!\! \!\!\! \!\!\! \!\!\! \!\!\!\Big(y_3 (y_1 \, \sin \theta_L \cos \theta_R + y_2 \, \cos \theta_L \, \sin \theta_R) \, \sin \phi_{m,\alpha} - \sqrt{2} \, y_1 \, y_2 \, \cos \phi_{m,\alpha} \Big) \, \sin 2 \, \theta_R \, \sin 3 \, \phi_s .
\label{eq:CDEGalphareduced_moddseven}
\end{eqnarray}

\vspace{0.2in}
\noindent The flavoured washout parameter $f_\alpha$, shown in Eq.~(\ref{eq:flvw}), is independent of the choice of $s$ and $m$ being even or odd and reads
\begin{equation}
\label{eq:flavouredwashoutratioCase3}
f_\alpha= \frac 13 \, \left( 1 + \left( \frac{\Delta y_{13}^2-\Delta y_{12}^2 \, \sin^2 \theta_L}{\Sigma y^2} \right) \, \cos 2 \, \phi_{m, \alpha} - \sqrt{2} \, \left( \frac{\Delta y_{12}^2}{\Sigma y^2}\right) \, \sin 2 \, \theta_L \, \cos \phi_{m,\alpha} \, \cos 3 \, \phi_s \right) \, .
\end{equation}
Furthermore, we compute the combination $\sum_\alpha C_{\mathrm{LFV},\alpha} \, f_\alpha$ for $s$ and $m$ both even. Since the result turns out to be rather lengthy, we only display it for $\lambda=0$ and $m=\frac{n}{2}$, as preferred in Case 3 b.1). Employing Eqs.~(\ref{eq:CLFValphasmevenCase3}) and (\ref{eq:flavouredwashoutratioCase3}), we have
\begin{equation}
\label{eq:sumalphaCLFValphafalphaCase3}
\sum_\alpha C_{\mathrm{LFV},\alpha} \, f_\alpha = - M^2 \left( \frac{y_1 \, y_2 \, (\Delta y_{12}^2)^2}{3 \, \Sigma y^2} \right) \, \kappa \, (2 + \kappa) \, \sin 2 \, \theta_L \, tcs_{R,-}  \, tcs_{R,+} \, \sin 6 \, \phi_s \, .
\end{equation}
As $C_{\mathrm{LNV},\alpha}=-C_{\mathrm{LFV},\alpha}$ for this choice of $s$ and $m$, the result for the combination $\sum_\alpha C_{\mathrm{LNV},\alpha} \, f_\alpha$ is the same up to an overall sign.
Likewise, we can consider the combination $\sum_\alpha C_{\mathrm{DEG},\alpha} \, f_\alpha$ for $s$ odd and $m$ even and find
\begin{eqnarray}\nonumber
\!\!\! \!\!\! \!\!\! \!\!\! \!\!\! \!\! \sum_\alpha C_{\mathrm{DEG},\alpha} \, f_\alpha &=&\frac{1}{3 \, \sqrt{2}} \, y_1 \, y_2 \,  (\Delta y^2_{12})^2 \,
\left(  \left( \frac{\Delta y_{13}^2-\Delta y_{12}^2 \, \sin^2 \theta_L}{\Sigma y^2} \right) \, \cos 3 \, \phi_{m}
\right.
\\
&&\left.\;\;\;\;\;\;\;\;\;\;\;\;\;\;\;\;\;\;\;\;\;\;\;\;\;\;\;\;\;\;\;\;\;\;\;\;- \sqrt{2} \, \left( \frac{\Delta y_{12}^2}{\Sigma y^2}\right) \, \sin 2 \, \theta_L \, \cos 3 \, \phi_s \right) \,  \sin 4 \, \theta_R \, \sin 3 \, \phi_s \, ,
\label{eq:sumalphaCDEGalphafalphaCase3}
\end{eqnarray}
using the expressions given in Eqs.~(\ref{eq:CDEGalpha_soddmeven_Case3}) and (\ref{eq:flavouredwashoutratioCase3}). Fixing $m=\frac{n}{2}$, see Case 3 b.1), we arrive at
\begin{equation}
\label{eq:sumalphaCDEGalphafalphamn2Case3}
\sum_\alpha C_{\mathrm{DEG},\alpha} \, f_\alpha = - \left( \frac{y_1 \, y_2 \, (\Delta y_{12}^2)^3}{6 \, \Sigma y^2} \right) \, \sin 2 \, \theta_L \, \sin 4 \, \theta_R \, \sin 6 \, \phi_s \, .
\end{equation}
For $s$ even and $m$ odd we find the same result up to an overall sign.

\section{Enhancement of residual symmetries}
\label{sec4}

For particular values of $\theta_L$ and $\theta_R$, the residual symmetry $G_\nu=Z_2 \times CP$, preserved by the neutrino Yukawa coupling matrix $Y_D$, can be enhanced.
Such values are interesting, since a small deviation from these can be related with different phenomenological aspects, e.g.~in the case of $\theta_L$ it can be the smallness of the reactor  mixing
angle $\theta_{13}$, see Eqs.~(\ref{eq:thetaLCase1}) and (\ref{eq:mixinganglesCase1}) and the detailed discussion in~\cite{Hagedorn:2014wha}, while for $\theta_R$ this can be the possibility
to have large couplings $y_f$, see Eqs.~(\ref{eq:strongNOCase1}) and (\ref{eq:strongIOCase1}).

In the following, we focus on the possible enhancement of the flavour symmetry. We consider the combinations $Y_D Y_D^\dagger$ and $Y_D^\dagger Y_D$ in order to analyse enhanced
symmetries due to particular values of $\theta_L$ and $\theta_R$, respectively. We constrain ourselves to the search of enhanced symmetries that are part of the flavour symmetry $G_f$
and leave aside the possibility that these enhanced symmetries are emergent/accidental. We also note that we usually expect larger symmetries to be preserved by the combination
$Y_D^\dagger Y_D$ at particular values of $\theta_R$ than by $Y_D Y_D^\dagger$ at particular values of $\theta_L$, since RH neutrinos transform as the real representation ${\bf 3^\prime}$ of $G_f$ that is unfaithful (for $n > 2$). The latter  can be seen from the fact that e.g.~the elements $c^2$ and $d^2$ are represented by the identity matrix in the representation ${\bf 3^\prime}$, compare Eq.~(\ref{abcd3prime}) in appendix~\ref{appA}. As a
consequence, the symmetry $Z_n \times Z_n$ comprised in $\Delta (6 \, n^2)$ acts as a Klein group in the case of the representation ${\bf 3^\prime}$ such that only 24 distinct representation matrices
can be obtained in total in this representation. When discussing enhanced symmetries for $Y_D^\dagger Y_D$, we only refer to the representation matrices in ${\bf 3^\prime}$ and the enhanced symmetry
which is generated by these.

We discuss the enhanced symmetries for Case 1) and Case 2) together as well as those for Case 3 a) and Case 3 b.1).

\subsection{Case 1) and Case 2)}
\label{sec41}

For Case 1) and Case 2) the residual $Z_2$ symmetry, which is part of $G_\nu$, is generated by
$Z=c^{n/2}$, compare Eqs.~(\ref{Z2Case1}) and (\ref{Z2Case2}), and the value of $\theta_L$, preferred by the global fit data on lepton mixing
angles, is close to $\theta_L \approx 0$ or $\theta_L \approx \pi$, compare Eq.~(\ref{eq:thetaLCase1}) and Tab~\ref{tab:Case2n14}.
Interesting values for $\theta_R$ are given by $\sin 2 \, \theta_R=0$, i.e.~$\theta_R=0, \pi/2, \pi, 3 \pi/2$, as well as by $\cos 2 \, \theta_R=0$, i.e.~$\theta_R=\pi/4, 3 \pi/4, 5 \pi/4, 7 \pi/4$,
as can be seen in section~\ref{subsec:case2numass} and section~\ref{subsec:case1numass}, respectively.

If $\theta_L=0, \pi$ and, for Case 2), the parameter $u$ also vanishes (or is a multiple of the index $n$), the combination $Y_D Y_D^\dagger$ becomes invariant under a further $Z_2$ group of $G_f$, which is generated by
\begin{equation}
Z^{L}_{\mbox{\footnotesize{add}}} = a \, b \;\;\; \mbox{and} \;\;\; Z^{L}_{\mbox{\footnotesize{add}}} ({\bf 3})=
\left( \begin{array}{ccc}
1 & 0 & 0\\
0 & 0 & 1\\
0 & 1 & 0
\end{array}
\right)
\, .
\end{equation}
The $Z_2$ symmetry associated with the generator $ Z^{L}_{\mbox{\footnotesize{add}}}$ forms a Klein group together with the residual $Z_2$ symmetry that is generated by $Z=c^{n/2}$.

Similarly, if $\theta_R=0, \pi/2, \pi, 3 \pi/2$ for Case 1), we find that in addition to the representation matrix
\begin{equation}
\frac 13 \, \left(
\begin{array}{ccc}
-1 & 2 & 2\\
2 & -1 & 2\\
2 & 2 &-1
\end{array}
\right)
\;\;\; \mbox{also} \;\;\; Z^{R}_{\mbox{\footnotesize{add}}} ({\bf 3^\prime})=
\left( \begin{array}{ccc}
1 & 0 & 0\\
0 & 0 & 1\\
0 & 1 & 0
\end{array}
\right)
\end{equation}
leaves the combination $Y_D^\dagger Y_D$ invariant. The symmetry associated with these two representation matrices is a Klein group.
This is also true for Case 2) for the choice $t$ even, where $s$ can be either even or odd.
For Case 2) with $t$ odd and any value of $s$ we have instead that the choice $\theta_R=0, \pi/2, \pi, 3 \pi/2$ leads to the combination $Y_D^\dagger Y_D$ being also invariant under the representation matrix
\begin{equation}
Z^{R}_{\mbox{\footnotesize{add}}} ({\bf 3^\prime})=
\frac 13 \, \left(
\begin{array}{ccc}
-1 & 2 \, \omega & 2 \, \omega^2\\
2 \, \omega & 2 \, \omega^2 & -1\\
2 \, \omega^2 & -1 & 2 \, \omega
\end{array}
\right)
\end{equation}
that corresponds to a $Z_4$ symmetry. Note that the square of this matrix is the representation matrix associated with the element $Z=c^{n/2}$.
These representation matrices and consequently the $Z_4$ symmetry are also preserved by the combination $Y_D^\dagger Y_D$
for $\theta_R=\pi/4, 3 \pi/4, 5 \pi/4, 7 \pi/4$ for Case 1), independently of the choice of $s$.
For Case 2) and for all possible choices of $s$ and $t$, we have that for $\theta_R=\pi/4, 3 \pi/4, 5 \pi/4, 7 \pi/4$ the representation matrix
\begin{equation}
	Z^{R}_{\mbox{\footnotesize{add}}} ({\bf 3^\prime})=
	\frac 13 \, \left(
		\begin{array}{ccc}
			-1 & 2 \, \omega & 2 \, \omega^2\\
			2 \, \omega^2 & -1 & 2 \, \omega\\
			2 \, \omega & 2 \, \omega^2 & -1
		\end{array}
	\right)
\end{equation}
also leaves the combination $Y_D^\dagger Y_D$ unchanged. This matrix together with the representation matrix for $Z=c^{n/2}$ gives rise to a Klein group.

In summary, the enhanced symmetries of the combinations $Y_D Y_D^\dagger$ and $Y_D^\dagger Y_D$ achieved for particular values of the angles $\theta_L$ and $\theta_R$, respectively,
do not depend on the choice of $s$ for Case 1), while for Case 2) we have to distinguish, at least for the combination $Y_D^\dagger Y_D$, between $t$ being even and $t$ being odd.

\subsection{Case 3 a) and Case 3 b.1)}
\label{sec42}

For Case 3 a) we have $m \approx 0$ ($m \approx n$) and $\theta_L \approx 0, \pi$, while for Case 3 b.1) it is $m \approx \frac n2$ and $\theta_L \approx \pi/2$, compare Tabs.~\ref{tab:Case3an16}, \ref{tab:Case3b1n8}, \ref{tab:Case3an34seven}-\ref{tab:Case3b1n20m10}.

We first discuss the results for the combination $Y_D Y_D^\dagger$.
For all choices of $s$ and independently of whether $m$ is even or odd, we find for Case 3 a) that additionally to the $Z_2$ symmetry generated by $Z = b$ in the limit of vanishing $m$ (or $m$ being the index $n$)
there is a $Z_2$ symmetry coming from
\begin{equation}
Z^{L}_{\mbox{\footnotesize{add}}} = c^{n/2} \, d^{n/2} \;\;\; \mbox{and} \;\;\; Z^{L}_{\mbox{\footnotesize{add}}} ({\bf 3})=
\frac 13 \, \left(
\begin{array}{ccc}
-1 & 2 \, \omega^2 & 2 \, \omega\\
2 \, \omega & -1 & 2 \, \omega^2\\
2 \, \omega^2 & 2 \, \omega & -1
\end{array}
\right)
\end{equation}
for $\theta_L=0, \pi$. These two symmetries give rise to a Klein group. Similarly, for all choices of $s$ and both types of $m$, $m$ even and $m$ odd,
we have for Case 3 b.1) that, beyond the $Z_2$ symmetry due to $Z=b \, c^{n/2} \, d^{n/2}$ in the limit of $m=n/2$, the $Z_2$ symmetry corresponding to the generator
\begin{equation}
Z^{L}_{\mbox{\footnotesize{add}}} = b \;\;\; \mbox{and} \;\;\; Z^{L}_{\mbox{\footnotesize{add}}} ({\bf 3})=
\left( \begin{array}{ccc}
1 & 0 & 0\\
0 & 0 & \omega^2\\
0 & \omega & 0
\end{array}
\right)
\end{equation}
leaves the combination $Y_D Y_D^\dagger$ for $\theta_L=\pi/2$ invariant. Checking the symmetry due to these two generators, we thus arrive at the same Klein group
as for Case 3 a).

For the combination $Y_D^\dagger Y_D$ we first note that it does not depend explicitly on the parameter $m$. Furthermore, from sections~\ref{subsec:case3numass} and~\ref{subsec:case3numass2} we know that the particular values for $\theta_R$ are $\theta_R=0, \pi/2, \pi, 3 \pi/2$ due to $\sin 2 \, \theta_R=0$ 
 as well as $\theta_R=\pi/4, 3 \pi/4, 5 \pi/4, 7 \pi/4$ due to $\cos 2 \, \theta_R=0$.
So, the enhanced symmetries for  $Y_D^\dagger Y_D$ at the particular  values of $\theta_R$
are the same for Case 3 a) and Case 3 b.1). We only need to distinguish between $m$ being even and $m$ being odd.
For $m$ even and all possible choices of $s$ we have that the $Z_2$ symmetry generated by the representation matrix
\begin{equation}
\left( \begin{array}{ccc}
1 & 0 & 0\\
0 & 0 & \omega^2\\
0 & \omega & 0
\end{array}
\right)
\end{equation}
in ${\bf 3^\prime}$ is part of $G_\nu$ and additionally for $\theta_R=0, \pi/2, \pi, 3 \pi/2$
the representation matrix
\begin{equation}
Z^{R}_{\mbox{\footnotesize{add}}} ({\bf 3^\prime})=
\frac 13 \, \left(
\begin{array}{ccc}
-1 & 2 \, \omega^2 & 2 \, \omega\\
2 \, \omega & -1 & 2 \, \omega^2\\
2 \, \omega^2 & 2 \, \omega & -1
\end{array}
\right)
\end{equation}
leaves the combination $Y_D^\dagger Y_D$ invariant. This representation matrix corresponds to the generator of a $Z_2$ symmetry and leads together with the
already present $Z_2$ symmetry to a Klein group.
For $m$ odd, we arrive at the same Klein group as enhanced symmetry. However, the symmetry comprised in $G_\nu$ is given by the representation matrix
\begin{equation}
\frac 13 \, \left(
\begin{array}{ccc}
-1 & 2 \, \omega^2 & 2 \, \omega\\
2 \, \omega & 2 & -\omega^2\\
2 \, \omega^2 & -\omega & 2
\end{array}
\right)
\end{equation}
in ${\bf 3^\prime}$, while the representation matrix
\begin{equation}
Z^{R}_{\mbox{\footnotesize{add}}} ({\bf 3^\prime})=
\left( \begin{array}{ccc}
1 & 0 & 0\\
0 & 0 & \omega^2\\
0 & \omega & 0
\end{array}
\right)
\end{equation}
is only a symmetry of $Y_D^\dagger Y_D$ for $\theta_R=0, \pi/2, \pi, 3 \pi/2$.
For the particular values $\theta_R=\pi/4, 3 \pi/4, 5 \pi/4, 7 \pi/4$ we find that there is no (non-trivial) representation matrix beyond the one associated with the generator of the
$Z_2$ symmetry, contained in $G_\nu$, leaving the combination  $Y_D^\dagger Y_D$ invariant.

In summary, we always find that the enhanced symmetry for the combination $Y_D Y_D^\dagger$ is a Klein group, generated by the elements $b$ and $c^{n/2} \, d^{n/2}$.
For the combination $Y_D^\dagger Y_D$ we also always find a Klein group for $\theta_R=0, \pi/2, \pi, 3 \pi/2$ and no enhancement of symmetry for
$\theta_R=\pi/4, 3 \pi/4, 5 \pi/4, 7 \pi/4$.

\vspace{0.2in}
\noindent We expect that the deviation of $\theta_R$ from one of its particular values is smaller than the one of $\theta_L$ from one of the identified particular values, since the symmetry of the combination $Y_D^\dagger Y_D$
is usually larger than the one of  $Y_D Y_D^\dagger$, simply because the representation ${\bf 3^\prime}$ is unfaithful for all groups $\Delta (6 \, n^2)$ with $n> 2$.

In one class of explicit models~\cite{Ishimori:2010au,King:2013eh,Feruglio:2019ybq},
the flavour (and CP) symmetries are spontaneously broken to the residual symmetries $G_\nu$ and $G_l$
with the help of flavour (and CP) symmetry breaking fields and a peculiar alignment of their VEVs, achieved with a potential of a particular form.
Depending on the choice of these fields and the form of their VEVs, a symmetry larger than $G_\nu$ and/or $G_l$ can be preserved by the mass matrices at leading order.
Contributions from higher-dimensional operators then induce small deviations so that only $G_\nu$ and $G_l$ remain intact, thus explaining the sizes of the angles $\theta_L$ and $\theta_R$.
An example can be found in \cite{Feruglio:2013hia},
where the correct size of $\theta_L$ and, consequently, of the reactor mixing angle $\theta_{13}$ is justified in this way.

\section{Summary and outlook}
\label{summ}

We have considered the low-scale type-I seesaw framework endowed with a flavour symmetry, belonging to the series of groups $\Delta (3 \, n^2)$ and $\Delta (6 \, n^2)$, $n\geq 2$, and a CP symmetry.
These symmetries are broken to different residual groups $G_l$ and $G_\nu$ in the charged lepton and neutrino sector, respectively, which 
strongly constrain the form of the mass matrices in these two sectors.
Since the Majorana mass matrix of the three RH neutrinos does not break the flavour and CP symmetry, they are (almost) degenerate in mass in this scenario. At the same time, the Dirac neutrino mass 
 matrix carries all non-trivial flavour structure  in the neutrino sector, fixed by $G_\nu$. For a given flavour and CP symmetry and residual groups $G_l$ and $G_\nu$, this matrix depends on five real parameters, see also~\cite{Hagedorn:2016lva},  that determine the light neutrino mass spectrum and give 
rise to one free (effective) parameter in the PMNS mixing matrix. They are, together with possible small splittings among the RH neutrino masses due to
the breaking of the flavour and CP symmetry, relevant for the generation of the BAU through low-scale leptogenesis. 

We have scrutinised the generation of the BAU and its dependence on the parameters, contained in the Dirac neutrino mass
 matrix, and on the RH neutrino masses, i.e.~their mass scale $M$ and the possible splittings $\kappa$ and $\lambda$, both numerically and analytically. In doing so, we have considered the four different viable lepton mixing patterns, 
 called Case 1) through Case 3 b.1),~\cite{Hagedorn:2014wha}. We have chosen  representative examples that allow for a
successful explanation of the measured values of the three lepton mixing angles and, possibly, the indication of the value of the CP phase $\delta$.
By numerically solving the quantum kinetic equations,
we have found that low-scale leptogenesis in both the freeze-in and freeze-out regime remains viable for a wide range of the RH neutrino mass scale $M$ and mixing angles $U^2_\alpha$
for all cases, Case 1) through Case 3 b.1).
Furthermore, we have supplemented the numerical findings with analytic expressions for different CP-violating combinations,  that allow for a qualitative understanding of the numerical results, such as signalling the absence of CP violation that leads to vanishing BAU. Indeed, in this scenario it is possible to reveal the parametric dependence of the BAU, e.g.~its dependence on the choice of the CP symmetry and thus its correlation with the Majorana phase $\alpha$ in Case 1), see Fig.~\ref{NO phi BAU 10 GeV combined}, discussion in section~\ref{sec53} and Eqs.~(\ref{eq:CLFValphaCase1seven}) and (\ref{eq:CLNValphaCase1seven}) in section~\ref{sec62}, and whether or not a non-zero splitting $\kappa$ (and $\lambda$) among the RH neutrino masses is necessary for the successful generation of the BAU, see Tab.~\ref{tab:overview}. This would be intractable in a scenario without flavour and CP symmetries.

In the following, we highlight the most striking results of this analysis. First of all, we find for each case, Case 1) through Case 3 b.1), instances in which it becomes possible to correlate the BAU with one or both Majorana phases $\alpha$ and $\beta$, present in the PMNS mixing matrix.
Secondly, for Case 2), $t$ odd, and Case 3 b.1), $m$ even and $s$ odd or vice versa, it is feasible to generate the BAU without the splittings $\kappa$ and $\lambda$, see Tab.~\ref{tab:overview} as well as the CP-violating combination $C_{\mathrm{DEG},\alpha}$ in section~\ref{sec6}. Interestingly enough, for Case 2), $t$ odd, a non-vanishing lightest neutrino mass $m_0$ is required for the successful generation of the BAU, while the variant with the least number of free parameters is given by Case 3 b.1), $m$ even and $s$ odd or vice versa, since it allows for non-zero BAU even if $m_0$ vanishes, as long as light neutrino masses follow NO.
Thirdly, we have demonstrated that the correct amount of BAU can also be generated in the limit, in which the splitting $\kappa$ is too large for a resonant enhancement, $\kappa \gg \Gamma/M$. This result depends on the case, Case 1) through Case 3 b.1), as well as on the choice of certain group theory parameters, see Tab.~\ref{tab:overview}, since 
the remaining subsystem of two RH neutrinos does not always contain CP violation necessary for  low-scale leptogenesis. 
In the limit of large $\kappa$, the dependence on the exact value of $\kappa$ can be neglected, and we have worked out analytic expressions that can reproduce the main parametric dependence of the BAU, see the CP-violating combination $C^{(23)}_{\mathrm{DEG},\alpha}$ in section~\ref{sec6}.
Finally, we note that for all cases, Case 1) through Case 3 b.1), active-sterile mixing angles $U_{\alpha i}^2$ much larger than those expected from the naive seesaw estimate, $U^2 \gg m_\nu/M$, can be obtained for specific choices of the group theory parameters and the angle $\theta_R$ being close to a special value, see Tab.~\ref{tab:overview}. These choices of parameters can permit, at the same time, the successful generation of the BAU. This observation is very important, since it leaves
open the tantalising possibility to find RH neutrinos at existing and planned direct search experiments~\cite{Atre:2009rg,Deppisch:2015qwa,Cai:2017mow,Agrawal:2021dbo,Abdullahi:2022jlv}.

\begin{table}
\begin{center}
\begin{tabular}{|l|c|c|c|}
\hline
Type of mixing pattern & BAU non-zero & BAU non-zero & Large total mixing\\
& for $\kappa=0$? & for large $\kappa$? & angle $U^2$ possible?\\
\hline\hline
Case 1) & No, see Fig.~\ref{NO kappa BAU Case I different masses}  & Yes, see Fig.~\ref{NO kappa BAU Case I different masses} & Yes, for $\cos 2 \, \theta_R \approx 0$\\[0.01in]
&  &  &  see Fig.~\ref{NO Mass U2 Case1}\\[0.01in]
\hline
Case 2), $t$ even & No, see Fig.~\ref{NO kappa BAU Case II different masses}  & No, see Fig.~\ref{NO kappa BAU Case II different masses}  & No\\[0.01in]
Case 2), $t$ odd & Yes, for $m_0 \neq 0$ & Yes, see Fig.~\ref{NO kappa BAU Case II todd different masses} & Yes, for $\sin 2 \, \theta_R \approx 0$\\[0.01in]
& see Fig.~\ref{NO thetaR/m0 BAU 100 GeV caseII odd no mass splittings}, plot (a) &  & see Fig.~\ref{NO Mass U2 Case2}\\[0.01in]
\hline
Case 3 b.1), $m$ and $s$ even & No, see Fig.~\ref{NO kappa BAU Case IIIb different masses massive lightest neutrino}  & No, see Fig.~\ref{NO kappa BAU Case IIIb different masses massive lightest neutrino}  & No\\[0.01in]
Case 3 b.1), $m$ even, $s$ odd & Yes, see Fig.~\ref{NO kappa BAU Case IIIb s odd different masses massless lightest neutrino}  & No, see Fig.~\ref{NO kappa BAU Case IIIb s odd different masses massless lightest neutrino}  & Yes, for $\cos 2 \, \theta_R \approx 0$\\[0.01in]
& except for strong IO &  & see Fig.~\ref{NO Mass U2 Case3}\\[0.01in]
Case 3 b.1), $m$ odd, $s$ even & Yes, see Fig.~\ref{NOIO BAU kappa caseIIIb odd phi4920} & Yes, see Fig.~\ref{NOIO BAU kappa caseIIIb odd phi4920} & Yes, for $\cos 2 \, \theta_R \approx 0$ \\[0.01in]
& except for strong IO &  & \\[0.01in]
Case 3 b.1), $m$ and $s$ odd & No & No  & No \\[0.01in]
\hline
\end{tabular}
\caption{{\small {\bf Overview over results for the different cases}, related to different mixing patterns for leptons. We highlight whether or not
it is possible to generate a non-zero value of the BAU for vanishing splitting $\kappa$ (and $\lambda$), whether or not we obtain a non-zero value of the BAU
for large values of $\kappa$, $\kappa \gg \Gamma/M$, 
and if it is possible to simultaneously have large total mixing angle $U^2$ and generate the observed amount of BAU.
}}
\label{tab:overview}
\end{center}
\end{table}

It is highly interesting to expand the current study in different directions.
First of all, the collider phenomenology of the three RH neutrinos as well as different LNV processes deserve a dedicated analysis in this scenario. 
While we have focused on a scenario in which the three RH neutrino masses are (almost) degenerate due to the imposed symmetries, it is equally
 interesting to consider another scenario in which the structure of the Dirac neutrino mass matrix is trivial, i.e.~this matrix is flavour-diagonal and flavour-universal,
 and the non-trivial flavour structure is encoded in the Majorana mass matrix of the RH neutrinos which preserves the residual symmetry $G_\nu$, see also~\cite{Hagedorn:2016lva}. As a consequence, the masses of the
 three RH neutrinos are non-degenerate. This is expected to lead to a different parameter space, allowing for the successful generation of the BAU, 
 especially no (further symmetry-breaking) source of small splittings among the RH neutrino masses is needed. 
 Also the collider phenomenology of the RH neutrinos as well as results for signals of LNV processes will in general differ from those of the discussed scenario.
 Apart from scenarios with three RH neutrinos, it is also worth to discuss extensions beyond the studied type-I seesaw framework such as the inverse seesaw mechanism 
 in which typically three more gauge singlet fermions are added, since also this mechanism of neutrino mass generation can be compatible with low-scale leptogenesis. At the same time, 
 it is well-known that inverse seesaw models can offer a variety of phenomenological imprints, such as collider signals, signals in precision flavour
 physics experiments as well as different LNV processes, see e.g.~\cite{Abada:2014kba}. Additionally, in both frameworks, the type-I seesaw and the inverse seesaw, a variation of the number of RH neutrino generations 
 and/or of the gauge singlet fermions can give rise to scenarios with distinct phenomenology, see e.g.~\cite{Abada:2014vea}. 
 Lastly, in view of a possible extension of the SM gauge group it is interesting to consider the effect of new gauge interactions, also involving
 RH neutrinos (and possibly gauge singlet fermions), see e.g.~\cite{Appelquist:2002mw,Pati:1973uk}, on the results for low-scale leptogenesis and their impact on different phenomenological aspects of the mentioned scenarios.

\section*{Acknowledgements}

We thank Garv Chauhan and Bhupal Dev for the exchange about their work~\cite{Chauhan:2021xus} that appeared during the final stage of our work.
Y.G. acknowledges the support of the French Community of Belgium through a FSR grant and the FRIA grant No.~1.E.063.22F. 
C.H. has been partly supported by the European Union's Horizon 2020 research and innovation programme under the Marie Sk\l{}odowska-Curie grant agreement No.~754496 (FELLINI programme) as well as  is supported by Spanish MINECO through the Ram\'o{}n y Cajal programme RYC2018-024529-I, by the national grant PID2020-113644GB-I00 and by the Generalitat Valenciana through PROMETEO/2021/083.
C.H. has also received support from the European Union's Horizon 2020 research and innovation programme under the Marie Sk\l{}odowska-Curie grant agreement No.~860881 (HIDDe$\nu$ network). 
 J.K. acknowledges the support of the Fonds de la Recherche Scientifique - FNRS under Grant No.~4.4512.10. Computational resources have been provided by the supercomputing facilities of the Université catholique de Louvain (CISM/UCL) and the Consortium des Équipements de Calcul Intensif en Fédération Wallonie Bruxelles (CÉCI) funded by the Fonds de la Recherche Scientifique de Belgique (F.R.S.-FNRS) under convention 2.5020.11 and by the Walloon Region.

\appendix

\mathversion{bold}
\section{Group theory of \texorpdfstring{$\Delta (6 \, n^2)$}{Delta (6, n\textasciicircum 2)} and representation matrices}
\mathversion{normal}
\label{appA}

In the following, we briefly describe the discrete groups $\Delta (3 \, n^2)$ and $\Delta (6 \, n^2)$ and some properties of their irreducible representations.
The groups $\Delta (3 \, n^2)$, $n \geq 2$ integer, can be given in terms of three generators $a$, $c$ and $d$ fulfilling the relations
\begin{equation}
a^3=e \; , \;\; c^n =e \; , \;\; d^n =e \; , \;\; c \, d =d \, c\; , \;\; a \, c \, a^{-1} = c^{-1} d^{-1} \; , \;\; a \, d \, a^{-1} = c
\end{equation}
with $e$ being the neutral element of the group~\cite{Luhn:2007uq}. As discussed in~\cite{Escobar:2008vc}, the groups $\Delta (6 \, n^2)$, $n \geq 2$ integer, are obtained
by adding a fourth generator $b$ to the set of $a$, $c$ and $d$. The relations involving $b$ are
\begin{equation}
b^2=e \; , \;\; (a \, b)^2= e \; , \;\; b \, c \, b^{-1} = d^{-1} \; , \;\; b \, d \, b^{-1} = c^{-1} \, .
\end{equation}
In the trivial representation ${\bf 1}$ all elements of the group are represented by the character $1$.
The representation matrices $g ({\bf 3})$ for $a$, $b$, $c$ and $d$ can be chosen in one irreducible, faithful, complex three-dimensional representation ${\bf 3}$
of $\Delta (6 \, n^2)$ as\footnote{The similarity transformation $U= \frac{1}{\sqrt{3}} \, \left(
\begin{array}{ccc}
1 & 1 & 1\\
\omega^2 & \omega & 1\\
\omega & \omega^2 & 1
\end{array}
\right)$ has to be applied to the generators $a ({\bf 3})$, $b ({\bf 3})$, $c ({\bf 3})$ and $d ({\bf 3})$ given in Eq.~(\ref{abcd3}) in order to obtain the form of the representation matrices
as given in~\cite{Escobar:2008vc}.}
\begin{eqnarray}
&&\label{abcd3}
a ({\bf 3}) =  \left( \begin{array}{ccc}
1 & 0 & 0\\
0 & \omega & 0\\
0 & 0 & \omega^2
\end{array}
\right)
\;\; , \;\;
b ({\bf 3}) =  \left( \begin{array}{ccc}
1 & 0 & 0\\
0 & 0 & \omega^2\\
0 & \omega & 0
\end{array}
\right)
\;\; , \\
&&\nonumber
c({\bf 3})= \frac 13 \, \left( \begin{array}{ccc}
1 + 2\cos\phi_n & 1 -\cos\phi_n - \sqrt{3} \sin \phi_n & 1-\cos\phi_n + \sqrt{3} \sin \phi_n \\
1-\cos\phi_n + \sqrt{3} \sin \phi_n &  1 + 2\cos\phi_n & 1 -\cos\phi_n - \sqrt{3} \sin \phi_n\\
1 -\cos\phi_n - \sqrt{3} \sin \phi_n &  1-\cos\phi_n + \sqrt{3} \sin \phi_n & 1 + 2\cos\phi_n
\end{array}
\right)
\end{eqnarray}
with $\omega=e^{2 \pi i/3}$ and $\phi_n = \frac{2 \pi}{n}$. The form of $d$ in ${\bf 3}$ can be computed via $d ({\bf 3})=a({\bf 3})^2 c ({\bf 3}) a ({\bf 3})$.
As mentioned in section~\ref{sec:symmetries}, the representation ${\bf 3}$ corresponds to ${\bf 3}_{\bf{1} \, (1)}$ in the nomenclature of~\cite{Escobar:2008vc}.

The existence of an irreducible, unfaithful, real three-dimensional representation ${\bf 3^\prime}$ requires that all its characters are real.
This cannot be fulfilled in all groups $\Delta (6 \, n^2)$, but only, if the index $n$ is even. In this case the form of the representation matrices of the generators is
\begin{equation}
\label{abcd3prime}
a ({\bf 3^\prime}) = a ({\bf 3}) \;\; , \;\; b ({\bf 3^\prime}) = b ({\bf 3}) \;\; , \;\;
c ({\bf 3^\prime}) = \frac 13 \, \left(
\begin{array}{ccc}
-1 & 2 & 2\\
2 & -1 & 2\\
2 & 2 &-1
\end{array}
\right)
\end{equation}
and $d ({\bf 3^\prime}) = a({\bf 3^\prime})^2 c ({\bf 3^\prime}) a ({\bf 3^\prime})$. Since these do not depend on the index $n$ of the group,
we find the same representation for all groups $\Delta (6 \, n^2)$ with even $n$. We observe that the group generated by the representation matrices $g({\bf 3^\prime})$ has 24 elements
and thus corresponds to the group $\Delta (6 \cdot 2^2)= \Delta (24)$. This group is isomorphic to the permutation group $S_4$. The  representation ${\bf 3^\prime}$ together with the one generated by
the representation matrices $a ({\bf 3^\prime})$, $c ({\bf 3^\prime})$, $d ({\bf 3^\prime})$ and $-b ({\bf 3^\prime})$ (i.e. the representation matrix $b ({\bf 3^\prime})$ acquires an overall sign, see~\cite{Escobar:2008vc}) are the only real three-dimensional representations in a generic group $\Delta (6 \,  n^2)$ with even $n$ and $3 \nmid n$. They are called ${\bf 3}_{\bf{1} \, (n/2)}$ and ${\bf 3}_{\bf{2} \, (n/2)}$ in~\cite{Escobar:2008vc}, respectively.

In order to check that ${\bf 3^\prime}$ and the other representation are the only real three-dimensional representations in a generic group $\Delta (6 \,  n^2)$ with even $n$ and $3 \nmid n$, we inspect the characters of the irreducible
three-dimensional representations. Following~\cite{Escobar:2008vc} we see that the characters
$\chi ({\bf 3_{\rm gen}})$ of a generic irreducible three-dimensional representation ${\bf 3_{\rm gen}}$ for a certain type of classes are given by $(-)\,\eta ^{- \rho \, l}$ with $\eta=e^{2 \pi i/n}$,
$\rho=0, 1, ..., n-1$ ($\rho$ labels this type of class of the group $\Delta (6 n^2)$) and $l=1, 2, ... , n-1$ ($l$ labels the different pairs of three-dimensional representations).
We have to require that all $(-)\,\eta ^{- \rho \, l}$ for a certain representation labelled by $l$ are real. This is ensured, if $\eta^{-l}$ is real for all powers $\rho$,
meaning $\eta^{-l}$ should be real itself. Hence, $2 \, l/n$ must be an integer. With the constraint on $l$, $1 \leq l \leq n-1$, we know that there is a
single solution to $2 \, l/n$ being an integer, namely $l=n/2$, i.~e.~there is a single pair of irreducible three-dimensional representations that are real.
In this case their characters are real for all classes, as can be explicitly checked with the help of the character table, shown in~\cite{Escobar:2008vc}.
Since the group, generated by $a ({\bf 3^\prime})$, $b ({\bf 3^\prime})$, $c ({\bf 3^\prime})$ and $d ({\bf 3^\prime})$ in Eq.~(\ref{abcd3prime}), only has 24 elements, the representation ${\bf 3^\prime}$
is unfaithful for $n>2$.

\mathversion{bold}
\section{Form of representation matrices for residual symmetries}
\mathversion{normal}
\label{appB}

We list the form of the representation matrices in the representations ${\bf 3}$ and ${\bf 3^\prime}$ for the generators of the different residual symmetries,
used in the discussion of Case 1), Case 2) and Case 3 a) and Case 3 b.1).

In all these cases, the residual flavour symmetry in the charged lepton sector involves the generator $a$ which corresponds to the representation matrix
\begin{equation}
a ({\bf 3}) = \left( \begin{array}{ccc}
1 & 0 & 0\\
0 & \omega & 0\\
0 & 0 & \omega^2
\end{array}
\right) \, .
\end{equation}
The residual flavour symmetry in the neutrino sector is generated by $Z$.
In Case 1) and Case 2) $Z$ is chosen as $c^{n/2}$ which is in the representation ${\bf 3}$ of the form
\begin{equation}
\label{eq:Zcn2in3}
Z ({\bf 3}) = \frac 13 \, \left(
\begin{array}{ccc}
-1 & 2 & 2\\
2 & -1 & 2\\
2 & 2 &-1
\end{array}
\right)
\end{equation}
independently of the index $n$, while the form of $Z=c^{n/2}$ in ${\bf 3^\prime}$ reads either
\begin{equation}
\label{eq:Zcn2_n2even}
Z ({\bf 3^\prime}) = \left(
\begin{array}{ccc}
1 & 0 & 0\\
0 & 1 & 0\\
0 & 0 & 1
\end{array}
\right)
\;\;\; \mbox{for} \;\;\; n/2 \;\; \mbox{even}
\end{equation}
or
\begin{equation}
\label{eq:Zcn2_n2odd}
Z ({\bf 3^\prime}) = \frac 13 \, \left(
\begin{array}{ccc}
-1 & 2 & 2\\
2 & -1 & 2\\
2 & 2 &-1
\end{array}
\right)  = Z ({\bf 3})
\;\;\; \mbox{for} \;\;\; n/2 \;\; \mbox{odd.}
\end{equation}

In Case 3 a) and Case 3 b.1) $Z$ is chosen as $b \, c^m d^m$ with $m=0, 1, ..., n-1$.
In the representation ${\bf 3}$ it is of the form
\small
\begin{equation}
\label{eq:Case3Zm3}
Z (m) ({\bf 3}) = \frac 13 \, \left(
\begin{array}{ccc}
1+2 \, \cos \gamma_m & \omega^2 \, \left(1-\cos\gamma_m + \sqrt{3}\, \sin \gamma_m \right) & \omega \, \left( 1- \cos \gamma_m -\sqrt{3} \, \sin \gamma_m \right)\\
\omega \, \left( 1-\cos \gamma_m +\sqrt{3}\, \sin \gamma_m \right) & 1- \cos \gamma_m -\sqrt{3} \, \sin \gamma_m & \omega^2 \, \left( 1+ 2\, \cos \gamma_m \right)\\
\omega^2 \, \left(  1- \cos \gamma_m -\sqrt{3} \, \sin \gamma_m  \right) & \omega \, \left( 1+ 2 \, \cos \gamma_m \right) & 1- \cos \gamma_m +\sqrt{3} \, \sin \gamma_m
\end{array}
\right)
\end{equation}
\normalsize
with $\gamma_m= \frac{2 \, \pi m}{n}$.
For the special values, $m = 0$, $m = n$ and $m=n/2$, the form of $Z (m) ({\bf 3})$ simplifies and we find
\begin{equation}
Z (m=0) ({\bf 3}) = Z (m=n) ({\bf 3}) = \left( \begin{array}{ccc}
1 & 0 & 0\\
0 & 0 & \omega^2\\
0 & \omega & 0
\end{array}
\right)
\end{equation}
and
\begin{equation}
Z (m=n/2) ({\bf 3}) = \frac 13 \, \left( \begin{array}{ccc}
-1 & 2 \, \omega^2 & 2 \, \omega \\
2 \, \omega & 2 & -\omega^2\\
2 \, \omega^2 & -\omega & 2
\end{array}
\right) \; .
\end{equation}
Similarly, we can compute the form of the representation matrix $Z (m) ({\bf 3^\prime})$. The decisive criterion for this form is whether $m$ is even or odd and
otherwise there is no further dependence on the parameter $m$ for $Z (m) ({\bf 3^\prime})$.
So, for $m$ being even we get
\begin{equation}
\label{Z3pmeven}
Z (m \, \mbox{even}) ({\bf 3^\prime}) =
\left( \begin{array}{ccc}
1 & 0 & 0\\
0 & 0 & \omega^2\\
0 & \omega & 0
\end{array}
\right)
\; ,
\end{equation}
while for $m$ odd we have
\begin{equation}
\label{Z3pmodd}
Z (m \, \mbox{odd}) ({\bf 3^\prime}) = \frac 13 \, \left(
\begin{array}{ccc}
-1 & 2 \, \omega^2 & 2 \, \omega\\
2 \, \omega & 2 & - \omega^2\\
2 \, \omega^2 & -\omega & 2
\end{array}
\right)
\; .
\end{equation}
We note that $Z (m \, \mbox{even}) ({\bf 3^\prime})$ coincides with $Z (m=0) ({\bf 3}) = Z (m=n) ({\bf 3})$
as well as $Z (m \, \mbox{odd}) ({\bf 3^\prime})$ coincides with $Z (m=n/2) ({\bf 3})$.

\mathversion{bold}
\section{CP symmetries and form of CP transformations}
\mathversion{normal}
\label{appC}

The CP symmetries correspond to automorphisms of the flavour symmetry $\Delta (6 \, n^2)$, see discussion  in~\cite{Feruglio:2012cw,Holthausen:2012dk,Chen:2014tpa,Grimus:1995zi,Ecker:1983hz,Ecker:1987qp,Neufeld:1987wa,Harrison:2002kp,Grimus:2003yn}.
In the present analysis we employ the ones used in~\cite{Hagedorn:2014wha}. These can be obtained as follows: consider the automorphism
\begin{equation}
\label{eq:XP23auto}
a \;\; \rightarrow \;\; a \;\; , \;\;  c \;\; \rightarrow \;\; c^{-1} \;\; , \;\;  d \;\; \rightarrow \;\; d^{-1} \;\; \mbox{and} \;\; b \;\; \rightarrow \;\; b
\end{equation}
in conjugation with an inner automorphism, see e.g.~above Eq.~(\ref{eq:Xs3Case1}). The automorphism in Eq.~(\ref{eq:XP23auto}) can be represented by $X_0 ({\bf 1})=1$ in the trivial representation ${\bf 1}$
and by the matrix
\begin{equation}
X_0 ({\bf 3}) = X_0 ({\bf 3^\prime}) = \left(
\begin{array}{ccc}
1 & 0 & 0\\
0 & 0 & 1\\
0 & 1 & 0
\end{array}
\right)
\end{equation}
in both three-dimensional representations ${\bf 3}$ and ${\bf 3^\prime}$.

In Case 1) the CP transformation $X (s) ({\bf 3})$ has the explicit form
\begin{equation}
\label{eq:case1Xin3}
X (s) ({\bf 3}) = \frac 13 \, e^{-4 \, i \, \phi_s} \, \left(
\begin{array}{ccc}
1+ 2 \, e^{6 \, i \, \phi_s} &  1 - e^{6 \, i \, \phi_s} & 1 - e^{6 \, i \, \phi_s}\\
1 - e^{6 \, i \, \phi_s} &  1+ 2 \, e^{6 \, i \, \phi_s} & 1 - e^{6 \, i \, \phi_s}\\
1 - e^{6 \, i \, \phi_s} & 1 - e^{6 \, i \, \phi_s} &  1+ 2 \, e^{6 \, i \, \phi_s}
\end{array}
\right)
\end{equation}
and $\phi_s=\frac{\pi \, s}{n}$.
The form of the CP transformation $X (s) ({\bf 3^\prime})$ in the representation ${\bf 3^\prime}$ depends on whether $s$ is even or odd, i.e.
\begin{equation}
\label{eq:case1Xin3prime_seven}
X(s) ({\bf 3^\prime}) = \left( \begin{array}{ccc}
1 & 0 & 0\\
0 & 1 & 0\\
0 & 0 & 1
\end{array}
\right) \;\;\; \mbox{for} \;\;\; s \;\; \mbox{even}
\end{equation}
and
\begin{equation}
\label{eq:case1Xin3prime_sodd}
X(s) ({\bf 3^\prime}) = \frac 13 \left( \begin{array}{ccc}
-1 & 2 & 2\\
2 & -1 & 2\\
2 & 2 &-1
\end{array}
\right) \;\;\; \mbox{for} \;\;\; s \;\; \mbox{odd.}
\end{equation}
We note that $X (s) ({\bf 3^\prime})$ for $s$ even coincides with $X (s=0) ({\bf 3})$ as well as that
$X (s) ({\bf 3^\prime})$ for $s$ odd equals $X (s=n/2) ({\bf 3})$. Note $n/2$ is an integer, since $n$ has to be even.

In Case 2) the form of the CP transformation $X(s,t)  ({\bf 3}) $ in the representation ${\bf 3}$
is more conveniently written in terms of the parameters $u$ and $v$
\begin{eqnarray}\nonumber
&&X (s, t) ({\bf 3}) = X (u, v) ({\bf 3}) \, =\frac 13 \, e^{- 2 \, i \, \phi_v/3} \\ \nonumber
&&\!\!\!\!\!\!\!\!\!\!\!\!\!\!\!\!\!\!
\times \left(
\begin{array}{ccc}
1 + 2 \, e^{i \, \phi_v} \, \cos \phi_u & 1 - e^{i \, \phi_v} \, (\cos \phi_u - \sqrt{3} \, \sin \phi_u) & 1 - e^{i \, \phi_v} \, (\cos \phi_u + \sqrt{3} \, \sin \phi_u)\\
1 - e^{i \, \phi_v} \, (\cos \phi_u - \sqrt{3} \, \sin \phi_u) & 1 - e^{i \, \phi_v} \, (\cos \phi_u + \sqrt{3} \, \sin \phi_u) & 1 + 2 \, e^{i \, \phi_v} \, \cos \phi_u\\
1 - e^{i \, \phi_v} \, (\cos \phi_u + \sqrt{3} \, \sin \phi_u)  & 1 + 2 \, e^{i \, \phi_v} \, \cos \phi_u & 1 - e^{i \, \phi_v} \, (\cos \phi_u - \sqrt{3} \, \sin \phi_u)
\end{array}
\right)
\\&&
\label{eq:case2Xin3}
\end{eqnarray}
with $\phi_u=\frac{\pi \, u}{n}$ and $\phi_v=\frac{\pi \, v}{n}$. For the definition of $u$ and $v$ see Eq.~(\ref{defuv}).
The form of the CP transformation $X (s,t) ({\bf 3^\prime})$ depends on whether $s$ and $t$ are even or odd.
The explicit form of $X (s,t) ({\bf 3^\prime})$, however, does neither contain $s$ nor $t$ as parameter. For $s$ and $t$ even we have
\begin{equation}
\label{eq:case2Xin3prime_steven}
X (s \, \mbox{even},t \, \mbox{even}) ({\bf 3^\prime}) = \left( \begin{array}{ccc}
1 & 0 & 0\\
0 & 0 & 1\\
0 & 1 & 0
\end{array}
\right) \; ,
\end{equation}
for $s$ even and $t$ odd we find
\begin{equation}
\label{eq:case2Xin3prime_seventodd}
X (s \, \mbox{even},t \, \mbox{odd}) ({\bf 3^\prime}) = \frac 13 \, \left( \begin{array}{ccc}
-1 & 2 \, \omega^2 & 2 \, \omega\\
2 \, \omega^2 & 2 \, \omega & -1\\
2 \, \omega & -1 & 2 \, \omega^2
\end{array}
\right) \; ,
\end{equation}
for $s$ odd and $t$ even we have
\begin{equation}
\label{eq:case2Xin3prime_soddteven}
X (s \, \mbox{odd},t \, \mbox{even}) ({\bf 3^\prime}) =  \frac 13 \, \left( \begin{array}{ccc}
-1 & 2 & 2\\
2 & 2 & -1\\
2 & -1 & 2
\end{array}
\right)
\end{equation}
and for $s$ and $t$ odd it is
\begin{equation}
\label{eq:case2Xin3prime_stodd}
X (s \, \mbox{odd},t \, \mbox{odd}) ({\bf 3^\prime})  = \frac 13 \, \left( \begin{array}{ccc}
-1 & 2 \, \omega & 2 \, \omega^2\\
2\, \omega & 2\, \omega^2 & -1\\
2 \, \omega^2  & -1 & 2 \, \omega
\end{array}
\right) \; .
\end{equation}
For completeness, we mention that the form of $X (s \, \mbox{even},t \, \mbox{even}) ({\bf 3^\prime})$ coincides with $X (s=0,t=0) ({\bf 3})$,
$X (s \, \mbox{even},t \, \mbox{odd}) ({\bf 3^\prime}) $ is of the same form as $X (s=0,t=n/2) ({\bf 3})$, while
$X (s \, \mbox{odd},t \, \mbox{even}) ({\bf 3^\prime}) $ agrees with $X (s=n/2,t=0) ({\bf 3})$
and
$X (s \, \mbox{odd},t \, \mbox{odd}) ({\bf 3^\prime}) $ with $X (s=n/2,t=n/2) ({\bf 3})$.
Note $n/2$ is an integer, since $n$ has to be even.

For Case 3 a) and Case 3 b.1) the form of the CP transformation $X (s) ({\bf 3})$ is given as
\begin{equation}
\label{eq:Case3Xs3}
X (s) ({\bf 3}) = \frac 13 \, e^{- i \, \phi_s} \, \left(
\begin{array}{ccc}
3 \, \cos 3 \, \phi_s + i \, \sin 3 \, \phi_s & -2 \, i \, \omega \,  \sin 3 \, \phi_s & -2 \, i \, \omega^2 \, \sin 3 \, \phi_s\\
-2 \, i \, \omega  \, \sin 3 \, \phi_s &    \omega^2 \, \left( 3 \, \cos 3 \, \phi_s + i \, \sin 3 \, \phi_s \right) & -2 \, i \, \sin 3 \, \phi_s\\
-2 \, i \, \omega^2  \, \sin 3 \, \phi_s & -2 \, i  \, \sin 3 \, \phi_s & \omega \, \left( 3 \, \cos 3 \, \phi_s + i \, \sin 3 \, \phi_s \right)
\end{array}
\right)
\end{equation}
with $\phi_s=\frac{\pi \, s}{n}$ and $\omega=e^{\frac{2 \pi i}{3}}$.
The form of the CP transformation $X (s) ({\bf 3^\prime})$ only depends on whether $s$ is even or odd. In particular, we have for $s$ even
\begin{equation}
\label{eq:Case3Xseven3prime}
X (s \, \mbox{even}) ({\bf 3^\prime}) = \left(
\begin{array}{ccc}
1 & 0 & 0\\
0 & \omega^2 & 0\\
0 & 0 & \omega
\end{array}
\right)
\end{equation}
and for $s$ odd
\begin{equation}
\label{eq:Case3Xsodd3prime}
X (s \, \mbox{odd}) ({\bf 3^\prime}) = \frac 13 \, \left(
\begin{array}{ccc}
-1 & 2 \, \omega & 2 \, \omega^2\\
2 \, \omega & - \omega^2 & 2 \\
2 \, \omega^2 & 2 & -\omega
\end{array}
\right) \; .
\end{equation}
We note that $X (s \, \mbox{even}) ({\bf 3^\prime})$ is of the same form as $X (s=0) ({\bf 3}) $, while $X (s \, \mbox{odd}) ({\bf 3^\prime}) $
equals $X (s=n/2) ({\bf 3}) $. Since $n$ has always to be even, also $n/2$ is an integer.

\section{Conventions of PMNS mixing matrix, neutrino masses and experimental data}
\label{appD}

\subsection{Conventions of PMNS mixing matrix}
\label{appD1}

As parametrisation of the PMNS mixing matrix we take
\begin{equation}
\label{eq:UPMNSdef}
U_{\mbox{\scriptsize{PMNS}}} = \tilde{U} (\theta_{ij}, \delta) \, {\rm diag}(1, e^{i \alpha/2}, e^{i (\beta/2 + \delta)})
\end{equation}
with $\tilde{U} (\theta_{ij}, \delta)$ being of the form of the Cabibbo-Kobayashi-Maskawa (CKM) matrix $V_{\mbox{\scriptsize{CKM}}}$ \cite{ParticleDataGroup:2020ssz}
\begin{equation}
\label{eq:UCKM}
\tilde{U} (\theta_{ij}, \delta) =
\begin{pmatrix}
c_{12} c_{13} & s_{12} c_{13} & s_{13} e^{- i \delta} \\
-s_{12} c_{23} - c_{12} s_{23} s_{13} e^{i \delta} & c_{12} c_{23} - s_{12} s_{23} s_{13} e^{i \delta} & s_{23} c_{13} \\
s_{12} s_{23} - c_{12} c_{23} s_{13} e^{i \delta} & -c_{12} s_{23} - s_{12} c_{23} s_{13} e^{i \delta} & c_{23} c_{13}
\end{pmatrix}
\end{equation}
and $s_{ij}=\sin\theta_{ij}$ and $c_{ij}=\cos\theta_{ij}$.  The mixing angles $\theta_{ij}$ range from $0$ to $\pi/2$, while the Majorana phases
$\alpha$, $\beta$ as well as the Dirac phase $\delta$ take values between $0$ and $2 \, \pi$.
Note one of the Majorana phases becomes unphysical, if the lightest neutrino mass $m_0$ vanishes.

\subsection{Neutrino masses}
\label{appD2}

For light neutrino masses following NO the three masses $m_i$ are parametrised as
\begin{equation}
\label{eq:massesNO}
m_1= m_0 \;\; , \;\;\; m_2= \sqrt{m_0^2 + \Delta m_{21}^2} \;\; , \;\;\; m_3= \sqrt{m_0^2 + \Delta m_{31}^2}
\end{equation}
with $m_0$ denoting the lightest neutrino mass.

For light neutrino masses with IO the masses $m_i$ are written as
\begin{equation}
\label{eq:massesIO}
m_1= \sqrt{m_0^2 + |\Delta m_{32}^2| - \Delta m_{21}^2 } \, , \;\; m_2= \sqrt{m_0^2 + |\Delta m_{32}^2| } \, ,
\;\; m_3= m_0
\end{equation}
where $m_0$ is the lightest neutrino mass.

\subsection{Experimental data}
\label{appD3}

We summarise the constraints on the lepton mixing parameters $\theta_{ij}$ and $\delta$ and on the mass squared differences from the current global fit performed by the NuFIT collaboration
(NuFIT 5.1, October 2021, without SK atmospheric data)~\cite{Esteban:2020cvm}. For light neutrino masses with NO, the constraints on the lepton mixing parameters read
\begin{eqnarray}\nonumber
&& \sin^2 \theta_{13} = 0.02220^{+ 0.00068} _{-0.00062} \;\; \mbox{and} \;\;\;\;\;\, 0.02034 \leq \sin^2 \theta_{13} \leq 0.02430 \; ,
\\ \nonumber
&& \sin^2 \theta_{12} = 0.304^{+ 0.013} _{-0.012} \;\;\;\;\;\;\;\;  \mbox{and} \;\;\; \;\;\; 0.269 \leq \sin^2 \theta_{12} \leq 0.343 \; ,
\\ \nonumber
&& \sin^2 \theta_{23} = 0.573 ^{+ 0.018} _{-0.023} \;\;\;\;\;\;\;\; \mbox{and} \;\;\;\;\;\; 0.405 \leq \sin^2 \theta_{23} \leq 0.620 \; ,
\\ \label{eq:anglesdeltabfappNO}
&& \delta=3.39^{+0.91}_{-0.44} \;\;\;\;\;\;\; \;\;\;\;\;\;\; \;\;\;\;\; \;\, \mbox{and} \;\;\;\;\;\; 1.83 \leq \delta \leq 7.07
\end{eqnarray}
for the best fit value, $1 \, \sigma$ level and $3 \, \sigma$ range, respectively, and the ones on the mass squared differences $\Delta m_{21}^2$ and $\Delta m_{31}^2$ are
\small
\begin{eqnarray}
\nonumber
\Delta m_{21}^2=m_2^2 -m_1^2 = \left( 7.42 _{-0.20}^{+0.21} \right) \, \times 10^{-5} \; \mathrm{eV}^2 \, &, &
\Delta m_{31}^2=m_3^2-m_1^2 = \left( 2.515 _{-0.028}^{+0.028} \right) \, \times 10^{-3} \; \mathrm{eV}^2 \, ,
\\ \nonumber
6.82 \, \times 10^{-5} \; \mathrm{eV}^2 \leq  &\Delta m_{21}^2& \leq 8.04 \, \times 10^{-5} \; \mathrm{eV}^2 \; ,
\\ \label{eq:masses3sappNO}
2.431 \, \times 10^{-3} \; \mathrm{eV}^2 \leq  &\Delta m_{31}^2& \leq 2.599 \, \times 10^{-3} \; \mathrm{eV}^2
\end{eqnarray}
\normalsize
at the $3 \, \sigma$ level. For light neutrino masses with IO, the constraints on the lepton mixing parameters are
\begin{eqnarray}\nonumber
&& \sin^2 \theta_{13} = 0.02238^{+ 0.00064} _{-0.00062} \;\;\; \mbox{and} \;\;\;\;\;\, 0.02053 \leq \sin^2 \theta_{13} \leq 0.02434 \; ,
\\ \nonumber
&& \sin^2 \theta_{12} = 0.304^{+ 0.012} _{-0.012} \;\;\;\;\;\;\;\;\;\, \mbox{and} \;\;\; \;\;\; 0.269 \leq \sin^2 \theta_{12} \leq 0.343 \; ,
\\ \nonumber
&& \sin^2 \theta_{23} = 0.578 ^{+ 0.017} _{-0.021} \;\;\;\;\;\;\;\;\;\, \mbox{and} \;\;\;\;\;\; 0.410 \leq \sin^2 \theta_{23} \leq 0.623 \; ,
\\ \label{eq:anglesdeltabfappIO}
&& \delta=5.01^{+0.47}_{-0.56}    \;\;\;\;\;\;\;\;\;\;\;\;\;\;\;\;\;\;\;\;\;\;\, \mbox{and} \;\;\;\;\;\; 3.35 \leq \delta \leq 6.30
\end{eqnarray}
for the best fit value, $1 \, \sigma$ level and $3 \, \sigma$ range, respectively, while the ones on the mass squared differences $\Delta m_{21}^2$ and $\Delta m_{32}^2$ read
\small
\begin{eqnarray}
\nonumber
\Delta m_{21}^2=m_2^2 -m_1^2 = \left( 7.42 _{-0.20}^{+0.21} \right) \, \times 10^{-5} \; \mathrm{eV}^2 \, &, &
\Delta m_{32}^2=m_3^2-m_2^2 = \left(  -2.498_{-0.029}^{+0.028} \right) \, \times 10^{-3} \; \mathrm{eV}^2
\\ \nonumber
6.82 \, \times 10^{-5} \; \mathrm{eV}^2 \leq  &\Delta m_{21}^2& \leq 8.04 \, \times 10^{-5} \; \mathrm{eV}^2 \; ,
\\ \label{eq:masses3sappIO}
-2.584 \, \times 10^{-3} \; \mathrm{eV}^2 \leq  &\Delta m_{32}^2& \leq -2.413 \, \times 10^{-3} \; \mathrm{eV}^2
\end{eqnarray}
\normalsize
at the $3 \, \sigma$ level. We note that NO corresponds to the best fit of the global fit, $\Delta \chi^2=\chi^2_\mathrm{IO}-\chi^2_\mathrm{NO}=2.6$.

Furthermore, the sum of the light neutrino masses is bounded by cosmological data (Planck TT,TE,EE+lowE +lensing+BAO)~\cite{Planck:2018vyg}
\begin{equation}
\sum\limits_{i=1}^3 m_i <  0.12\;\text{eV} \;\; \mbox{at the} \;\; 95 \% \; \mbox{C.L.} \; .
\end{equation}
We thus take for the lightest neutrino mass $m_0$ as upper bound
\begin{equation}
\label{eq:m0bound}
m_0 \lesssim  0.03 \;\text{eV}\;\; \mbox{for NO and} \;\; m_0 \lesssim  0.015 \;\text{eV}\;\; \mbox{for IO.}
\end{equation}

\section{Further tables for Case 3 a) and Case 3 b.1)}
\label{appE}

This appendix comprises tables with results for the lepton mixing angles for Case 3 a), $n=34$ and $m=2$ and all possible values of $s$, and for Case 3 b.1), $n=20$, $m=9$ and $m=10$ as well as all allowed values of $s$.

\begin{table}[h]
\begin{center}
\begin{tabular}{c}
$
\begin{array}{|l||c|c|c|c|c|c|c|c|c|}
\hline
\;\;\;\; s & 0 & 2 & 4 & 6 & 8 & 10 & 12 & 14 & 16\\
\hline
\;\;\;\; \theta_L & \ba1.93\\ \left[3.10\right]\ea & \ba 1.99\\\left[3.10\right]\ea & \ba 2.31 \\\left[3.05\right]\ea & 0.134 & \ba0.0656\\\left[0.997\right]\ea & \ba0.0413\\\left[1.18\right]\ea & \ba0.0392\\\left[1.20\right]\ea & \ba0.0528\\\left[1.09\right]\ea & \ba0.184\\\left[0.501\right]\ea \\
\hline
\;  \sin^{2}\theta_{12} & 0.304 & 0.304 & 0.304 & 0.335 & 0.304 & 0.304 & 0.304 & 0.304 & 0.304\\
\hline
\;  \Delta\chi^2_{12} & \sim 0 & \sim 0 & \sim 0 & 5.84 & \sim 0 & \sim 0 & \sim 0 & \sim 0 & \sim 0 \\
\hline
\end{array}
$
\end{tabular}
\\[0.1in]
\begin{tabular}{c}
$
\begin{array}{|l||c|c|c|c|c|c|c|c|}
\hline
\;\;\;\; s & 18 & 20 & 22 & 24 & 26 & 28 & 30 & 32\\
\hline
\;\;\;\; \theta_L
& \ba2.64\\ \left[2.96\right]\ea  & \ba2.05\\\left[3.09\right]\ea  & \ba1.94\\\left[3.10\right]\ea  & \ba1.96\\\left[3.10\right]\ea  & \ba2.14\\\left[3.08\right]\ea & 3.01 & \ba0.0921\\\left[0.83\right]\ea & \ba0.0456\\\left[1.15\right]\ea\\
\hline
\;  \sin^{2}\theta_{12} & 0.304 &  0.304 & 0.304 &  0.304 & 0.304 &  0.335 & 0.304 & 0.304\\
\hline
\;  \Delta\chi^2_{12} & \sim 0 & \sim 0 & \sim 0 & \sim 0 & \sim 0 & 5.84 & \sim 0 & \sim 0 \\
\hline
\end{array}
$
\end{tabular}
\caption{{\small  \textbf{Case 3 a)} Results for $n=34$, $m=2$ and $s$ even. The lepton mixing angles $\theta_{13}$ and $\theta_{23}$ are fixed to $\sin^2 \theta_{13} \approx 0.0225$
and $\sin^2 \theta_{23} \approx 0.607$.
We display $\theta_L$, $\sin^2 \theta_{12}$ and $\Delta \chi^2_{12}$ for light neutrino masses with NO. We write $\sim 0$ when $\Delta \chi^2_{12}<10^{-3}$. The presence of a second best fit value for $\theta_L$ is indicated in square brackets. These results can also be found in~\cite{Hagedorn:2014wha}, see Case 3 a) with $n=17$ and $m=1$.}
\label{tab:Case3an34seven}
}
\end{center}
\end{table}

\begin{table}
\begin{center}
\begin{tabular}{c}
$
\begin{array}{|l||c|c|c|c|c|c|c|c|c|}
\hline
\;\;\;\; s & 1 & 3 & 5 & 7 & 9 & 11 & 13 & 15 & 17\\
\hline
\;\;\;\; \theta_L & \ba1.95\\ \left[3.10\right]\ea & \ba 2.09\\ \left[3.08\right]\ea & 2.89 & \ba0.119\\ \left[0.703 \right]\ea & \ba0.0487\\ \left[1.12\right]\ea & \ba0.0387\\ \left[1.21\right]\ea & \ba0.0432\\ \left[1.17\right]\ea & \ba0.0761\\ \left[0.927\right]\ea & 0\\
\hline
\;  \sin^{2}\theta_{12} & 0.304 & 0.304 & 0.319 & 0.304 & 0.304 & 0.304 & 0.304 & 0.304 & 0.341\\
\hline
\;  \Delta\chi^2_{12} & \sim 0 & \sim 0 & 1.43 & \sim 0 & \sim 0 & \sim 0 & \sim 0 & \sim 0 & 8.09 \\
\hline
\end{array}
$
\end{tabular}
\\[0.1in]
\begin{tabular}{c}
$
\begin{array}{|l||c|c|c|c|c|c|c|c|}
\hline
\;\;\;\; s & 19 & 21 & 23 & 25 & 27 & 29 & 31 & 33\\
\hline
\;\;\;\; \theta_L & \ba2.21\\ \left[3.07\right]\ea  & \ba1.97\\ \left[3.10\right]\ea  & \ba1.93\\ \left[3.10\right]\ea  & \ba2.02\\ \left[3.09\right]\ea  & \ba2.44\\ \left[3.02\right]\ea & 0.251 & \ba0.0582\\ \left[1.05\right]\ea & \ba0.0401\\ \left[1.20\right]\ea\\
\hline
\;  \sin^{2}\theta_{12} & 0.304 &  0.304 & 0.304 &  0.304 & 0.304 &  0.319 & 0.304 & 0.304\\
\hline
\;  \Delta\chi^2_{12}& \sim 0 & \sim 0 & \sim 0 & \sim 0 & \sim 0 & 1.43 & \sim 0 & \sim 0 \\
\hline
\end{array}
$
\end{tabular}
\caption{{\small  \textbf{Case 3 a)} Results for $n=34$, $m=2$ and $s$ odd. The lepton mixing angles $\theta_{13}$ and $\theta_{23}$ are fixed to $\sin^2 \theta_{13} \approx 0.0225$ and $\sin^2 \theta_{23} \approx 0.607$. We display $\theta_L$, $\sin^2 \theta_{12}$ and $\Delta \chi^2_{12}$ for light neutrino masses with NO. We write $\sim 0$ when $\Delta \chi^2_{12}<10^{-3}$. The presence of a second best fit value for $\theta_L$ is indicated in square brackets.}
\label{tab:Case3an34sodd}
}
\end{center}
\end{table}

\begin{table}
\begin{tabular}{c}
$
\begin{array}{|l||c|c|c|c|c|c|c|c|c|c|c|c|c|c|}
\hline
\;\;\;\; s & 0 & 1 & 2 & 3 & 4 & 5 & 6 & 7 & 8 &  9 \\
\hline
\;\;\;\; \theta_L & 1.53 & 1.53 & 1.51 & 1.46 & 1.65 & 1.62 &  1.61 & 1.61 & 1.62 & 1.64  \\
\hline
\; \sin^{2}\theta_{23} & 0.440 & 0.439 & 0.437 & 0.423& 0.431 & 0.438 &  0.440 & 0.440 & 0.439 & 0.435 \\[0.01in]
\; \sin^{2}\theta_{12} & 0.335 & 0.335 & 0.335 & 0.335 & 0.335 & 0.335 &  0.335 & 0.335 & 0.335 & 0.335 \\[0.01in]
\; \sin^{2}\theta_{13} & 0.0223 & 0.0223 & 0.0223 & 0.0223 & 0.0223 & 0.0223 &  0.0223 & 0.0223 & 0.0223 & 0.0223 \\
\hline
\;  \Delta\chi^2  & 6.73 & 6.78 & 7.01 & 9.04 & 7.73 & 6.89 & 6.75 & 6.74 & 6.82 & 7.23 \\
\hline
\end{array}
$
\end{tabular}
\\[0.1in]
\begin{tabular}{c}
$
\begin{array}{|l||c|c|c|c|c|c|c|c|c|c|c|c|c|c|}
\hline
\;\;\;\; s & 10 & 11 & 12 & 13 & 14 & 15 & 16 & 17 & 18 &  19 \\
\hline
\;\;\;\; \theta_L & 1.43~\left[1.71\right] & 1.50 & 1.53 & 1.53 & 1.53 & 1.52 &  1.49 & 1.68 & 1.63 & 1.61  \\
\hline
\; \sin^{2}\theta_{23} & 0.410 & 0.435 & 0.439 & 0.440 & 0.440 & 0.438 &  0.431 & 0.423 & 0.437 & 0.439 \\[0.01in]
\; \sin^{2}\theta_{12} & 0.335 & 0.335 & 0.335 & 0.335 & 0.335 & 0.335 &  0.335 & 0.335 & 0.335 & 0.335 \\[0.01in]
\; \sin^{2}\theta_{13} &  0.0223 & 0.0223 & 0.0223 & 0.0223 & 0.0223 & 0.0223 &  0.0223 & 0.0223 & 0.0223 & 0.0223 \\
\hline
\;  \Delta\chi^2  & 12.64 & 7.23 & 6.82 & 6.74 & 6.75 & 6.89 & 7.74 & 9.04 & 7.01 & 6.78 \\
\hline
\end{array}
$
\end{tabular}
\caption{{\small  \textbf{Case 3 b.1)} Results for a larger value of $n$, $n=20$, together with $m=9$ and $0 \leq s \leq n-1=19$. The results for $m=11$ can be deduced
from the displayed ones, using the symmetry transformations given in~\cite{Hagedorn:2014wha}. For light neutrino masses with NO we give $\Delta \chi^2$ resulting from the fit to the global fit data~\cite{Esteban:2020cvm}. The presence of a second best fit value for $\theta_L$ is indicated in square brackets.}
\label{tab:Case3b1n20m9}
}
\end{table}

\begin{table}
\begin{center}
\begin{tabular}{c}
$
\begin{array}{|l||c|c|c|c|c|c|}
\hline
\;\;\;\; s & 2 & 3 & 4  & 9 & 10 \\
\hline
\;\;\;\; \theta_L & 1.31 & 1.31~\left[1.83\right] & 1.31~\left[1.83\right]  & 1.31~\left[1.83\right] & 1.31~\left[1.83\right] \\
\hline
\; \sin^{2}\theta_{23} & 0.622 & 0.533~\left[0.467\right] & 0.436~\left[0.564\right]  & 0.406~\left[0.595\right] & 0.5 \\[0.01in]
\; \sin^{2}\theta_{12} & 0.318 & 0.318 & 0.318 & 0.318 & 0.318 \\[0.01in]
\; \sin^{2}\theta_{13} & 0.0220 & 0.0222 & 0.0222  & 0.0221~\left[0.0222\right] & 0.0222 \\
\hline
\;  \Delta\chi^2 & 11.2 & 3.51~\left[2.36\right] & 2.68~\left[1.50\right] & 10.1~\left[2.89\right] & 4.12  \\
\hline
\end{array}
$
\end{tabular}
\\[0.1in]
\begin{tabular}{c}
$
\begin{array}{|l||c|c|c|c|c|}
\hline
\;\;\;\; s & 11 & 16 & 17 &  18 \\
\hline
\;\;\;\; \theta_L & 1.31~\left[1.83\right]  & 1.31~\left[1.83\right] & 1.31~\left[1.83\right] & 1.83 \\
\hline
\; \sin^{2}\theta_{23} & 0.594~\left[0.406\right]  & 0.564~\left[0.436\right] & 0.467~\left[0.533\right] & 0.621\\[0.01in]
\; \sin^{2}\theta_{12} & 0.318  & 0.318 & 0.318 & 0.318\\[0.01in]
\; \sin^{2}\theta_{13} & 0.0222~\left[0.0221\right]  & 0.0222 & 0.0222 & 0.0222 \\
\hline
\;  \Delta\chi^2  & 2.89~\left[10.1\right]  & 1.50~\left[2.68\right] & 2.36~\left[3.51\right] & 11.2 \\
\hline
\end{array}
$
\end{tabular}
\end{center}
\caption{{\small  \textbf{Case 3 b.1)} Results for a larger value of $n$, $n=20$, together with $m=10$ and all values of $s$ that lead to a viable description of the experimental data on lepton mixing angles~\cite{Esteban:2020cvm}. The goodness of this fit is encoded in the value of $\Delta\chi^2$, displayed for light neutrino masses with NO. The presence of a second best fit value for $\theta_L$ is indicated in square brackets.}
\label{tab:Case3b1n20m10}
}
\end{table}

\newpage

\section{Supplementary plots}
\label{appF}

In this appendix, we collect further plots that supplement the discussion on low-scale leptogenesis and might be of interest for the reader.

\subsection{Case 1)}
\label{appF1}

\begin{figure}
	\begin{subfigure}{.5\textwidth}
		\centering
		\includegraphics[width = 1.05\textwidth]{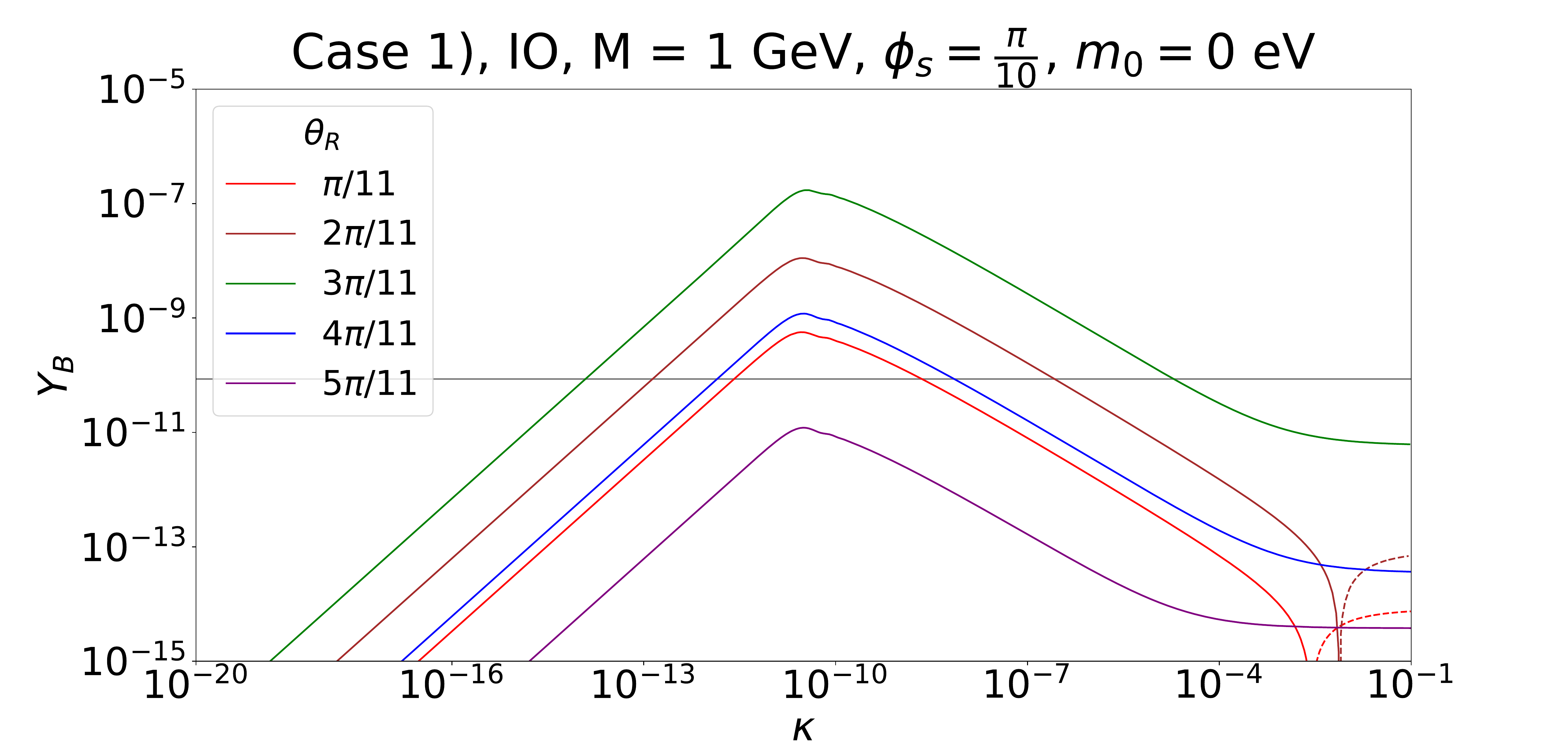}
		\captionof{figure}{Vanishing initial conditions.}
	\end{subfigure}
	\begin{subfigure}{.5\textwidth}
		\centering
		\includegraphics[width = 1.05\textwidth]{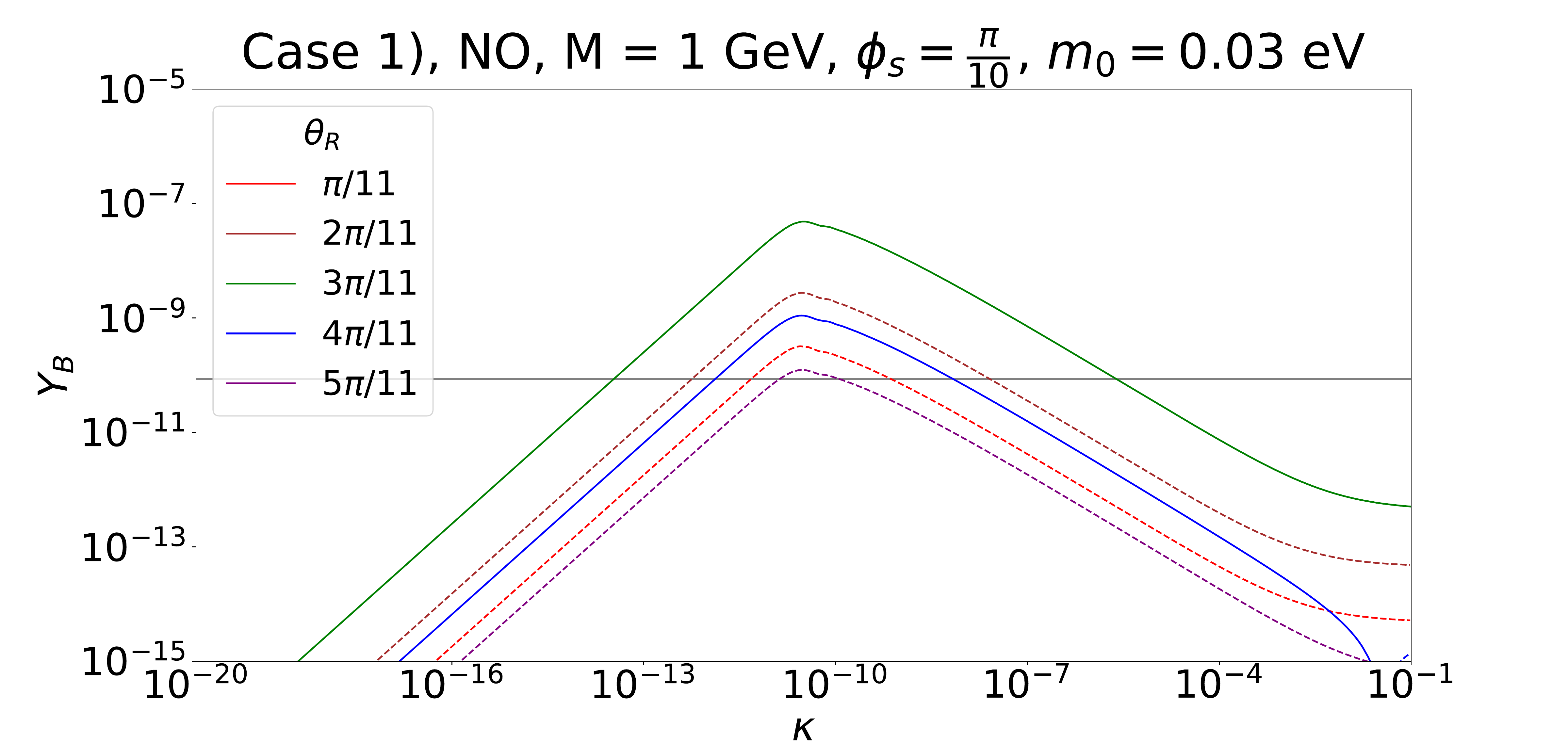}
		\caption{Vanishing initial conditions.}
	\end{subfigure}
	\begin{subfigure}{.5\textwidth}
		\includegraphics[width = 1.05\textwidth]{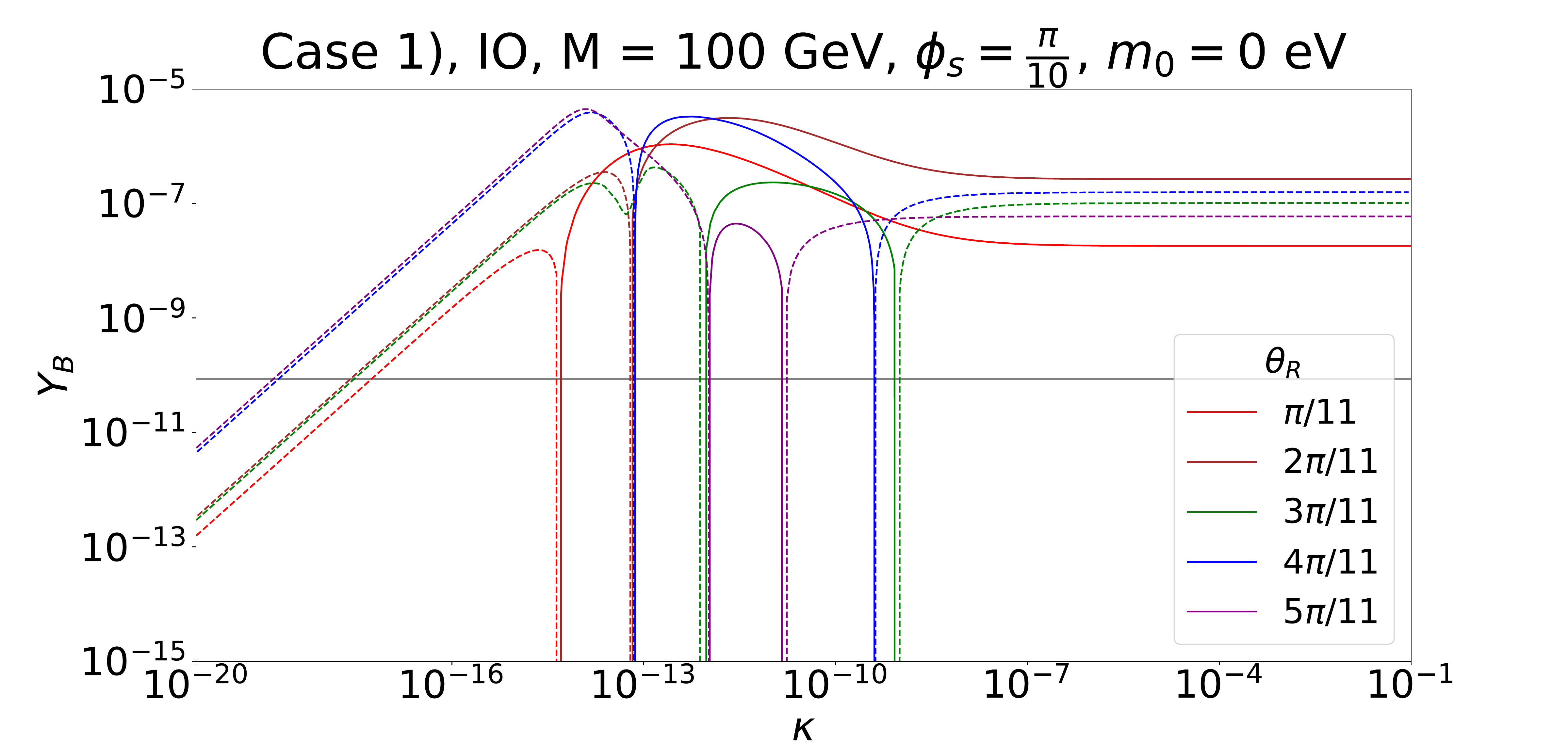}
		\caption{Vanishing initial conditions.}
	\end{subfigure}
	\begin{subfigure}{.5\textwidth}
		\includegraphics[width = 1.05\textwidth]{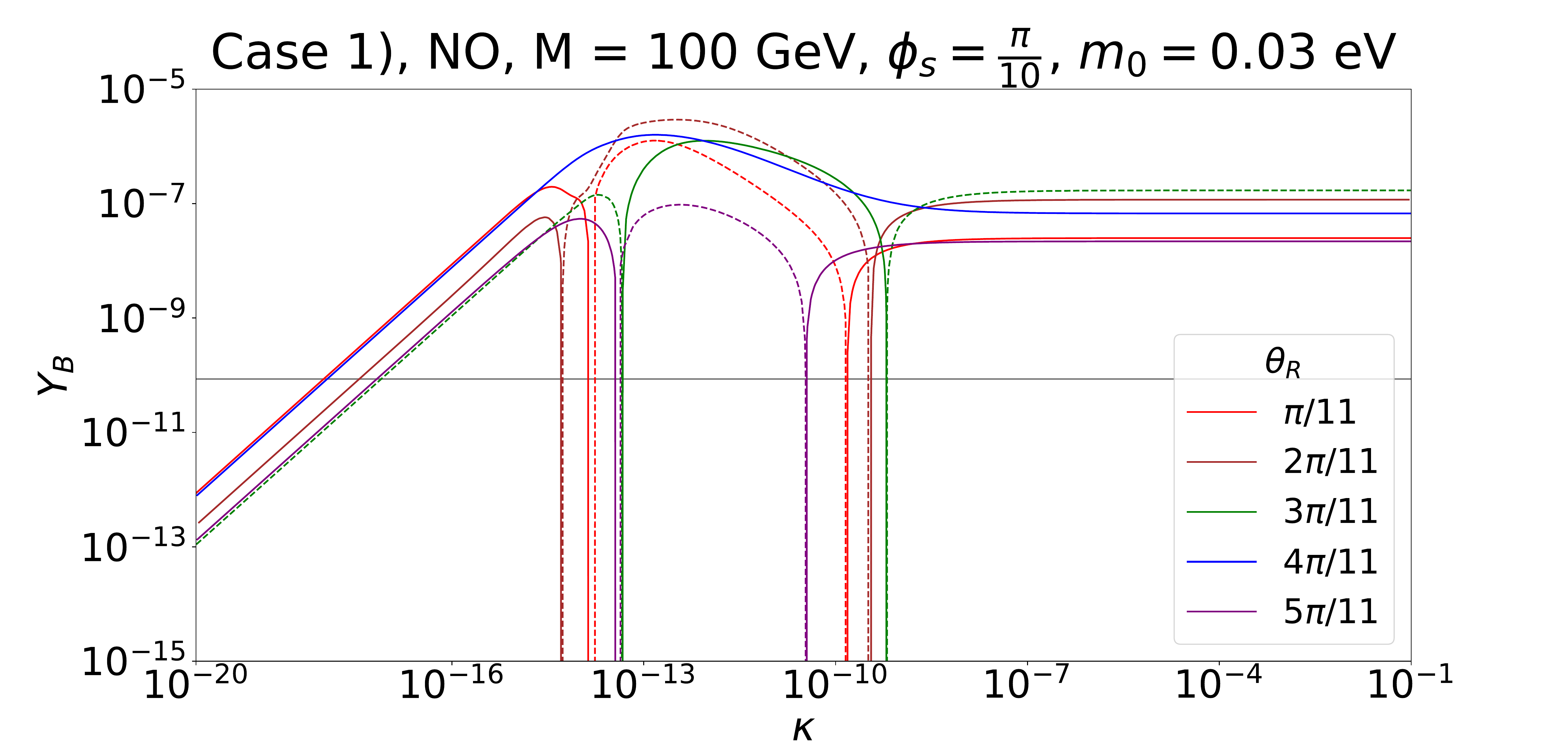}
		\caption{Vanishing initial conditions.}
	\end{subfigure}
	\begin{subfigure}{.5\textwidth}
		\includegraphics[width = 1.05\textwidth]{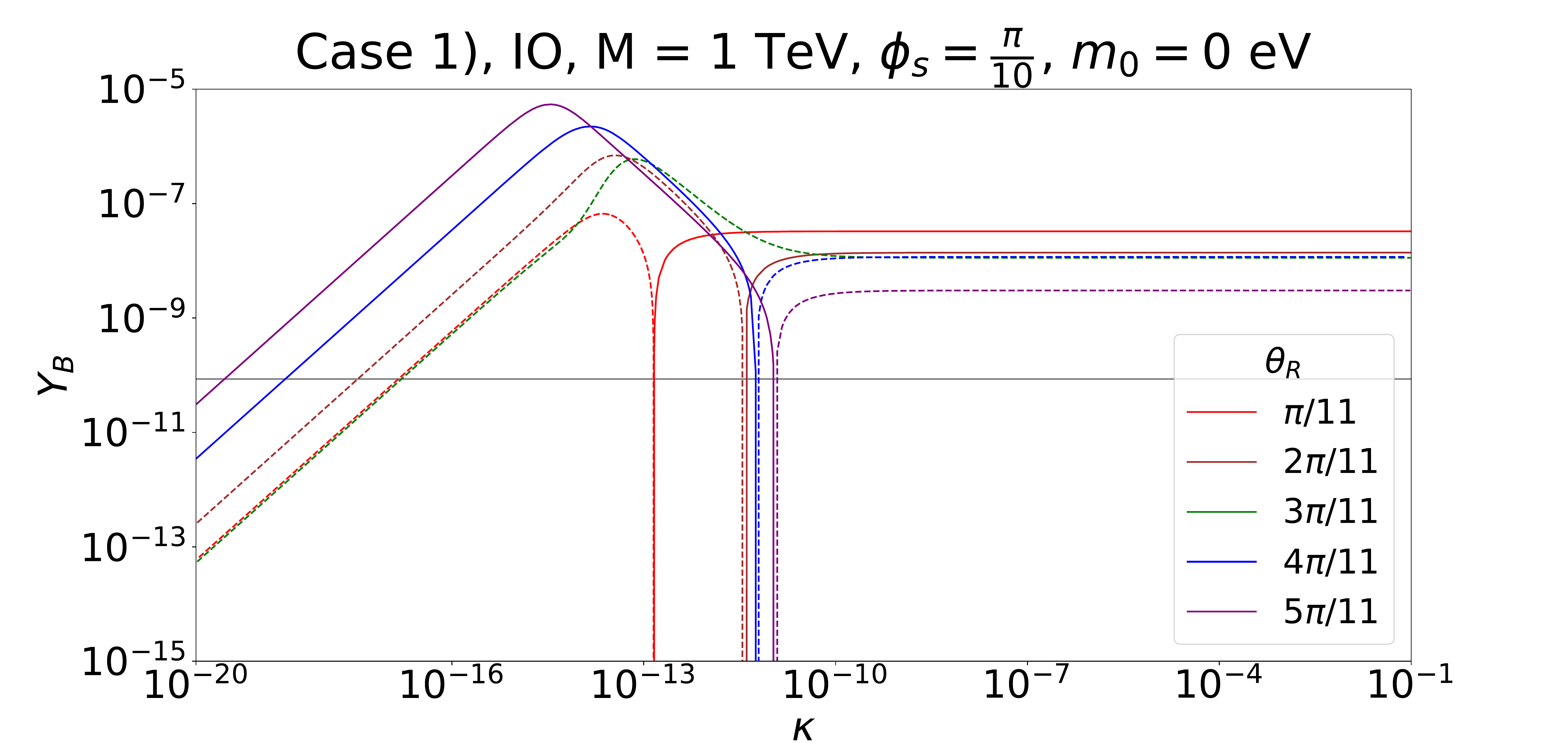}
		\caption{Vanishing initial conditions.}
	\end{subfigure}
	\begin{subfigure}{.5\textwidth}
		\includegraphics[width = 1.05\textwidth]{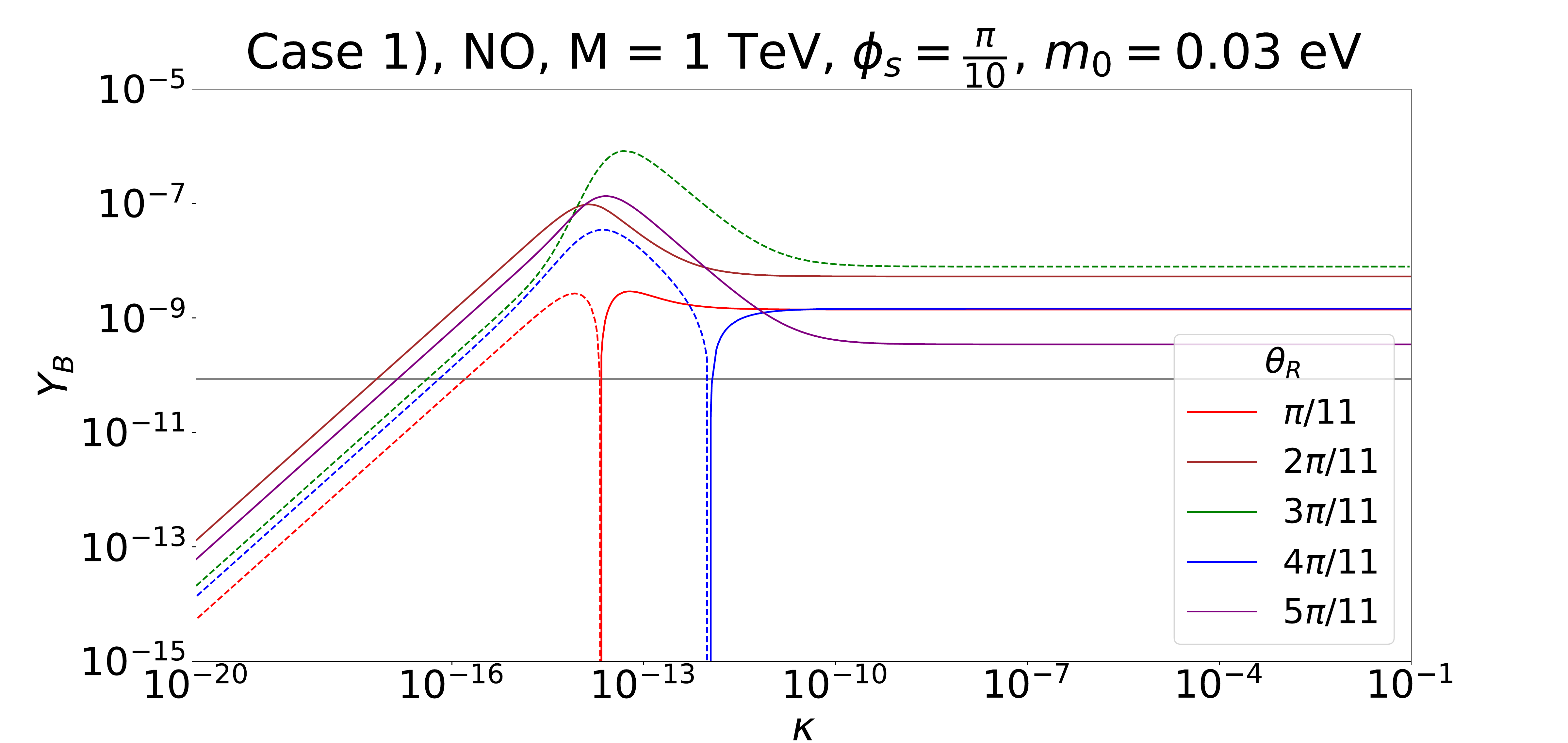}
		\caption{Vanishing initial conditions.}
	\end{subfigure}
\caption{{\small {\bf Case 1)} $Y_B$ as function of $\kappa$ for a Majorana mass $M=1$ GeV, $M=100$ GeV and $M=1$ TeV for light neutrino masses following strong IO (left plots) and NO with $m_0=0.03$ eV (right plots). The value of $s$ is fixed to $s=1$ and thus $\phi_s=\frac{\pi}{10}$. Results for different values of the angle $\theta_R$ are displayed.
 Both negative (dashed lines) as well as positive (continuous lines) values of the BAU are represented. The grey line indicates the observed value of the BAU, $Y_B\approx 8.6 \cdot 10^{-11}$.}}
\label{kappa BAU Case I different masses IO/Massive m0}
\end{figure}

\begin{figure}
	\begin{subfigure}{.5\textwidth}
		\centering
		\includegraphics[width = 1.1\textwidth]{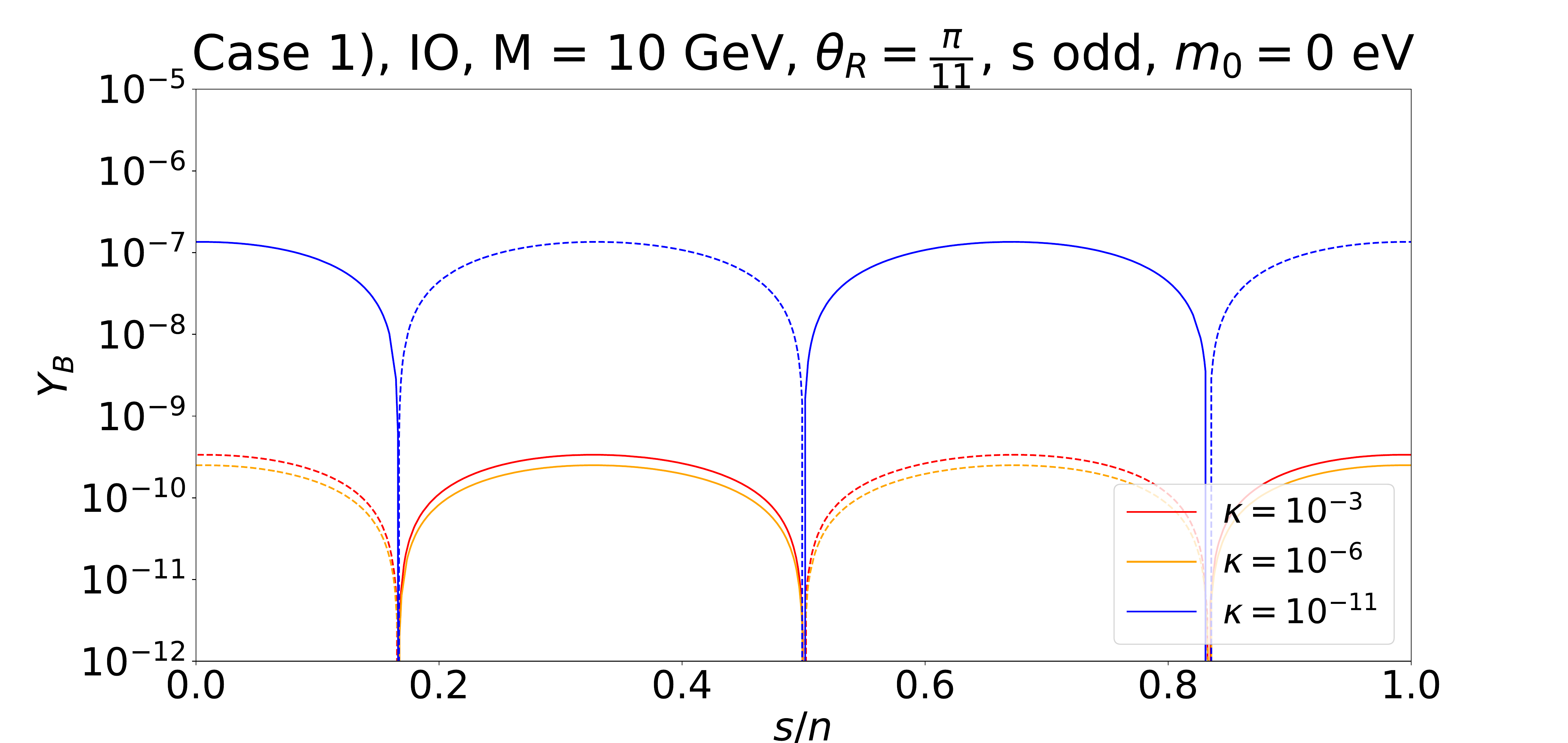}
		\caption{Vanishing initial conditions.}
	\end{subfigure}
	\begin{subfigure}{.5\textwidth}
		\centering
		\includegraphics[width = 1.1\textwidth]{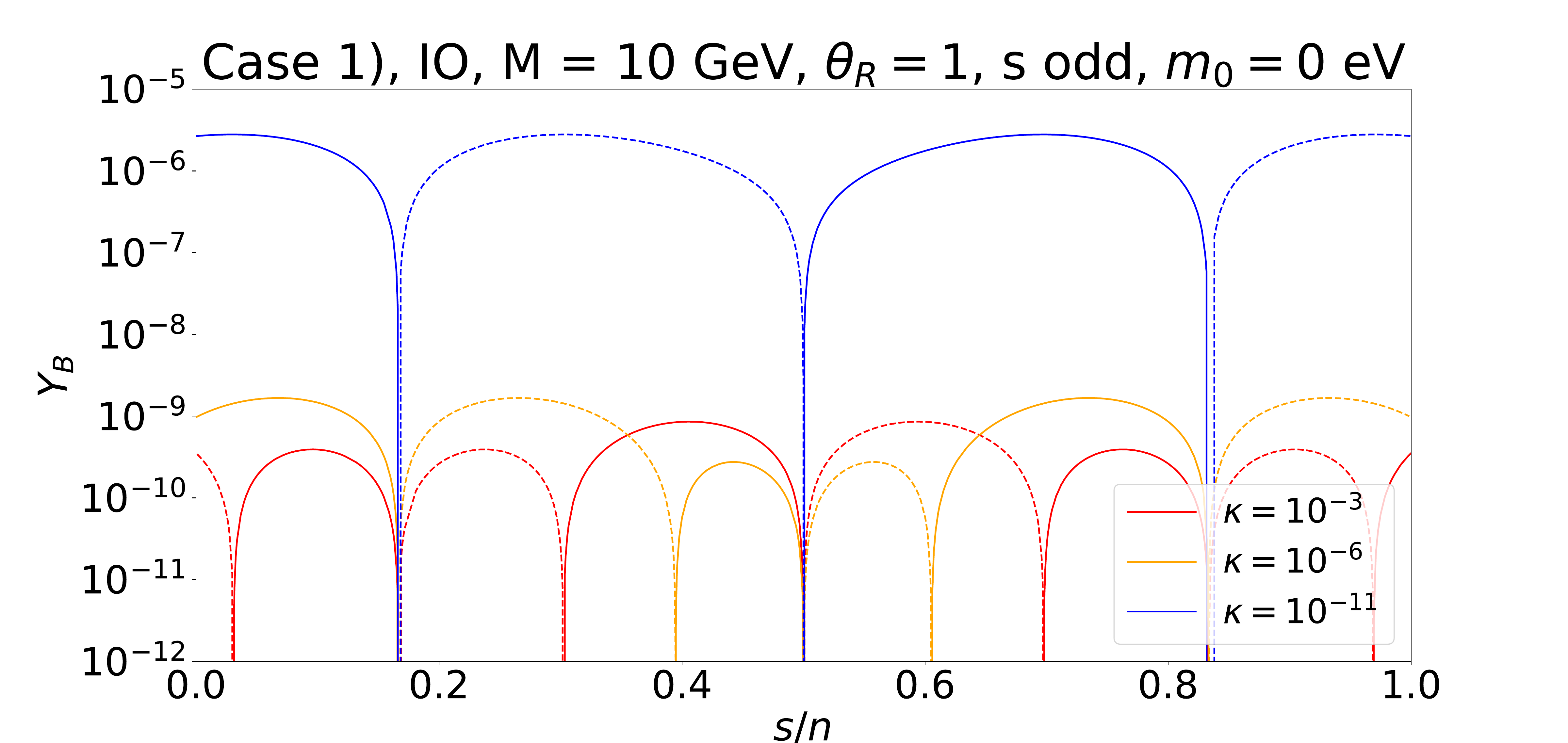}
		\caption{Vanishing initial conditions.}
	\end{subfigure}
\caption{{\small {\bf Case 1)} $Y_B$ as function of $\frac{s}{n}$, treated as continuous parameter, for a Majorana mass $M=10$ GeV and for $\theta_R=\frac{\pi}{11}$ (left plot) and $\theta_R=1$ (right plot). 
The light neutrino mass spectrum follows strong IO. The parameter $s$ is assumed to be odd. Different values of $\kappa$, $\kappa \in \{10^{-3},10^{-6},10^{-11}\}$, are considered. Both negative (dashed lines) as well as positive (continuous lines) values of the BAU are represented.}}
\label{IO phi BAU 10 GeV Case 1}
\end{figure}

\begin{figure}
	\centering
	\includegraphics[width=0.49\textwidth]{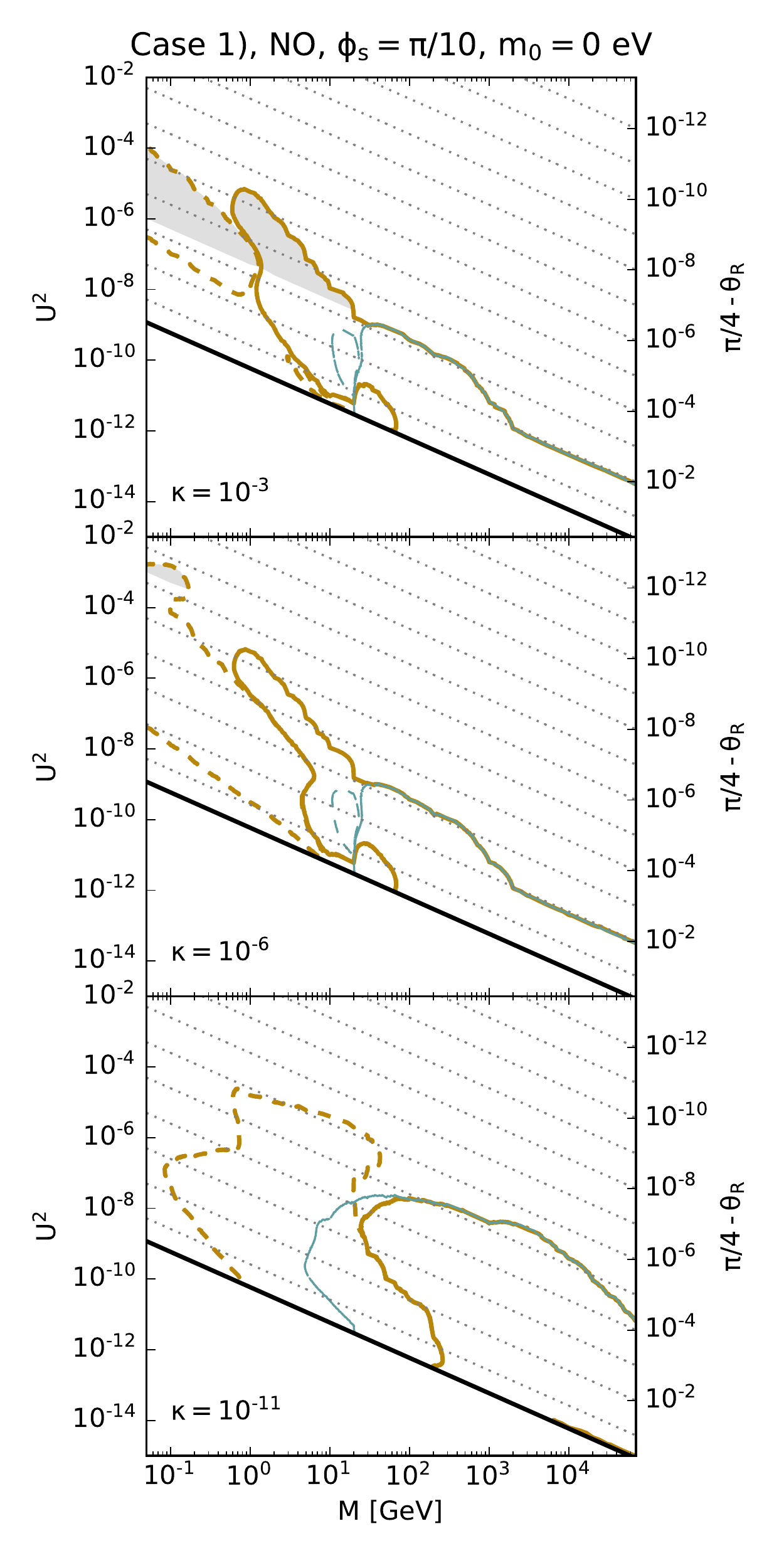}
	\includegraphics[width=0.49\textwidth]{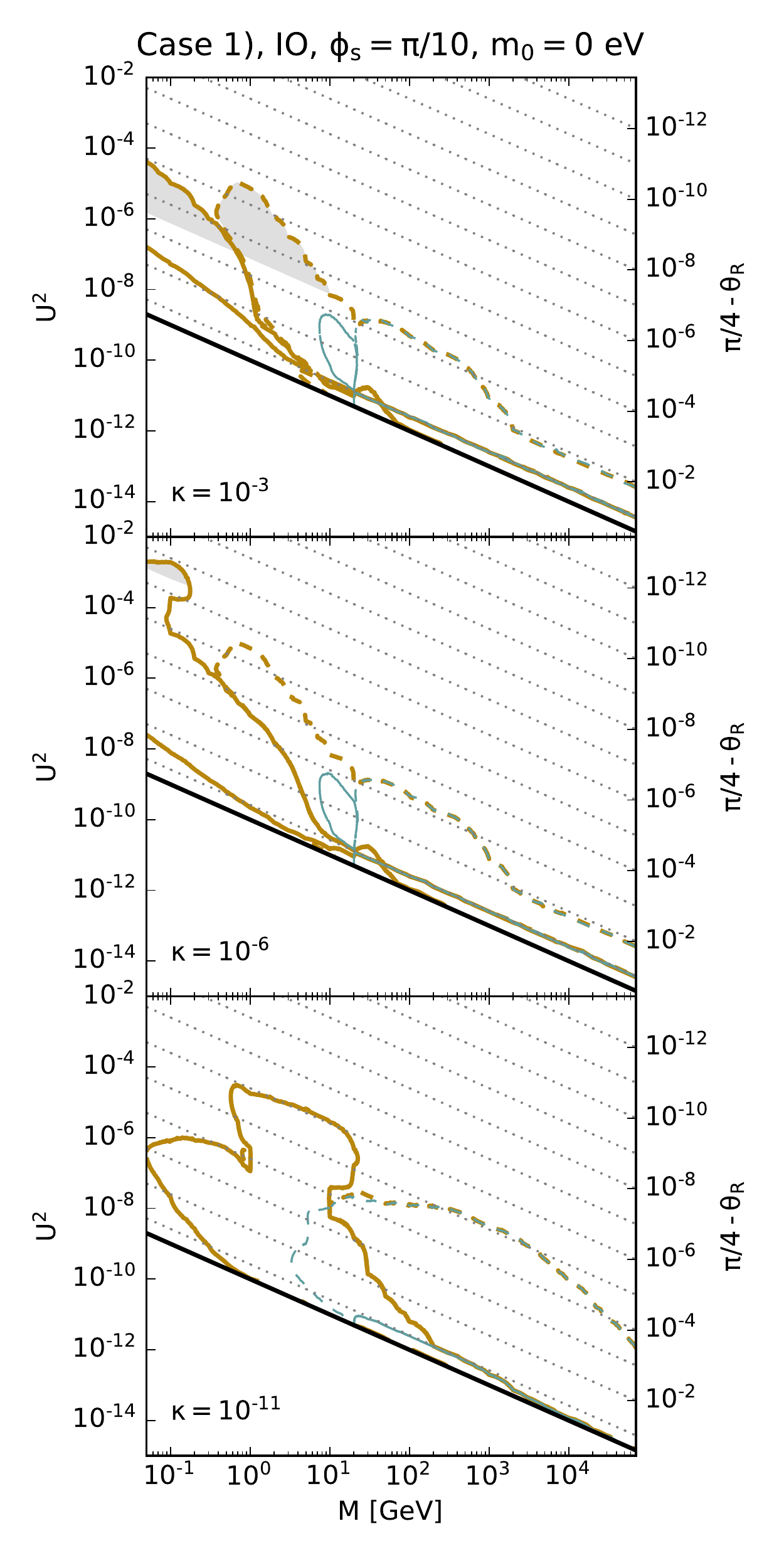}
	\caption{{\small {\bf Case 1)}
	Range of total mixing angle $U^2$ consistent with the observed amount of  BAU for heavy neutrino masses $M$ between $50$ MeV and $70$ TeV, as presented in Fig.~\ref{NO Mass U2 Case1}.
	The three rows correspond to different values of $\kappa$, $\kappa \in \{ 10^{-3}, 10^{-6}, 10^{-11} \}$, respectively.
	The vanishing and thermal initial conditions are shown by the ochre and turquoise lines, respectively.
	The continuous (dashed) lines indicate positive (negative) values of the BAU.
	The grey shaded areas show the regions in which the condition in Eq.~\eqref{eq:kappacorrectionscriterion} and alike is no longer satisfied. The value of the angle $\theta_R$ can be read off.
}}
\label{fig:Case1_MU2IC}
\end{figure}

For Case 1), we show in Fig.~\ref{kappa BAU Case I different masses IO/Massive m0} plots for the three further choices of the Majorana mass $M$, $M=1$ GeV, $M=100$ GeV and $M=1$ TeV, for light neutrino masses with strong IO as well as for a NO light neutrino mass spectrum with $m_0=0.03$ eV in addition to the two plots shown for $M=10$ GeV in Fig.~\ref{kappa BAU 10 GeV IO and NO massive case I} in the main text. Furthermore, we display in Fig.~\ref{IO phi BAU 10 GeV Case 1} two plots for the
BAU as function of $s/n$, treated as continuous parameter, assuming $s$ odd, for light neutrino masses with strong IO. These plots supplement the information found in Fig.~\ref{NO phi BAU 10 GeV combined} in the main text. Lastly, we show in Fig.~\ref{fig:Case1_MU2IC} dedicated plots for each value of the splitting
$\kappa$ in order to facilitate the comparison between the results for vanishing and thermal initial conditions. These plots are combined in
Fig.~\ref{NO Mass U2 Case1} in the main text, where the results for the different values of $\kappa$ are shown in different colours.

\newpage

\subsection{Case 2)}
\label{appF2}

\begin{figure}[h!]
	\begin{subfigure}{.5\textwidth}
		\centering
		\includegraphics[width = \textwidth]{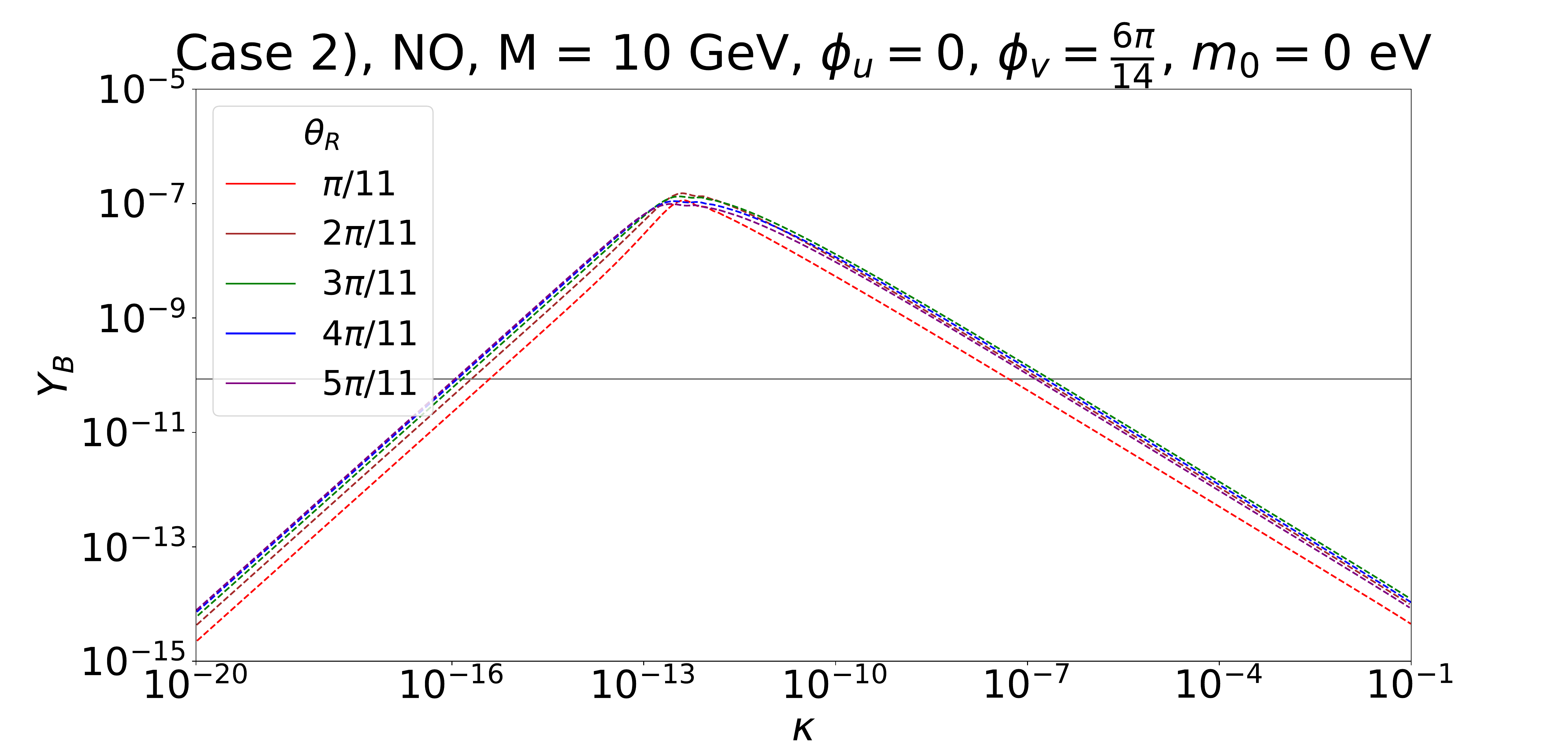}
		\caption{Vanishing initial conditions.}
	\end{subfigure}
	\begin{subfigure}{.5\textwidth}
		\centering
		\includegraphics[width = \textwidth]{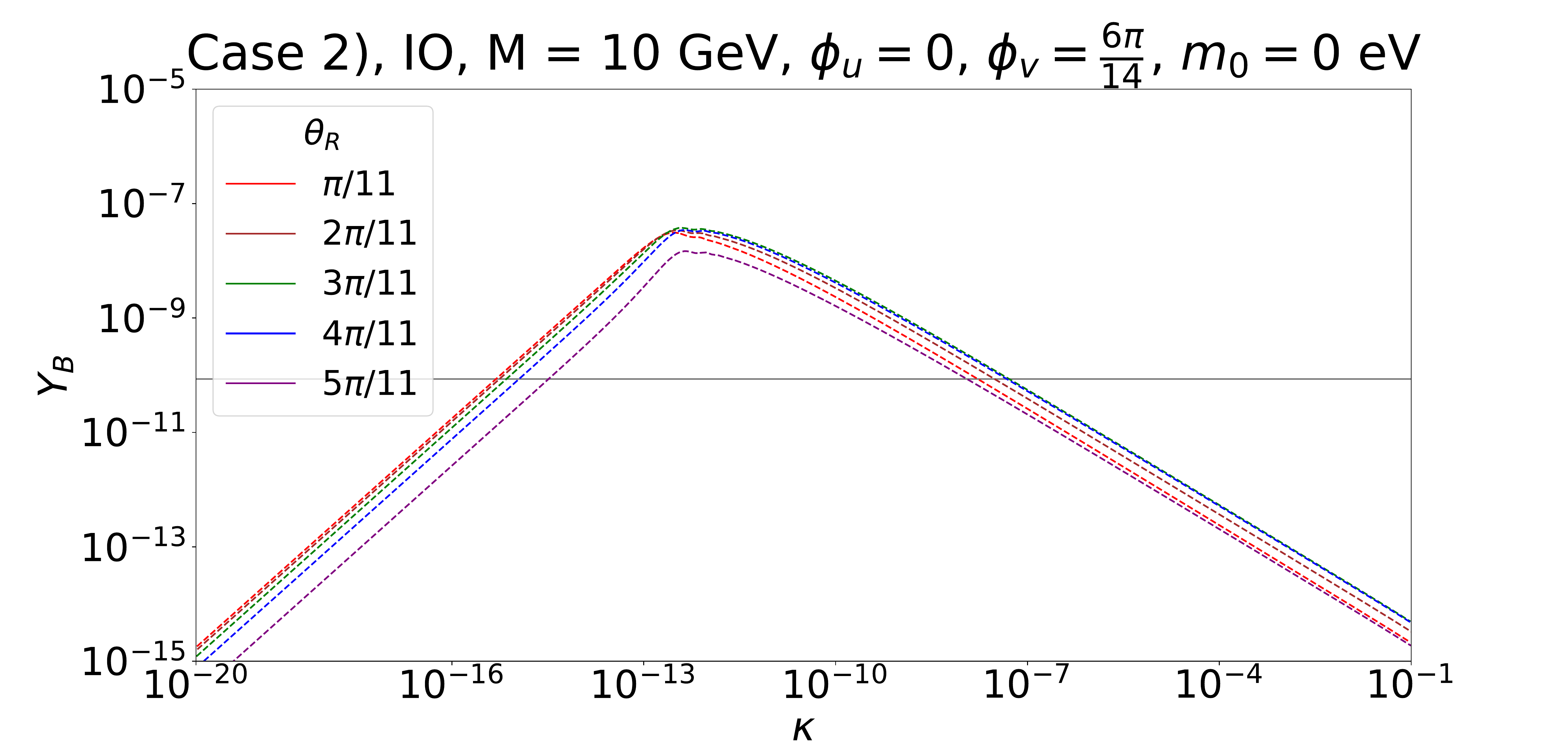}
		\caption{Vanishing initial conditions.}
	\end{subfigure}
\caption{{\small {\bf Case 2)} $Y_B$ as function of the splitting $\kappa$ for a Majorana mass $M=10$ GeV. Light neutrino masses have strong NO (left plot) and strong IO (right plot), respectively.
For the remaining choices see Fig.~\ref{NO kappa BAU Case II different masses}.
}}
\label{NOIO kappa BAU 10 GeV Case II massless}
\end{figure}

\begin{figure}
	\begin{subfigure}{.5\textwidth}
		\centering
		\includegraphics[width = \textwidth]{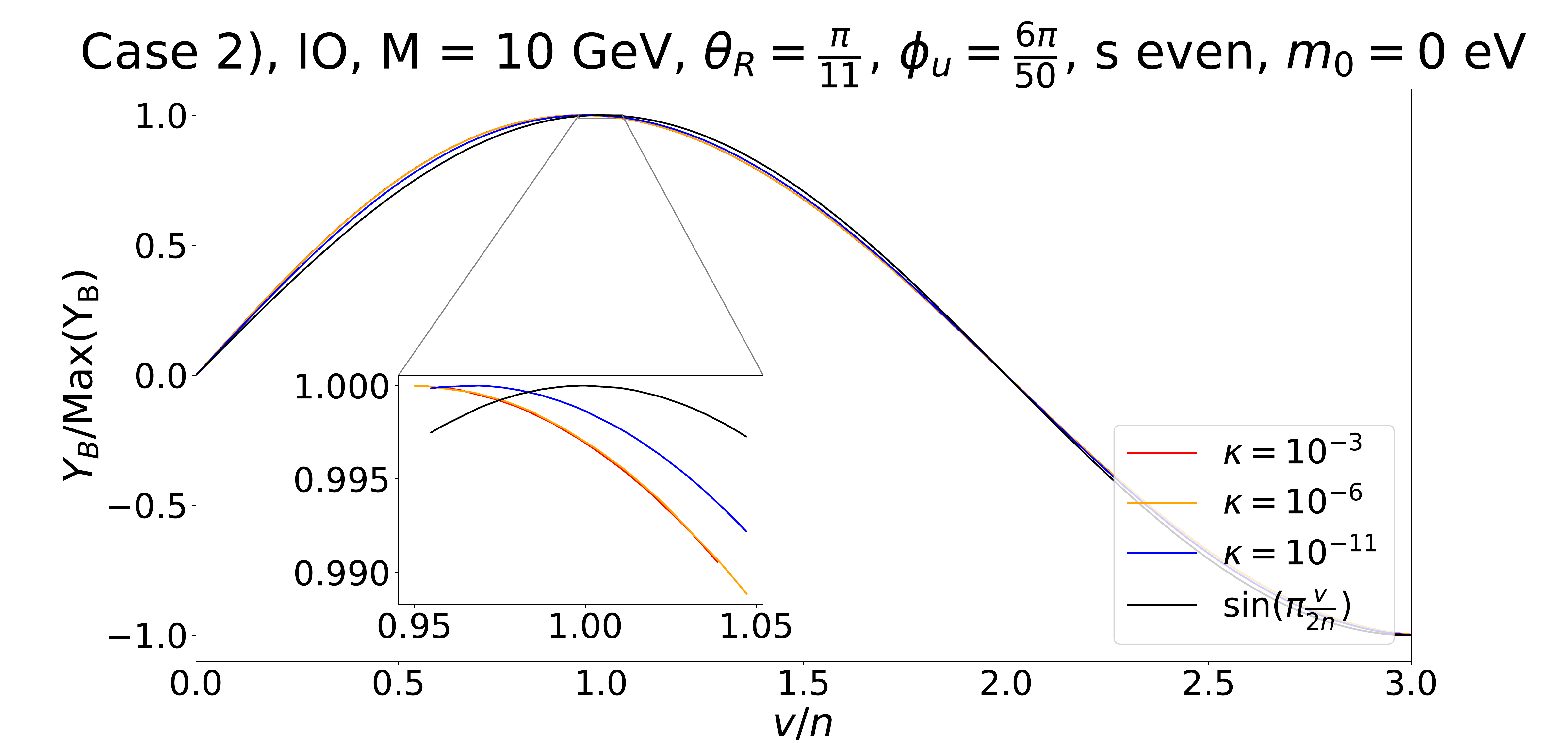}
		\caption{Vanishing initial conditions.}
	\end{subfigure}
	\begin{subfigure}{.5\textwidth}
		\centering
		\includegraphics[width = \textwidth]{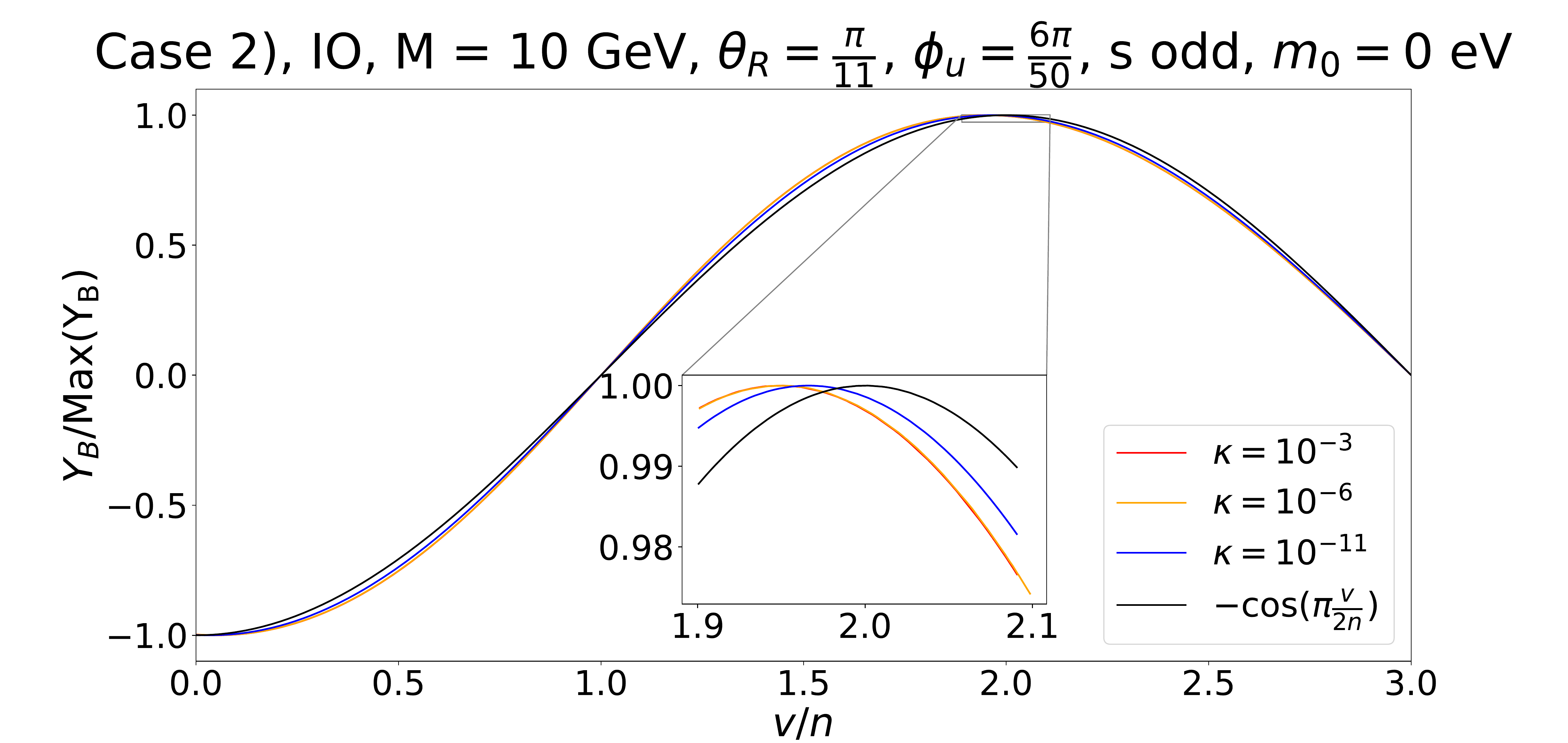}
		\caption{Vanishing initial conditions.}
\end{subfigure}
\caption{{\small {\bf Case 2)} $Y_B$ as function of $\frac{v}{n}$, treated as continuous parameter, for a Majorana mass $M=10$ GeV and different values of $\kappa$, $\kappa \in \{10^{-3},10^{-6},10^{-11}\}$. Light neutrino masses have strong IO. 
 The angle $\theta_R$ is fixed to $\theta_R=\frac{\pi}{11}$. The group theory parameters $n$ and $u$
are chosen as $n=50$ and $u=6$ so that $t$ is even and $s$ can be even (left plot) or odd (right plot), as in Fig.~\ref{NO BAU phiv e-3e-6e-11 caseII massless}. We compare the numerical results with the analytical expectation which is proportional to $\sin(\frac{\pi v}{2n})$ (for $s$  even) and to $-\cos(\frac{\pi v}{2n})$ (for $s$ odd), compare Eq.~\eqref{eq:CLFValphaCase2stevenIO}.
}}
\label{IO BAU phiv e-3e-6e-11 caseII massless}
\end{figure}

\begin{figure}
	\begin{subfigure}{.5\textwidth}
		\centering
		\includegraphics[width = \textwidth]{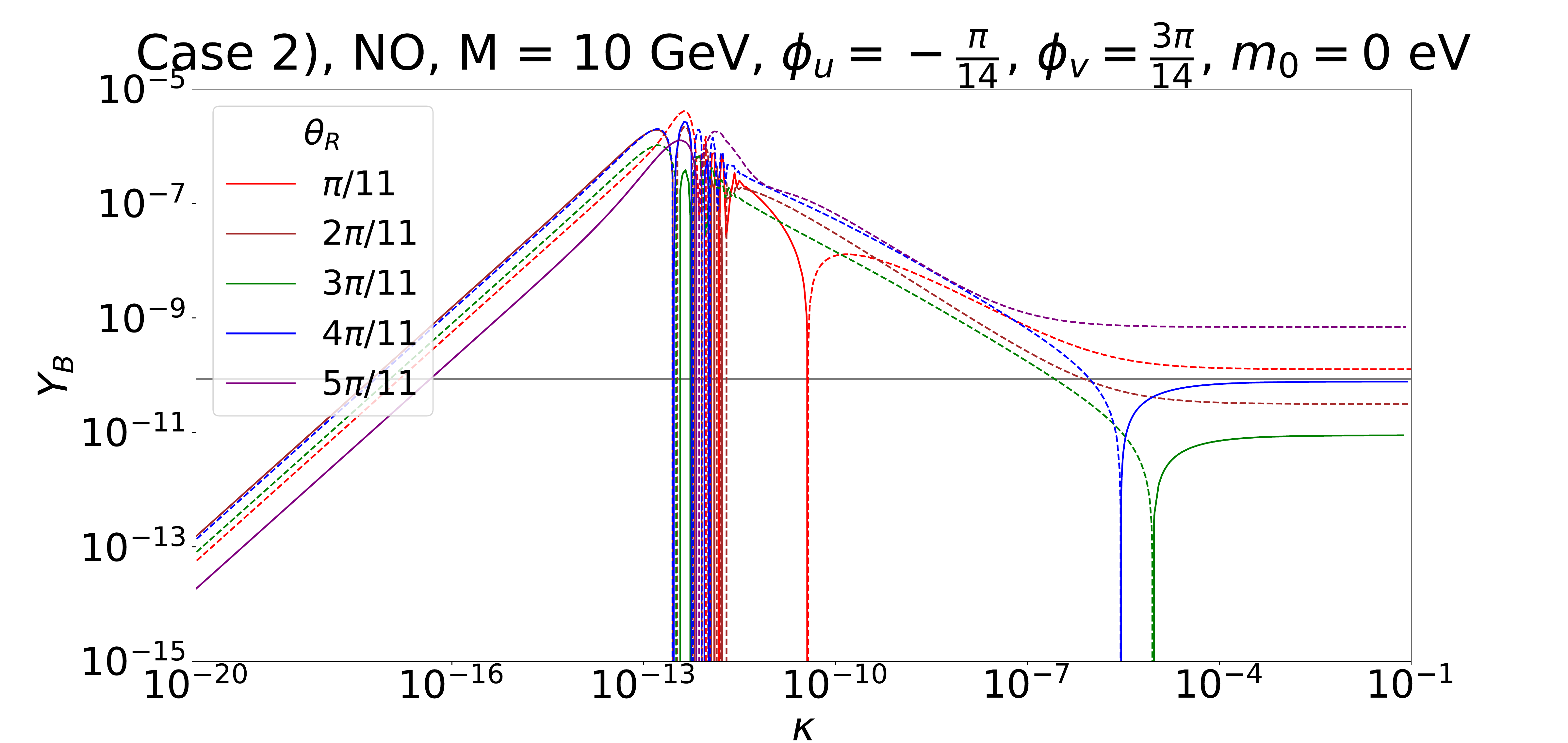}
		\caption{Vanishing initial conditions.}
	\end{subfigure}
	\begin{subfigure}{.5\textwidth}
		\includegraphics[width = \textwidth]{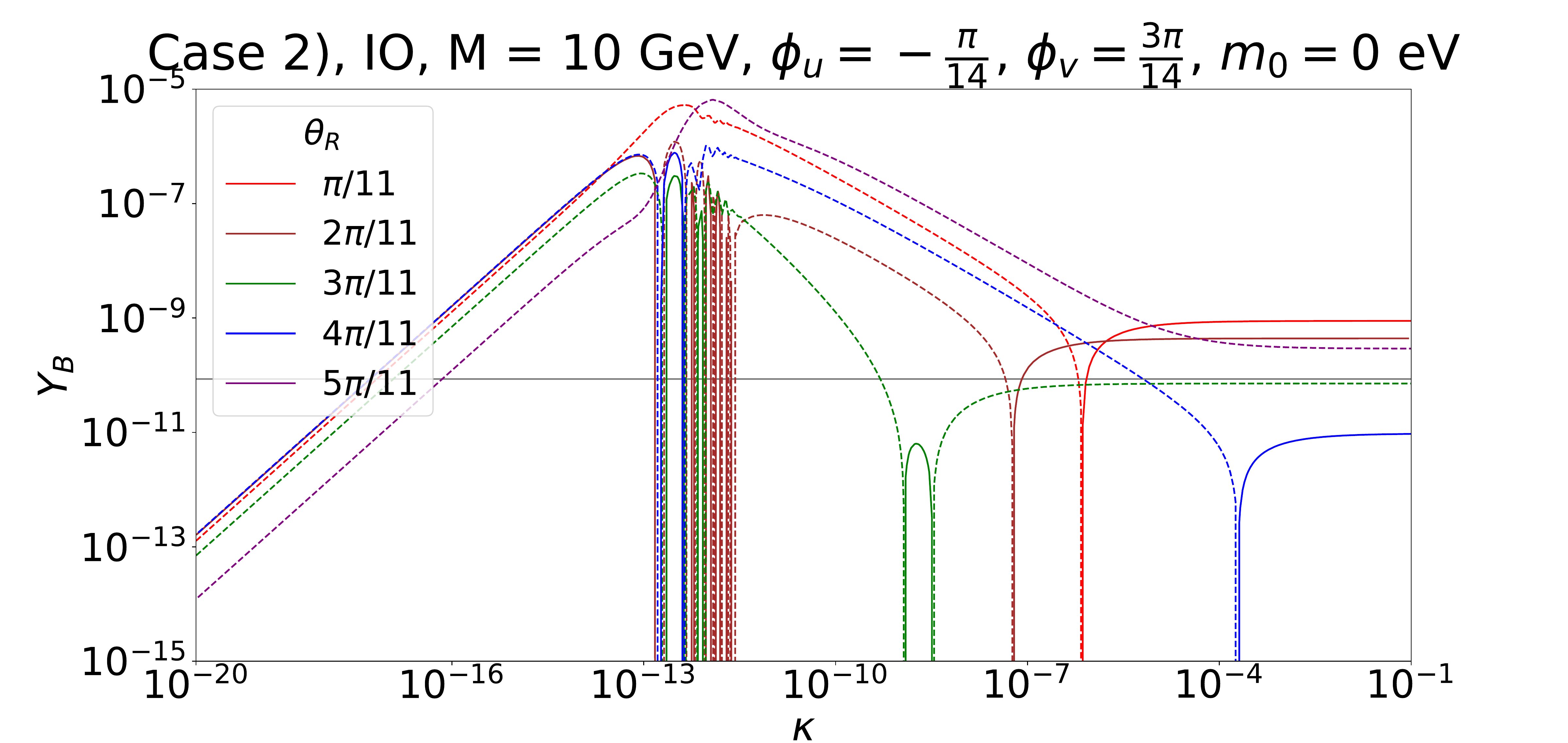}
		\caption{Vanishing initial conditions.}
	\end{subfigure}
	\caption{{\small {\bf Case 2)} $Y_B$ as function of $\kappa$ for a Majorana mass $M=10$ GeV. Light neutrino masses follow strong NO (left plot) and strong IO (right plot), respectively. The parameters $(s,t)$ are fixed to $(0,1)$ such that $\phi_u=-\frac{\pi}{14}$ and $\phi_v=\frac{3 \, \pi}{14}$.
For the remaining choices see Fig.~\ref{NO kappa BAU Case II todd different masses}.
}}
\label{NOIO kappa BAU 10 GeV Case II todd u-1}
\end{figure}

\begin{figure}
	\begin{subfigure}{.5\textwidth}
		\centering
		\includegraphics[width = \textwidth]{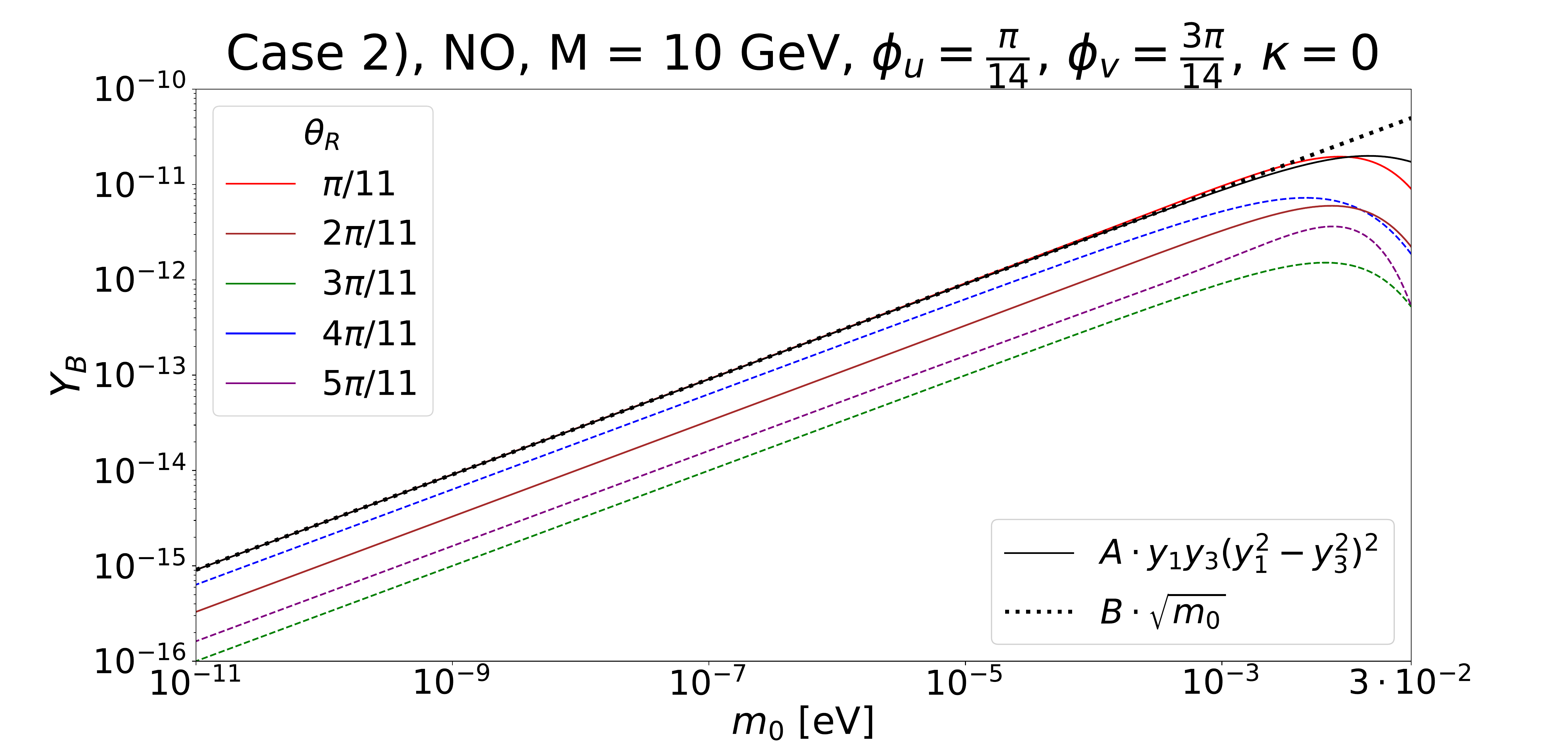}
		\caption{Vanishing initial conditions.}
	\end{subfigure}
	\begin{subfigure}{.5\textwidth}
		\centering
		\includegraphics[width = \textwidth]{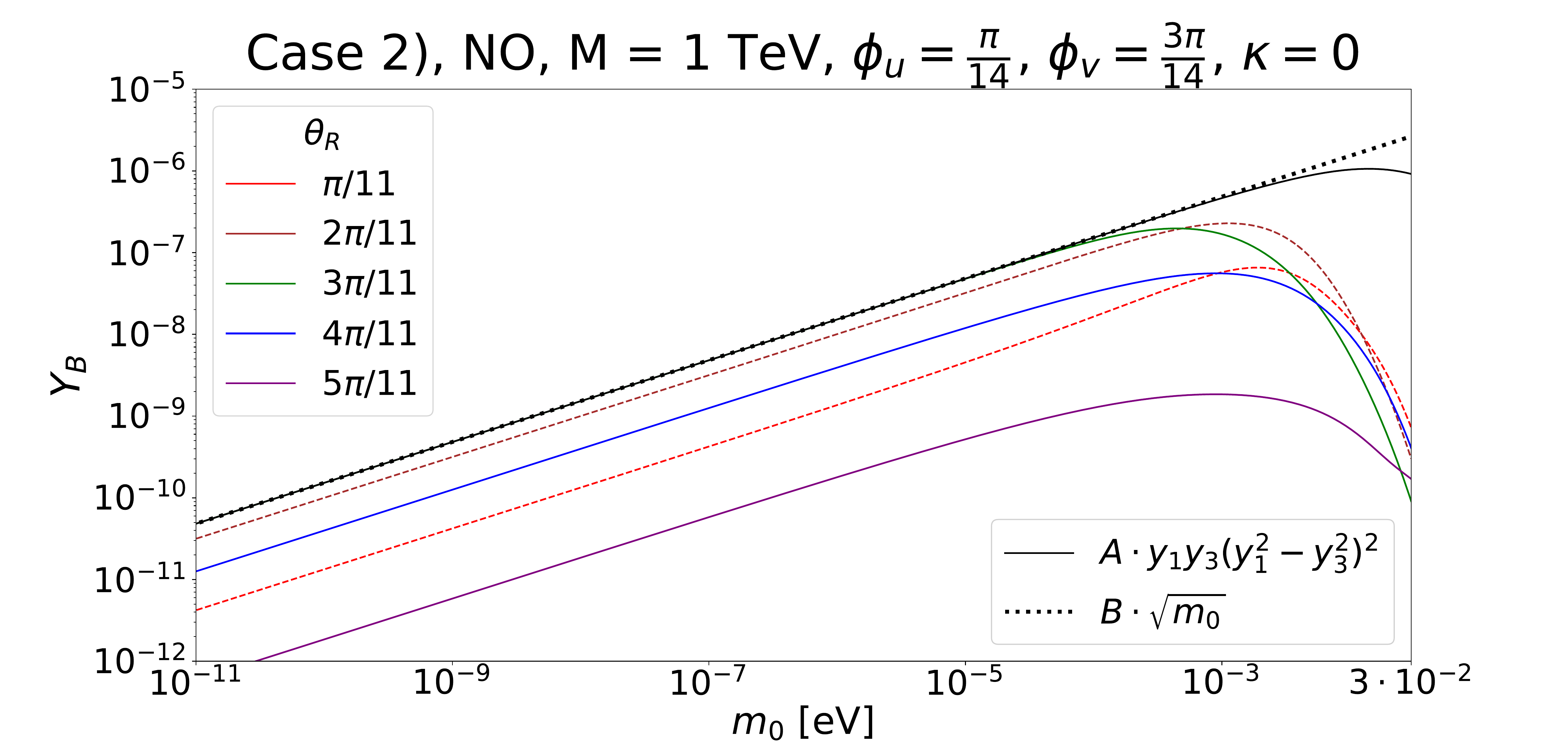}
		\caption{Vanishing initial conditions.}
	\end{subfigure}
\caption{{\small {\bf Case 2)} $Y_B$ as function of the lightest neutrino mass $m_0$ in the absence of splittings, $\kappa=0$ (and $\lambda=0$), for different values of $\theta_R$. We set the Majorana mass $M$ to $10$ GeV (left plot) and $1$ TeV (right plot), respectively, and fix the group theory parameters $n$, $s$ and $t$ to $n=14$, $s=1$ and $t=1$ such that $\phi_u=\frac{\pi}{14}$ and $\phi_v=\frac{3 \, \pi}{14}$.
Light neutrino masses follow NO. In addition, we compare, for $\theta_R=\frac{\pi}{11}$ (left plot) and $\theta_R=\frac{3\pi}{11}$ (right plot), the numerical results with the following two analytical expressions: $A \cdot y_1 \, y_3 \, (y_1^2-y_3^2)^2$ with $A \approx 4.14 \cdot 10^{30}$ (left plot) and $A\approx 7.37 \cdot 10^{29}$ (right plot) and $B \cdot \sqrt{m_0}$ with $B \approx 2.87 \cdot 10^{-10} \mbox{eV}^{-1/2}$ (left plot) and $B \approx 1.53 \cdot 10^{-5} \mbox{eV}^{-1/2}$ (right plot). The first (second) expression corresponds to the continuous (dotted) black line.
Both negative (dashed lines) as well as positive (continuous lines) values of the BAU are represented.}}
\label{NO m0 BAU 10 GeV/1 TeV caseII odd no mass splittings}
\end{figure}

\begin{figure}
	\centering
	\includegraphics[width=0.49\textwidth]{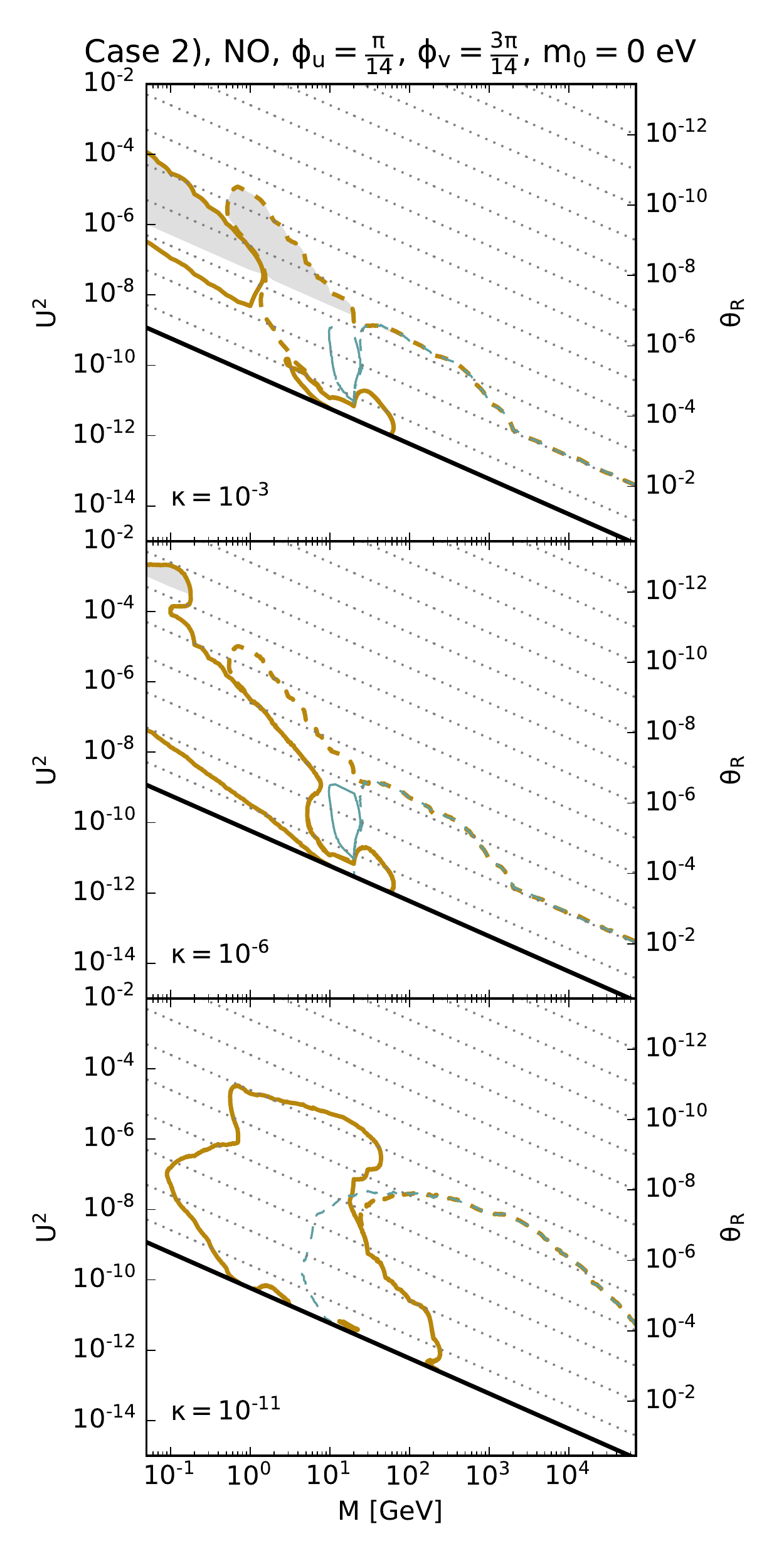}
	\includegraphics[width=0.49\textwidth]{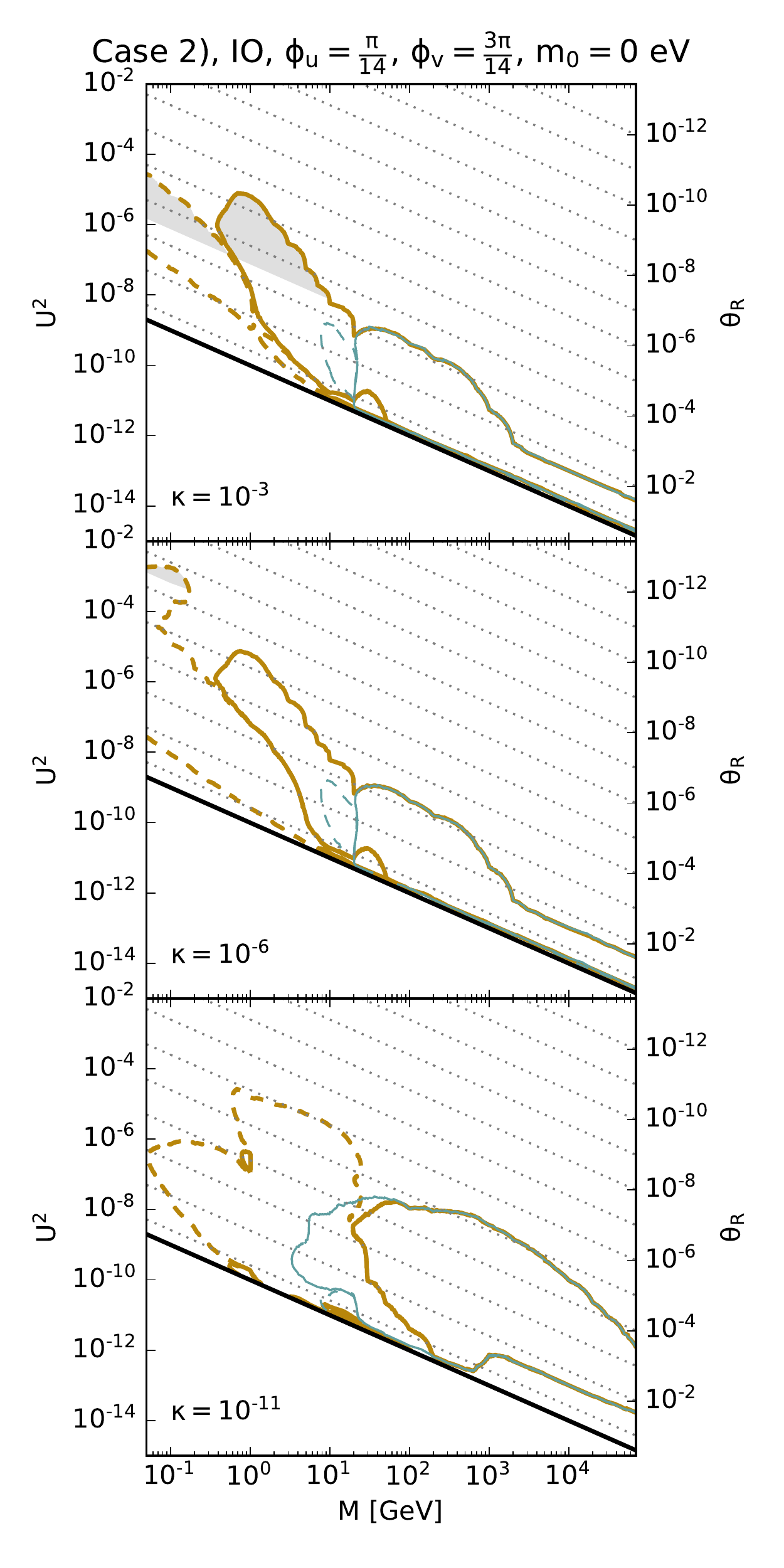}
	\caption{{\small {\bf Case 2)}
	Range of total mixing angle $U^2$ consistent with the observed BAU for heavy neutrino masses between $50$ MeV and $70$ TeV, as presented in Fig.~\ref{NO Mass U2 Case2}.
	The three rows correspond to different values of $\kappa$, $\kappa \in \{ 10^{-3}, 10^{-6}, 10^{-11} \}$, respectively.
	The vanishing and thermal initial conditions are shown by the ochre and turquoise lines, respectively.
	The continuous (dashed) lines indicate positive (negative) values of the BAU.
	The grey shaded areas show the regions in which a condition like the one in  Eq.~\eqref{eq:kappacorrectionscriterion} is no longer satisfied. The angle $\theta_R$ can be read off.
}}
\label{fig:Case2_MU2IC}
\end{figure}

\begin{figure}
	\centering
	\includegraphics[width = \textwidth]{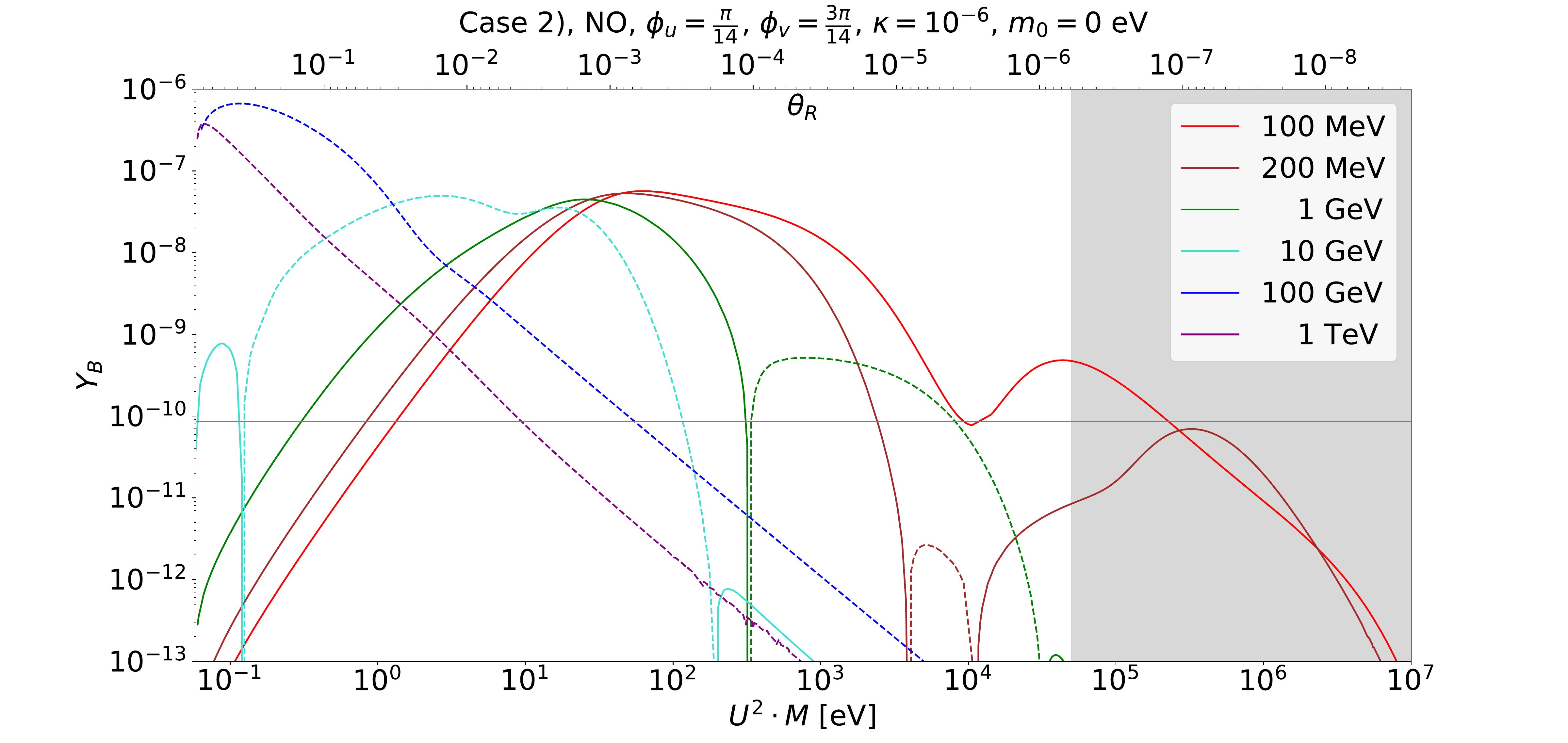}
	\caption{{\small \textbf{Case 2)} $Y_B$ as function of $U^2\cdot M$ for different values of the heavy neutrino mass $M$. The splitting $\kappa$ is fixed to $10^{-6}$. Light neutrino masses follow strong NO. The group theory parameters $n$, $u$ and $v$ are chosen as $n=14$,
		$u=1$ and $v=3$
corresponding to $\phi_u=\frac{\pi}{14}$ and $\phi_v=\frac{3\, \pi}{14}$. Both negative (dashed lines) as well as positive (continuous lines) values of the BAU are represented. The grey line indicates the observed value of the BAU, $Y_B\approx 8.6 \cdot 10^{-11}$. The grey shaded area shows the region in which a criterion similar to the one in  Eq.~\eqref{eq:kappacorrectionscriterion} is not fulfilled.}}
\label{NO BAU vs U2 10 GeV e-6 multiple curves caseII}
\end{figure}

For Case 2), we add the following plots to those already present in the main text: two plots, found in Fig.~\ref{NOIO kappa BAU 10 GeV Case II massless}, of the BAU with respect to the splitting $\kappa$  for the choice $t$ even (more concretely, we have $n=14$, $u=0$ and $v=6$) for light neutrino masses with strong NO and with strong IO, respectively, for a fixed value of the Majorana mass $M$, $M=10$ GeV, and several values of the angle $\theta_R$. These, in particular the plot for strong NO, can be compared with Fig.~\ref{NO kappa BAU Case II different masses}, especially plot (b) of this figure. Furthermore, we show in this appendix in Fig.~\ref{IO BAU phiv e-3e-6e-11 caseII massless} two plots for light neutrino masses with strong IO corresponding to those shown in Fig.~\ref{NO BAU phiv e-3e-6e-11 caseII massless} in the main text for neutrinos with strong NO. Also for strong IO the BAU depends on $\phi_v$ ($v/n$ treated as continuous parameter), as expected from the CP-violating combinations. For the choice $t$ odd, we display in Fig.~\ref{NOIO kappa BAU 10 GeV Case II todd u-1} two plots for $u=-1$ (we use $n=14$, $s=0$ and $t=1$) which should be
compared with the plots in the main text for $u=1$, see e.g.~Fig.~\ref{NO kappa BAU Case II todd different masses}. In order to illustrate the
possible strong enhancement of the BAU depending on the Majorana mass $M$, we provide in addition to plot (a) in Fig.~\ref{NO thetaR/m0 BAU 100 GeV caseII odd no mass splittings} for $M=100$ GeV in the main text two plots for $M=10$ GeV and $M=1$ TeV and otherwise the same choices of parameters as in the figure of the main text, see Fig.~\ref{NO m0 BAU 10 GeV/1 TeV caseII odd no mass splittings}. In Fig.~\ref{fig:Case2_MU2IC}
  we also show dedicated plots for each value of the splitting $\kappa$ in order to facilitate the comparison between the results for vanishing and thermal initial conditions. These plots are combined in
Fig.~\ref{NO Mass U2 Case2} in the main text, where the results for the different values of $\kappa$ are shown in different colours.
Lastly, we show in Fig.~\ref{NO BAU vs U2 10 GeV e-6 multiple curves caseII} the corresponding plot for Case 2) and $t$ odd ($n=14$, $u=1$ and $v=3$), as given in Fig.~\ref{NO BAU vs U2 10 GeV e-6 multiple curves} for Case 1). We note  for these plots as well the striking similarity, up to the overall sign of the BAU (exchange of continuous and dashed lines).

\newpage

\subsection{Case 3 a) and Case 3 b.1)}
\label{appF3}

\begin{figure}
	\begin{subfigure}{.5\textwidth}
		\centering
		\includegraphics[width = \textwidth]{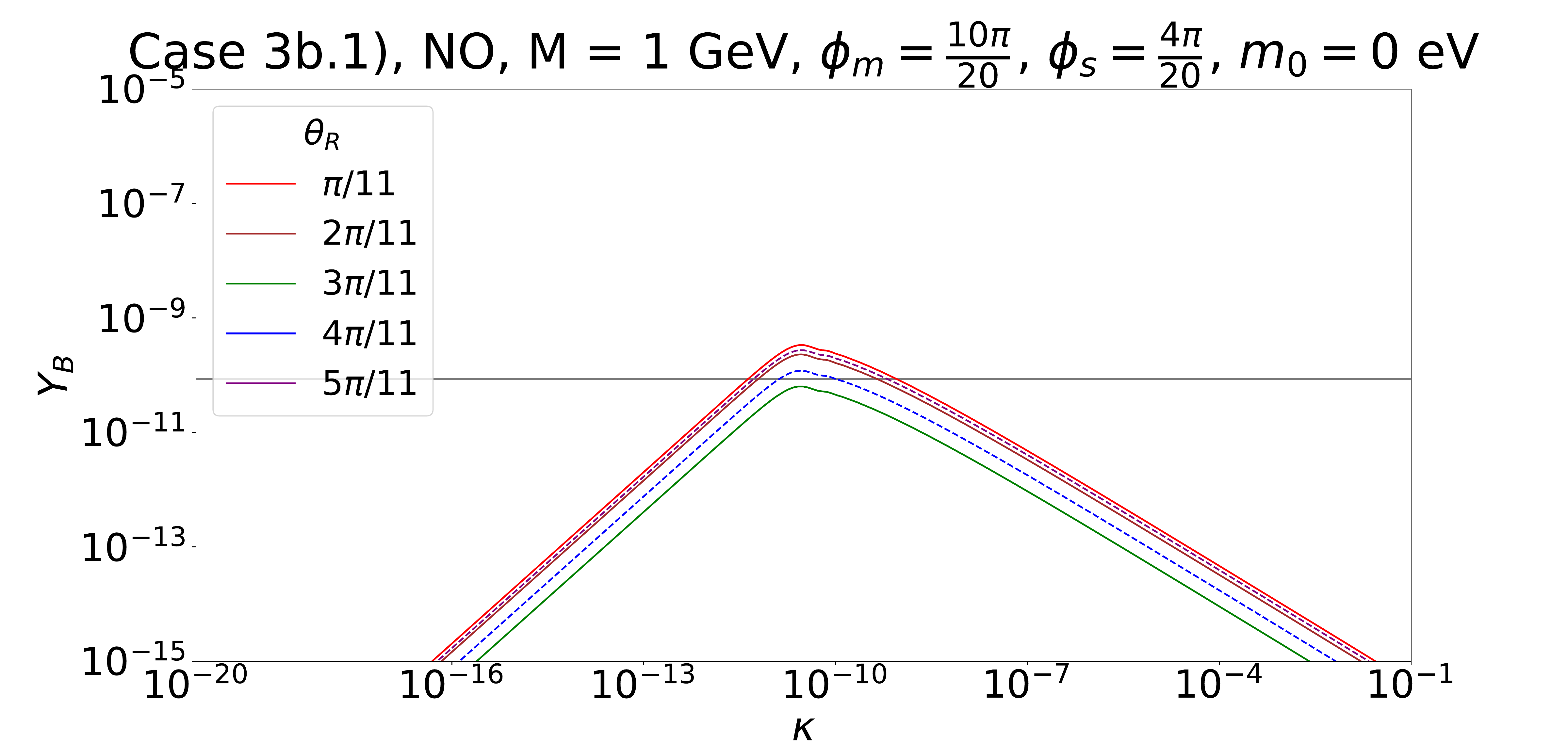}
		\caption{Vanishing initial conditions.}
	\end{subfigure}
	\begin{subfigure}{.5\textwidth}
		\centering
		\includegraphics[width = \textwidth]{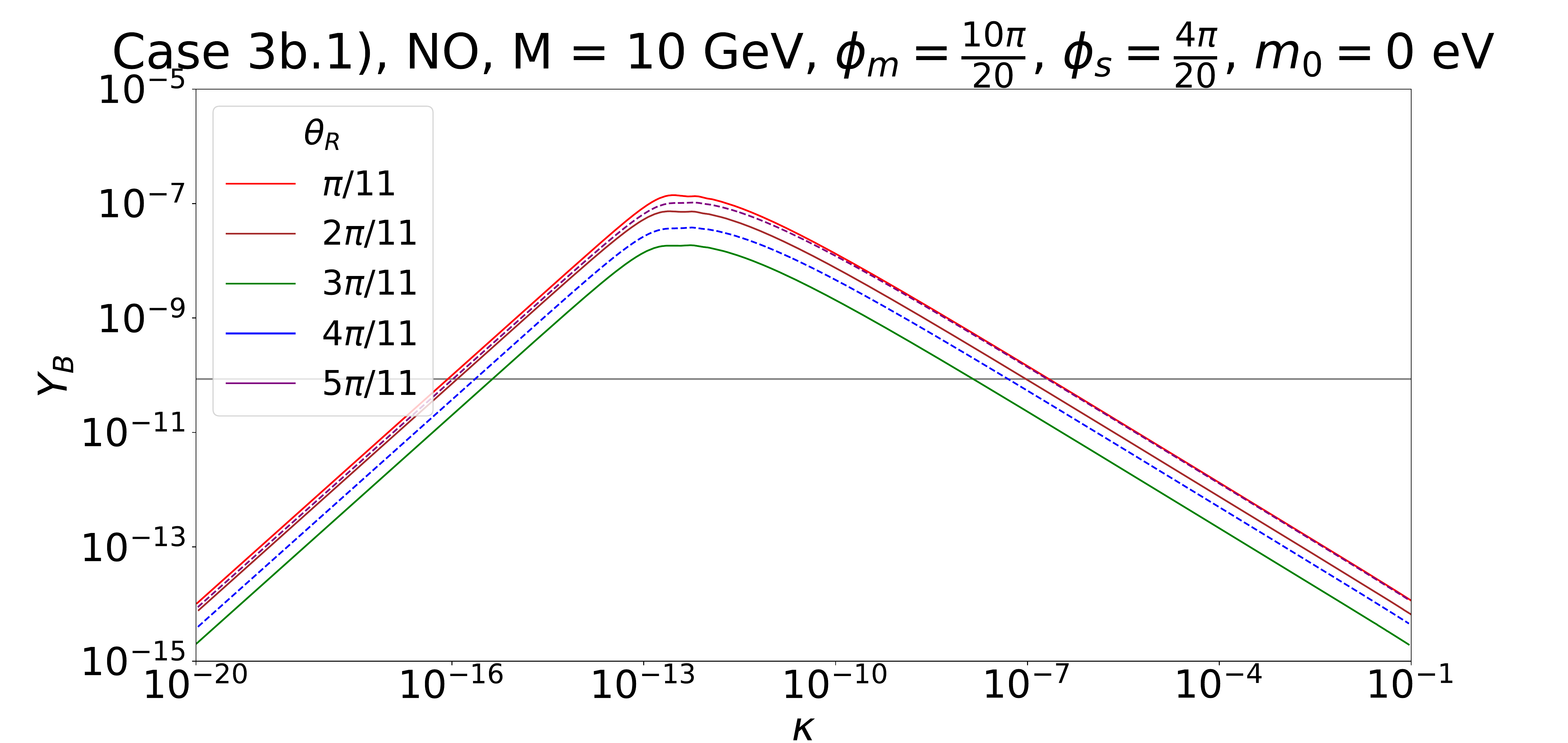}
		\caption{Vanishing initial conditions.}
		\label{NO kappa BAU 10 GeV case IIIb massless}
	\end{subfigure}
	\begin{subfigure}{.5\textwidth}
		\includegraphics[width = \textwidth]{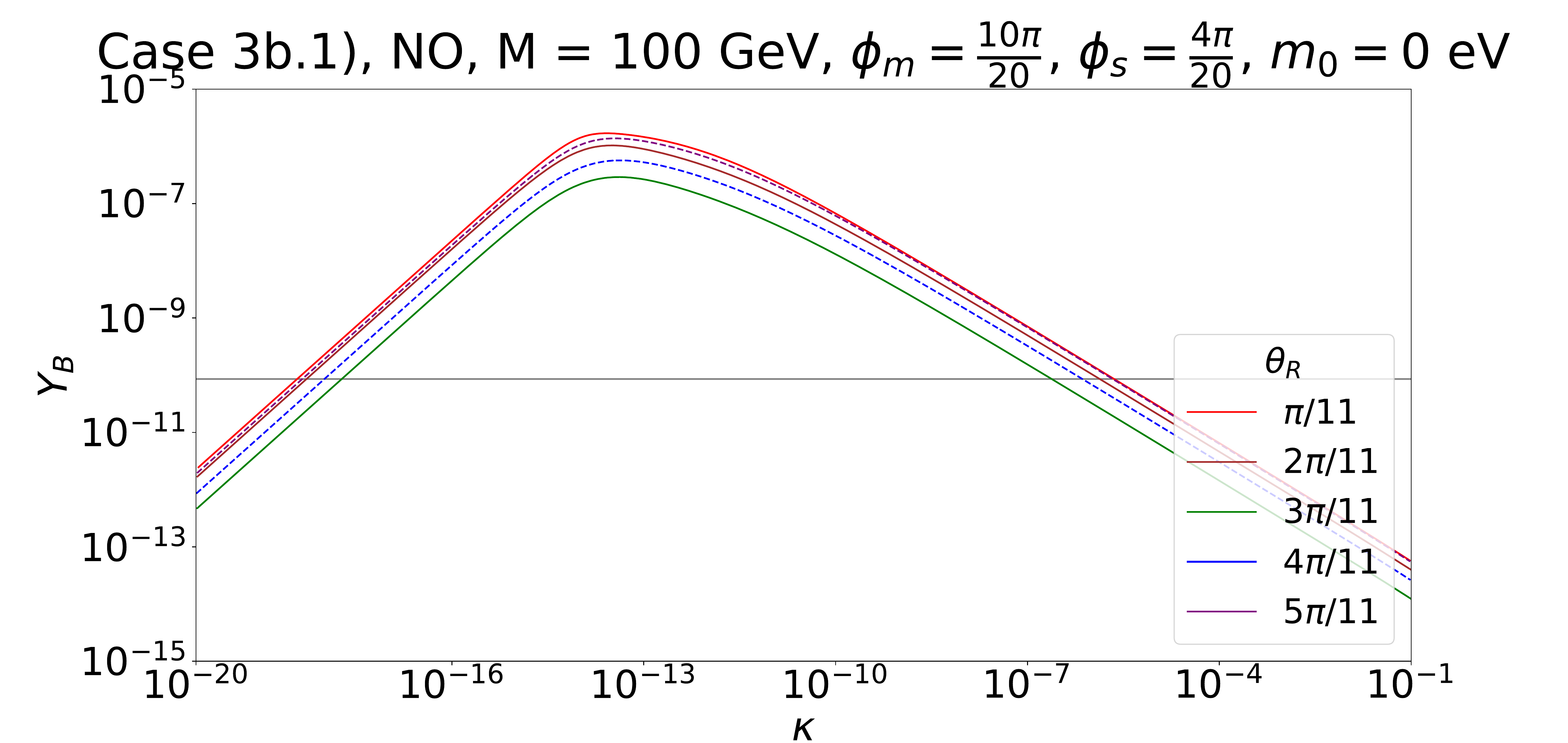}
		\caption{Vanishing initial conditions.}
	\end{subfigure}
	\begin{subfigure}{.5\textwidth}
		\includegraphics[width = \textwidth]{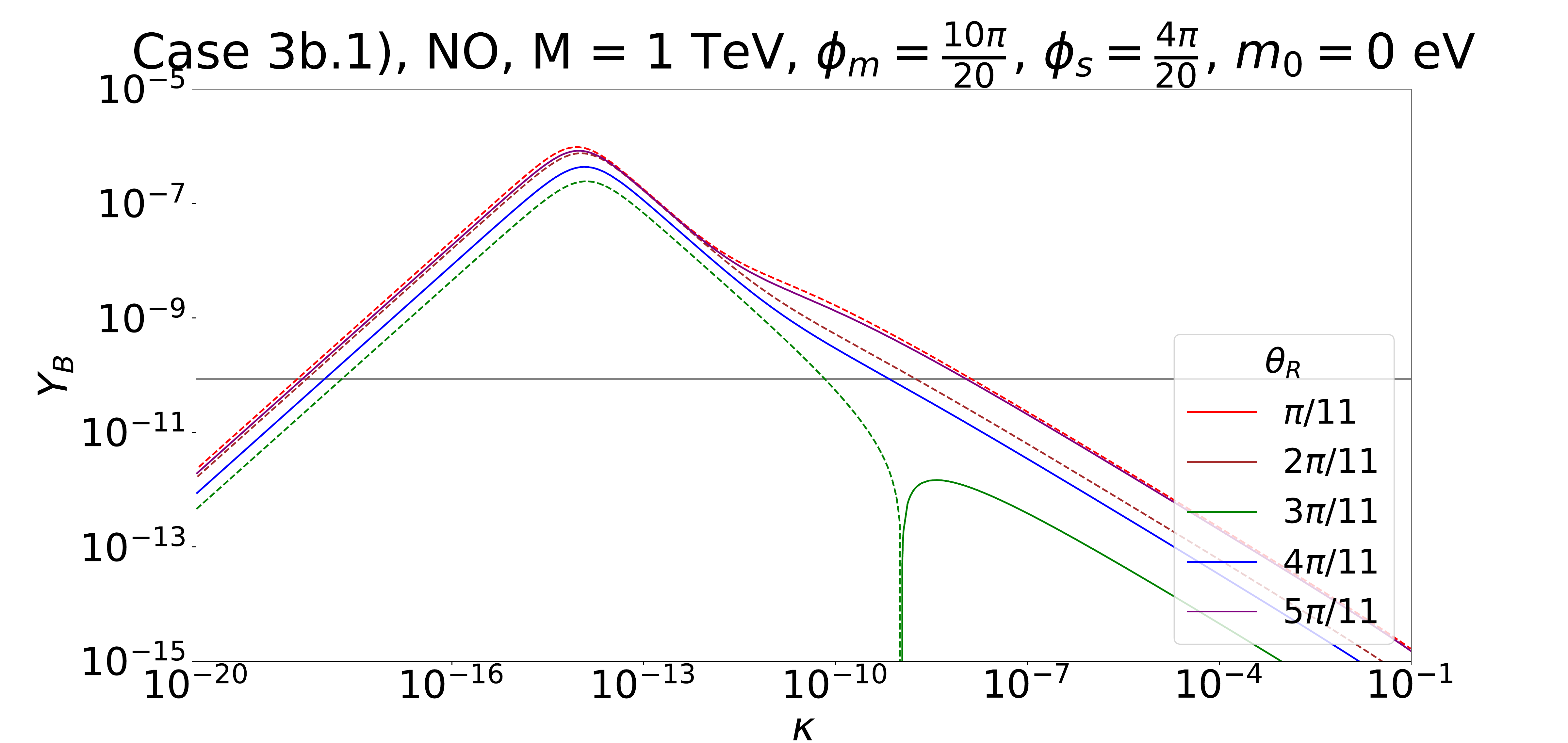}
		\caption{Vanishing initial conditions.}
	\end{subfigure}
	\caption{{\small {\bf Case 3 b.1)} $Y_B$ as function of the splitting $\kappa$ for a Majorana mass $M=1$ GeV (upper left plot), $M=10$ GeV (upper right plot), $M=100$ GeV (lower left plot) and $M=1$ TeV (lower right plot). For each of these choices, different values of $\theta_R$ have been studied. The group theory parameters are set to $n=20$, $m=10$ and $s=4$. Light neutrino masses have strong NO. For the rest of the choices
		see Fig.~\ref{NO kappa BAU Case IIIb different masses massive lightest neutrino}.
}}
\label{NO kappa BAU Case IIIb different masses massless lightest neutrino}
\end{figure}

\begin{figure}
	\begin{subfigure}{.5\textwidth}
		\centering
		\includegraphics[width = \textwidth]{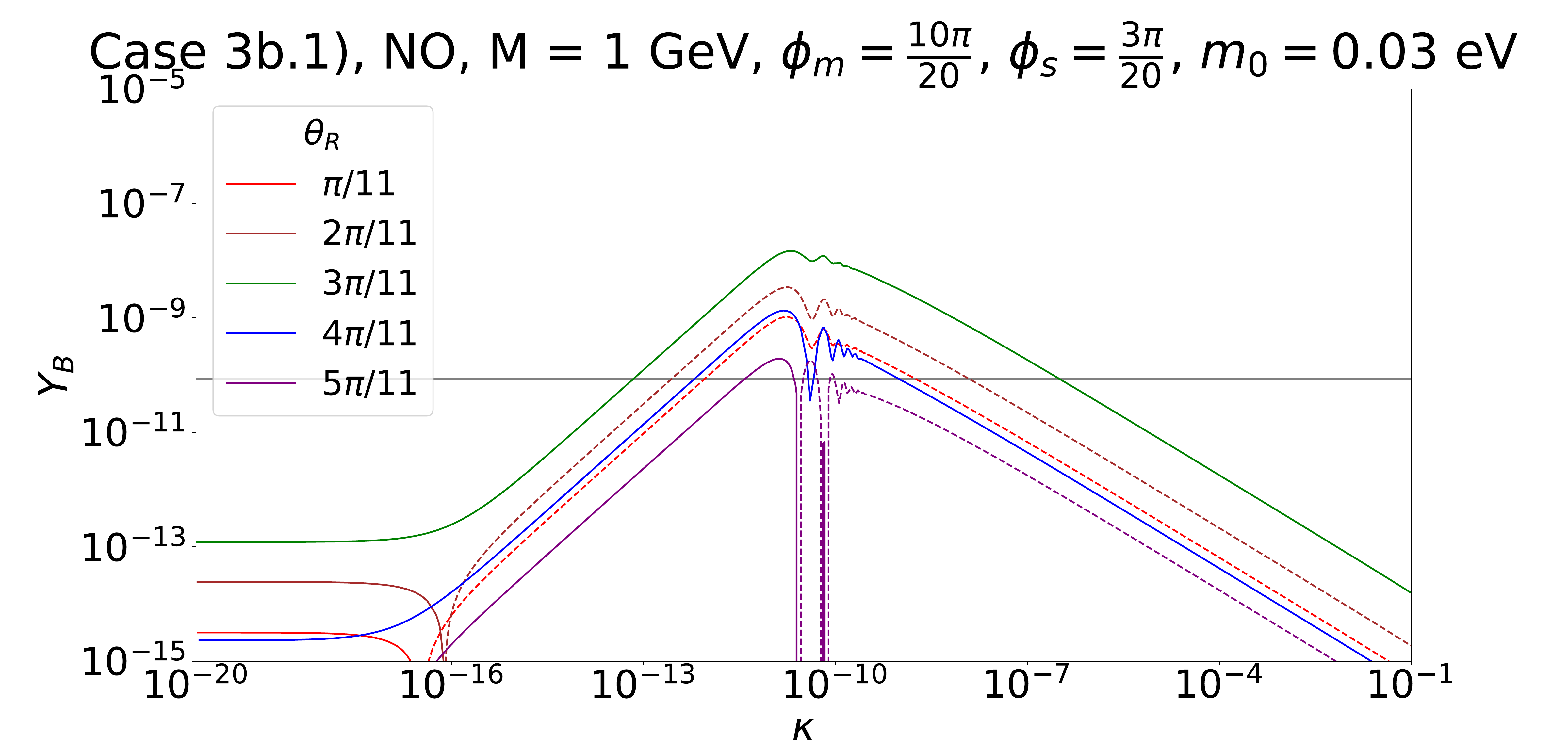}
		\caption{Vanishing initial conditions.}
	\end{subfigure}
	\begin{subfigure}{.5\textwidth}
		\centering
		\includegraphics[width = \textwidth]{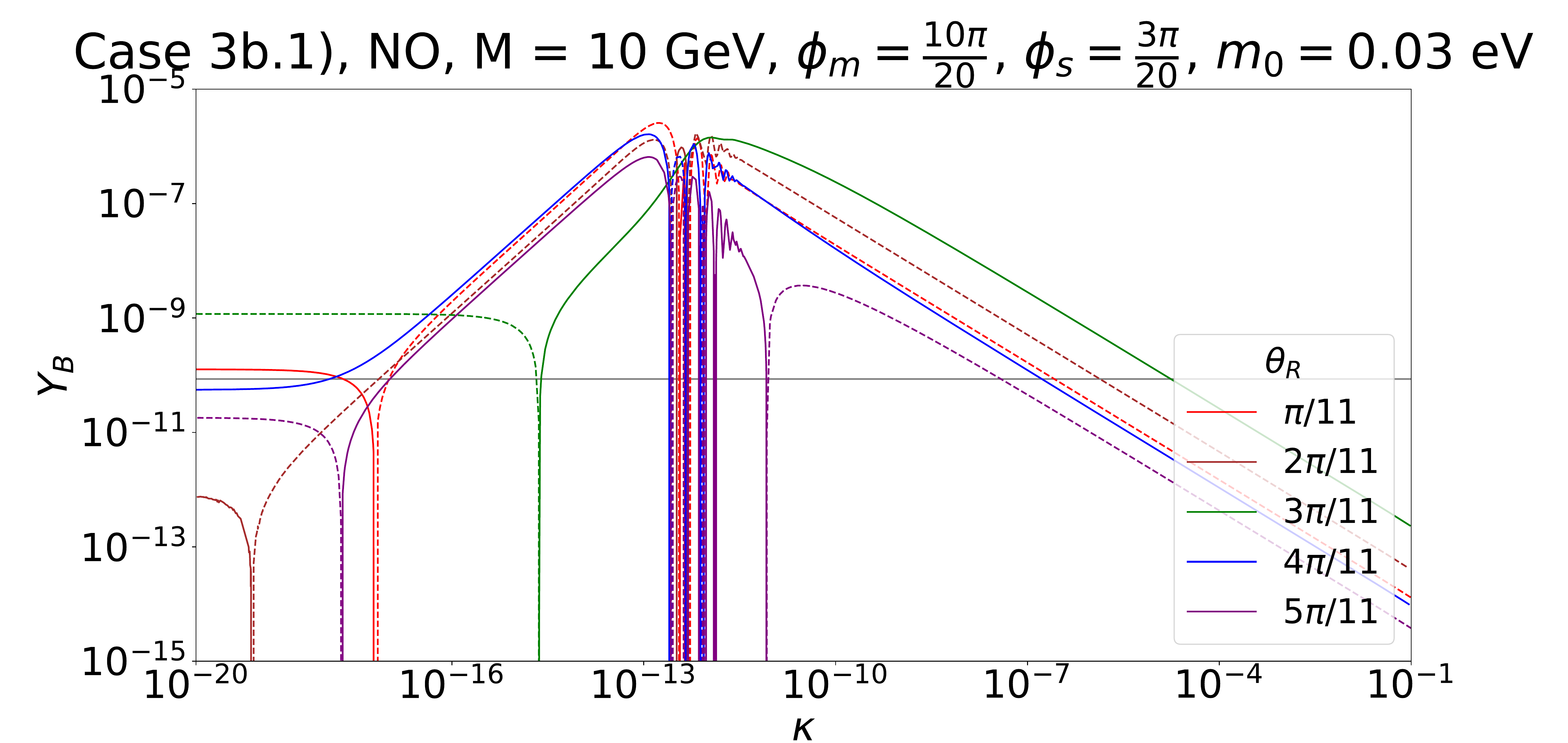}
		\caption{Vanishing initial conditions.}
		\label{NO kappa BAU 10 GeV IIIb odd massive}
	\end{subfigure}
	\begin{subfigure}{.5\textwidth}
		\includegraphics[width = \textwidth]{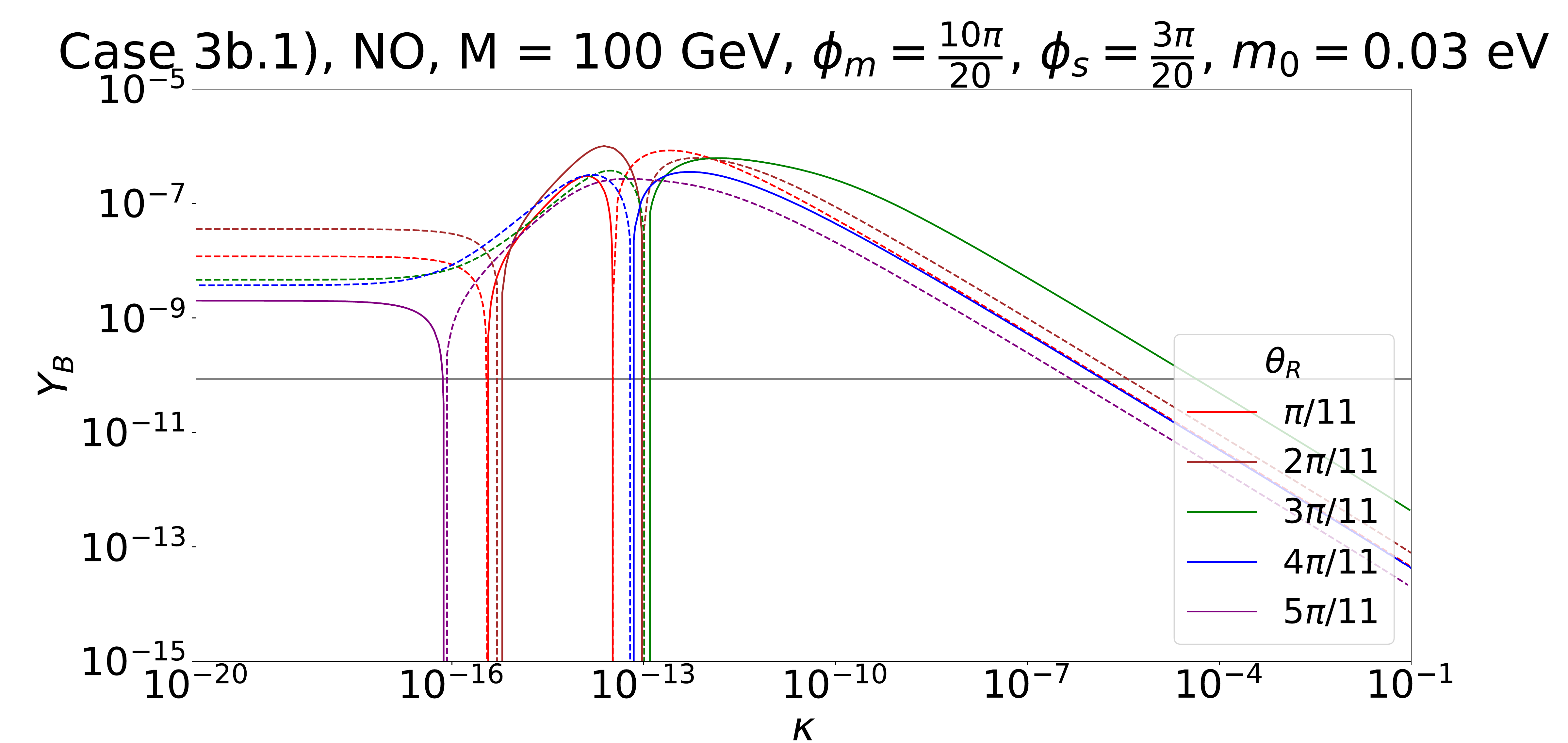}
		\caption{Vanishing initial conditions.}
	\end{subfigure}
	\begin{subfigure}{.5\textwidth}
		\includegraphics[width = \textwidth]{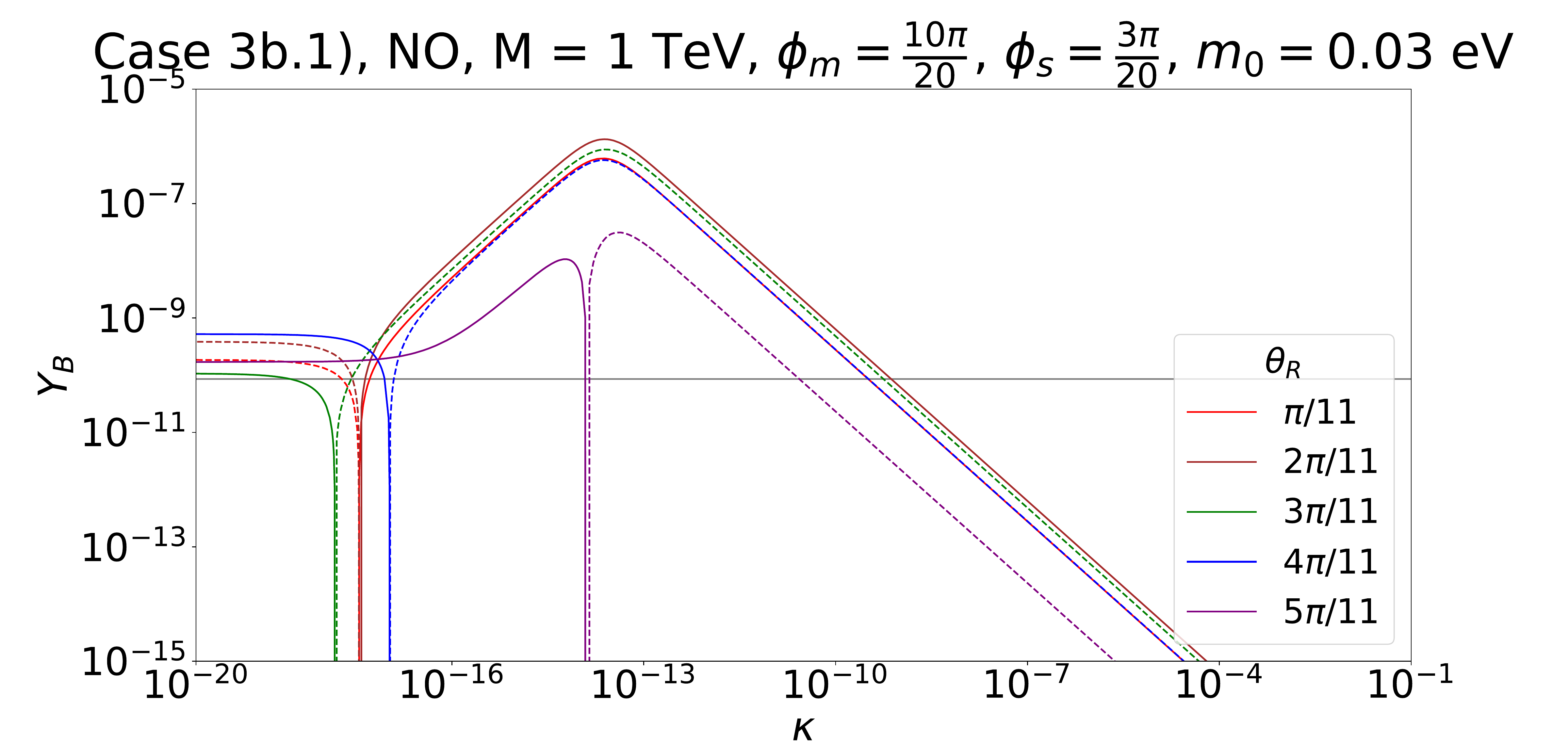}
		\caption{Vanishing initial conditions.}
	\end{subfigure}
	\caption{{\small {\bf Case 3 b.1)} $Y_B$ as function of $\kappa$ for a Majorana mass $M=1$ GeV (upper left plot), $M=10$ GeV (upper right plot), $M=100$ GeV (lower left plot) and $M=1$ TeV (lower right plot). For each of these choices, different values of $\theta_R$ have been
considered. The group theory parameters are fixed to $n=20$, $m=10$ and $s=3$. Light neutrino masses have NO with $m_0=0.03$ eV.
For the rest of the choices
see Fig.~\ref{NO kappa BAU Case IIIb s odd different masses massless lightest neutrino}.
}}
\label{NO kappa BAU Case IIIb s odd different masses massive lightest neutrino}
\end{figure}

\begin{figure}
	\centering
	\includegraphics[width = \textwidth]{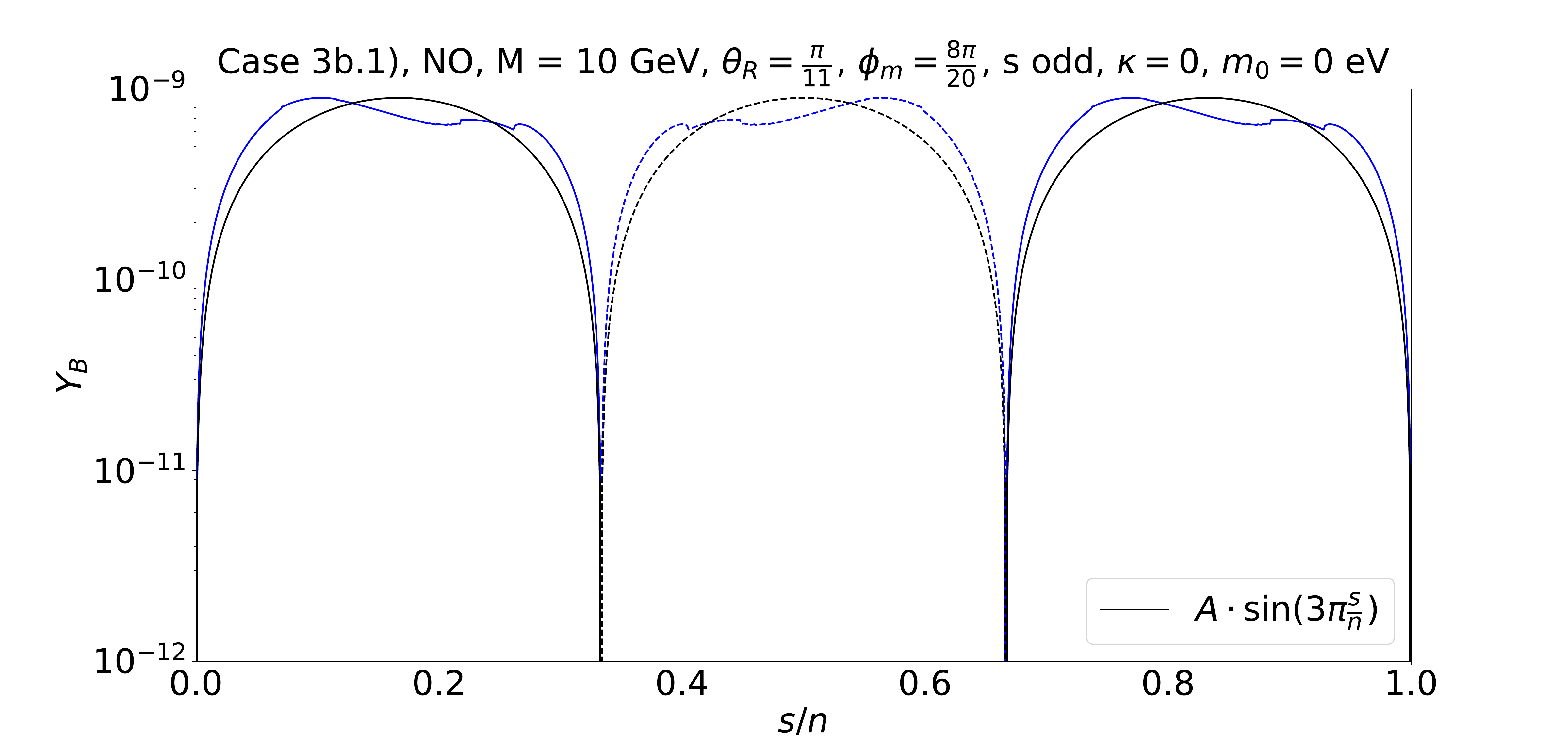}
	\caption{{\small {\bf Case 3 b.1)} $Y_B$ as function of $\frac{s}{n}$, treated as continuous parameter, for a Majorana mass $M=10$ GeV
		and a fixed value of $\theta_R$, $\theta_R=\frac{\pi}{11}$. The splitting $\kappa$ (and $\lambda$) vanishes.
		Light neutrino masses have strong NO.
		We assume $s$ being odd. The group theory parameters are fixed to $n=20$ and $m=8$.
		The analytic expectation, arising from the CP-violating combination $C_{\mathrm{DEG},\alpha}$,
		see Eq.~(\ref{eq:CDEGalpha_soddmeven_Case3}), is shown in black. The coefficient $A$ is given by $\mbox{Max}\big[|Y_B|\big]$.
		Both negative (dashed lines) as well as positive (continuous lines) values of the BAU are represented.
}}
\label{NO phis BAU 10 GeV phim820 caseIIIb}
\end{figure}

\begin{figure}
	\centering
	\includegraphics[width = \textwidth]{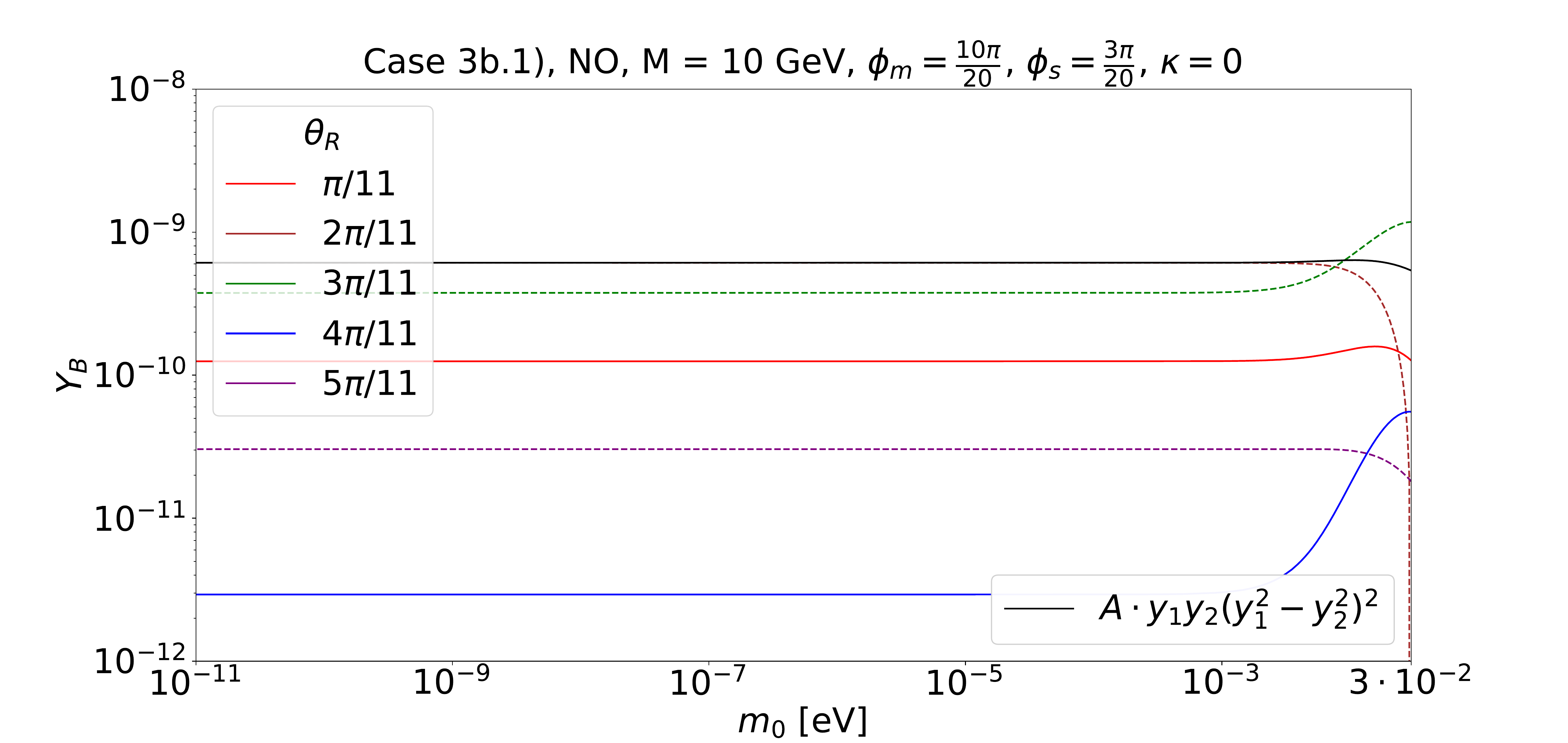}  
	\caption{{\small {\bf Case 3 b.1)} $Y_B$ as function of the lightest neutrino mass $m_0$ for a Majorana mass $M=10$ GeV and for different values of $\theta_R$ in absence of the splitting $\kappa$ (and $\lambda$). Light neutrino masses have NO. The group theory parameters are fixed to $n=20$, $m=10$ and $s=3$. For $\theta_R=\frac{2\pi}{11}$, we compare the numerical results (brown dashed curve) with the analytical expression $A\cdot y_1y_2(y_1^2-y_2^2)$ (continuous black line) with $A \approx -8.16 \cdot 10^{31}$. This expression reflects the dependence of the BAU, due to the CP-violating combination $C_{\mathrm{DEG},\alpha}$, on the couplings $y_f$, compare Eq.~\eqref{eq:CDEGalpha_soddmeven_Case3}.
		Both negative (dashed lines) as well as positive (continuous lines) values of the BAU are represented.
}}
\label{NO m0 BAU 10 GeV caseIIIb}
\end{figure}

\begin{figure}
	\centering
	\includegraphics[width=0.49\textwidth]{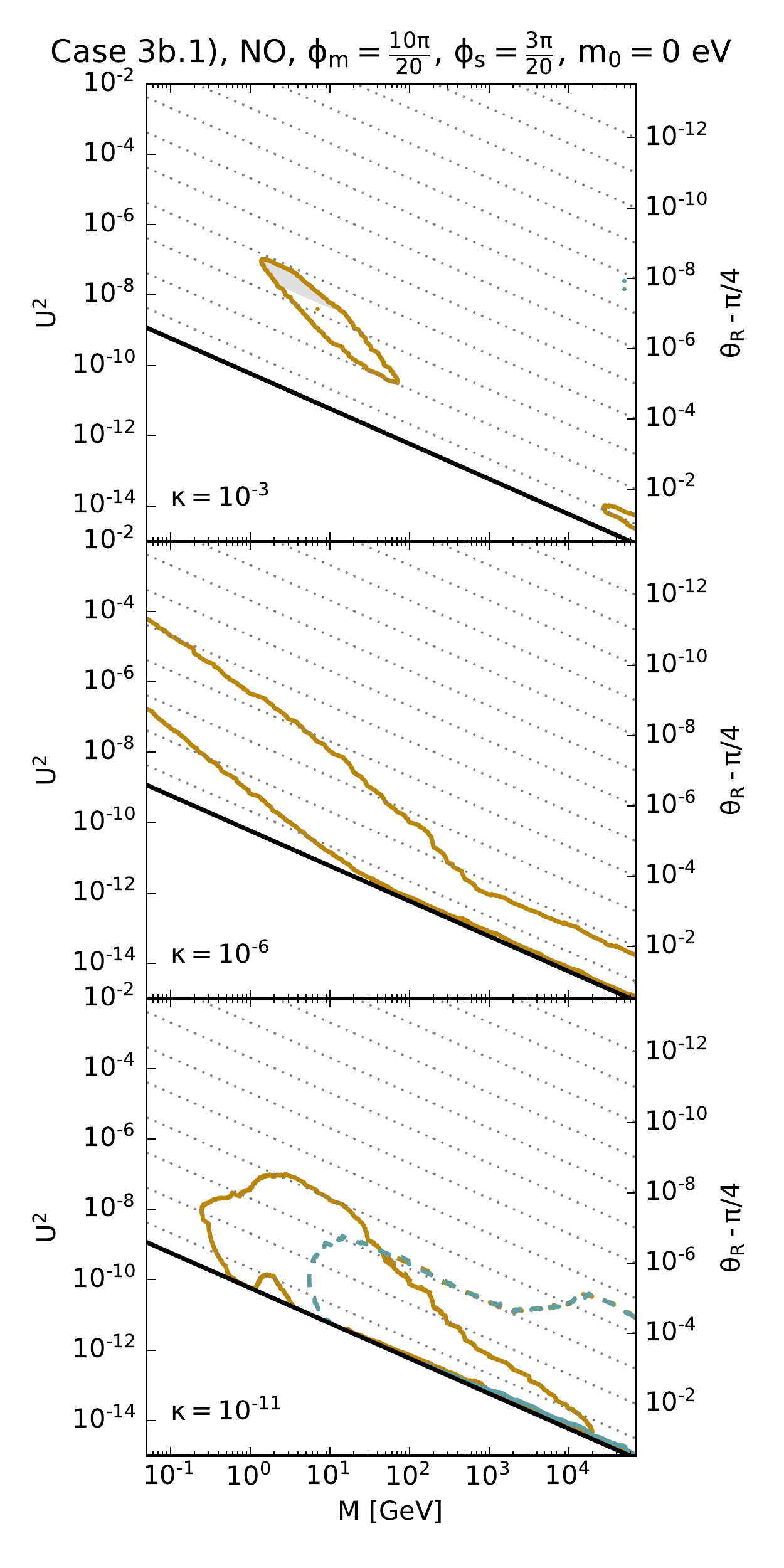}
	\includegraphics[width=0.49\textwidth]{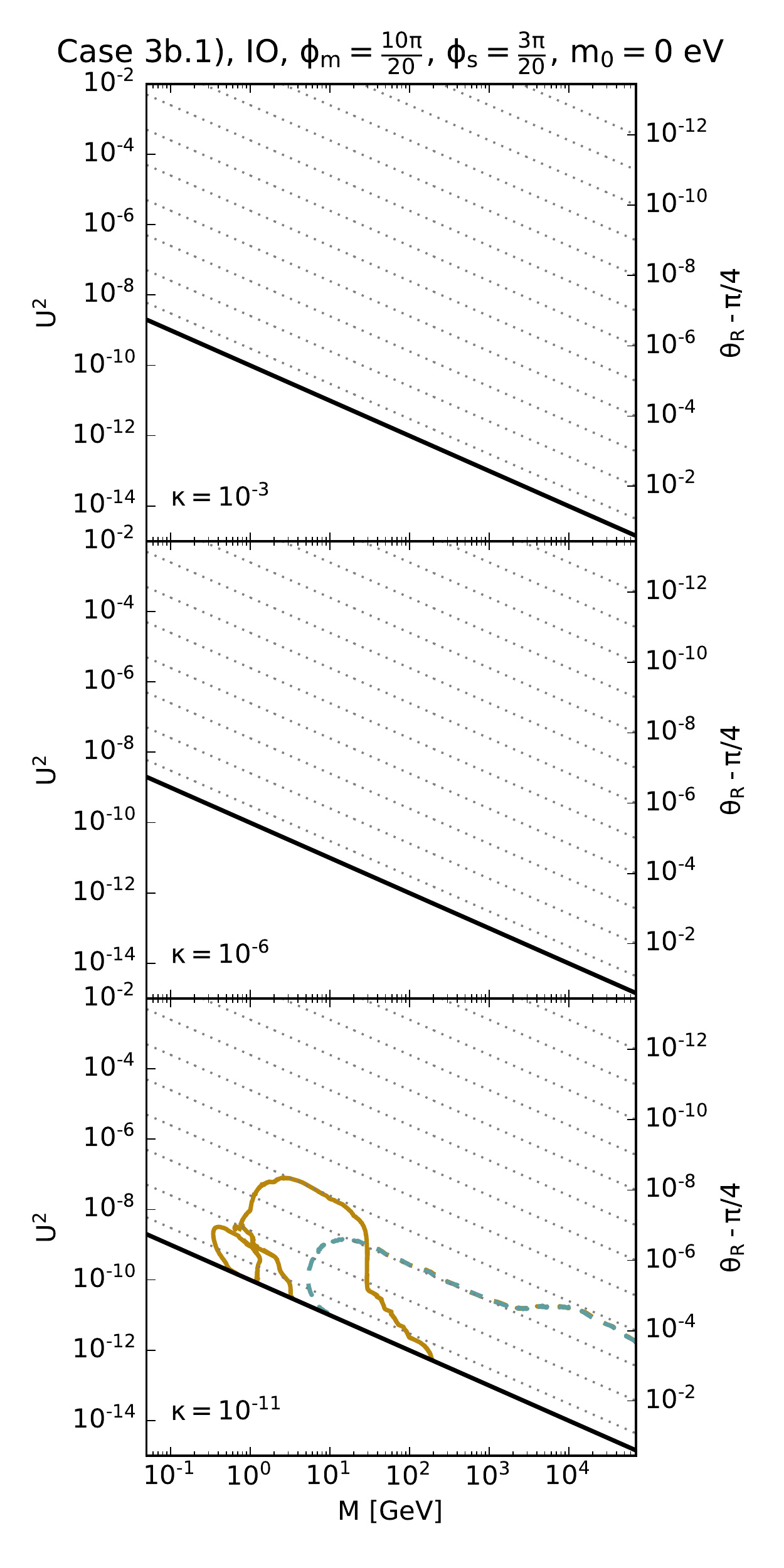}
	\caption{{\small {\bf Case 3 b.1)}
	Range of total mixing angle $U^2$ consistent with the observed BAU for heavy neutrino masses between $50$ MeV and $70$ TeV, as presented in Fig.~\ref{NO Mass U2 Case3}.
	The three rows correspond to different values of $\kappa$, $\kappa \in \{ 10^{-3}, 10^{-6}, 10^{-11} \}$, respectively. 
	The vanishing and thermal initial conditions are shown by the ochre and turquoise lines, respectively.
	The continuous (dashed) lines indicate positive (negative) values of the BAU.
	The grey shaded area shows the region in which a condition like the one in Eq.~\eqref{eq:kappacorrectionscriterion} is no longer satisfied.
}}
\label{fig:Case3b_MU2IC}
\end{figure}

\begin{figure}
	\centering
	\includegraphics[width = \textwidth]{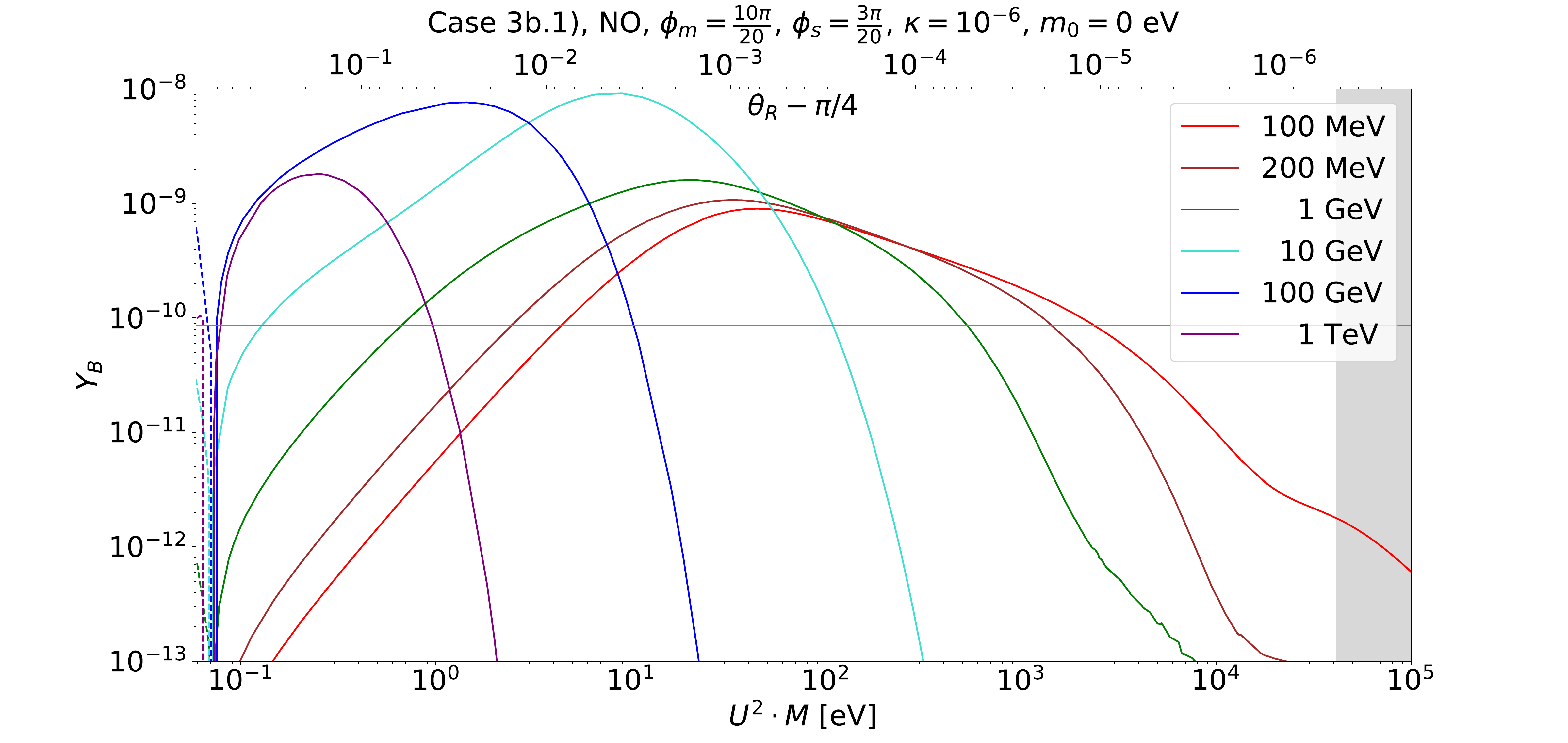}
	\caption{{\small \textbf{Case 3 b.1)} $Y_B$ as function of $U^2\cdot M$ for different values of the Majorana mass $M$.
		The splitting $\kappa$ is set to $10^{-6}$. Light neutrino masses follow strong NO.
		The group theory parameters $n$, $m$ and $s$ are fixed to $n=20$, $m=10$ and $s=3$.
Both negative (dashed lines) as well as positive (continuous lines) values of the BAU are represented. The grey line indicates the observed value of the BAU, $Y_B\approx 8.6 \cdot 10^{-11}$. The grey shaded area represents the region in which a criterion similar to the one in Eq.~\eqref{eq:kappacorrectionscriterion} is not fulfilled.}}
\label{NO BAU vs U2 10 GeV e-6 multiple curves caseIIIb odd}
\end{figure}

For Case 3 b.1), we also provide further plots. In Fig.~\ref{NO kappa BAU Case IIIb different masses massless lightest neutrino}, we show the BAU with respect to the splitting $\kappa$ for different values of the Majorana mass $M$ and of the angle $\theta_R$ and
for fixed values of the group theory parameters $n$, $m$ and $s$, $n=20$, $m=10$ and $s=4$. Light neutrino masses follow strong NO. These plots correspond to those shown in
Fig.~\ref{NO kappa BAU Case IIIb different masses massive lightest neutrino} in the main text for light neutrino masses with NO, but $m_0$ non-zero. Similarly, we show in Fig.~\ref{NO kappa BAU Case IIIb s odd different masses massive lightest neutrino} the BAU as function of $\kappa$ for an odd value of $s$, namely $s=3$,
and light neutrino masses with NO and $m_0=0.03$ eV. These plots should be compared to those in Fig.~\ref{NO kappa BAU Case IIIb s odd different masses massless lightest neutrino} in the main text, where we use the same choice of parameters, but assume that
the lightest neutrino mass vanishes. Fig.~\ref{NO phis BAU 10 GeV phim820 caseIIIb} illustrates the behaviour of the BAU with $s/n$, treated as continuous parameter, for a fixed value of $M$, $\theta_R$, light neutrino masses
with strong NO and vanishing splitting $\kappa$ (and $\lambda$). We assume $s$ to be odd.
Note that we use in this figure, $n=20$ and $m=8$ in contrast to $n=20$ and $m=10$ as well as $n=50$ and $m=24$, as
employed in Fig.~\ref{NO BAU phis caseIIIb odd} in the main text. Fig.~\ref{NO m0 BAU 10 GeV caseIIIb} captures the dependence of the BAU on the lightest neutrino mass $m_0$ for a fixed value of $M$, $n$, $m$ and $s$ as well as $\kappa=0$
($\lambda$ vanishes anyway) and five different values of $\theta_R$. Light neutrino masses follow NO. This figure can be compared to Fig.~\ref{NO thetaR/m0 BAU 100 GeV caseII odd no mass splittings}, plot (a), for Case 2), found in the main text, and completes the discussion of Case 3 b.1), $m$ even and $s$
odd and no splitting. We also show dedicated plots for each value of the splitting $\kappa$ in Fig.~\ref{fig:Case3b_MU2IC} in order to facilitate the comparison between the results for vanishing and thermal initial conditions. These plots are combined in
Fig.~\ref{NO Mass U2 Case3} in the main text, where the results for the different values of $\kappa$ are shown in different colours.
 Lastly, in Fig.~\ref{NO BAU vs U2 10 GeV e-6 multiple curves caseIIIb odd}, we display the resulting BAU as function of $U^2 \cdot M$ for several different values of the Majorana mass $M$ and fixed values of $\kappa$ as well
as the group theory parameters $n$, $m$ and $s$ and also a light neutrino mass spectrum with strong NO. This figure complements the left plot in Fig.~\ref{NO Mass U2 Case3} in the main text. Furthermore,
it can be compared to corresponding figures for Case 1), see Fig.~\ref{NO BAU vs U2 10 GeV e-6 multiple curves} in the main text, and for Case 2), see Fig.~\ref{NO BAU vs U2 10 GeV e-6 multiple curves caseII} in appendix~\ref{appF2}.

\newpage

\section{Additional formulae for CP-violating combinations}
\label{appG}

In this appendix, some of the formulae for CP-violating combinations for Case 2) and Case 3 a) and Case 3 b.1), respectively, are collected
which turn out to be rather lengthy.

\subsection{Case 2)}
\label{appG1}

For $s$ even and $t$ odd, the expression for the CP-violating combination $C_{\mathrm{LNV},\alpha}$ reads
\small
\begin{eqnarray}\nonumber
&&C_{\mathrm{LNV},\alpha} = \frac{1}{9} \, M^2 \, \Big( 2 \, y_2 \, \Delta \sigma_{12}
\\ \nonumber
&&\Big[(y_3 \, (\Delta y^2_{12}+\Delta y^2_{13}\, \sin 2 \, \theta_R)\, \sin\theta_L\, cs_{R,-} + y_1 \, (\Delta y^2_{23}-\Delta y^2_{13}\, \sin 2 \, \theta_R)\, \cos\theta_L\, cs_{R,+})  \, \cos \frac{\phi_{u,\alpha}}{2} \, \cos \frac{\phi_v}{2}
\\  \nonumber
&&- (y_1 \, (\Delta y^2_{23}+\Delta y^2_{13}\, \sin 2 \, \theta_R)\,\sin\theta_L \, cs_{R,-} + y_3 \, (\Delta y^2_{12}-\Delta y^2_{13}\, \sin 2 \, \theta_R)\, \cos\theta_L\, cs_{R,+})  \, \sin \frac{\phi_{u,\alpha}}{2} \, \cos \frac{\phi_v}{2}
\\  \nonumber
&&+(y_1 \, (\Delta y^2_{23}+\Delta y^2_{13}\, \sin 2 \, \theta_R)\, \cos\theta_L\, cs_{R,-}  - y_3 \, (\Delta y^2_{12}-\Delta y^2_{13}\, \sin 2 \, \theta_R)\, \sin\theta_L\, cs_{R,+})  \, \cos \frac{\phi_{u,\alpha}}{2} \, \sin \frac{\phi_v}{2}
\\  \nonumber
&&- (y_3 \, (\Delta y^2_{12}+\Delta y^2_{13}\, \sin 2 \, \theta_R)\, \cos\theta_L\, cs_{R,-}  - y_1 \, (\Delta y^2_{23}-\Delta y^2_{13}\, \sin 2 \, \theta_R)\, \sin\theta_L\, cs_{R,+}) \, \sin \frac{\phi_{u,\alpha}}{2} \, \sin \frac{\phi_v}{2}\Big]
\\ \nonumber
&&-  \Delta y^2_{13} \, [\Delta\sigma_{12}+ 6 \, \Delta\sigma_{23}] \, [((y_1^2+y_3^2)\, \cos 2 \, \theta_L \, \sin 4 \, \theta_R- 2 \, y_1 \, y_3 \, \sin 2 \, \theta_L\, \cos 4 \, \theta_R) \, \cos \phi_{u,\alpha} + \Delta y^2_{13} \, \sin 4 \, \theta_R] \Big).\\&&
\end{eqnarray}

\normalsize

\subsection{Case 3 a) and Case 3 b.1)}
\label{appG2}

For $s$ odd and $m$ even, the form of the CP-violating combination $C_{\mathrm{LNV},\alpha}$ is given by
\begin{eqnarray}\nonumber
&&\!\!\!\!\!\!\!\!\!C_{\mathrm{LNV},\alpha} = \frac{1}{9\, \sqrt{2}} \, M^2
\\ \nonumber
&&\!\!\!\!\!\!\!\!\!\;\Big(
\Delta y^2_{12} \, [\Delta \sigma_{12}+3\, \Delta \sigma_{23}] \, [-2 \, y_1 \, y_2\,\sin 2 \, \theta_L\, \cos 4 \, \theta_R +(\Delta y^2_{12}+(y_1^2+y_2^2)\, \cos 2 \, \theta_L)\, \sin 4 \, \theta_R] \, \cos 2 \, \phi_{m,\alpha}
\\  \nonumber
&&\!\!\!\!\!\!\!\!\!+6 \, \sqrt{2} \, y_3 \, \Delta\sigma_{23} \, [y_1\, \cos \theta_L \, (\Delta y^2_{23}+\Delta y^2_{12}\, \cos 2 \, \theta_R)\, \cos \theta_R \\ \nonumber
&&\;\;\;\;\;\;\;\;\;\;\;\;\;\;\;\;\;\;\;\;\;\;\;\;\;\;\;\;\;\;\;\;\;\;\;\;\;\;\;\;\;\;\;\;\;\;\;\;\;\;\;\;\;\;\;\;\;\;\;\;\;\;\;\;\;\;\;+ y_2 \, \sin \theta_L\, (\Delta y^2_{13}+\Delta y^2_{12}\, \cos 2 \, \theta_R) \, \sin \theta_R] \, \sin 2 \, \phi_{m,\alpha}
\\  \nonumber
&&\!\!\!\!\!\!\!\!\!+12 \, \sqrt{2}\, y_3 \, \Delta \sigma_{23} \,
[y_2 \, \cos \theta_L \, (\Delta y^2_{13}-\Delta y^2_{12}\, \cos 2 \, \theta_R) \, \cos \theta_R \\ \nonumber
&&\;\;\;\;\;\;\;\;\;\;\;\;\;\;\;\;\;\;\;\;\;\;\;\;\;\;\;\;\;\;\;\;\;\;\;\;\;\;\;\;\;\;\;\;\;\;\;\;\;\;\;\;\;\;\;\;\;+y_1 \,  \sin \theta_L \, (\Delta y^2_{23}-\Delta y^2_{12}\, \cos 2 \, \theta_R) \,\sin \theta_R] \, \sin \phi_{m,\alpha} \, \sin 3 \, \phi_s
\\  \nonumber
&&\!\!\!\!\!\!\!\!\!+12\, y_3 \, \Delta \sigma_{23} \,
[-y_1 \, \sin \theta_L \, (\Delta y^2_{23}+\Delta y^2_{12}\, \cos 2 \, \theta_R) \, \cos \theta_R
\\ \nonumber
&&\;\;\;\;\;\;\;\;\;\;\;\;\;\;\;\;\;\;\;\;\;\;\;\;\;\;\;\;\;\;\;\;\;\;\;\;\;\;\;\;\;\;\;\;\;\;\;\;\;\;\;\;\;\;\;\;+y_2 \, \cos \theta_L \, (\Delta y^2_{13}+\Delta y^2_{12}\, \cos 2 \, \theta_R)\, \sin \theta_R]\, \sin \phi_{m,\alpha} \, \cos 3 \, \phi_s
\\  \nonumber
&&\!\!\!\!\!\!\!\!\!- 2 \, \sqrt{2}  \, \Delta y^2_{12} \, [\Delta \sigma_{12}+3\, \Delta \sigma_{23}] \, [2 \, y_1 \, y_2 \, \cos 2 \, \theta_L \, \cos 4 \, \theta_R+(y_1^2+y_2^2)\, \sin 2 \, \theta_L\, \sin 4 \, \theta_R]\, \cos \phi_{m,\alpha} \, \cos 3 \, \phi_s
\\
&&\!\!\!\!\!\!\!\!\!+ 2 \, y_1 \, y_2 \, \Delta y^2_{12} \, [\Delta \sigma_{12}+3\, \Delta \sigma_{23}]\,  \sin 2 \, \theta_R \, \cos \phi_{m,\alpha} \, \sin 3 \, \phi_s
+ 2 \, (\Delta y^2_{12})^2 \, [\Delta \sigma_{12}+3\, \Delta \sigma_{23}] \, \sin 4 \, \theta_R
\Big).
\end{eqnarray}

\bibliographystyle{JHEP}
\bibliography{refs}

\providecommand{\href}[2]{#2}\begingroup\raggedright\begin{thebibliography}{100}

\bibitem{ParticleDataGroup:2020ssz}
{\scshape Particle Data Group} collaboration, P.~A. Zyla et~al., \emph{{Review
  of Particle Physics}},
  \href{http://dx.doi.org/10.1093/ptep/ptaa104}{\emph{PTEP} {\bf 2020} (2020)
  083C01}.

\bibitem{Cai:2017jrq}
Y.~Cai, J.~Herrero-Garc\'\i{}a, M.~A. Schmidt, A.~Vicente and R.~R. Volkas,
  \emph{{From the trees to the forest: a review of radiative neutrino mass
  models}}, \href{http://dx.doi.org/10.3389/fphy.2017.00063}{\emph{Front. in
  Phys.} {\bf 5} (2017) 63}, [\href{https://arxiv.org/abs/1706.08524}{{\tt
  1706.08524}}].

\bibitem{Asaka:2005pn}
T.~Asaka and M.~Shaposhnikov, \emph{{The $\nu$MSM, dark matter and baryon
  asymmetry of the universe}},
  \href{http://dx.doi.org/10.1016/j.physletb.2005.06.020}{\emph{Phys. Lett. B}
  {\bf 620} (2005) 17--26}, [\href{https://arxiv.org/abs/hep-ph/0505013}{{\tt
  hep-ph/0505013}}].

\bibitem{Asaka:2005an}
T.~Asaka, S.~Blanchet and M.~Shaposhnikov, \emph{{The $\nu$MSM, dark matter and
  neutrino masses}},
  \href{http://dx.doi.org/10.1016/j.physletb.2005.09.070}{\emph{Phys. Lett. B}
  {\bf 631} (2005) 151--156}, [\href{https://arxiv.org/abs/hep-ph/0503065}{{\tt
  hep-ph/0503065}}].

\bibitem{Chun:2017spz}
E.~J. Chun et~al., \emph{{Probing Leptogenesis}},
  \href{http://dx.doi.org/10.1142/S0217751X18420058}{\emph{Int. J. Mod. Phys.
  A} {\bf 33} (2018) 1842005}, [\href{https://arxiv.org/abs/1711.02865}{{\tt
  1711.02865}}].

\bibitem{Abdullahi:2022jlv}
A.~M. Abdullahi et~al., \emph{{The Present and Future Status of Heavy Neutral
  Leptons}},  in \emph{{2022 Snowmass Summer Study}}, 3, 2022.
\newblock \href{https://arxiv.org/abs/2203.08039}{{\tt 2203.08039}}.

\bibitem{Ishimori:2010au}
H.~Ishimori, T.~Kobayashi, H.~Ohki, Y.~Shimizu, H.~Okada and M.~Tanimoto,
  \emph{{Non-Abelian Discrete Symmetries in Particle Physics}},
  \href{http://dx.doi.org/10.1143/PTPS.183.1}{\emph{Prog. Theor. Phys. Suppl.}
  {\bf 183} (2010) 1--163}, [\href{https://arxiv.org/abs/1003.3552}{{\tt
  1003.3552}}].

\bibitem{King:2013eh}
S.~F. King and C.~Luhn, \emph{{Neutrino Mass and Mixing with Discrete
  Symmetry}},
  \href{http://dx.doi.org/10.1088/0034-4885/76/5/056201}{\emph{Rept. Prog.
  Phys.} {\bf 76} (2013) 056201}, [\href{https://arxiv.org/abs/1301.1340}{{\tt
  1301.1340}}].

\bibitem{Feruglio:2019ybq}
F.~Feruglio and A.~Romanino, \emph{{Lepton flavor symmetries}},
  \href{http://dx.doi.org/10.1103/RevModPhys.93.015007}{\emph{Rev. Mod. Phys.}
  {\bf 93} (2021) 015007}, [\href{https://arxiv.org/abs/1912.06028}{{\tt
  1912.06028}}].

\bibitem{Grimus:2011fk}
W.~Grimus and P.~O. Ludl, \emph{{Finite flavour groups of fermions}},
  \href{http://dx.doi.org/10.1088/1751-8113/45/23/233001}{\emph{J. Phys. A}
  {\bf 45} (2012) 233001}, [\href{https://arxiv.org/abs/1110.6376}{{\tt
  1110.6376}}].

\bibitem{Feruglio:2012cw}
F.~Feruglio, C.~Hagedorn and R.~Ziegler, \emph{{Lepton Mixing Parameters from
  Discrete and CP Symmetries}},
  \href{http://dx.doi.org/10.1007/JHEP07(2013)027}{\emph{JHEP} {\bf 07} (2013)
  027}, [\href{https://arxiv.org/abs/1211.5560}{{\tt 1211.5560}}].

\bibitem{Holthausen:2012dk}
M.~Holthausen, M.~Lindner and M.~A. Schmidt, \emph{{CP and Discrete Flavour
  Symmetries}}, \href{http://dx.doi.org/10.1007/JHEP04(2013)122}{\emph{JHEP}
  {\bf 04} (2013) 122}, [\href{https://arxiv.org/abs/1211.6953}{{\tt
  1211.6953}}].

\bibitem{Chen:2014tpa}
M.-C. Chen, M.~Fallbacher, K.~T. Mahanthappa, M.~Ratz and A.~Trautner,
  \emph{{CP Violation from Finite Groups}},
  \href{http://dx.doi.org/10.1016/j.nuclphysb.2014.03.023}{\emph{Nucl. Phys. B}
  {\bf 883} (2014) 267--305}, [\href{https://arxiv.org/abs/1402.0507}{{\tt
  1402.0507}}].

\bibitem{Grimus:1995zi}
W.~Grimus and M.~N. Rebelo, \emph{{Automorphisms in gauge theories and the
  definition of CP and P}},
  \href{http://dx.doi.org/10.1016/S0370-1573(96)00030-0}{\emph{Phys. Rept.}
  {\bf 281} (1997) 239--308}, [\href{https://arxiv.org/abs/hep-ph/9506272}{{\tt
  hep-ph/9506272}}].

\bibitem{Ecker:1983hz}
G.~Ecker, W.~Grimus and H.~Neufeld, \emph{{Spontaneous {CP} Violation in
  Left-right Symmetric Gauge Theories}},
  \href{http://dx.doi.org/10.1016/0550-3213(84)90373-0}{\emph{Nucl. Phys. B}
  {\bf 247} (1984) 70--82}.

\bibitem{Ecker:1987qp}
G.~Ecker, W.~Grimus and H.~Neufeld, \emph{{A Standard Form for Generalized {CP}
  Transformations}},
  \href{http://dx.doi.org/10.1088/0305-4470/20/12/010}{\emph{J. Phys. A} {\bf
  20} (1987) L807}.

\bibitem{Neufeld:1987wa}
H.~Neufeld, W.~Grimus and G.~Ecker, \emph{{Generalized {CP} Invariance, Neutral
  Flavor Conservation and the Structure of the Mixing Matrix}},
  \href{http://dx.doi.org/10.1142/S0217751X88000254}{\emph{Int. J. Mod. Phys.
  A} {\bf 3} (1988) 603--616}.

\bibitem{Harrison:2002kp}
P.~F. Harrison and W.~G. Scott, \emph{{Symmetries and generalizations of
  tri-bimaximal neutrino mixing}},
  \href{http://dx.doi.org/10.1016/S0370-2693(02)01753-7}{\emph{Phys. Lett. B}
  {\bf 535} (2002) 163--169}, [\href{https://arxiv.org/abs/hep-ph/0203209}{{\tt
  hep-ph/0203209}}].

\bibitem{Grimus:2003yn}
W.~Grimus and L.~Lavoura, \emph{{A Nonstandard CP transformation leading to
  maximal atmospheric neutrino mixing}},
  \href{http://dx.doi.org/10.1016/j.physletb.2003.10.075}{\emph{Phys. Lett. B}
  {\bf 579} (2004) 113--122}, [\href{https://arxiv.org/abs/hep-ph/0305309}{{\tt
  hep-ph/0305309}}].

\bibitem{Luhn:2007uq}
C.~Luhn, S.~Nasri and P.~Ramond, \emph{{The Flavor group $\Delta(3\,n^2)$}},
  \href{http://dx.doi.org/10.1063/1.2734865}{\emph{J. Math. Phys.} {\bf 48}
  (2007) 073501}, [\href{https://arxiv.org/abs/hep-th/0701188}{{\tt
  hep-th/0701188}}].

\bibitem{Escobar:2008vc}
J.~A. Escobar and C.~Luhn, \emph{{The Flavor Group $\Delta(6\,n^2)$}},
  \href{http://dx.doi.org/10.1063/1.3046563}{\emph{J. Math. Phys.} {\bf 50}
  (2009) 013524}, [\href{https://arxiv.org/abs/0809.0639}{{\tt 0809.0639}}].

\bibitem{Hagedorn:2014wha}
C.~Hagedorn, A.~Meroni and E.~Molinaro, \emph{{Lepton mixing from
  $\Delta(3\,n^2)$ and $\Delta(6\,n^2)$ and CP}},
  \href{http://dx.doi.org/10.1016/j.nuclphysb.2014.12.013}{\emph{Nucl. Phys. B}
  {\bf 891} (2015) 499--557}, [\href{https://arxiv.org/abs/1408.7118}{{\tt
  1408.7118}}].

\bibitem{Fukugita:1986hr}
M.~Fukugita and T.~Yanagida, \emph{{Baryogenesis Without Grand Unification}},
  \href{http://dx.doi.org/10.1016/0370-2693(86)91126-3}{\emph{Phys. Lett. B}
  {\bf 174} (1986) 45--47}.

\bibitem{Kuzmin:1985mm}
V.~A. Kuzmin, V.~A. Rubakov and M.~E. Shaposhnikov, \emph{{On the Anomalous
  Electroweak Baryon Number Nonconservation in the Early Universe}},
  \href{http://dx.doi.org/10.1016/0370-2693(85)91028-7}{\emph{Phys. Lett. B}
  {\bf 155} (1985) 36}.

\bibitem{Davidson:2008bu}
S.~Davidson, E.~Nardi and Y.~Nir, \emph{{Leptogenesis}},
  \href{http://dx.doi.org/10.1016/j.physrep.2008.06.002}{\emph{Phys. Rept.}
  {\bf 466} (2008) 105--177}, [\href{https://arxiv.org/abs/0802.2962}{{\tt
  0802.2962}}].

\bibitem{Akhmedov:1998qx}
E.~K. Akhmedov, V.~A. Rubakov and A.~Y. Smirnov, \emph{{Baryogenesis via
  neutrino oscillations}},
  \href{http://dx.doi.org/10.1103/PhysRevLett.81.1359}{\emph{Phys. Rev. Lett.}
  {\bf 81} (1998) 1359--1362},
  [\href{https://arxiv.org/abs/hep-ph/9803255}{{\tt hep-ph/9803255}}].

\bibitem{Pilaftsis:2003gt}
A.~Pilaftsis and T.~E.~J. Underwood, \emph{{Resonant leptogenesis}},
  \href{http://dx.doi.org/10.1016/j.nuclphysb.2004.05.029}{\emph{Nucl. Phys. B}
  {\bf 692} (2004) 303--345}, [\href{https://arxiv.org/abs/hep-ph/0309342}{{\tt
  hep-ph/0309342}}].

\bibitem{Pilaftsis:2005rv}
A.~Pilaftsis and T.~E.~J. Underwood, \emph{{Electroweak-scale resonant
  leptogenesis}},
  \href{http://dx.doi.org/10.1103/PhysRevD.72.113001}{\emph{Phys. Rev. D} {\bf
  72} (2005) 113001}, [\href{https://arxiv.org/abs/hep-ph/0506107}{{\tt
  hep-ph/0506107}}].

\bibitem{Hagedorn:2017wjy}
C.~Hagedorn, R.~N. Mohapatra, E.~Molinaro, C.~C. Nishi and S.~T. Petcov,
  \emph{{CP Violation in the Lepton Sector and Implications for Leptogenesis}},
  \href{http://dx.doi.org/10.1142/S0217751X1842006X}{\emph{Int. J. Mod. Phys.
  A} {\bf 33} (2018) 1842006}, [\href{https://arxiv.org/abs/1711.02866}{{\tt
  1711.02866}}].

\bibitem{Jenkins:2008rb}
E.~E. Jenkins and A.~V. Manohar, \emph{{Tribimaximal Mixing, Leptogenesis, and
  $\theta_{13}$}},
  \href{http://dx.doi.org/10.1016/j.physletb.2008.08.028}{\emph{Phys. Lett. B}
  {\bf 668} (2008) 210--215}, [\href{https://arxiv.org/abs/0807.4176}{{\tt
  0807.4176}}].

\bibitem{Bertuzzo:2009im}
E.~Bertuzzo, P.~Di~Bari, F.~Feruglio and E.~Nardi, \emph{{Flavor symmetries,
  leptogenesis and the absolute neutrino mass scale}},
  \href{http://dx.doi.org/10.1088/1126-6708/2009/11/036}{\emph{JHEP} {\bf 11}
  (2009) 036}, [\href{https://arxiv.org/abs/0908.0161}{{\tt 0908.0161}}].

\bibitem{Hagedorn:2009jy}
C.~Hagedorn, E.~Molinaro and S.~T. Petcov, \emph{{Majorana Phases and
  Leptogenesis in See-Saw Models with $A_4$ Symmetry}},
  \href{http://dx.doi.org/10.1088/1126-6708/2009/09/115}{\emph{JHEP} {\bf 09}
  (2009) 115}, [\href{https://arxiv.org/abs/0908.0240}{{\tt 0908.0240}}].

\bibitem{AristizabalSierra:2009ex}
D.~Aristizabal~Sierra, F.~Bazzocchi, I.~de~Medeiros~Varzielas, L.~Merlo and
  S.~Morisi, \emph{{Tri-Bimaximal Lepton Mixing and Leptogenesis}},
  \href{http://dx.doi.org/10.1016/j.nuclphysb.2009.10.009}{\emph{Nucl. Phys. B}
  {\bf 827} (2010) 34--58}, [\href{https://arxiv.org/abs/0908.0907}{{\tt
  0908.0907}}].

\bibitem{Mohapatra:2015gwa}
R.~N. Mohapatra and C.~C. Nishi, \emph{{Implications of
  \ensuremath{\mu}-\ensuremath{\tau} flavored CP symmetry of leptons}},
  \href{http://dx.doi.org/10.1007/JHEP08(2015)092}{\emph{JHEP} {\bf 08} (2015)
  092}, [\href{https://arxiv.org/abs/1506.06788}{{\tt 1506.06788}}].

\bibitem{Hagedorn:2016lva}
C.~Hagedorn and E.~Molinaro, \emph{{Flavor and CP symmetries for leptogenesis
  and 0\ensuremath{\nu}\ensuremath{\beta}\ensuremath{\beta} decay}},
  \href{http://dx.doi.org/10.1016/j.nuclphysb.2017.03.015}{\emph{Nucl. Phys. B}
  {\bf 919} (2017) 404--469}, [\href{https://arxiv.org/abs/1602.04206}{{\tt
  1602.04206}}].

\bibitem{Chen:2016ptr}
P.~Chen, G.-J. Ding and S.~F. King, \emph{{Leptogenesis and residual CP
  symmetry}}, \href{http://dx.doi.org/10.1007/JHEP03(2016)206}{\emph{JHEP} {\bf
  03} (2016) 206}, [\href{https://arxiv.org/abs/1602.03873}{{\tt 1602.03873}}].

\bibitem{Fong:2021tqj}
C.~S. Fong, M.~H. Rahat and S.~Saad, \emph{{Low-scale resonant leptogenesis in
  SU(5) GUT with $T_{13}$ family symmetry}},
  \href{http://dx.doi.org/10.1103/PhysRevD.104.095028}{\emph{Phys. Rev. D} {\bf
  104} (2021) 095028}, [\href{https://arxiv.org/abs/2103.14691}{{\tt
  2103.14691}}].

\bibitem{Chauhan:2021xus}
G.~Chauhan and P.~S.~B. Dev, \emph{{Resonant Leptogenesis, Collider Signals and
  Neutrinoless Double Beta Decay from Flavor and CP Symmetries}},
  \href{https://arxiv.org/abs/2112.09710}{{\tt 2112.09710}}.

\bibitem{Klaric:2020phc}
J.~Klari\'c, M.~Shaposhnikov and I.~Timiryasov, \emph{{Uniting Low-Scale
  Leptogenesis Mechanisms}},
  \href{http://dx.doi.org/10.1103/PhysRevLett.127.111802}{\emph{Phys. Rev.
  Lett.} {\bf 127} (2021) 111802},
  [\href{https://arxiv.org/abs/2008.13771}{{\tt 2008.13771}}].

\bibitem{Ghiglieri:2017csp}
J.~Ghiglieri and M.~Laine, \emph{{GeV-scale hot sterile neutrino oscillations:
  a numerical solution}},
  \href{http://dx.doi.org/10.1007/JHEP02(2018)078}{\emph{JHEP} {\bf 02} (2018)
  078}, [\href{https://arxiv.org/abs/1711.08469}{{\tt 1711.08469}}].

\bibitem{Esteban:2020cvm}
I.~Esteban, M.~C. Gonzalez-Garcia, M.~Maltoni, T.~Schwetz and A.~Zhou,
  \emph{{The fate of hints: updated global analysis of three-flavor neutrino
  oscillations}}, \href{http://dx.doi.org/10.1007/JHEP09(2020)178}{\emph{JHEP}
  {\bf 09} (2020) 178}, [\href{https://arxiv.org/abs/2007.14792}{{\tt
  2007.14792}}].

\bibitem{Minkowski:1977sc}
P.~Minkowski, \emph{{$\mu \to e\gamma$ at a Rate of One Out of $10^{9}$ Muon
  Decays?}}, \href{http://dx.doi.org/10.1016/0370-2693(77)90435-X}{\emph{Phys.
  Lett. B} {\bf 67} (1977) 421--428}.

\bibitem{Glashow:1979nm}
S.~L. Glashow, \emph{{The Future of Elementary Particle Physics}},
  \href{http://dx.doi.org/10.1007/978-1-4684-7197-7_15}{\emph{NATO Sci. Ser. B}
  {\bf 61} (1980) 687}.

\bibitem{Gell-Mann:1979vob}
M.~Gell-Mann, P.~Ramond and R.~Slansky, \emph{{Complex Spinors and Unified
  Theories}}, {\emph{Conf. Proc. C} {\bf 790927} (1979) 315--321},
  [\href{https://arxiv.org/abs/1306.4669}{{\tt 1306.4669}}].

\bibitem{Mohapatra:1979ia}
R.~N. Mohapatra and G.~Senjanović, \emph{{Neutrino Mass and Spontaneous Parity
  Nonconservation}},
  \href{http://dx.doi.org/10.1103/PhysRevLett.44.912}{\emph{Phys. Rev. Lett.}
  {\bf 44} (1980) 912}.

\bibitem{Yanagida:1980xy}
T.~Yanagida, \emph{{Horizontal Symmetry and Masses of Neutrinos}},
  \href{http://dx.doi.org/10.1143/PTP.64.1103}{\emph{Prog. Theor. Phys.} {\bf
  64} (1980) 1103}.

\bibitem{Schechter:1980gr}
J.~Schechter and J.~W.~F. Valle, \emph{{Neutrino Masses in $SU(2) \times U(1)$
  Theories}}, \href{http://dx.doi.org/10.1103/PhysRevD.22.2227}{\emph{Phys.
  Rev. D} {\bf 22} (1980) 2227}.

\bibitem{Canetti:2012zc}
L.~Canetti, M.~Drewes and M.~Shaposhnikov, \emph{{Matter and Antimatter in the
  Universe}}, \href{http://dx.doi.org/10.1088/1367-2630/14/9/095012}{\emph{New
  J. Phys.} {\bf 14} (2012) 095012},
  [\href{https://arxiv.org/abs/1204.4186}{{\tt 1204.4186}}].

\bibitem{Drewes:2013gca}
M.~Drewes, \emph{{The Phenomenology of Right Handed Neutrinos}},
  \href{http://dx.doi.org/10.1142/S0218301313300191}{\emph{Int. J. Mod. Phys.
  E} {\bf 22} (2013) 1330019}, [\href{https://arxiv.org/abs/1303.6912}{{\tt
  1303.6912}}].

\bibitem{Garbrecht:2018mrp}
B.~Garbrecht, \emph{{Why is there more matter than antimatter? Calculational
  methods for leptogenesis and electroweak baryogenesis}},
  \href{http://dx.doi.org/10.1016/j.ppnp.2019.103727}{\emph{Prog. Part. Nucl.
  Phys.} {\bf 110} (2020) 103727},
  [\href{https://arxiv.org/abs/1812.02651}{{\tt 1812.02651}}].

\bibitem{Bodeker:2020ghk}
D.~Bodeker and W.~Buchmuller, \emph{{Baryogenesis from the weak scale to the
  grand unification scale}},
  \href{http://dx.doi.org/10.1103/RevModPhys.93.035004}{\emph{Rev. Mod. Phys.}
  {\bf 93} (2021) 035004}, [\href{https://arxiv.org/abs/2009.07294}{{\tt
  2009.07294}}].

\bibitem{Atre:2009rg}
A.~Atre, T.~Han, S.~Pascoli and B.~Zhang, \emph{{The Search for Heavy Majorana
  Neutrinos}},
  \href{http://dx.doi.org/10.1088/1126-6708/2009/05/030}{\emph{JHEP} {\bf 05}
  (2009) 030}, [\href{https://arxiv.org/abs/0901.3589}{{\tt 0901.3589}}].

\bibitem{Deppisch:2015qwa}
F.~F. Deppisch, P.~S. Bhupal~Dev and A.~Pilaftsis, \emph{{Neutrinos and
  Collider Physics}},
  \href{http://dx.doi.org/10.1088/1367-2630/17/7/075019}{\emph{New J. Phys.}
  {\bf 17} (2015) 075019}, [\href{https://arxiv.org/abs/1502.06541}{{\tt
  1502.06541}}].

\bibitem{Cai:2017mow}
Y.~Cai, T.~Han, T.~Li and R.~Ruiz, \emph{{Lepton Number Violation: Seesaw
  Models and Their Collider Tests}},
  \href{http://dx.doi.org/10.3389/fphy.2018.00040}{\emph{Front. in Phys.} {\bf
  6} (2018) 40}, [\href{https://arxiv.org/abs/1711.02180}{{\tt 1711.02180}}].

\bibitem{Agrawal:2021dbo}
P.~Agrawal et~al., \emph{{Feebly-interacting particles: FIPs 2020 workshop
  report}}, \href{http://dx.doi.org/10.1140/epjc/s10052-021-09703-7}{\emph{Eur.
  Phys. J. C} {\bf 81} (2021) 1015},
  [\href{https://arxiv.org/abs/2102.12143}{{\tt 2102.12143}}].

\bibitem{Antusch:2016ejd}
S.~Antusch, E.~Cazzato and O.~Fischer, \emph{{Sterile neutrino searches at
  future $e^-e^+$, $pp$, and $e^-p$ colliders}},
  \href{http://dx.doi.org/10.1142/S0217751X17500786}{\emph{Int. J. Mod. Phys.
  A} {\bf 32} (2017) 1750078}, [\href{https://arxiv.org/abs/1612.02728}{{\tt
  1612.02728}}].

\bibitem{FCC:2018evy}
{\scshape FCC} collaboration, A.~Abada et~al., \emph{{FCC-ee: The Lepton
  Collider}: {Future Circular Collider Conceptual Design Report Volume 2}},
  \href{http://dx.doi.org/10.1140/epjst/e2019-900045-4}{\emph{Eur. Phys. J. ST}
  {\bf 228} (2019) 261--623}.

\bibitem{CEPCStudyGroup:2018ghi}
{\scshape CEPC Study Group} collaboration, M.~Dong et~al., \emph{{CEPC
  Conceptual Design Report: Volume 2 - Physics \& Detector}},
  \href{https://arxiv.org/abs/1811.10545}{{\tt 1811.10545}}.

\bibitem{Drewes:2012ma}
M.~Drewes and B.~Garbrecht, \emph{{Leptogenesis from a GeV Seesaw without Mass
  Degeneracy}}, \href{http://dx.doi.org/10.1007/JHEP03(2013)096}{\emph{JHEP}
  {\bf 03} (2013) 096}, [\href{https://arxiv.org/abs/1206.5537}{{\tt
  1206.5537}}].

\bibitem{Canetti:2014dka}
L.~Canetti, M.~Drewes and B.~Garbrecht, \emph{{Probing leptogenesis with
  GeV-scale sterile neutrinos at LHCb and Belle II}},
  \href{http://dx.doi.org/10.1103/PhysRevD.90.125005}{\emph{Phys. Rev. D} {\bf
  90} (2014) 125005}, [\href{https://arxiv.org/abs/1404.7114}{{\tt
  1404.7114}}].

\bibitem{Garbrecht:2014bfa}
B.~Garbrecht, \emph{{More Viable Parameter Space for Leptogenesis}},
  \href{http://dx.doi.org/10.1103/PhysRevD.90.063522}{\emph{Phys. Rev. D} {\bf
  90} (2014) 063522}, [\href{https://arxiv.org/abs/1401.3278}{{\tt
  1401.3278}}].

\bibitem{Shuve:2014zua}
B.~Shuve and I.~Yavin, \emph{{Baryogenesis through Neutrino Oscillations: A
  Unified Perspective}},
  \href{http://dx.doi.org/10.1103/PhysRevD.89.075014}{\emph{Phys. Rev. D} {\bf
  89} (2014) 075014}, [\href{https://arxiv.org/abs/1401.2459}{{\tt
  1401.2459}}].

\bibitem{Hernandez:2015wna}
P.~Hern\'andez, M.~Kekic, J.~L\'opez-Pav\'on, J.~Racker and N.~Rius,
  \emph{{Leptogenesis in GeV scale seesaw models}},
  \href{http://dx.doi.org/10.1007/JHEP10(2015)067}{\emph{JHEP} {\bf 10} (2015)
  067}, [\href{https://arxiv.org/abs/1508.03676}{{\tt 1508.03676}}].

\bibitem{Abada:2018oly}
A.~Abada, G.~Arcadi, V.~Domcke, M.~Drewes, J.~Klaric and M.~Lucente,
  \emph{{Low-scale leptogenesis with three heavy neutrinos}},
  \href{http://dx.doi.org/10.1007/JHEP01(2019)164}{\emph{JHEP} {\bf 01} (2019)
  164}, [\href{https://arxiv.org/abs/1810.12463}{{\tt 1810.12463}}].

\bibitem{Drewes:2021nqr}
M.~Drewes, Y.~Georis and J.~Klari\'c, \emph{{Mapping the Viable Parameter Space
  for Testable Leptogenesis}},
  \href{http://dx.doi.org/10.1103/PhysRevLett.128.051801}{\emph{Phys. Rev.
  Lett.} {\bf 128} (2022) 051801},
  [\href{https://arxiv.org/abs/2106.16226}{{\tt 2106.16226}}].

\bibitem{Dodelson:1993je}
S.~Dodelson and L.~M. Widrow, \emph{{Sterile-neutrinos as dark matter}},
  \href{http://dx.doi.org/10.1103/PhysRevLett.72.17}{\emph{Phys. Rev. Lett.}
  {\bf 72} (1994) 17--20}, [\href{https://arxiv.org/abs/hep-ph/9303287}{{\tt
  hep-ph/9303287}}].

\bibitem{Shi:1998km}
X.-D. Shi and G.~M. Fuller, \emph{{A New dark matter candidate: Nonthermal
  sterile neutrinos}},
  \href{http://dx.doi.org/10.1103/PhysRevLett.82.2832}{\emph{Phys. Rev. Lett.}
  {\bf 82} (1999) 2832--2835},
  [\href{https://arxiv.org/abs/astro-ph/9810076}{{\tt astro-ph/9810076}}].

\bibitem{Drewes:2016upu}
M.~Drewes et~al., \emph{{A White Paper on keV Sterile Neutrino Dark Matter}},
  \href{http://dx.doi.org/10.1088/1475-7516/2017/01/025}{\emph{JCAP} {\bf 01}
  (2017) 025}, [\href{https://arxiv.org/abs/1602.04816}{{\tt 1602.04816}}].

\bibitem{Boyarsky:2018tvu}
A.~Boyarsky, M.~Drewes, T.~Lasserre, S.~Mertens and O.~Ruchayskiy,
  \emph{{Sterile neutrino Dark Matter}},
  \href{http://dx.doi.org/10.1016/j.ppnp.2018.07.004}{\emph{Prog. Part. Nucl.
  Phys.} {\bf 104} (2019) 1--45}, [\href{https://arxiv.org/abs/1807.07938}{{\tt
  1807.07938}}].

\bibitem{Hernandez:2016kel}
P.~Hern\'andez, M.~Kekic, J.~L\'opez-Pav\'on, J.~Racker and J.~Salvado,
  \emph{{Testable Baryogenesis in Seesaw Models}},
  \href{http://dx.doi.org/10.1007/JHEP08(2016)157}{\emph{JHEP} {\bf 08} (2016)
  157}, [\href{https://arxiv.org/abs/1606.06719}{{\tt 1606.06719}}].

\bibitem{Drewes:2016jae}
M.~Drewes, B.~Garbrecht, D.~Gueter and J.~Klaric, \emph{{Testing the low scale
  seesaw and leptogenesis}},
  \href{http://dx.doi.org/10.1007/JHEP08(2017)018}{\emph{JHEP} {\bf 08} (2017)
  018}, [\href{https://arxiv.org/abs/1609.09069}{{\tt 1609.09069}}].

\bibitem{Shaposhnikov:2008pf}
M.~Shaposhnikov, \emph{{The $\nu$MSM, leptonic asymmetries, and properties of
  singlet fermions}},
  \href{http://dx.doi.org/10.1088/1126-6708/2008/08/008}{\emph{JHEP} {\bf 08}
  (2008) 008}, [\href{https://arxiv.org/abs/0804.4542}{{\tt 0804.4542}}].

\bibitem{Drewes:2019byd}
M.~Drewes, J.~Klari\'c and P.~Klose, \emph{{On lepton number violation in heavy
  neutrino decays at colliders}},
  \href{http://dx.doi.org/10.1007/JHEP11(2019)032}{\emph{JHEP} {\bf 11} (2019)
  032}, [\href{https://arxiv.org/abs/1907.13034}{{\tt 1907.13034}}].

\bibitem{Shaposhnikov:2006nn}
M.~Shaposhnikov, \emph{{A Possible symmetry of the $\nu$MSM}},
  \href{http://dx.doi.org/10.1016/j.nuclphysb.2006.11.003}{\emph{Nucl. Phys. B}
  {\bf 763} (2007) 49--59}, [\href{https://arxiv.org/abs/hep-ph/0605047}{{\tt
  hep-ph/0605047}}].

\bibitem{Kersten:2007vk}
J.~Kersten and A.~Y. Smirnov, \emph{{Right-Handed Neutrinos at CERN LHC and the
  Mechanism of Neutrino Mass Generation}},
  \href{http://dx.doi.org/10.1103/PhysRevD.76.073005}{\emph{Phys. Rev. D} {\bf
  76} (2007) 073005}, [\href{https://arxiv.org/abs/0705.3221}{{\tt
  0705.3221}}].

\bibitem{Moffat:2017feq}
K.~Moffat, S.~Pascoli and C.~Weiland, \emph{{Equivalence between massless
  neutrinos and lepton number conservation in fermionic singlet extensions of
  the Standard Model}},  \href{https://arxiv.org/abs/1712.07611}{{\tt
  1712.07611}}.

\bibitem{Baur:2019iai}
A.~Baur, H.~P. Nilles, A.~Trautner and P.~K.~S. Vaudrevange, \emph{{A String
  Theory of Flavor and $\mathscr {CP}$}},
  \href{http://dx.doi.org/10.1016/j.nuclphysb.2019.114737}{\emph{Nucl. Phys. B}
  {\bf 947} (2019) 114737}, [\href{https://arxiv.org/abs/1908.00805}{{\tt
  1908.00805}}].

\bibitem{Altarelli:2005yx}
G.~Altarelli and F.~Feruglio, \emph{{Tri-bimaximal neutrino mixing, $A_4$ and
  the modular symmetry}},
  \href{http://dx.doi.org/10.1016/j.nuclphysb.2006.02.015}{\emph{Nucl. Phys. B}
  {\bf 741} (2006) 215--235}, [\href{https://arxiv.org/abs/hep-ph/0512103}{{\tt
  hep-ph/0512103}}].

\bibitem{deMedeirosVarzielas:2005qg}
I.~de~Medeiros~Varzielas, S.~F. King and G.~G. Ross, \emph{{Tri-bimaximal
  neutrino mixing from discrete subgroups of $SU(3)$ and $SO(3)$ family
  symmetry}},
  \href{http://dx.doi.org/10.1016/j.physletb.2006.11.015}{\emph{Phys. Lett. B}
  {\bf 644} (2007) 153--157}, [\href{https://arxiv.org/abs/hep-ph/0512313}{{\tt
  hep-ph/0512313}}].

\bibitem{Lin:2009bw}
Y.~Lin, \emph{{Tri-bimaximal Neutrino Mixing from $A_4$ and $\theta_{13}\sim
  \theta_C$}},
  \href{http://dx.doi.org/10.1016/j.nuclphysb.2009.08.018}{\emph{Nucl. Phys. B}
  {\bf 824} (2010) 95--110}, [\href{https://arxiv.org/abs/0905.3534}{{\tt
  0905.3534}}].

\bibitem{Ding:2012xx}
G.-J. Ding, \emph{{TFH Mixing Patterns, Large $\theta_{13}$ and $\Delta(96)$
  Flavor Symmetry}},
  \href{http://dx.doi.org/10.1016/j.nuclphysb.2012.04.002}{\emph{Nucl. Phys. B}
  {\bf 862} (2012) 1--42}, [\href{https://arxiv.org/abs/1201.3279}{{\tt
  1201.3279}}].

\bibitem{Feruglio:2013hia}
F.~Feruglio, C.~Hagedorn and R.~Ziegler, \emph{{A realistic pattern of lepton
  mixing and masses from $S_4$ and CP}},
  \href{http://dx.doi.org/10.1140/epjc/s10052-014-2753-2}{\emph{Eur. Phys. J.
  C} {\bf 74} (2014) 2753}, [\href{https://arxiv.org/abs/1303.7178}{{\tt
  1303.7178}}].

\bibitem{Luhn:2013lkn}
C.~Luhn, \emph{{Trimaximal TM$_{1}$ neutrino mixing in $S_{4}$ with spontaneous
  CP violation}},
  \href{http://dx.doi.org/10.1016/j.nuclphysb.2013.07.003}{\emph{Nucl. Phys. B}
  {\bf 875} (2013) 80--100}, [\href{https://arxiv.org/abs/1306.2358}{{\tt
  1306.2358}}].

\bibitem{Hagedorn:2018gpw}
C.~Hagedorn and J.~K\"onig, \emph{{Lepton and quark mixing from a stepwise
  breaking of flavor and $CP$}},
  \href{http://dx.doi.org/10.1103/PhysRevD.100.075036}{\emph{Phys. Rev. D} {\bf
  100} (2019) 075036}, [\href{https://arxiv.org/abs/1811.07750}{{\tt
  1811.07750}}].

\bibitem{Hagedorn:2018bzo}
C.~Hagedorn and J.~K\"onig, \emph{{Lepton and quark masses and mixing in a SUSY
  model with $\Delta(384)$ and CP}},
  \href{http://dx.doi.org/10.1016/j.nuclphysb.2020.114953}{\emph{Nucl. Phys. B}
  {\bf 953} (2020) 114953}, [\href{https://arxiv.org/abs/1811.09262}{{\tt
  1811.09262}}].

\bibitem{Hagedorn:2011un}
C.~Hagedorn and M.~Serone, \emph{{Leptons in Holographic Composite Higgs Models
  with Non-Abelian Discrete Symmetries}},
  \href{http://dx.doi.org/10.1007/JHEP10(2011)083}{\emph{JHEP} {\bf 10} (2011)
  083}, [\href{https://arxiv.org/abs/1106.4021}{{\tt 1106.4021}}].

\bibitem{Hagedorn:2011pw}
C.~Hagedorn and M.~Serone, \emph{{General Lepton Mixing in Holographic
  Composite Higgs Models}},
  \href{http://dx.doi.org/10.1007/JHEP02(2012)077}{\emph{JHEP} {\bf 02} (2012)
  077}, [\href{https://arxiv.org/abs/1110.4612}{{\tt 1110.4612}}].

\bibitem{Fischer:2021nha}
O.~Fischer, M.~Lindner and S.~van~der Woude, \emph{{Robustness of ARS
  leptogenesis in scalar extensions}},
  \href{http://dx.doi.org/10.1007/JHEP05(2022)149}{\emph{JHEP} {\bf 05} (2022)
  149}, [\href{https://arxiv.org/abs/2110.14499}{{\tt 2110.14499}}].

\bibitem{Flood:2021qhq}
I.~Flood, R.~Porto, J.~Schlesinger, B.~Shuve and M.~Thum, \emph{{Hidden-sector
  neutrinos and freeze-in leptogenesis}},
  \href{http://dx.doi.org/10.1103/PhysRevD.105.095025}{\emph{Phys. Rev. D} {\bf
  105} (2022) 095025}, [\href{https://arxiv.org/abs/2109.10908}{{\tt
  2109.10908}}].

\bibitem{Ding:2014ora}
G.-J. Ding, S.~F. King and T.~Neder, \emph{{Generalised CP and $\Delta(6\,n^2)$
  family symmetry in semi-direct models of leptons}},
  \href{http://dx.doi.org/10.1007/JHEP12(2014)007}{\emph{JHEP} {\bf 12} (2014)
  007}, [\href{https://arxiv.org/abs/1409.8005}{{\tt 1409.8005}}].

\bibitem{Dev:2017wwc}
B.~Dev, M.~Garny, J.~Klaric, P.~Millington and D.~Teresi, \emph{{Resonant
  enhancement in leptogenesis}},
  \href{http://dx.doi.org/10.1142/S0217751X18420034}{\emph{Int. J. Mod. Phys.
  A} {\bf 33} (2018) 1842003}, [\href{https://arxiv.org/abs/1711.02863}{{\tt
  1711.02863}}].

\bibitem{Wyler:1982dd}
D.~Wyler and L.~Wolfenstein, \emph{{Massless Neutrinos in Left-Right Symmetric
  Models}}, \href{http://dx.doi.org/10.1016/0550-3213(83)90482-0}{\emph{Nucl.
  Phys. B} {\bf 218} (1983) 205--214}.

\bibitem{Mohapatra:1986aw}
R.~N. Mohapatra, \emph{{Mechanism for Understanding Small Neutrino Mass in
  Superstring Theories}},
  \href{http://dx.doi.org/10.1103/PhysRevLett.56.561}{\emph{Phys. Rev. Lett.}
  {\bf 56} (1986) 561--563}.

\bibitem{Mohapatra:1986bd}
R.~N. Mohapatra and J.~W.~F. Valle, \emph{{Neutrino Mass and Baryon Number
  Nonconservation in Superstring Models}},
  \href{http://dx.doi.org/10.1103/PhysRevD.34.1642}{\emph{Phys. Rev. D} {\bf
  34} (1986) 1642}.

\bibitem{Bernabeu:1987gr}
J.~Bernabeu, A.~Santamaria, J.~Vidal, A.~Mendez and J.~W.~F. Valle,
  \emph{{Lepton Flavor Nonconservation at High-Energies in a Superstring
  Inspired Standard Model}},
  \href{http://dx.doi.org/10.1016/0370-2693(87)91100-2}{\emph{Phys. Lett. B}
  {\bf 187} (1987) 303--308}.

\bibitem{Davidson:2002qv}
S.~Davidson and A.~Ibarra, \emph{{A Lower bound on the right-handed neutrino
  mass from leptogenesis}},
  \href{http://dx.doi.org/10.1016/S0370-2693(02)01735-5}{\emph{Phys. Lett. B}
  {\bf 535} (2002) 25--32}, [\href{https://arxiv.org/abs/hep-ph/0202239}{{\tt
  hep-ph/0202239}}].

\bibitem{Dev:2017trv}
P.~S.~B. Dev, P.~Di~Bari, B.~Garbrecht, S.~Lavignac, P.~Millington and
  D.~Teresi, \emph{{Flavor effects in leptogenesis}},
  \href{http://dx.doi.org/10.1142/S0217751X18420010}{\emph{Int. J. Mod. Phys.
  A} {\bf 33} (2018) 1842001}, [\href{https://arxiv.org/abs/1711.02861}{{\tt
  1711.02861}}].

\bibitem{Hambye:2001eu}
T.~Hambye, \emph{{Leptogenesis at the TeV scale}},
  \href{http://dx.doi.org/10.1016/S0550-3213(02)00293-6}{\emph{Nucl. Phys. B}
  {\bf 633} (2002) 171--192}, [\href{https://arxiv.org/abs/hep-ph/0111089}{{\tt
  hep-ph/0111089}}].

\bibitem{Flanz:1994yx}
M.~Flanz, E.~A. Paschos and U.~Sarkar, \emph{{Baryogenesis from a lepton
  asymmetric universe}},
  \href{http://dx.doi.org/10.1016/0370-2693(94)01555-Q}{\emph{Phys. Lett. B}
  {\bf 345} (1995) 248--252}, [\href{https://arxiv.org/abs/hep-ph/9411366}{{\tt
  hep-ph/9411366}}].

\bibitem{Covi:1996wh}
L.~Covi, E.~Roulet and F.~Vissani, \emph{{CP violating decays in leptogenesis
  scenarios}},
  \href{http://dx.doi.org/10.1016/0370-2693(96)00817-9}{\emph{Phys. Lett. B}
  {\bf 384} (1996) 169--174}, [\href{https://arxiv.org/abs/hep-ph/9605319}{{\tt
  hep-ph/9605319}}].

\bibitem{Sakharov:1967dj}
A.~D. Sakharov, \emph{{Violation of CP Invariance, C asymmetry, and baryon
  asymmetry of the universe}},
  \href{http://dx.doi.org/10.1070/PU1991v034n05ABEH002497}{\emph{Pisma Zh.
  Eksp. Teor. Fiz.} {\bf 5} (1967) 32--35}.

\bibitem{Klaric:2021cpi}
J.~Klari\'c, M.~Shaposhnikov and I.~Timiryasov, \emph{{Reconciling resonant
  leptogenesis and baryogenesis via neutrino oscillations}},
  \href{http://dx.doi.org/10.1103/PhysRevD.104.055010}{\emph{Phys. Rev. D} {\bf
  104} (2021) 055010}, [\href{https://arxiv.org/abs/2103.16545}{{\tt
  2103.16545}}].

\bibitem{Sigl:1993ctk}
G.~Sigl and G.~Raffelt, \emph{{General kinetic description of relativistic
  mixed neutrinos}},
  \href{http://dx.doi.org/10.1016/0550-3213(93)90175-O}{\emph{Nucl. Phys. B}
  {\bf 406} (1993) 423--451}.

\bibitem{Biondini:2017rpb}
S.~Biondini et~al., \emph{{Status of rates and rate equations for thermal
  leptogenesis}}, \href{http://dx.doi.org/10.1142/S0217751X18420046}{\emph{Int.
  J. Mod. Phys. A} {\bf 33} (2018) 1842004},
  [\href{https://arxiv.org/abs/1711.02864}{{\tt 1711.02864}}].

\bibitem{Laine:2022pgk}
M.~Laine, \emph{{Sterile neutrino rates for general $M$, $T$, $\mu$, $k$:
  review of a theoretical framework}},
  \href{http://dx.doi.org/10.1016/j.aop.2022.169022}{\emph{Annals Phys.} {\bf
  444} (2022) 169022}, [\href{https://arxiv.org/abs/2203.05772}{{\tt
  2203.05772}}].

\bibitem{Ghiglieri:2017gjz}
J.~Ghiglieri and M.~Laine, \emph{{GeV-scale hot sterile neutrino oscillations:
  a derivation of evolution equations}},
  \href{http://dx.doi.org/10.1007/JHEP05(2017)132}{\emph{JHEP} {\bf 05} (2017)
  132}, [\href{https://arxiv.org/abs/1703.06087}{{\tt 1703.06087}}].

\bibitem{Buchmuller:2005eh}
W.~Buchmuller, R.~D. Peccei and T.~Yanagida, \emph{{Leptogenesis as the origin
  of matter}},
  \href{http://dx.doi.org/10.1146/annurev.nucl.55.090704.151558}{\emph{Ann.
  Rev. Nucl. Part. Sci.} {\bf 55} (2005) 311--355},
  [\href{https://arxiv.org/abs/hep-ph/0502169}{{\tt hep-ph/0502169}}].

\bibitem{Garbrecht:2019zaa}
B.~Garbrecht, P.~Klose and C.~Tamarit, \emph{{Relativistic and spectator
  effects in leptogenesis with heavy sterile neutrinos}},
  \href{http://dx.doi.org/10.1007/JHEP02(2020)117}{\emph{JHEP} {\bf 02} (2020)
  117}, [\href{https://arxiv.org/abs/1904.09956}{{\tt 1904.09956}}].

\bibitem{Bezrukov:2008ut}
F.~Bezrukov, D.~Gorbunov and M.~Shaposhnikov, \emph{{On initial conditions for
  the Hot Big Bang}},
  \href{http://dx.doi.org/10.1088/1475-7516/2009/06/029}{\emph{JCAP} {\bf 06}
  (2009) 029}, [\href{https://arxiv.org/abs/0812.3622}{{\tt 0812.3622}}].

\bibitem{Bezrukov:2012sa}
F.~Bezrukov, M.~Y. Kalmykov, B.~A. Kniehl and M.~Shaposhnikov, \emph{{Higgs
  Boson Mass and New Physics}},
  \href{http://dx.doi.org/10.1007/JHEP10(2012)140}{\emph{JHEP} {\bf 10} (2012)
  140}, [\href{https://arxiv.org/abs/1205.2893}{{\tt 1205.2893}}].

\bibitem{Domcke:2020quw}
V.~Domcke, K.~Kamada, K.~Mukaida, K.~Schmitz and M.~Yamada, \emph{{Wash-In
  Leptogenesis}},
  \href{http://dx.doi.org/10.1103/PhysRevLett.126.201802}{\emph{Phys. Rev.
  Lett.} {\bf 126} (2021) 201802},
  [\href{https://arxiv.org/abs/2011.09347}{{\tt 2011.09347}}].

\bibitem{Anisimov:2010gy}
A.~Anisimov, D.~Besak and D.~Bodeker, \emph{{Thermal production of relativistic
  Majorana neutrinos: Strong enhancement by multiple soft scattering}},
  \href{http://dx.doi.org/10.1088/1475-7516/2011/03/042}{\emph{JCAP} {\bf 03}
  (2011) 042}, [\href{https://arxiv.org/abs/1012.3784}{{\tt 1012.3784}}].

\bibitem{Eijima:2017anv}
S.~Eijima and M.~Shaposhnikov, \emph{{Fermion number violating effects in low
  scale leptogenesis}},
  \href{http://dx.doi.org/10.1016/j.physletb.2017.05.068}{\emph{Phys. Lett. B}
  {\bf 771} (2017) 288--296}, [\href{https://arxiv.org/abs/1703.06085}{{\tt
  1703.06085}}].

\bibitem{Hambye:2016sby}
T.~Hambye and D.~Teresi, \emph{{Higgs doublet decay as the origin of the baryon
  asymmetry}},
  \href{http://dx.doi.org/10.1103/PhysRevLett.117.091801}{\emph{Phys. Rev.
  Lett.} {\bf 117} (2016) 091801},
  [\href{https://arxiv.org/abs/1606.00017}{{\tt 1606.00017}}].

\bibitem{DOnofrio:2015gop}
M.~D'Onofrio and K.~Rummukainen, \emph{{Standard model cross-over on the
  lattice}}, \href{http://dx.doi.org/10.1103/PhysRevD.93.025003}{\emph{Phys.
  Rev. D} {\bf 93} (2016) 025003},
  [\href{https://arxiv.org/abs/1508.07161}{{\tt 1508.07161}}].

\bibitem{Bodeker:2019ajh}
D.~B\"odeker and D.~Schr\"oder, \emph{{Equilibration of right-handed
  electrons}},
  \href{http://dx.doi.org/10.1088/1475-7516/2019/05/010}{\emph{JCAP} {\bf 05}
  (2019) 010}, [\href{https://arxiv.org/abs/1902.07220}{{\tt 1902.07220}}].

\bibitem{Roy:2010xq}
A.~Roy and M.~Shaposhnikov, \emph{{Resonant production of the sterile neutrino
  dark matter and fine-tunings in the $\nu$MSM}},
  \href{http://dx.doi.org/10.1103/PhysRevD.82.056014}{\emph{Phys. Rev. D} {\bf
  82} (2010) 056014}, [\href{https://arxiv.org/abs/1006.4008}{{\tt
  1006.4008}}].

\bibitem{Antusch:2002rr}
S.~Antusch, J.~Kersten, M.~Lindner and M.~Ratz, \emph{{Neutrino mass matrix
  running for nondegenerate seesaw scales}},
  \href{http://dx.doi.org/10.1016/S0370-2693(02)01960-3}{\emph{Phys. Lett. B}
  {\bf 538} (2002) 87--95}, [\href{https://arxiv.org/abs/hep-ph/0203233}{{\tt
  hep-ph/0203233}}].

\bibitem{Hernandez:2014fha}
P.~Hernandez, M.~Kekic and J.~Lopez-Pavon, \emph{{$N_{\rm eff}$ in low-scale
  seesaw models versus the lightest neutrino mass}},
  \href{http://dx.doi.org/10.1103/PhysRevD.90.065033}{\emph{Phys. Rev. D} {\bf
  90} (2014) 065033}, [\href{https://arxiv.org/abs/1406.2961}{{\tt
  1406.2961}}].

\bibitem{Vincent:2014rja}
A.~C. Vincent, E.~F. Martinez, P.~Hern\'andez, M.~Lattanzi and O.~Mena,
  \emph{{Revisiting cosmological bounds on sterile neutrinos}},
  \href{http://dx.doi.org/10.1088/1475-7516/2015/04/006}{\emph{JCAP} {\bf 04}
  (2015) 006}, [\href{https://arxiv.org/abs/1408.1956}{{\tt 1408.1956}}].

\bibitem{Sabti:2020yrt}
N.~Sabti, A.~Magalich and A.~Filimonova, \emph{{An Extended Analysis of Heavy
  Neutral Leptons during Big Bang Nucleosynthesis}},
  \href{http://dx.doi.org/10.1088/1475-7516/2020/11/056}{\emph{JCAP} {\bf 11}
  (2020) 056}, [\href{https://arxiv.org/abs/2006.07387}{{\tt 2006.07387}}].

\bibitem{Boyarsky:2020dzc}
A.~Boyarsky, M.~Ovchynnikov, O.~Ruchayskiy and V.~Syvolap, \emph{{Improved big
  bang nucleosynthesis constraints on heavy neutral leptons}},
  \href{http://dx.doi.org/10.1103/PhysRevD.104.023517}{\emph{Phys. Rev. D} {\bf
  104} (2021) 023517}, [\href{https://arxiv.org/abs/2008.00749}{{\tt
  2008.00749}}].

\bibitem{Domcke:2020ety}
V.~Domcke, M.~Drewes, M.~Hufnagel and M.~Lucente, \emph{{MeV-scale Seesaw and
  Leptogenesis}}, \href{http://dx.doi.org/10.1007/JHEP01(2021)200}{\emph{JHEP}
  {\bf 01} (2021) 200}, [\href{https://arxiv.org/abs/2009.11678}{{\tt
  2009.11678}}].

\bibitem{Mastrototaro:2021wzl}
L.~Mastrototaro, P.~D. Serpico, A.~Mirizzi and N.~Saviano, \emph{{Massive
  sterile neutrinos in the early Universe: From thermal decoupling to
  cosmological constraints}},
  \href{http://dx.doi.org/10.1103/PhysRevD.104.016026}{\emph{Phys. Rev. D} {\bf
  104} (2021) 016026}, [\href{https://arxiv.org/abs/2104.11752}{{\tt
  2104.11752}}].

\bibitem{Mastrototaro:2019vug}
L.~Mastrototaro, A.~Mirizzi, P.~D. Serpico and A.~Esmaili, \emph{{Heavy sterile
  neutrino emission in core-collapse supernovae: Constraints and signatures}},
  \href{http://dx.doi.org/10.1088/1475-7516/2020/01/010}{\emph{JCAP} {\bf 01}
  (2020) 010}, [\href{https://arxiv.org/abs/1910.10249}{{\tt 1910.10249}}].

\bibitem{Arguelles:2021dqn}
C.~A. Arg\"uelles, N.~Foppiani and M.~Hostert, \emph{{Heavy neutral leptons
  below the kaon mass at hodoscopic neutrino detectors}},
  \href{http://dx.doi.org/10.1103/PhysRevD.105.095006}{\emph{Phys. Rev. D} {\bf
  105} (2022) 095006}, [\href{https://arxiv.org/abs/2109.03831}{{\tt
  2109.03831}}].

\bibitem{Kelly:2021xbv}
K.~J. Kelly and P.~A.~N. Machado, \emph{{MicroBooNE experiment, NuMI absorber,
  and heavy neutral leptons}},
  \href{http://dx.doi.org/10.1103/PhysRevD.104.055015}{\emph{Phys. Rev. D} {\bf
  104} (2021) 055015}, [\href{https://arxiv.org/abs/2106.06548}{{\tt
  2106.06548}}].

\bibitem{Bondarenko:2021cpc}
K.~Bondarenko, A.~Boyarsky, J.~Klaric, O.~Mikulenko, O.~Ruchayskiy, V.~Syvolap
  et~al., \emph{{An allowed window for heavy neutral leptons below the kaon
  mass}}, \href{http://dx.doi.org/10.1007/JHEP07(2021)193}{\emph{JHEP} {\bf 07}
  (2021) 193}, [\href{https://arxiv.org/abs/2101.09255}{{\tt 2101.09255}}].

\bibitem{Chrzaszcz:2019inj}
M.~Chrzaszcz, M.~Drewes, T.~E. Gonzalo, J.~Harz, S.~Krishnamurthy and
  C.~Weniger, \emph{{A frequentist analysis of three right-handed neutrinos
  with GAMBIT}},
  \href{http://dx.doi.org/10.1140/epjc/s10052-020-8073-9}{\emph{Eur. Phys. J.
  C} {\bf 80} (2020) 569}, [\href{https://arxiv.org/abs/1908.02302}{{\tt
  1908.02302}}].

\bibitem{Alimena:2019zri}
J.~Alimena et~al., \emph{{Searching for long-lived particles beyond the
  Standard Model at the Large Hadron Collider}},
  \href{http://dx.doi.org/10.1088/1361-6471/ab4574}{\emph{J. Phys. G} {\bf 47}
  (2020) 090501}, [\href{https://arxiv.org/abs/1903.04497}{{\tt 1903.04497}}].

\bibitem{Planck:2018vyg}
{\scshape Planck} collaboration, N.~Aghanim et~al., \emph{{Planck 2018 results.
  VI. Cosmological parameters}},
  \href{http://dx.doi.org/10.1051/0004-6361/201833910}{\emph{Astron.
  Astrophys.} {\bf 641} (2020) A6},
  [\href{https://arxiv.org/abs/1807.06209}{{\tt 1807.06209}}].

\bibitem{Drewes:2016gmt}
M.~Drewes, B.~Garbrecht, D.~Gueter and J.~Klaric, \emph{{Leptogenesis from
  Oscillations of Heavy Neutrinos with Large Mixing Angles}},
  \href{http://dx.doi.org/10.1007/JHEP12(2016)150}{\emph{JHEP} {\bf 12} (2016)
  150}, [\href{https://arxiv.org/abs/1606.06690}{{\tt 1606.06690}}].

\bibitem{Canetti:2012kh}
L.~Canetti, M.~Drewes, T.~Frossard and M.~Shaposhnikov, \emph{{Dark Matter,
  Baryogenesis and Neutrino Oscillations from Right Handed Neutrinos}},
  \href{http://dx.doi.org/10.1103/PhysRevD.87.093006}{\emph{Phys. Rev. D} {\bf
  87} (2013) 093006}, [\href{https://arxiv.org/abs/1208.4607}{{\tt
  1208.4607}}].

\bibitem{Canetti:2010aw}
L.~Canetti and M.~Shaposhnikov, \emph{{Baryon Asymmetry of the Universe in the
  $\nu$MSM}},
  \href{http://dx.doi.org/10.1088/1475-7516/2010/09/001}{\emph{JCAP} {\bf 09}
  (2010) 001}, [\href{https://arxiv.org/abs/1006.0133}{{\tt 1006.0133}}].

\bibitem{Garbrecht:2011aw}
B.~Garbrecht and M.~Herranen, \emph{{Effective Theory of Resonant Leptogenesis
  in the Closed-Time-Path Approach}},
  \href{http://dx.doi.org/10.1016/j.nuclphysb.2012.03.009}{\emph{Nucl. Phys. B}
  {\bf 861} (2012) 17--52}, [\href{https://arxiv.org/abs/1112.5954}{{\tt
  1112.5954}}].

\bibitem{Antusch:2017pkq}
S.~Antusch, E.~Cazzato, M.~Drewes, O.~Fischer, B.~Garbrecht, D.~Gueter et~al.,
  \emph{{Probing Leptogenesis at Future Colliders}},
  \href{http://dx.doi.org/10.1007/JHEP09(2018)124}{\emph{JHEP} {\bf 09} (2018)
  124}, [\href{https://arxiv.org/abs/1710.03744}{{\tt 1710.03744}}].

\bibitem{Blanchet:2006ch}
S.~Blanchet, P.~Di~Bari and G.~G. Raffelt, \emph{{Quantum Zeno effect and the
  impact of flavor in leptogenesis}},
  \href{http://dx.doi.org/10.1088/1475-7516/2007/03/012}{\emph{JCAP} {\bf 03}
  (2007) 012}, [\href{https://arxiv.org/abs/hep-ph/0611337}{{\tt
  hep-ph/0611337}}].

\bibitem{Abada:2014kba}
A.~Abada, M.~E. Krauss, W.~Porod, F.~Staub, A.~Vicente and C.~Weiland,
  \emph{{Lepton flavor violation in low-scale seesaw models: SUSY and non-SUSY
  contributions}}, \href{http://dx.doi.org/10.1007/JHEP11(2014)048}{\emph{JHEP}
  {\bf 11} (2014) 048}, [\href{https://arxiv.org/abs/1408.0138}{{\tt
  1408.0138}}].

\bibitem{Abada:2014vea}
A.~Abada and M.~Lucente, \emph{{Looking for the minimal inverse seesaw
  realisation}},
  \href{http://dx.doi.org/10.1016/j.nuclphysb.2014.06.003}{\emph{Nucl. Phys. B}
  {\bf 885} (2014) 651--678}, [\href{https://arxiv.org/abs/1401.1507}{{\tt
  1401.1507}}].

\bibitem{Appelquist:2002mw}
T.~Appelquist, B.~A. Dobrescu and A.~R. Hopper, \emph{{Nonexotic Neutral Gauge
  Bosons}}, \href{http://dx.doi.org/10.1103/PhysRevD.68.035012}{\emph{Phys.
  Rev. D} {\bf 68} (2003) 035012},
  [\href{https://arxiv.org/abs/hep-ph/0212073}{{\tt hep-ph/0212073}}].

\bibitem{Pati:1973uk}
J.~C. Pati and A.~Salam, \emph{{Unified Lepton-Hadron Symmetry and a Gauge
  Theory of the Basic Interactions}},
  \href{http://dx.doi.org/10.1103/PhysRevD.8.1240}{\emph{Phys. Rev. D} {\bf 8}
  (1973) 1240--1251}.

\end{thebibliography}\endgroup

\end{document}